\documentclass[twocolumn]{aastex63}

\usepackage{booktabs}
\usepackage{multirow}

\usepackage{longtable}
\usepackage{rotating}
\usepackage{float} 
\usepackage{textcomp} 
\usepackage{amsmath}
\usepackage[normalem]{ulem}
\usepackage{enumitem}

\newcommand{\coonezero}{CO(1\textrm{--}0)\ }
\newcommand{\cotwoone}{CO(2\textrm{--}1)\ }
\newcommand{\cionezero}{[CI](1\textrm{--}0)\ }
\newcommand{\vzgal}{V\textit{z}{--}GAL\ }
\newcommand{\zgal}{\textit{z}{--}GAL\ }

\usepackage{hyperref}

\begin{document}

\title{V\textit{z}$-$GAL: Probing Cold Molecular Gas in Dusty Star-forming Galaxies at $\bf \textit{z}=1-6$}

\correspondingauthor{Prachi Prajapati}
\email{prajapati@ph1.uni-koeln.de}

\author[0000-0002-3094-1077]{Prachi Prajapati}\thanks{Doctoral Researcher at International Max Planck Research \\ School (IMPRS) for Astronomy \& Astrophysics}
\affiliation{Institute for Astrophysics, University of Cologne, Z{\"u}lpicher Straße 77, 50937 Cologne, Germany}
\affiliation{Max Planck Institute for Radio Astronomy (MPIfR), Auf dem H{\"u}gel 69, 53121 Bonn, Germany}

\author[0000-0001-9585-1462]{Dominik Riechers}
\affiliation{Institute for Astrophysics, University of Cologne, Z{\"u}lpicher Straße 77, 50937 Cologne, Germany}

\author[0000-0003-2027-8221]{Pierre Cox}
\affiliation{Sorbonne Université, Université Paris 6 and CNRS, Institut d’Astrophysique de Paris, 98bis boulevard Arago, 75014 Paris, France}

\author[0000-0003-4678-3939]{Axel Weiß}
\affiliation{Max Planck Institute for Radio Astronomy (MPIfR), Auf dem H{\"u}gel 69, 53121 Bonn, Germany}

\author[0000-0003-4357-3450]{Am\'elie Saintonge}
\affiliation{Max Planck Institute for Radio Astronomy (MPIfR), Auf dem H{\"u}gel 69, 53121 Bonn, Germany}

\author[0000-0002-0675-0078]{Bethany Jones}
\affiliation{Institute for Astrophysics, University of Cologne, Z{\"u}lpicher Straße 77, 50937 Cologne, Germany}

\author[0000-0002-5268-2221]{Tom J.~L.~C. Bakx}
\affiliation{Departement of Space, Earth \& Environement, Chalmers University of Technology, Chalmersplatsen 4, Gothenburg SE-412 96, Sweden}

\author[0000-0002-0320-1532]{Stefano Berta}
\affiliation{Institut de Radioastronomie Millim\'etrique (IRAM), 300 rue de la Piscine, 38406 Saint-Martin-d'H{\`e}res, France}

\author[0000-0001-5434-5942]{Paul van der Werf}
\affiliation{Leiden Observatory, Leiden University, PO Box 9513, 2300 RA Leiden, The Netherlands}

\author[0000-0002-7176-4046]{Roberto Neri}
\affiliation{Institut de Radioastronomie Millim\'etrique (IRAM), 300 rue de la Piscine, 38406 Saint-Martin-d'H{\`e}res, France}

\author[0000-0001-7387-0558]{Kirsty M. Butler}
\affiliation{Departement of Space, Earth \& Environement, Chalmers University of Technology, Chalmersplatsen 4, Gothenburg SE-412 96, Sweden}

\author[0000-0002-3892-0190]{Asantha Cooray}
\affiliation{University of California Irvine, Department of Physics \& Astronomy, FRH 2174, Irvine, CA 92697, USA}

\author[0009-0007-2281-4944]{Diana Ismail}
\affiliation{Université de Strasbourg, CNRS, Observatoire astronomique de Strasbourg, UMR 7550, 67000 Strasbourg, France}

\author[0000-0002-7892-396X]{Andrew J. Baker}
\affiliation{Department of Physics and Astronomy, Rutgers University---New Brunswick, NJ 08854-8019, USA}
\affiliation{Department of Physics and Astronomy, University of the Western Cape, Bellville 7535, Cape Town, South Africa}

\author[0009-0009-9483-8763]{Edoardo Borsato}
\affiliation{Dipartimento di Fisica e Astronomia "G. Galilei," Universit\`a di Padova, vicolo dell’Osservatorio 3, I-35122 Padova, Italy}

\author[0000-0001-6159-9174]{Andrew Harris}
\affiliation{Department of Astronomy, University of Maryland, College Park 20742 MD, USA}

\author[0000-0001-5118-1313]{Rob Ivison}
\affiliation{Institute for Astronomy, University of Edinburgh, Royal Observatory, Blackford Hill, Edinburgh EH9 3HJ, UK}
\affiliation{School of Cosmic Physics, Dublin Institute for Advanced Studies, 31 Fitzwilliam Pl., Dublin D02 XF86, Ireland}

\author[0000-0003-1939-5885]{Matthew Lehnert}
\affiliation{Sorbonne Université, Université Paris 6 and CNRS, Institut d’Astrophysique de Paris, 98bis boulevard Arago, 75014 Paris, France}
\affiliation{Universit\'e Lyon 1, ENS de Lyon, Centre de Recherche Astrophysique de Lyon (UMR5574), 69230 Saint-Genis-Laval, France}

\author[0000-0003-3948-7621]{Lucia Marchetti}
\affiliation{Department of Astronomy, University of Cape Town, 7701 Rondebosch, Cape Town, South Africa}
\affiliation{INAF - Istituto di Radioastronomia, via Gobetti 101, I-40129 Bologna, Italy}
\affiliation{The Inter-University Institute for Data Intensive Astronomy (IDIA), Department of Astronomy, University of Cape Town, 7701 Rondebosch, Cape Town, South Africa}

\author[0000-0002-2985-7994]{Hugo Messias}
\affiliation{European Southern Observatory, Alonso de C\'ordova 3107, Casilla 19001, Vitacura, Santiago, Chile}
\affiliation{Joint ALMA Observatory, Vitacura, Santiago 763-0355, Chile}

\author[0000-0002-4721-3922]{Alain Omont}
\affiliation{Sorbonne Université, Université Paris 6 and CNRS, Institut d’Astrophysique de Paris, 98bis boulevard Arago, 75014 Paris, France}

\author[0000-0003-3745-4228]{Catherine Vlahakis}
\affiliation{National Radio Astronomy Observatory, 520 Edgemont Rd., Charlottesville, VA 22901, USA }

\author[0000-0002-8117-9991]{Chentao Yang}
\affiliation{Department of Space, Earth \& Environment, Chalmers University of Technology, Gothenburg SE-412 96, Sweden}


\begin{abstract}

{We present the first results of V\textit{z}-GAL, a high-redshift CO($J=1-0$) large survey with the Karl G. Jansky Very Large Array, targeting 92 \textit{Herschel}-selected, infrared-luminous, dusty star-forming galaxies (DSFGs) at redshifts 1 to 6. These sources are selected based on having redshifts and mid/high-\textit{J} CO transitions from the NOrthern Extended Millimeter Array \zgal survey. We successfully detect CO($J=1-0$) emission in 90/92 galaxies at the expected positions and redshifts, including 9 tentative detections at $2\sigma - 3\sigma$ significance, and CO($J=2-1$) emission in 10 of these galaxies. The CO($J=1-0$) luminosities suggest apparent gas masses in the range $\mu {M}_{\rm H_2}$ = $(2-20) \times {10}^{11}~(\alpha_{CO}/{4.0})~\mathrm{M_{\odot}}$, which implies gas depletion times of $(50-600)$~Myr. These timescales show similar spread as local ULIRGs, suggesting a self-regulatory mechanism that maintains a consistent SFR per unit gas mass in starbursts across redshifts. To quantify the contribution of ``excitation correction" factors to gas mass estimates, we calculate median CO line brightness temperature ratios of $r_{21}=0.88\pm0.25$, $r_{31}=0.61\pm0.22$, $r_{41}=0.49\pm0.15$, $r_{51}=0.47\pm0.13$, and $r_{61}=0.28\pm0.13$. Accounting for these corrections results in a reduced scatter in `gas mass--star formation rate' relations. We also find a median log(${L}^{\prime}_{\mathrm{[CI]}(^{3}P_1 - ^{3}P_0)}/{L}^{\prime}_{\mathrm{CO}(J=1-0)})=-0.71\pm0.12$ for a subsample of 23 sources, consistent with the ratios derived for local star-forming galaxies. Together, our findings are in agreement with common conditions in the cold gas reservoirs among star-forming galaxies over a broad range in star formation modes, efficiencies, and scales.}

\end{abstract}

\keywords{high-redshift galaxies --- galaxy evolution --- starbursts --- local galaxies --- CO line emission: molecular gas --- [CI] line emission: atomic carbon --- star formation: interstellar medium}

\section{Introduction} \label{sec:intro}

Sub-millimeter or infrared selected dusty star-forming galaxies (DSFGs) in the early Universe play a vital role in understanding galaxy formation and evolution. Many of these massive, dust-enshrouded systems --- undergoing intense star formation --- host infrared luminosities (${L}_\mathrm{8-1000 \mu m}$) above 10$^{12}$$-$10$^{13}$ $\mathrm{L}_\mathrm{\odot}$. While such infrared-luminous DSFGs are rare in the local Universe, they are relatively more abundant at high-redshifts (high-\textit{z}), contributing significantly to the comoving star formation rate density (SFRD) at its peak at redshift \textit{z} = $1-3$ \citep[e.g.,][]{magnelli+2013,madau_dickinson2014,bourne+2017,hatsukade+2018,zavala+2021}. The understanding of these galaxies is {therefore} essential for exploring the processes that drive star formation and the role of dust in the early Universe \citep{blain+2002,casey+2014,hodgedacunha2020}. 

{Remarkably, the brightest DSFGs are typically triggered by major mergers \citep[e.g.,][]{tacconi+2008,riechers+2017}, that are short-lived, with starburst phases lasting about 100 Myr. During this period, their star formation rates (SFRs) can exceed 1000~$\mathrm{M}_\mathrm{\odot} \ \mathrm{yr}^{-1}$ in the most exceptional cases, far surpassing most intensely star-forming galaxies in the local Universe; i.e., ultra-luminous infrared galaxies \citep[ULIRGs;][]{sandersmirabel1996}. The significant number of DSFGs at high redshifts, along with their diverse nature, makes it crucial to characterize their properties within the broader context of galaxy formation and evolution.}

{Observing the far-infrared spectral transitions of multi-phase gas and the underlying continuum of DSFGs is of particular importance for understanding their obscured star formation. Star-forming activity is typically fueled by cold molecular gas reservoirs in galaxies \citep[e.g.,][]{gaosolomon2004,saintongebarbara2022,koyama2017}. Alongside regulating the star formation, extended cold gas environments also influence the gas inflow and feedback processes within and around the galaxies. Molecular gas in local galaxies has an excellent probe of both the total star-forming gas masses extended over large spatial scales and the evolutionary mechanisms \citep[e.g.,][]{Leroy+2013, Genzel_2015, Saintonge+2017}. Furthermore, the cold molecular gas history of the Universe has already been shown to reflect the evolution of the cosmic SFRD up to redshift 7 \citep[see e.g.,][for an overview]{riechers2019coldz,decarli2020,walter+2020}. Therefore, studying the cold molecular gas of DSFGs at high-\textit{z} is essential.} 

{The most abundant molecule after molecular hydrogen, $^{12}$C$^{16}$O (carbon monoxide; hereafter, CO) is a widely used tracer of molecular gas (i.e., $\mathrm{H_2}$). The rotational ground state transition, CO($J=1$ $\rightarrow$ $J=0$), hereafter referred to as $\mathrm {CO(1-0)}$, is emitted by any molecular gas with densities $\gtrsim {10}^{2}~\mathrm{{cm}^{-3}}$. Even if the Boltzmann tail in the velocity distribution of the molecules ensures the presence of some molecules in higher-$J$ states, these levels are significantly less populated than the ground state at lower densities. However, this may not be true for dense, highly excited environments, where the partition function is shifted towards higher-\textit{J} levels. At high-\textit{z}, the cosmic microwave background (CMB) also contributes in driving the partition function away from the prevailing conditions found in the local Universe \citep[see details in][]{dacunha2013,harrington2021}.}

{Although challenging and limited to very few galaxies \citep{carilli2002,greve04,riechers06,riechers09,riechers+2010} until the advent of the Karl G. Jansky Very Large Array (VLA) upgrade \citep[e.g.,][]{ivison+2011,riechers+2011sled2,riechers+2011lensed,riechers+2011sled1,harris+2012,sharon+2016}, observations of the \coonezero line (${\nu}_{\mathrm{rest}} = 115.271~\mathrm{GHz}$) are crucial to robustly calibrate total molecular gas masses using an CO-$\mathrm{{H}_{2}}$ conversion factor, $\alpha_{\rm CO}$ (see Appendix~\ref{app:a} for more details). This conversion avoids an additional uncertainty related to the line excitation, which affects the higher-\textit{J} CO lines. Furthermore, combining \coonezero data with higher-$J$ CO transitions provides the anchor for studies of the gas excitation via the CO spectral line energy distribution (SLED) from which one can determine the physical properties of the molecular gas including its kinetic temperature and density \citep[e.g.,][]{weiss2005m82,weiss2007,riechers06,riechers+2010,riechers+2013,spilker2014,yang2017,harrington2021,harrington2025}. Such modeling can further benefit from observations of the atomic carbon fine structure transitions, namely [CI](1$-$0) and [CI](2$-$1), to better constrain the excitation conditions in these extreme sources \citep[e.g.,][]{Gururanjan2023,friascastillo2024}.} 

{Most studies of the molecular gas in high-\textit{z} galaxies to date are based on mid- to high-\textit{J} CO lines \citep[e.g.,][]{valentino2020,berta+2023,hagimoto2023}, which limits our understanding of the cold gas reservoirs due to the lack of systematic \coonezero observations. Following a pilot program that targeted 14 galaxies \citep{stanley+2023}, we have therefore taken up a \coonezero large survey, VLA V\textit{z}--GAL, targeting 92 \textit{Herschel}-selected DSFGs at redshifts between 1 to 6. All 106 of these galaxies have higher-\textit{J} CO line observations and robust spectroscopic redshifts from the NOrthern Extended Millimeter Array (NOEMA) \zgal program \citep{neri+2020,cox+2023}.}

In Section~\ref{sec:sample}, we describe the observations and data reduction. In Section~\ref{sec:meausrements}, we report the \coonezero spectra, total intensity maps, and integrated properties of the line and underlying continuum. This section also includes additional \cotwoone results of ten targets. Section~\ref{sec:results} discusses the implications of these results, particularly, addressing total molecular gas masses, depletion times, gas-to-dust mass comparison, and CO line ratios. We also compare [CI](1$-$0) to \coonezero line luminosities for 23 DSFGs with available [CI] observations from the \zgal \citep{neri+2020,cox+2023}. Finally, Section~\ref{sec:conclusion} summarizes the main findings of this paper and highlights the scope to utilize this large sample for a number of studies in the future. 

A spatially flat Lambda Cold Dark Matter ($\Lambda$CDM) cosmology with $ {H}_\mathrm{0}$ = 67.4 $\mathrm{km} \ \mathrm{s}^{-1} \mathrm{Mpc}^{-1}$ and ${\Omega}_\mathrm{M}$ = 0.315, and ${\Omega}_\mathrm{\Lambda}$ = $\left(1-{\Omega}_\mathrm{M}\right)$ has been adopted in the calculation of luminosity distances ($D_L$) throughout this paper \citep{planckcosmo2020}.

\section{Sample, Observations, and Data} \label{sec:sample}

{Large surveys over the past fifteen years have significantly increased the number of detections of DSFGs. In particular, the \textit{Herschel Space Observatory} \citep{pilbratt+2010} conducted extensive surveys at $70-500 \mu \mathrm{m}$, covering over $1000~\mathrm{deg}^{2}$, leading to the discovery of several hundred of thousands DSFGs \citep{eales+2010, oliver+2012, viero+2014, lutz+2011, nayyeri+2016, bakx+2018, ward+2022}. In addition, other surveys such as the \textit{all-sky Planck-HFI} \citep{canameras+2015,planckcollab2015} and the \textit{South Pole Telescope} (SPT) surveys \citep{vieira+2010,carlstrom+2011,weiss2013} have uncovered additional sources at the bright end of the same population.}
 
{Accurate spectroscopic redshifts are essential for conducting detailed follow-up studies of multi-phase gas in high-\textit{z} DSFGs. Since the first detections of sub-millimeter galaxies (SMGs) and thanks to extensive efforts from various research groups using ground-based sub-millimeter facilities, spectroscopic redshifts for about 400 DSFGs are now available, spanning a wide range from $1.5<z<7$ \citep[e.g.,][and references therein]{greve2005,weiss+2009,walter+2012,riechers+2013,riechers2010redshift,riechers+2017,riechers2021gadotOH+,canameras+2015,neri+2020,harrington2021,urquhart2022,reuter2020,cox+2023}.} 

\subsection{NOEMA \zgal Spectroscopic Survey}
\label{sec:noemazgal}

\begin{figure}[h]
\centering
\includegraphics[width=0.45\textwidth]{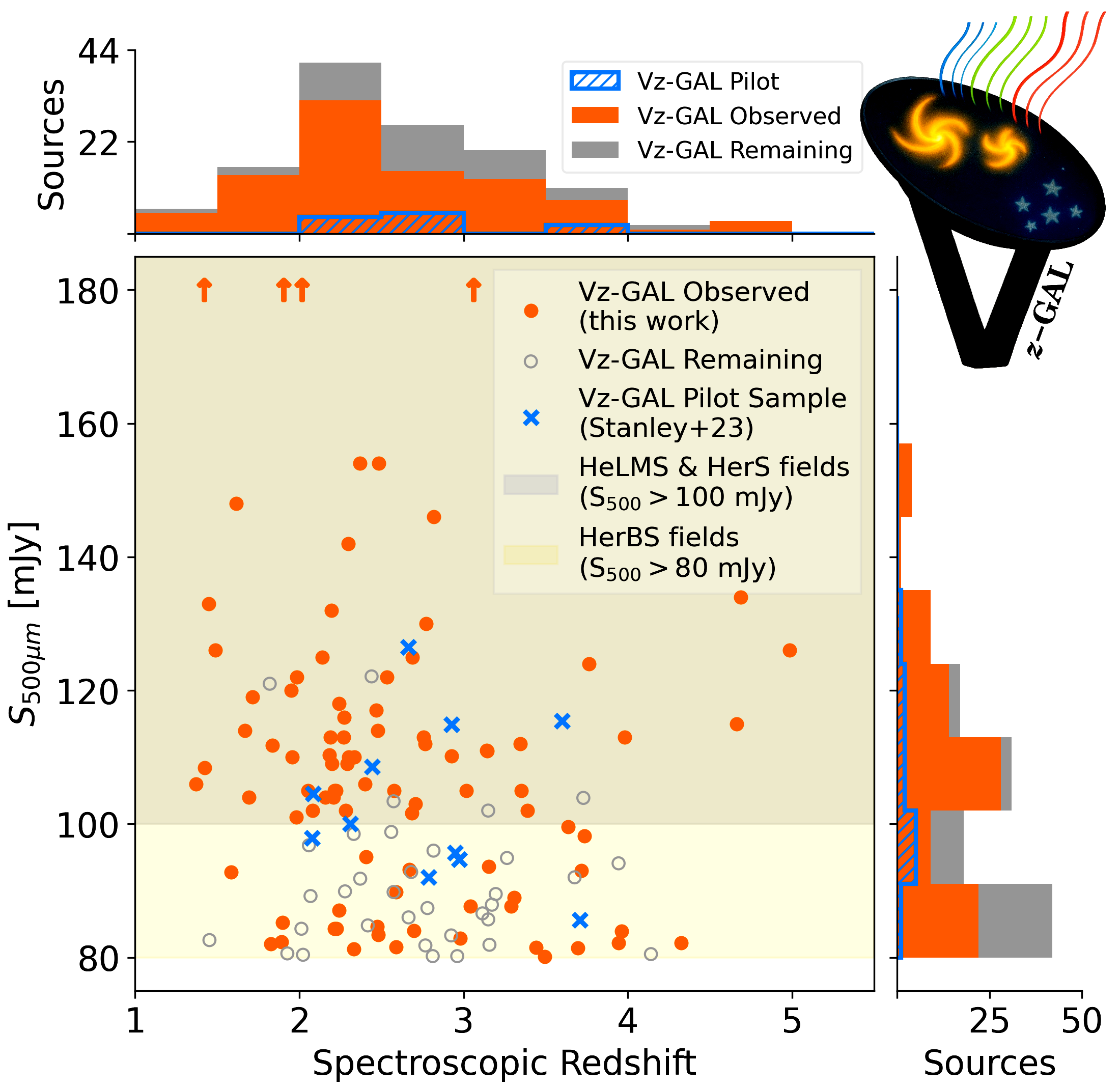}
\caption{\textit{Herschel}/SPIRE $500\mu m$ flux densities ($S_{500\mu m}$) of the \vzgal DSFGs (this work) compared to the pilot program \citep{stanley+2023} with respect to their spectroscopic redshifts based on the \zgal project. {The histograms show fluxes of the \vzgal sample with shown in gray the remaining observations not executed to date.} In the central portion, the filled area in silver color ($S_{500\mu m}$ $\mathrm{>100~mJy}$) indicates the flux selection criteria for the HeLMS \& HerS targets. The area in yellow (i.e., $S_{500\mu m}$ $\mathrm{>80~mJy}$) shows the lower flux cut that was used to select the HerBS sources. Orange data points and arrows (the latter for $S_{500\mu m}$ $\mathrm{>180~mJy}$) represent sources from this work in comparison to the DSFGs from pilot sample shown as blue crosses. {Sources not yet observed are mostly in the $S_{500\mu m} = \mathrm{80-100~mJy}$ regime and hence selected from the GAMA fields. This incompleteness is largely due to scheduling restrictions in certain LST ranges and thus random beyond the $S_{500\mu m}$ distribution. As a result, based on the homogeneous selection of the parent sample, the survey is representative and nearly complete down to $S_{500\mu m}$ = $\mathrm{100~mJy}$, while the redshift--$S_{500\mu m}$ space is relatively evenly incomplete in the $S_{500\mu m} = \mathrm{80-100~mJy}$ regime based on the observations not yet executed.}} 
\label{fig:vzgal_summary}
\end{figure}

The NOEMA \zgal project \citep{neri+2020,cox+2023} is to date the largest spectroscopic redshift survey of high-\textit{z} DSFGs, having measured redshifts for 135 DSFGs based on mid to high-$J$ CO transitions, mainly from CO(2$-$1) to CO(5$-$4), with some sources seen in [C{\small I}] and $\rm H_2O$ emission lines. {These sub-millimeter/infrared bright galaxies (${L}_\mathrm{8-1000 \mu m}$ $\gtrsim$ 10$^{12.5}$ $\mathrm{L}_\mathrm{\odot}$) were originally selected based on simple flux cuts of the \textit{Herschel} $500 \, \mu m$ flux density ($S_{500 \mu m}$; Figure~\ref{fig:vzgal_summary}) of the HerMES Large Mode Survey and Stripe 82 Survey \citep[HeLMS and HerS:][]{nayyeri+2016} and the {\it Herschel} Bright Sources sample (HerBS) detected in the H-ATLAS survey \citep{bakx+2018}. These selection criteria were $S_{500 \mu m} \geq 100 \, \rm mJy$ for the HeLMS and HerS galaxies, and $S_{500 \mu m} \geq 80 \, \rm mJy$ for the HerBS sources, respectively, after the removal of known low-redshift and non-thermal sources. The sample is complete in the selected fields, with the exception of sources for which redshifts were already known at the time \zgal was initiated.}

{Detailed studies of selected sources in the parent sample from which the \zgal sample was drawn have revealed a large diversity, e.g., numerous lensed galaxies \citep[e.g.,][]{negrello+2010,negrello+2017,conley+2011,riechers+2011lensed,bussmann+2013,bussmann+2015,wardlow+2013,nayyeri+2016lensed,bakx+2020lensed1}, active galactic nuclei \citep[AGNs; e.g.,][]{stanley+2023}, hyper-luminous galaxies (HyLIRGs) with luminosities exceeding 10$^{13}$~$\mathrm{L}_\mathrm{\odot}$ \citep[e.g.,][]{ivison+2013,riechers+2013,riechers+2017,oteo+2016}, and plausible galaxy mergers and protocluster environments \citep[e.g.,][]{riechers+2014,bussmann+2015,oteo+2018,gomezgujjaro+2019,bakx+2024}. Similar discoveries were made in the Planck and SPT samples \citep[e.g.,][]{canameras+2015,spilker+2016,miller2018,harrington17,niki2025}, with differences in the relative distributions which can likely be explained due to the different selection functions.}

The spectroscopic redshifts of the \zgal sources are in the range $1 < z < 6$ with a median value of $z = 2.56\pm0.10$ \citep{cox+2023}, centered approximately on the peak epoch of `galaxy growth' with the highest star formation rate per unit co-moving volume. These data served as a foundation for studies reporting on the dust properties of the DSFGs and their evolution with redshift \citep{ismail+2023}, and the physical properties of the galaxies using higher-\textit{J} CO lines \citep{berta+2023}. However, in order to obtain a better understanding of the physical conditions of the cold molecular gas in these DSFGs, measurements of the \coonezero emission line are essential as reported here.

\subsection{VLA \vzgal \coonezero Sample Selection, Observations, and Data} \label{sec:vlaobs}

{Using the NSF’s K. G. Jansky VLA, a first study of the \coonezero emission in 14 DSFGs (\texttt{VLA/20A-083}; PI: Riechers) in the redshift range $2<z<4$, from the \zgal pilot sample \citep{neri+2020}, was conducted by \cite{stanley+2023}. We here present the results of the continuation of this effort as a VLA Large Program, V\textit{z}$-$GAL\footnote{\href{https://vzgal.uni-koeln.de/}{\vzgal website: https://vzgal.uni-koeln.de/}}, to measure molecular gas in the entire \zgal sample as traced by \coonezero transition, and also observe \cotwoone for the \zgal targets beyond $z=3.6$.}

The observations presented here as part of the \vzgal large program (\texttt{VLA/23B-169} and \texttt{VLA/25A-099}; PI: Riechers) were carried out during semesters 2023B and 2025A. {In total, we observed 92/135 bright, dusty galaxies.} Together with the 14 DSFGs from the pilot sample \citep{stanley+2023}, this covers 106/135 \zgal sources in \coonezero emission line and the underlying continuum. {Observations of the remaining sources were not executed to date, and hence will be the focus of future efforts. The \vzgal targets were selected from the \zgal sample, therefore follow the same selection criteria, with the only additional constraint of the sources not yet observed (see Figure~\ref{fig:vzgal_summary}).} 
 
Together with the targets of the pilot project, the sample of high-$z$ galaxies presented in this paper quintuples the existing \coonezero detections of \textit{Herschel}-selected bright DSFGs with $S_{500 \mu m} \geq 80 \, \rm mJy$, including lensed galaxies \citep[22 sources in total:][]{riechers+2011lensed,harris+2012,fu2013,ivison+2013,riechers+2013,leung19,bakx+2020lensed2} and is equivalent to the total number of available \coonezero detections for high-\textit{z} DSFGs spread across more heterogeneous samples \citep[e.g.,][and references therein]{ivison+2011,riechers2011carma,riechers11cloverleaf,aravena+2016,sharon+2016,harrington19,friascastillo23}. Many from such heterogeneous samples are JCMT/SCUBA or SCUBA-2 selected DSFGs \citep[e.g., unlensed dusty galaxies with $S_{850 \mu m} > 7.5 \, \rm mJy$ from][]{friascastillo23}. 

To maximize detection efficiency by minimizing phase noise and source resolution, we used the most compact D configuration\footnote{\url{https://science.nrao.edu/facilities/vla/docs/manuals/oss/performance/resolution}} of the VLA to observe the \coonezero emission line in the selected dusty galaxies. Their robust spectroscopic redshifts of $1.4 < z <5.4$ are covered using the \texttt{Ku} (2~cm), \texttt{K} (1~cm), \texttt{Ka} (0.9~cm), and \texttt{Q} (0.7~cm) band receivers of the VLA. We also utilized our remaining, approved time for 2025A in its subsequent semester (with C configuration) to repeat the \coonezero observations for two of the sources, namely HeLMS-57 and HerBS-204, to improve their signal-to-noise ratio (SNR) after combination with the previously observed D array data.

Furthermore, we observed ten of these \vzgal galaxies at redshifts $z>3.6$ in \cotwoone (${\nu}_{\mathrm{rest}} = 230.542~\mathrm{GHz}$), using the VLA \texttt{Ka/Q} band receivers. These are all \zgal sources where such observations are possible with the VLA at reasonable system temperatures. The WIDAR correlator was configured in 8-bit mode to maximize line sensitivity, achieving 2~MHz spectral resolution ($\sim12-36~\mathrm{km} \ \mathrm{s}^{-1}$ per channel across different bands) in dual polarization over a 2~GHz bandwidth. Exposure times were optimized for line detections and not for the underlying continuum. The desired spectral line in each observation set was typically centered in one of the 128~MHz sub-bands of 1~GHz intermediate frequency (IF) bands/sessions, with the second IF band placed either adjacent to the first one for enhancing continuum sensitivity or at a sufficiently different frequency for measuring the spectral index or covering other faint lines. Throughout this paper, we use the acronym IF1 for the IF band at lower frequencies and IF2 for the other one. 

Standard VLA flux/bandpass calibrators (3C48, 3C138, 3C147, and 3C286) were used to perform flux calibration. Note that during the observations, quasar 3C138 was in a flaring state, which affected one set of observations (see Section~\ref{subsec:datareduction}). Phase calibrators in the proximity of respective science targets were selected from the VLA calibrator catalog.

\section{Data Reduction and Measurements} \label{sec:meausrements}

\subsection{Data Reduction} \label{subsec:datareduction}

The data were reduced, calibrated, and imaged using \texttt{CASA v6.4.1} \citep{McMullin+2007casa} on two high performance computers in Cologne, namely CHEOPS\footnote{\url{https://itcc.uni-koeln.de/hpc/hpc/technische-details}} and RAMSES\footnote{\url{https://itcc.uni-koeln.de/hpc/hpc/ramses}}. Manual flagging was necessary to remove radio frequency interference (RFI), bad baselines, and non-functional antennas for all sources. The measured flux densities of the flux calibrators agreed with their \citet{Purley2017} models to within 10–15\% accuracy. However, the flaring flux calibrator 3C138 affected \coonezero observation of HerS-11, one of the DSFGs from our sample. To correct for this, we rescaled its observed flux densities using \texttt{getcalmodvla\footnote{\url{https://casadocs.readthedocs.io/en/latest/api/tt/casatasks.calibration.getcalmodvla.html}}} in \texttt{CASA v6.7}. 

Following the calibration process, we split apart the calibrated target data from calibrators using the \texttt{CASA} task \texttt{split}. We generated spectral cubes for each of the targets and manually cleaned their integrated-intensity (moment-0) maps using \texttt{tclean}. The default baseline weighting scheme, \texttt{natural}, was used in \texttt{tclean} and the data were averaged to a resolution of 6~MHz (on average $\sim$60 $\mathrm{km} \ \mathrm{s}^{-1}$). The CO emission of our targets is mostly unresolved or only marginally resolved (see Section~\ref{subsec:spec}), with angular resolutions between  $1.2^{\prime\prime}$and $5.2^{\prime\prime}$. Our observations for objects that required long exposure times were divided into multiple sets, which we combined using \texttt{concat} to improve the SNR while imaging. For HeLMS-57 and HerBS-204, the D- and additional C-configuration data were combined using \texttt{concat} after being \texttt{split} with a common \textit{uv}-range of $0-125~k\lambda$ and $0-82~k\lambda$, respectively. This step ensured that the spatial scales were matched between the two array setups and the SNR was improved across the combined datasets. Although retaining the full \textit{uv}-range of C-array data might have led to better angular resolution, we noticed the increased noise due to phase instability for the longer baselines, leading us to combine only the common \textit{uv}-range.

Continuum subtraction was performed using \texttt{uvcontsub}, and clean continuum maps were generated in multi-frequency synthesis (\texttt{mfs}) mode of \texttt{tclean} using line-free channels for the sources with continuum detections over 2${\sigma}_{c}$, where ${\sigma}_{c}$ is the root mean squared (rms) noise across the averaged bandwidth. For targets with continuum detections in both IF1 and IF2, we combined the line-free channels from both IFs to create a continuum map at the mean central frequency of both using the \texttt{tclean mfs} mode. We further use the \texttt{CASA} 2D Gaussian fitting task \texttt{imfit} to estimate the total continuum flux densities for cases with continuum above 2${\sigma}_{c}$ (see the tables in Appendix~\ref{app:a}). For non-detections with SNR below 2, we provide 3${\sigma}_{c}$ upper-limits using the clean continuum maps, assuming the emission to be unresolved. Please refer to Appendix~\ref{app:a} for details on continuum measurements. 

\vspace{1cm}
\subsection{CO Spectral Line Measurements} \label{subsec:spec}

We partially resolve some lensed DSFGs and systems with multiple galaxies at the same spectroscopic redshift in the VLA data -- see list of these sources and details in Appendix~\ref{app:b}. Therefore, we choose two different methods to extract the spectra: (i) peak pixel spectrum extraction for unresolved targets, and (ii) 2${\sigma}$ contours based extraction for resolved systems --- ${\sigma}$ being the rms noise of the moment-0 map --- as this level is high enough to limit noise contamination while still inclusive enough to capture faint, extended line emission that would be systematically missed at more conservative thresholds. For case (ii), we iterated the flux extraction process for defining the extraction region used. Here, the initial spectra are extracted from a large circular aperture centered on the \zgal based target coordinates, encompassing the source emission. We defined line channels to be used for the moment-0 maps from here. Having the map created using \texttt{CARTA v4.1.0} \citep{cartaref2024}, we redefined the extraction regions based on its 2${\sigma}$ contours and repeated the steps until the 2${\sigma}$ contours converged. Finally using \texttt{CASA tclean}, we created total intensity (moment-0) maps directly from the \texttt{split} target visibilities for all DSFGs, regardless of their different spectral extraction methods.

  The \coonezero and \cotwoone moment-0 maps of all the \vzgal sources are presented in  Appendix~\ref{app:b}. {Of the 92 DSFGs covered by the observations, \coonezero emission is detected at $>$5$\sigma$ significance for 57 targets, at $>$3$\sigma$ significance for 81 targets, and $>$2$\sigma$ significance for 90 targets. We consider the 9 sources recovered at $2\sigma - 3\sigma$ significance to be tentative detections, but due to the known redshifts and positions, the probability of false positives is low compared to random $2\sigma - 3\sigma$ peaks in the data cubes. Statistically, we expect less than one false detection among these tentative identifications, but more sensitive observations are required to confirm them.} The remaining two sources --- HeLMS-44 ($z = 1.37$) and HerBS-53 ($z = 1.42$) --- show no significant emission, with \coonezero signal levels below 2$\sigma$ in the VLA \texttt{Q}-band. These non-detections are due to high noise at their redshifted \coonezero frequencies near the upper edge of the $40-50$ GHz band. 
  
  Three sources with multiple spatial components (HerBS-109, HerBS-187, HerBS-194) show only a partial \coonezero detection; i.e., we detected only one of their components above 2$\sigma$ level. Please note that HerBS-109 does not have available infrared luminosity \citep{berta+2023} and out of its three spatial components, we only detected the southern one. Although we provide the \coonezero line properties of its southern component in Appendix~\ref{app:a}, we avoid using this target in the analyses presented in Section~\ref{sec:results}. Otherwise, the statistical results in Section~\ref{sec:results} incorporate limits from all the partial- and non-detections, wherever relevant.
  
Furthermore, all the ten DSFGs with additional \cotwoone data (HeLMS-19, HeLMS-24, HeLMS-27, HeLMS-36, HeLMS-45, HerS-11, HerBS-78, HerBS-177, HerBS-179, and HerBS-185) were successfully detected with at least SNR of 3, with only HeLMS-27 and HerBS-179 showing tentative $2\sigma - 3\sigma$ detections.  

Besides moment-0 maps, we show in Appendix~\ref{app:b} the CO line spectra and their Gaussian fits for all the 92 targets. We compare these spectra to higher-\textit{J} CO lines detected with NOEMA \citep{cox+2023} for each source. As no mid/high-\textit{J} CO lines were previously detected for HeLMS-49, we instead compare the \coonezero line profile with that of [CI](1$-$0). Approximately {30\%} of the sources show double-peaked line profiles. Among these, the sources with significant overlap between two fitted Gaussian profiles show the extracted flux values to be consistent (within error-bars) with those derived using only a single Gaussian fit to the same profile. We choose to use the results of a single Gaussian fit in these cases to avoid larger uncertainties on the integrated line fluxes due to more fitting parameters. However, for sources with well-separated peaks --- velocity separation ($v_2 - v_1$) $\geq 1.25 \times \sqrt{{{{\sigma}_{1}}^2}+{{{\sigma}_{2}}^2}}$) --- we fit them with two separate Gaussian profiles and sum their fluxes to derive the total line flux. Here ${\sigma}_{1}$ and ${\sigma}_{2}$ are the standard deviations of two Gaussians fitted to two different line components of a double-peaked profile. The integrated properties of the measured \coonezero and \cotwoone spectra are listed in Appendix~\ref{app:a}. 

Overall, the \coonezero line profiles have consistent widths and shapes with those seen at higher-\textit{J} CO levels. Qualitative notes on individual galaxies with differences in this comparison (e.g., HerS-13) and with newly resolved spatial components (e.g., HeLMS-32) are given in Appendix~\ref{app:b}.

\subsubsection{CO Line Luminosity and Gas Mass Derivation} \label{subsec:lum}

To date, integrated \coonezero line luminosities, ${{{L}^{\prime}}_{\rm CO(1-0)}}$, have been used {in the literature} for robustly calibrating total molecular gas masses, $M_{\rm H_2}$, for targets with unknown excitation conditions. With \coonezero observations, we avoid uncertainties related to the conversion from higher-$J$ CO emission lines to the ground-state. Relations between the observables and these physical quantities used in our calculation are as follows \citep[see][for details]{solomon97}:
\begin{equation}
    {{L}^{\prime}}_{\rm CO} = 3.25 \times {10}^{7} \times \frac{{{I}_{\rm CO}} \cdot {{{D}_{L}}^{2}}}{{{\nu}^{2}_{\rm CO,rest}} \cdot {\left(1+z\right)}}~\mathrm{\left[K~km~{s}^{-1}~{pc}^{2} \right]} 
    \label{eqn:LprimeCO}
\end{equation}
\begin{equation}
    {M}_{\rm H_2} = {\alpha}_{\rm CO} \times {{L}^{\prime}}_{\rm CO(1-0)}~\mathrm{[M_{\odot}]}
    \label{eqn:gasmass}
\end{equation}
where, ${I}_{\rm CO} = {S}_{\rm peak}~{\Delta v}$ is the measured CO line flux in [Jy km $\mathrm{{s}^{-1}}$], $D_L$ is the luminosity distance in [Mpc], ${\nu}_{\rm CO,rest}$ is the rest frequency of the CO line in [GHz], and ${\alpha}_{\rm CO}$ is the CO-$\mathrm{{H}_{2}}$ conversion factor. Here ${\Delta v}$ and ${S}_{\rm peak}$ are the full width at half maximum (FWHM) and peak intensity of the spectral line, respectively. For double peaked spectra, we adopt the net FWHM as following:

\begin{equation}
    \mathrm{net~FWHM} = \Delta s + \left[0.5 \times (\Delta v_1 + \Delta v_2)\right]
\end{equation}

\noindent
where, $\Delta s$ is the separation between two fitted-Gaussians of a double peaked line. Further, $\Delta v_1$ and $\Delta v_2$ are their respective FWHM. In such cases, the ${S}_{peak}$ of the double Gaussian line fit corresponds to the derived total line flux divided by the net FWHM.

Using this robustly calibrated relation, we list the ${{L}^{\prime}}_{\rm CO(1-0)}$ and the derived gas masses adopting a Milky Way-like ${\alpha}_{\rm CO}$ of $\mathrm{4 ~{M}_{\odot}~{\left(K~km~{s}^{-1}~{pc}^{2} \right)}^{-1}}$ in Appendix~\ref{app:a}. From now on, we will refer to the ${\alpha}_{\rm CO}$ value without the units attached, for the ease of reading. We discuss the choice of adopting the Milky Way-like conversion factor in Appendix~\ref{app:a}. 

{As we do not have derived values for the lensing magnification factor $\mu$ for the full sample, we report the line luminosities and gas masses of all the sources without applying a correction for $\mu$. However, we indicate the `candidate' lensed DSFGs with a special marker in the tables in the Appendix~\ref{app:a}.}

The \vzgal DSFGs have \coonezero line luminosities in the range $\mu {L}^{\prime}_{\mathrm{CO(1-0)}} \sim (0.5-5) \times {10}^{11}~\mathrm{K~km~{s}^{-1}~{pc}^{2}}$, and have gas masses $\mu {M}_{\rm H_2}$ ranging from $2 \times {10}^{11}~(\alpha_{CO}/{4.0})~\mathrm{M_{\odot}}$ to $20 \times {10}^{11}~(\alpha_{CO}/{4.0})~\mathrm{M_{\odot}}$. For the sake of completeness, we also tabulate these properties for the DSFGs from the \vzgal pilot program \citep{stanley+2023}, adopting the same ${\alpha}_{\rm CO}$ conversion factor (see Appendix~\ref{app:a}).

\section{Results and Discussion} \label{sec:results}

In this section, we present the implications of the \coonezero results discussing first the correlation between the total infrared and CO line luminosities (Sections~\ref{subsec:correlLirLco} and \ref{subsec:sfe}), followed by the redshift evolution of the gas depletion time presented in Section~\ref{subsec:tdepl}. Further, we combine our \coonezero and \cotwoone results with the mid/high-$J$ CO transitions from \citet{cox+2023} to calculate the CO line ratios (Section~\ref{subsec:colineratio}). Here we also derive empirical relations between the CO(1$-$0), CO(2$-$1), and CO(3$-$2) line luminosities to compare our statistical trends in scaling relations with those derived using the \zgal data. Finally, Section~\ref{subsec:gasdustratio} presents the gas-to-dust mass ratios of our sample and Section~\ref{subsec:CIcomparison} explores a correlation between the line luminosities of [CI](1$-$0) and CO(1$-$0). We summarize this discussion by listing the physical conditions of the interstellar medium (ISM) under which the calibration of total molecular gas masses, based only on \coonezero observations, might be challenging (Section~\ref{subsec:co10caveats}).

\subsection{${{L}^{\prime}}_{\rm CO(1-0)}-L_{\rm IR}$ Correlation}\label{subsec:correlLirLco}

The {Kennicutt-Schmidt (KS) relation} or {star formation law} in galaxies is defined as the relation between the gas surface density and the SFR surface density \citep{schmidt1959,kennicutt1989}. For most of the high-\textit{z} galaxies with unavailable spatially resolved observations, a comparison between the integrated ${L}^{\prime}_{\rm CO}$ (as a proxy for $M_{\rm gas}$) and total infrared luminosity ($L_{\rm IR}$, a proxy for SFR) serves as a proxy for the KS-relation. Although the ${L}^{\prime}_{\rm CO}-L_{\rm IR}$ correlation is not strictly the same as the KS-law, it is beneficial to check the scaling relation over a large redshift range. Based solely on observables, the relative positions of different galaxy populations in this diagram provide qualitative insights into underlying physical conditions --- such as variations in gas and dust temperatures and densities. The trends in CO-$\mathrm{H_2}$ conversion factor (${\alpha}_{\rm CO}$), hypothesized to depend on these physical properties, can also be qualitatively discussed using this correlation. In the literature, ${\alpha}_{\rm CO}$ has commonly been adopted with a bimodal distribution (see details in Appendix~\ref{app:a}). 

The total infrared luminosities ($L_{\rm IR}$; $8-1000 \mu$m) used in this work are derived by integrating a modified black body (MBB) function. In a preceding \zgal paper, \citet{berta+2023} transformed the \textit{Herschel}- and NOEMA-observed $\rm 50-1000 \mu m$ luminosities \citep{ismail+2023} of our DSFGs into $8-1000~\mu$m luminosities using $L_{50-1000 \mu m}/L_{8-1000 \mu m}$ = $0.7\pm0.1$. These $8-1000 \mu$m infrared luminosities ($L_{\rm IR}$), which are connected to SFRs and/or dust temperatures, have been used in all the upcoming analyses.

  \begin{figure*}
\begin{center}
\includegraphics[width=0.575\textwidth]{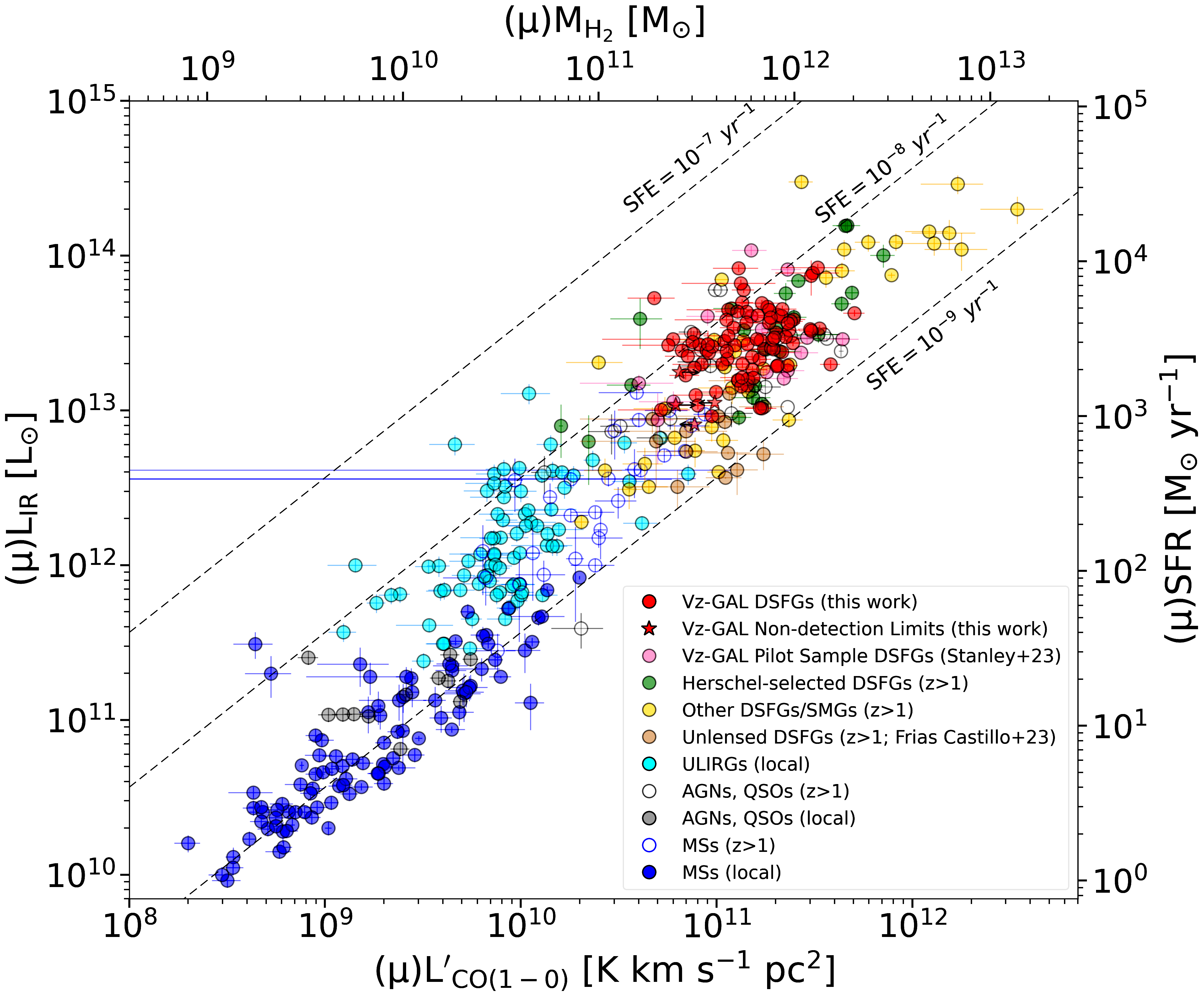}
\includegraphics[width=0.575\textwidth]{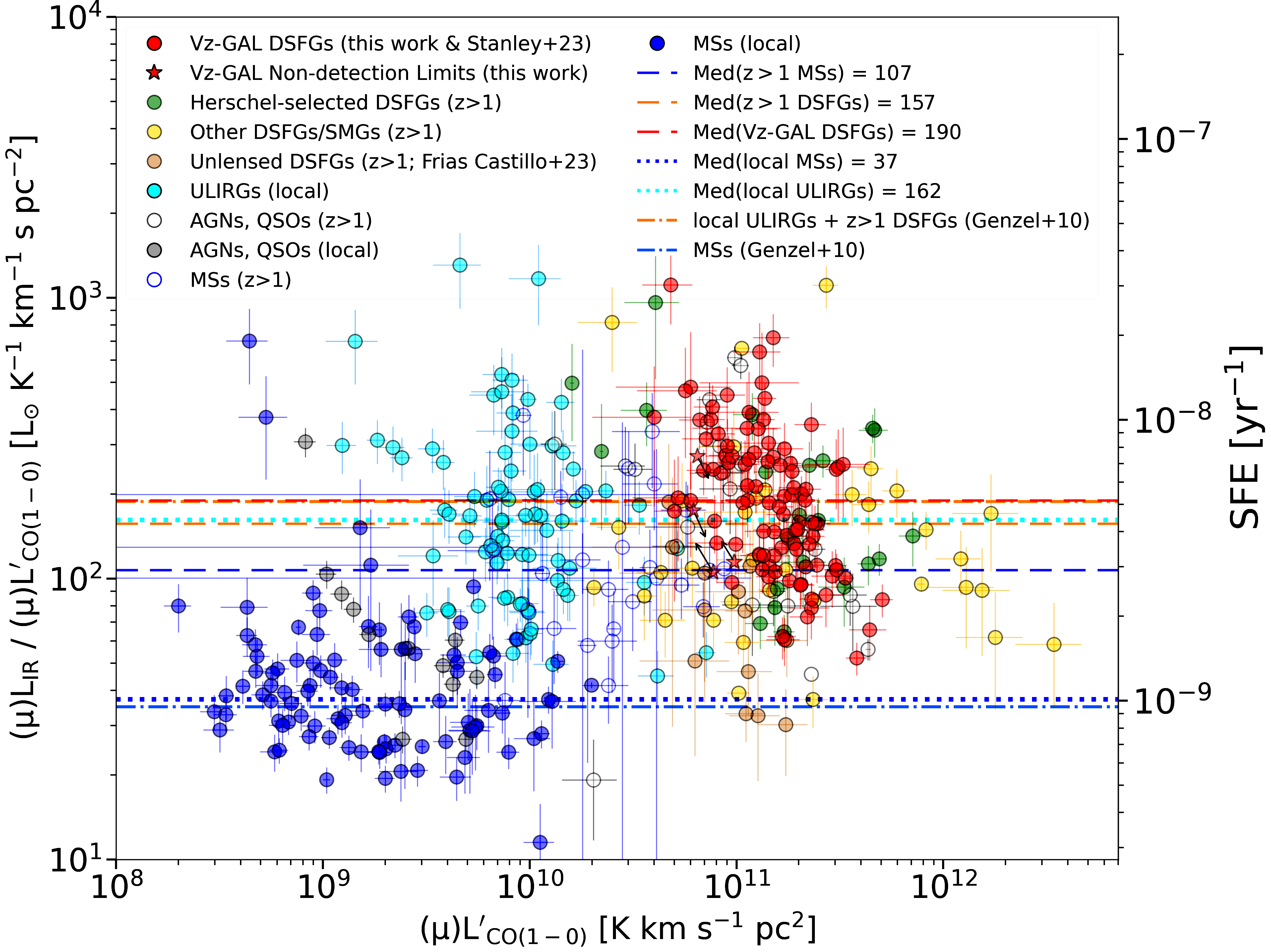} 
\end{center}
  \caption{\textit{(Upper panel)} A proxy for the Kennicutt-Schmidt (KS) relation of the \vzgal sample in terms of the infrared ($\rm 8-1000\, \mu m$) luminosity \citep[$L_{\rm IR}$;][]{berta+2023} and the \coonezero line luminosity \citep[${{L}^{\prime}}_{\rm CO(1-0)}$; this work and][]{stanley+2023}. Dashed diagonals present constant star formation efficiency (SFE) lines with SFR [${\rm M}_{\odot}~{\rm year}^{-1}$] = $1.09 \times {10}^{-10}$ ${L}_{\rm IR} [{\rm L}_{\odot}]$ and ${\alpha}_{\rm CO} = \mathrm{4 ~{M}_{\odot}~{\left(K~km~{s}^{-1}~{pc}^{2} \right)}^{-1}}$. Here, we highlight the \vzgal pilot sample from \citet{stanley+2023} separately to illustrate its consistency with our measurements. However, in the subsequent statistical analyses and corresponding figures, the 14 pilot galaxies are combined with our 92 DSFGs and treated as a single, unified \vzgal sample. \textit{(Lower panel)} Ratio of the infrared to \coonezero luminosity, a proxy for star formation efficiency (SFE), versus the CO line luminosity (without lensing correction) as a probe for molecular gas mass. Relations derived by \citet{genzel2010} are shown with dash-dotted lines for comparison, up-scaling their far-infrared luminosities $L_{\rm FIR}$ ($\rm 50-300 \, \mu m$) to $L_{\rm IR}$ ($\rm 8-1000 \, \mu m$) using $L_{\rm IR}$/$L_{\rm FIR}$  $\sim 1.3$ \citep{graciacarpio2008} to match our plot. In both panels, we also show literature data points with available \coonezero observations: high-\textit{z} \textit{Herschel}-selected DSFGs \citep{riechers+2011lensed,harris+2012,fu2013,ivison+2013,harrington19,leung19,bakx+2020lensed2}, other distant DSFGs and sub-millimeter galaxies (SMGs) \citep{greve04,riechers08,carilli10,harris10,ivison+2011,emonts11,riechers+2011sled1,riechers+2011sled2,swinbank11,thomson12,riechers+2013,sharon13,emonts14,sharon15,aravena+2016,sharon+2016,harrington17,dannerhr19,friascastillo23}, high-\textit{z} AGNs/QSOs \citep{carilli02,lewis02,riechers06,aravena08,riechers09,wagg10,lestrade11,riechers11,sharon+2016,Riechers2020}, distant MS galaxies \citep{aravena12,aravena14,rudnick17,gomezgujjaro+2019,kaasinen19,Riechers2020}, local ULIRGs \citep{solomon97,chung09,combes11,gowardhan18}, and nearby massive MS galaxies \citep[${M}_{\ast} > {10}^{10}~{\rm M}_{\odot}$;][]{geach11,Saintonge+2017,villanueva17,dunne21}. }
  \label{fig:KSrelation}
\end{figure*}

Figure~\ref{fig:KSrelation} presents the distribution in the ${L}^{\prime}_{\rm CO(1-0)}-L_{\rm IR}$ plane of all the \vzgal DSFGs, including the sources from the pilot program \citep{stanley+2023}. For comparison, we added nearby and high-$z$ galaxies from the literature with available \coonezero observations (see caption of Figure~\ref{fig:KSrelation} for references). There appears to be a systematic offset in the $L_{\rm IR}/{{L}^{\prime}}_{\rm CO(1-0)}$ ratios between SFR$-M_{\ast}$ main-sequence (MS) galaxies at low and high redshifts, local ULIRGs, and  distant DSFGs. This ratio being a proxy for SFR per unit gas mass; i.e., star formation efficiencies (SFEs), such an offset may suggest differences in their SFEs \citep[and/or dust temperatures assuming the same gas mass; see][]{magdis2012}. On the other hand, the overall trend across statistically significant samples remains relatively smooth without evidence for a clear bimodality. 
High-\textit{z} DSFGs follow similar SFE lines {and spread} as local ULIRGs (upper panel of Figure~\ref{fig:KSrelation}), perhaps implying comparable gas/dust temperatures and densities, and conditions that are more extreme than those in MS galaxies. This likely reflects their denser star-forming regions and more intense radiation fields associated with starburst activity, which result in warmer dust temperatures, higher $L_{\rm IR}$, and elevated SFEs compared to local disks \citep[e.g.,][]{riechers2021gadotOH+}. Thus, the observed distributions in Figure~\ref{fig:KSrelation} (upper panel) do not provide evidence for assuming a bimodal conversion factor rather than continuous variation in ${\alpha}_{\rm CO}$ across these populations, driven by their underlying physical conditions. They, however, also cannot rule out a constant ${\alpha}_{\rm CO}$.

In Figure~\ref{fig:KSrelation} (lower panel), we examine the $L_{\rm IR}$/${{L}^{\prime}}_{\rm CO(1-0)}$ ratio, a proxy for SFE, as a function of ${{L}^{\prime}}_{\rm CO(1-0)}$ to further explore the observed offset. This ratio reflects variations in gas density and temperature across galaxy populations, which in turn may affect the ${\alpha}_{\rm CO}$. Here although $L_{\rm IR}$/${{L}^{\prime}}_{\rm CO(1-0)}$ ratio is independent of the lensing magnification factor ($\mu$) under the assumption of no differential lensing between the emission coming from dust and gas, the apparent ${L}^{\prime}_{\rm CO(1-0)}$ of the high-\textit{z} DSFGs are $\sim 1$ dex higher than local ULIRGs. This can be at least partially explained by the lensed nature of the majority of the high-\textit{z} galaxies. It is however worth noting that 79\% of our sample shows $\mu {L}^{\prime}_{\rm CO(1-0)}$ values in agreement with those explored by \citet{friascastillo23} for unlensed DSFGs at $z=2-5$. Assuming an average magnification factor of $\mu = 7$ \citep[see e.g.,][]{bussmann+2015} for our sample, leading to ${L}^{\prime}_{\rm CO(1-0)} \sim (0.7-7) \times {10}^{10}~\mathrm{K~km~{s}^{-1}~{pc}^{2}}$, we still infer marginally larger cold gas reservoirs in our high-\textit{z} DSFGs than those seen in local ULIRGs. 

While the local disks have a median ratio $\sim3\times$ lower than high-\textit{z} MS galaxies \citep[see also][]{scoville2016}, local starbursts ($z\sim0$ ULIRGs) and high-\textit{z} DSFGs are observed to exhibit similar ratios. However, local ULIRGs are known to be more compact than high-\textit{z} DSFGs and possibly have higher IR surface brightness \citep[e.g.,][]{solomon1992,gaosolomon2004}. The fact that, despite these differences, their observed $L_{\rm IR}$/${{L}^{\prime}}_{\rm CO(1-0)}$ ratios are consistent with distant DSFGs suggests that the star formation processes operating in these two populations are self-regulated. In other words, an internally regulated mechanism is likely responsible for maintaining the star formation rate per unit molecular gas mass within a small range across these two populations.

Both the highly star-forming populations (local ULIRGs and high-\textit{z} DSFGs) have $\sim4.3\times$ higher $L_{\rm IR}$/${{L}^{\prime}}_{\rm CO(1-0)}$ ratios than local MS galaxies. However, these ratios are only about $1.5\times$ higher than those of MS galaxies from the early Universe.  Our results confirm this trend, also observed by \citet{genzel2010}, with improved statistics. \citet{genzel2010} interpreted such higher ratios as an evidence for more rapid or efficient star formation (see their Figure 2). 

It is also important to note that this offset may arise from warmer dust temperatures ($T_{\rm d}$) in the highly star-forming galaxies, both at low and high redshifts \citep[see e.g.,][]{magdis2012,bethermin2013}. For instance, assuming a constant gas and dust mass across local disks and ULIRGs, we can model the galaxy’s emission as a black body, where $L_{\rm IR} \propto{{T_{\rm d}}^4}$, when optically thick at all wavelengths \citep{magdis2012,jin2022}. Alternatively, if we treat the emission as that of a gray body in the thin limit, this relationship approximately becomes $L_{\rm IR} \propto{{T_{\rm d}}^{4.5}}$. Under this assumption, the observed offset in the $L_{\rm IR}$/${{L}^{\prime}}_{\rm CO(1-0)}$ ratios between these two populations could be explained by increasing the dust temperature in starbursts $(3-8)\times$ compared to local disks --- as the $L_{\rm IR}$ of starbursts are on average $\sim1.5-4$ dex higher. This temperature difference may also be connected to a more efficient star formation process.\footnote{In addition to thermal and photoelectric heating, possibilities of other dust heating mechanisms cannot be neglected; e.g., cosmic ray heating, shock heating, gas-dust collisional heating.}

\subsection{Redshift Evolution of $L_{\rm IR}/{{L}^{\prime}}_{\rm CO(1-0)}$ Ratio, a Proxy for Star Formation Efficiency} \label{subsec:sfe}

The integrated $L_{\rm IR}/{{L}^{\prime}}_{\rm CO(1-0)}$ ratio is a proxy for the star formation efficiency (SFE) of galaxies. Given an available molecular gas reservoir, SFE measures the efficiency at which stars form out of this material. This is formulated as SFE = (SFR/$M_{\rm gas}$), where the star formation rate (SFR) is related to the rest-frame $8-1000~\mu \rm m$ total infrared luminosity (${L}_{\rm IR}$) using a relation from \citet{kennicutt1989} as shown below, corrected for a \citet{chabrier2003} initial mass function. 

\begin{equation}
    \mathrm{SFR} [{\rm M}_{\odot}~{\rm yr}^{-1}] = 1.09 \times {10}^{-10} {L}_{\rm IR} [{\rm L}_{\odot}]
    \label{eqn:sfr_lir}
\end{equation}

\noindent
We caution the reader that the integrated measurements presented here average over the smaller scale variations in physical conditions within each galaxy, providing a global SFR.

\begin{figure*}
\begin{center}
\includegraphics[width=0.58\textwidth]{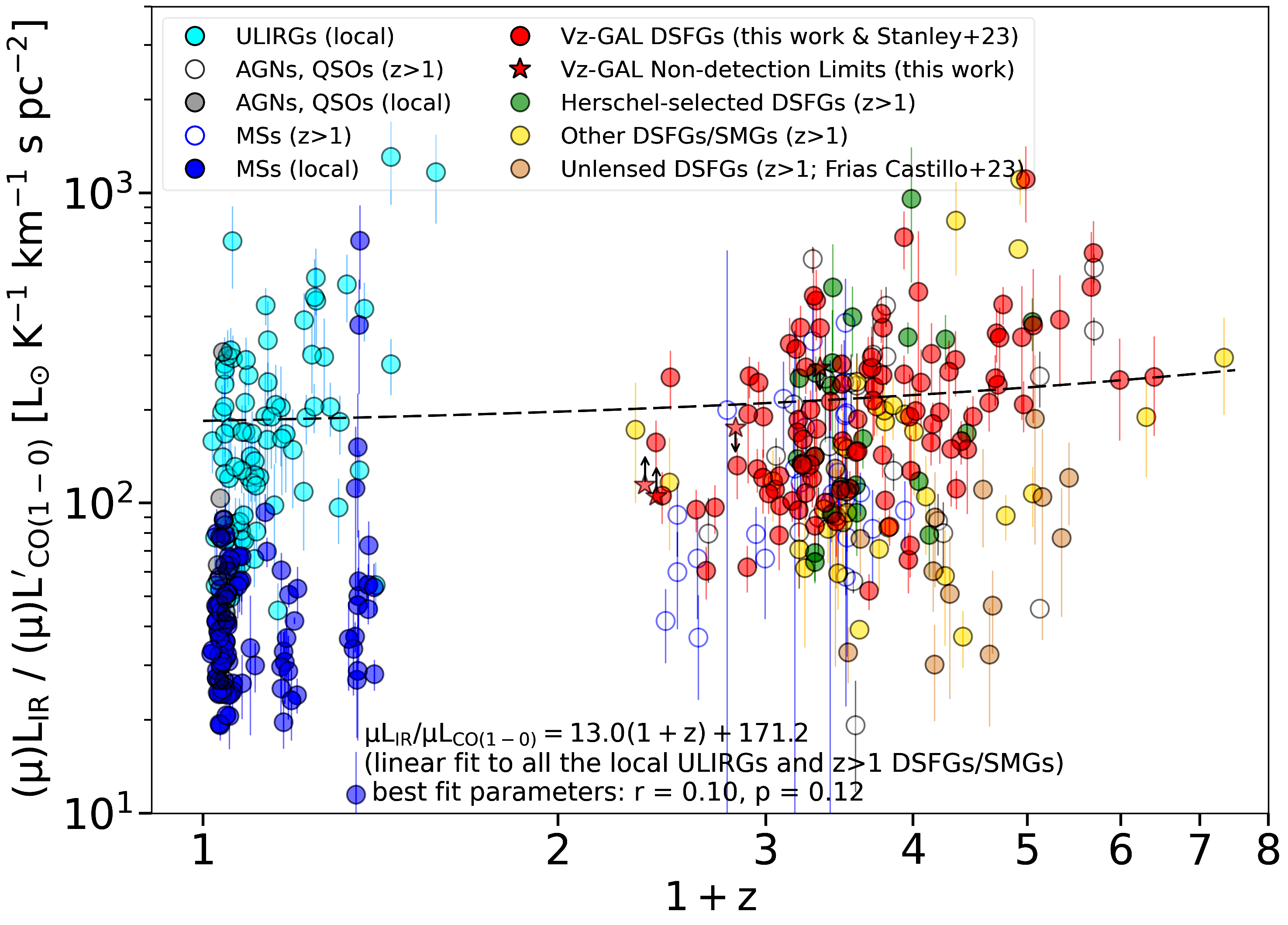}
\includegraphics[width=0.58\textwidth]{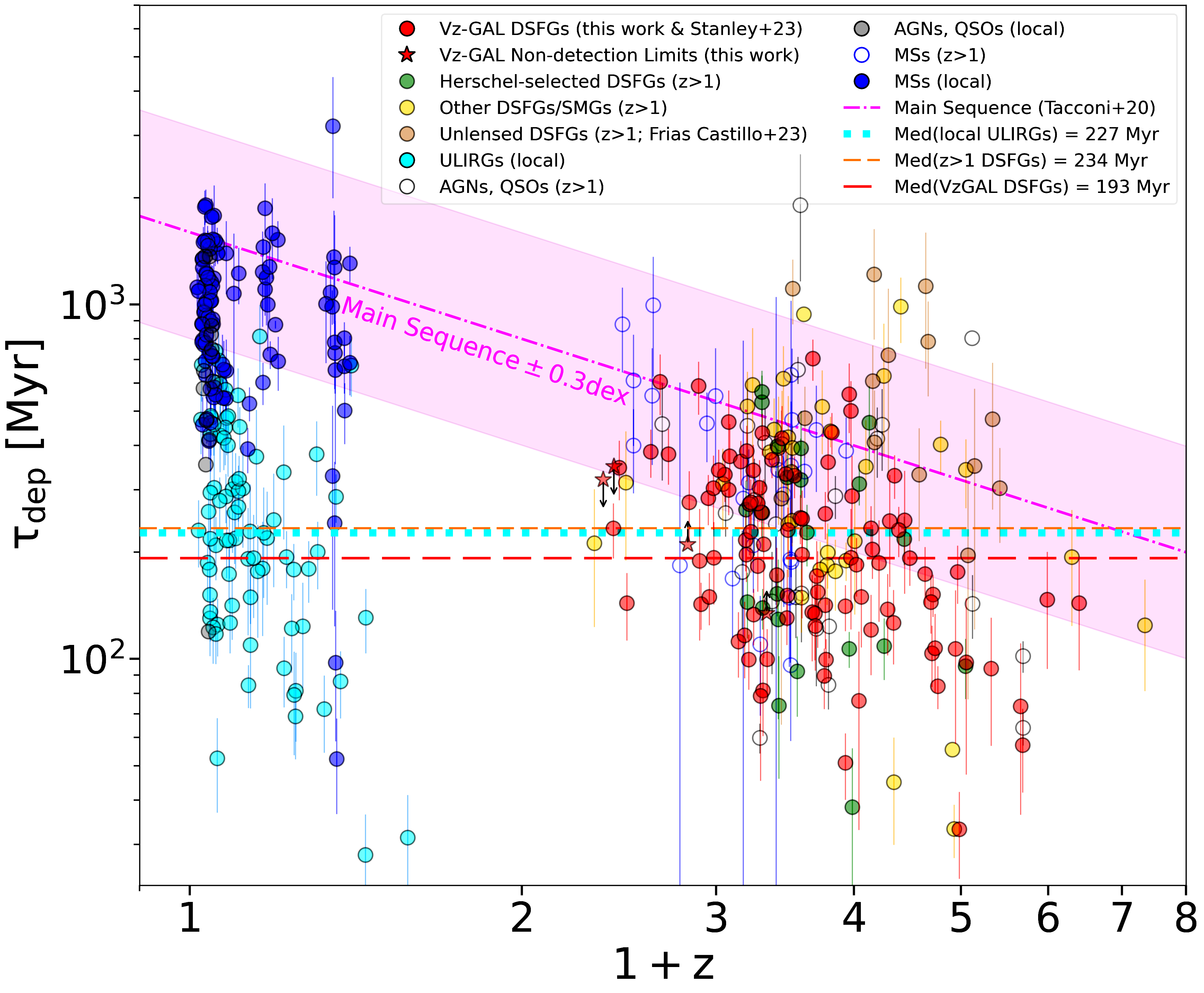} 
\end{center}
  \caption{\textit{(Upper panel)} Redshift evolution of the total infrared ($8-1000 \mu m$) to \coonezero luminosity ratio $L_{\rm IR}/L^\prime_{\rm CO(1-0)}$ as a proxy for SFE. The linear fit shown with a dashed black line considers all the high-\textit{z} DSFGs and local ULIRGs, which shows no confirmation of a linear trend. We choose to plot the redshift in log-scale here to highlight the variation across MS galaxies and ULIRGs at $z\sim0$. \textit{(Lower panel)} Variation of the gas depletion time ($\tau_{\rm dep}$) as a function of redshift when assuming ${M}_{\rm H_2} = 4 {{L}^{\prime}}_{\rm CO(1-0)}$ (see Section~\ref{subsec:lum}) and SFR = $1.09 \times {10}^{-10}$ ${L}_{\rm IR}$ (see Section~\ref{subsec:sfe}). The dashed red line is the median $\tau_{dep}$ for the \vzgal sources, which is compared to that of local ULIRGs (cyan dotted line) and all DSFGs including the \vzgal sample (dashed orange line). The SFR$-M_{\ast}$ main-sequence (MS) shown as magenta dashed line is based on Figure 7 from \citet{tacconi2020}, which varies as $\tau_{\rm dep}=1.6 \times {(1+z)}^{-1}$. Both panels include the literature data points mentioned in the caption of Figure~\ref{fig:KSrelation}. Please note that a gap in \coonezero observations (rest frame $115~{\rm GHz}$) around $z=1$ is due to the current observational limitation at corresponding redshifted frequencies.}
  \label{fig:z_evol_sfe_tdepl}
\end{figure*}

With our comprehensive \coonezero observations, we examine the redshift evolution of the $L_{\rm IR}/{{L}^{\prime}}_{\rm CO(1-0)}$ ratio, a direct observational proxy for SFE (see upper panel of Figure~\ref{fig:z_evol_sfe_tdepl}). This approach avoids assumptions about ${\alpha}_{\rm CO}$, which affect SFE estimates through uncertainties in ${M}_{\rm gas}$. The $L_{\rm IR}/{{L}^{\prime}}_{\rm CO(1-0)}$ ratio shows no significant trend with redshift when comparing high-\textit{z} DSFGs to local ($z \sim 0$) ULIRGs, though the scatter increases at $z > 2$, and the number statistics become limited at $z > 4$. Notably, the \vzgal sample and other \textit{Herschel}-selected sources exhibit relatively less scatter {than the rest} of the distant DSFGs collected from the literature. Our 500~$\mu m$-selected DSFGs appear to be {marginally shifted} towards the higher end of this ratio compared to other distant, dusty galaxies. This could reflect source selection effects, plausibly leading to different dust temperatures for a fixed gas mass.\footnote{To avoid the $T_{\rm d}-z$ degeneracy \citep[see][]{casey+2014}, we consider the galaxies around cosmic noon ($z=2-3$). This includes all the DSFGs with different selection criteria, and MS galaxies in the given redshift range. We plot their $L_{\rm IR}$ versus $L_{\rm IR}/{{L}^{\prime}}_{\rm CO(1-0)}$ ratios in Appendix~\ref{app:b}, which do not show significant variation across various samples, despite the differences in (i) the selection criteria, (ii) observed IR luminosities (hence, $T_{\rm d}$) and/or (iii) lensing factors. The median ratios for these galaxy populations are comparable within a factor of 2, except for a few outliers (see Appendix~\ref{app:b}).}

As discussed in Section~\ref{subsec:correlLirLco}, the \vzgal sources show consistency with the local ULIRG samples over a wide redshift range, owing to their high SFRs per available gas mass and/or elevated dust temperatures. This result confirms the similarity in their star-forming environments. Furthermore, a clear distinction between MS galaxies and starbursts is evident in the local Universe, but this separation is poorly constrained beyond $z \sim 2$, where MS galaxy data are sparse. Therefore, it remains uncertain whether DSFGs and MS galaxies at early cosmic epochs have (1) similar SFEs or gas depletion times, and/or (2) comparable dust heating mechanisms.

\subsection{Redshift Evolution of Gas Depletion Timescales}
\label{subsec:tdepl}

The inverse of the star formation efficiency (SFE) is defined as the molecular gas depletion time, $\tau_{\rm dep}=(M_{\rm gas}/\rm SFR)$, which is a measure of the duration it would take for the gas reservoir to be depleted by the ongoing star formation rate (SFR). This definition assumes an absence of external processes (e.g., gas accretion, outflows or AGN feedback) affecting the present gas reservoir and a constant SFR throughout the star-forming period. This timescale ($\tau_{\rm dep}$) is useful to investigate galaxy evolution in the early Universe. In Table~\ref{tab:co10}, we list the depletion times of our targeted DSFGs.

Figure~\ref{fig:z_evol_sfe_tdepl} (lower panel) shows the redshift evolution of the gas depletion time ($\tau_{\rm dep}$). The observed $\tau_{\rm dep}$ values ($\sim 50-600$~Myr) align with those reported for other DSFGs in the literature for the adopted ${\alpha}_{\rm CO}$ \citep[e.g.,][]{carilli10,sharon+2016,berta+2023,stanley+2023}, and are consistent with typical values found in local ULIRGs undergoing intense starbursts, as also reflected in the scaling relations in Figure~\ref{fig:KSrelation}. The median $\tau_{\rm dep}$ for high-\textit{z} DSFGs is comparable to that of local ULIRGs, suggesting that these high-\textit{z} galaxies are equally efficient at star formation. In other words, the $\tau_{\rm dep}$ in high-\textit{z} dusty systems are not disproportionately higher than local ULIRGs, although they seem to have more molecular gas available. As discussed in Section~\ref{subsec:correlLirLco}, this infers a plausible balance of internal processes within these galaxy populations that regulates the SFR per available unit molecular gas mass. Notably, at higher redshifts ($z>3.5$), the \vzgal DSFGs exhibit somewhat shorter depletion timescales --- indicative of elevated SFEs --- compared to both local MS galaxies and ULIRGs. However, this apparent trend remains tentative due to limited sample size at these redshifts. 

About 42\% of our sample lies within $0.3$~dex of the empirical main sequence (MS) defined by \citet{tacconi2020} at corresponding redshifts. The majority fall below the MS, consistent with their observed higher SFRs (upper panel of Figure~\ref{fig:KSrelation}) and enhanced SFEs (upper panel of Figure~\ref{fig:z_evol_sfe_tdepl}) compared to local MS galaxies. On the other hand, high-\textit{z} MS galaxies from the literature follow the empirical trend from \citet{tacconi2020} when compared with their local counterparts. We note that many of the local MS galaxies have known observational ${\alpha}_{\rm CO}$ values that are higher \citep[e.g.,][]{Saintonge+2017} than the assumed conversion factor of $\mathrm{4 ~{M}_{\odot}~{\left(K~km~{s}^{-1}~{pc}^{2} \right)}^{-1}}$ in Figure~\ref{fig:z_evol_sfe_tdepl} (lower panel), making them evenly spread about the MS if corrected. 

Overall, Figure~\ref{fig:z_evol_sfe_tdepl} confirms that high-\textit{z} DSFGs plausibly represent a heterogeneous population encompassing a broad range of dusty, star-forming systems. This population likely includes both galaxies experiencing extreme starburst activity, analogous to local ULIRGs, as well as systems that lie within the scatter of the empirical star-forming MS. It may also be possible that DSFGs at high redshifts are more massive {versions} of local, giant MS galaxies that are undergoing starburst episodes, making their integrated SFEs comparable to local ULIRGs. The high-\textit{z} MS galaxies are already expected to exhibit relatively shorter gas depletion times compared to local MS galaxies \citep{tacconi2020}. However, measurements of stellar masses to derive the distance from the star-forming MS ($\Delta \mathrm{MS}$) and better estimates of the ${\alpha}_{\rm CO}$ conversion factor for our sample will be necessary to extract detailed inferences \citep[see also discussion in][]{berta+2023}. In addition, better statistics at $z>4$ will be beneficial to confirm the consistency plausible decreasing trend in $\tau_{\rm dep}$ with redshift, which would indicate that infrared-luminous, dusty galaxies at earlier epochs had increasingly higher SFEs than seen in local massive MS galaxies and ULIRGs.

\subsection{CO Line Ratios} \label{subsec:colineratio}

The relative line luminosities of various rotational transitions of the CO molecule, tracing the molecular gas ($\rm H_2$), depend on the gas excitation conditions. As seen in Equation~\ref{eqn:gasmass} and existing literature, the \coonezero transition is robustly calibrated as an anchor point to derive the total molecular gas masses, and is also important to measure precise CO line ratios. Also known as CO brightness temperature ratios, these are defined as follows:
\begin{equation}
 {{r}}_{{J1}} = \frac{{{{L}}^{\prime}}_{{\rm CO}({J} \rightarrow {J-1})}}{{{{L}}^{\prime}}_{{\rm CO(1-0)}}}
 \label{eqn:lineratio}
\end{equation}
\noindent
where, \textit{J} is an index of the upper level in CO($J \rightarrow J-1$) transition, whose line luminosity is ${{{{L}}^{\prime}}_{{\rm CO}({J} \rightarrow {J-1})}}$. 

For optically thick and thermalized gas, all the CO rotational levels are equally excited and the brightness temperature ($T_{\rm b}$) of all these levels is the same. This leads to ${{r}}_{{J1}}=1$ for all the values of \textit{J} (hence, ${I}_{\rm CO} \propto {J}^{2}$; see Figure~\ref{fig:co_sled_avg}). In practice, these ratios exhibit significant variation --- particularly as a function of the gas excitation temperature ($T_{\rm ex}$) and the molecular hydrogen number density \citep[$n_{\rm H_2}$; e.g.,][]{weiss2005m82,riechers+2011sled2,riechers+2011sled1,harrington2021}. For example, the brightness temperature ratios of high-\textit{J} CO levels (say, $J\gtrsim{6}$) are typically elevated when the system has intense radiation field and/or shock-excited hot/dense gas due to stellar feedback processes, such as supernovae or outflows. In contrast, the population/occupancy of CO molecules in lower-\textit{J} levels dominate in Milky Way like normal star-forming disks \citep{fixsen1999,niki2021,harrington2025}. In this way, the observed CO line ratios inform us about the physical properties, namely temperature and density, of the star-forming gas.

\begin{figure*}
\centering
\includegraphics[width=0.6\textwidth]
{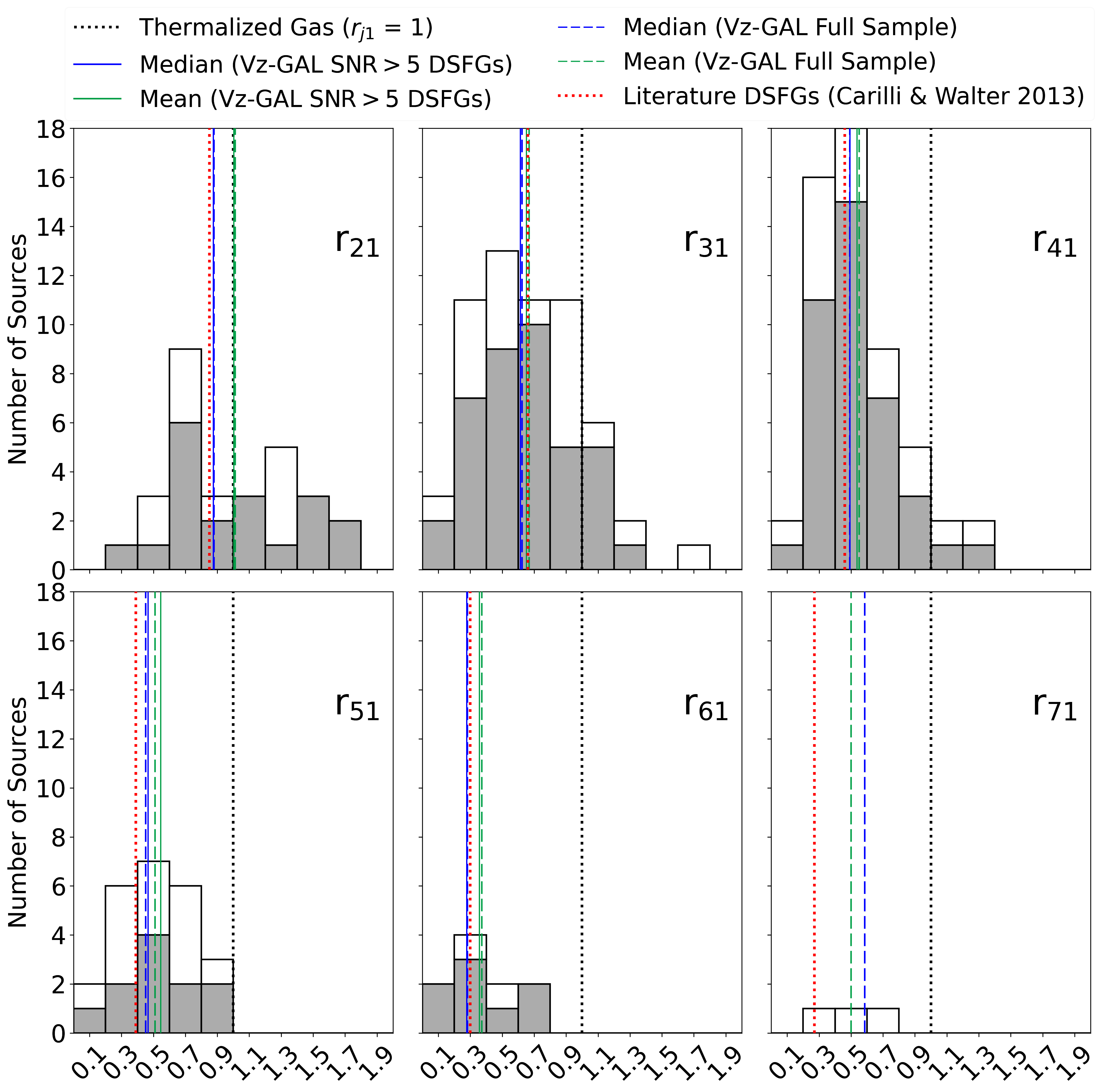}
\caption{CO-line ratio histograms for the \vzgal sample. The shaded histograms in gray are with the sources detected above 5$\sigma$ significance and the white histograms show the rest of our targets that have SNR$>$2. The ratios show mostly sub-thermal excitation of the mid/high-$J$ CO lines, indicating the importance of \coonezero as an anchor point and as a robustly calibrated molecular gas mass tracer. The values compiled by \citet{carilliwalter2013} are plotted for comparison (red dotted lines). Optically thick and thermalized gas with ${r}_{J1}=1$ is shown as black dashed lines in each subplots.}
\label{fig:lineratio_stat}
\end{figure*}

\renewcommand{\arraystretch}{1.2}

\begin{deluxetable}{ccccccc}
\tabletypesize{\scriptsize} 
\tablecaption{CO line brightness temperature ratios
\label{tab:lineratios}}
\tablehead{
    \colhead{${{J}_{\rm up}}$} & \multicolumn{2}{c}{\vzgal (SNR$>$5)} &
    \multicolumn{2}{c}{\vzgal (full)} &
     \colhead{CW13 } &
    \colhead{H21 LPs} \\
    & Median & Mean & Median & Mean &  DSFGs & Mean 
}
\startdata
1  &  1&1&1&1&1&1    \\
2  &  0.88$\pm$0.25 & 1.00$\pm$0.25  &0.88&0.99& 0.85  &  0.88$\pm$0.07   \\
3  &  0.61$\pm$0.22 & 0.65$\pm$0.22  &0.61&0.66& 0.66  &  0.69$\pm$0.12   \\
4  &  0.49$\pm$0.15 & 0.54$\pm$0.15  &0.49&0.55& 0.46  &  0.52$\pm$0.14   \\
5  &  0.47$\pm$0.13 & 0.54$\pm$0.13  &0.45&0.51& 0.39  &  0.37$\pm$0.15  \\
6  &  0.28$\pm$0.13 & 0.36$\pm$0.13  &0.28&0.37& 0.30  &  0.25$\pm$0.14   \\
7  &  -- & --  &0.59$^{\ast}$&0.50$^{\ast}$&  0.27 &  0.17$\pm$0.12  \\
\enddata
\tablecomments{Ratios for the \vzgal sample in the first two columns are only with the DSFGs detected above 5$\sigma$ significance in both CO lines. Their statistical error-bars are the (unscaled) median absolute deviation (MAD) of the respective distributions. Subsequent two columns with line ratios of the full \vzgal sample are also consistent. CW13 DSFGs are from \citet{carilliwalter2013}. H21 LPs are the Planck-identified lensed galaxies from \citet{harrington2021}. \\ $^{\ast}$ $r_{71}$ is based on only three source statistics and all these have \coonezero detection with $2<\mathrm{SNR}<5$ (see the text for details).}
\end{deluxetable}

In Figure~\ref{fig:lineratio_stat}, we show the statistical distribution of CO line ratios of the \vzgal sample up to \textit{J} = 7. We observe mostly subthermal excitation (${{r}}_{{J1}}<1$) for our targets. The median line ratios for the targets detected above $5\sigma$ confidence in both the CO rotational levels involved with the respective line ratio are presented in Table~\ref{tab:lineratios}. Our values with improved statistics and high SNR measurements confirm those derived by \citet{carilliwalter2013} and \citet{harrington2021}. The derived $r_{31}$ here is also in agreement with the values derived by \citet{sharon+2016} and \citet{yang2017}. Our median $r_{71}$ is $\sim2.5\times$ higher than the value presented by \citet{carilliwalter2013}, albeit this is using only three \vzgal DSFGs (HeLMS-19, HerS-11, and HerBS-185) with available CO(7$-$6) spectra. These galaxies, with \coonezero detection below $5\sigma$ significance, have $r_{71}$ of $0.59\pm0.17$, $0.65\pm0.36$, and $0.26\pm0.11$, respectively. Therefore, the derived median will need better statistics and deeper observations to be confirmed (see the bottom-right panel of Figure~\ref{fig:lineratio_stat}). Further, we find no apparent trend between any of the CO line ratios and redshift within the \vzgal sample ($1<z<6$) as also observed by \citet{sharon+2016} for $r_{31}$ in a sample of 14 galaxies at $z=2-3$.

These brightness temperature ratios reiterate the importance of \coonezero observations, given that the subthermal line ratios can lead to additional uncertainties in the total molecular gas masses derived based on higher-\textit{J} CO levels. For example, gas masses estimated from the CO(3$-$2) observations may lead to an underestimation by $(40\pm22)\%$, under unknown excitation, compared to the values derived based on the \coonezero line. CO(2$-$1), seems to serve as a better ${{{{L}}^{\prime}}_{{\rm CO}({1} - {0})}}$ tracer than CO(3$-$2) as it shows line ratios ($r_{21}$) close to unity, which has also been noticed in integrated measurements of massive local galaxies \citep[e.g., xCOLD-GASS sources from][]{Saintonge+2017}. We discuss this further in Section~\ref{subsubsec:co21co10}.

In Appendix~\ref{app:b}, we list the sources with unusually low ratios (${{r}}_{{J1}}<0.2$). Further, we observed superthermal line ratios (${{r}}_{{J1}}>1$ for $2<J<4$) for a few \vzgal DSFGs {(see statistics in Figure~\ref{fig:lineratio_stat})}; however, these ratios are consistent with being ${{r}}_{{J1}} \leq 1$ (subthermal) when considering the associated uncertainties in the flux calibration of line measurements (see Appendix~\ref{app:b}). Superthermal ratios are expected to arise if the gas has a highly excited partition function. This implies that, in such environments, the relative columns in the $J=0$ and $J=1$ levels are lower than those under Milky Way-like conditions because a significant fraction {of the gas} is in higher excited states. Under high excitation conditions, since the opacity in the \coonezero line is expected to be lower than that of the higher-\textit{J} transitions (at least securely up to $J\sim4$), the \coonezero line is apparently the first to get optically thin making the line ratios superthermal. The confirmation of such ratios will need detailed investigation using semi-empirical models and/or better observations of low SNR targets.\footnote{One of the plausible reasons that could contribute to the scatter observed in histograms presented in Figure~\ref{fig:lineratio_stat} might be gravitational lensing. Although differential magnification may not have a significant impact, the placement of caustics relative to the gas distribution might affect what we see while observing the regions of highest surface infrared brightness, given that these DSFGs are amongst the brightest high-\textit{z} sources in the \textit{Herschel} catalogs.}

\begin{figure*}
\begin{center}
\includegraphics[width=0.58\textwidth]{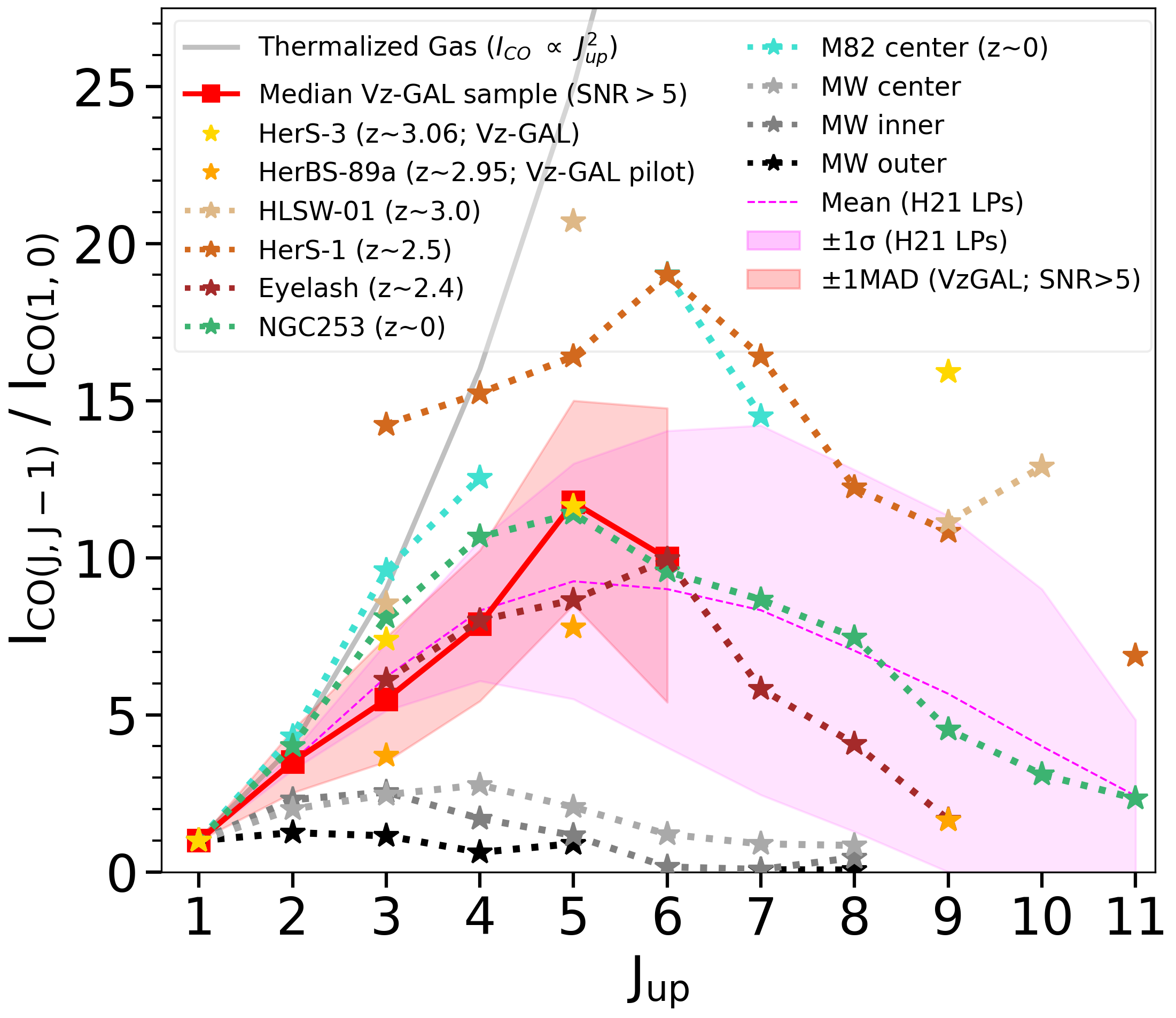}
\includegraphics[width=0.58\textwidth]{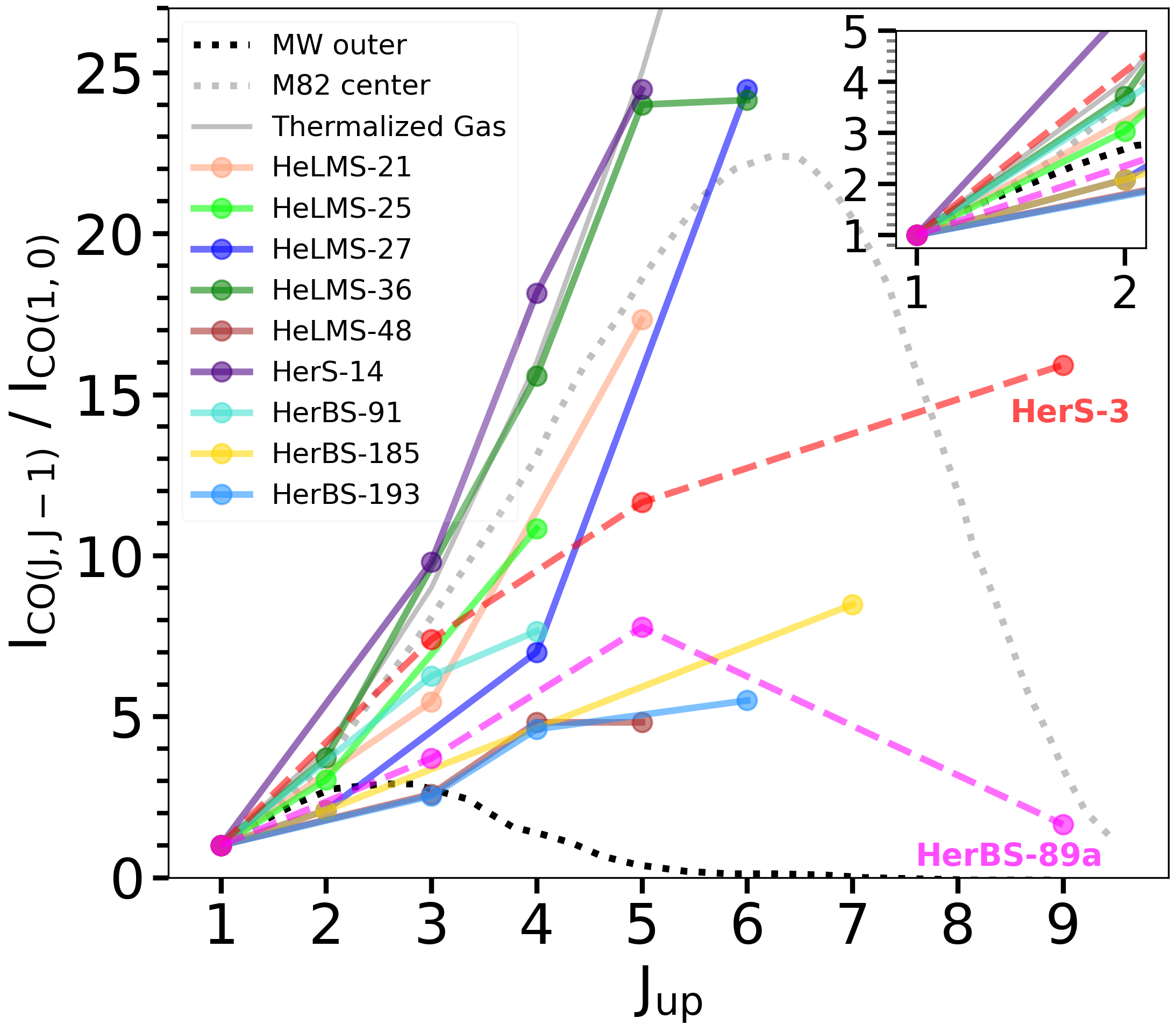} 
\end{center}
\caption{\textit{(Upper panel)} Median CO SLED of the \vzgal sample (solid red curve) compared to sources from the literature including: five high-$z$ DSFGs, HerS-3 at $z=3.061$ \citep[this work and][]{cox2025hers3}, HerBS-89a at $z=2.95$ \citep{berta2021}, HLSW-01 at $z=2.96$ \citep{riechers+2011lensed,scott2011}, HerS-1 at $z=2.55$ \citep{harrington19,lui2022}, and the Cosmic Eyelash at $z=2.32$ \citep{danielson2011,swinbank11}; local starbursts NGC~253 \citep{israel1995,rosenberg2014,perezbeaupuits2018} and M82 center \citep{weiss2005m82}; and the Milky Way, where we adopted the line fluxes of three different regions from \citet{fixsen1999}, namely the central star-forming zone (MW center), the inner disk (MW inner), and the outer disk (MW outer). Magenta-shaded region is based on the median line ratios of Planck-identified lensed galaxies \citep[H21 LPs;][]{harrington2021}, which is in good agreement with our \vzgal median values (region shaded in red). \textit{(Lower panel)} CO SLEDs of 11 individual high-\textit{z} DSFGs from our \vzgal sample in comparison with the Milky Way and local starburst M82. This plot highlights the variety of excitation conditions across the \vzgal large sample. SLED modeling for two of these galaxies, namely HerS-3 and HerBS-89a, has already been presented by \citet{cox2025hers3} and \citet{berta2021}, respectively. The gray solid curve in both panels shows optically thick and thermalized gas condition of the CO SLED. All the line fluxes ($I_{\rm CO}$) used here are in the units of Jy km ${\rm s^{-1}}$.}
  \label{fig:co_sled_avg}
\end{figure*}

In the upper panel of Figure~\ref{fig:co_sled_avg}, we compare the median CO line ratios of the \vzgal sample to SLEDs of five high-\textit{z} DSFGs (HerS-3, HerBS-89a, HLSW-01, HerS-1, and the Cosmic Eyelash), two local starburst galaxies (NGC~253 and M82), and the Milky Way --- all with available \coonezero observations. These results summarize that the median CO SLED of the \vzgal sample ($1<z<6$), peaking at $J\sim5$, follows a similar behavior as depicted by the local starburst NGC~253 and the Cosmic Eyelash, one of the extreme DSFGs. \citet{harrington2021} also {observe SLEDs peaking between $J=4$ and $J=6$ for most of their {\it Planck}-identified high-\textit{z} lensed DSFGs}. Such a peak at $J>4$ confirms a presence of warm/dense gas component in our targeted DSFGs along with the cold gas traced by \coonezero and other low-\textit{J} levels. This warm/dense gas seem to dominate the CO SLEDs of these DSFGs. Such dense star-forming environments are also found close to the Galactic central molecular zone \citep[see e.g.,][]{kohno2024}. Note that each median CO line ratio contributing to the \vzgal CO SLED in the upper panel of Figure~\ref{fig:co_sled_avg} is based on different sample sizes, depending on the availability of higher-\textit{J} CO observations across the NOEMA bands. As a result, the median CO SLED does not necessarily reflect the excitation properties of any individual \vzgal\ source and might be affected by redshift-dependent selection biases.

Overall, we notice a wide range of excitation conditions within high-\textit{z} DSFGs, some also resembling the dense cores/center of local starbursts like M82. To better represent this variety, we depict the observed, well-sampled CO SLEDs of 11 \vzgal DSFGs in the lower panel of Figure~\ref{fig:co_sled_avg}. It is clear from this presentation that the \vzgal sample includes high-\textit{z} dusty galaxies with a noticeable spread in excitation conditions.

\subsubsection{A Comparison to \zgal Results: Empirical Relations Between CO(1$-$0), CO(2$-$1), and CO(3$-$2) Line Luminosity} \label{subsubsec:co21co10}

\vzgal is a systematic \coonezero large survey of dusty galaxies in the early Universe, spanning {the redshift range} $1<z<6$. For this sample, \citet{berta+2023} investigated trends in the scaling relations based on the detected lowest-\textit{J} level of the higher-\textit{J} CO lines using NOEMA. In the absence of \coonezero observations, they consistently used either \cotwoone or CO(3--2) to estimate gas masses and depletion timescales for most of these high-\textit{z} DSFGs. The line ratios they adopted from \citet{carilliwalter2013} are consistent with our findings (see Table~\ref{tab:lineratios}). Similarly, \citet{hagimoto2023} examined {a sample} of 71 \textit{Herschel}-selected DSFGs (\textit{Herschel}-BEARS) at $1.5<z<4.5$ in the southern sky, utilizing mid/high-\textit{J} CO lines and \citet{harrington2021} line ratios. The statistical trends in the scaling relation shown in the upper panel of our Figure~\ref{fig:KSrelation} confirm the results from these studies, given that the ratios they assumed are in agreement {with} our robustly measured CO line ratios. To further analyze these low/mid-\textit{J} CO levels, here we derive correlations with ${L}^{\prime}_{\rm CO(1-0)}$.

\begin{figure*}
  \gridline{
    \fig{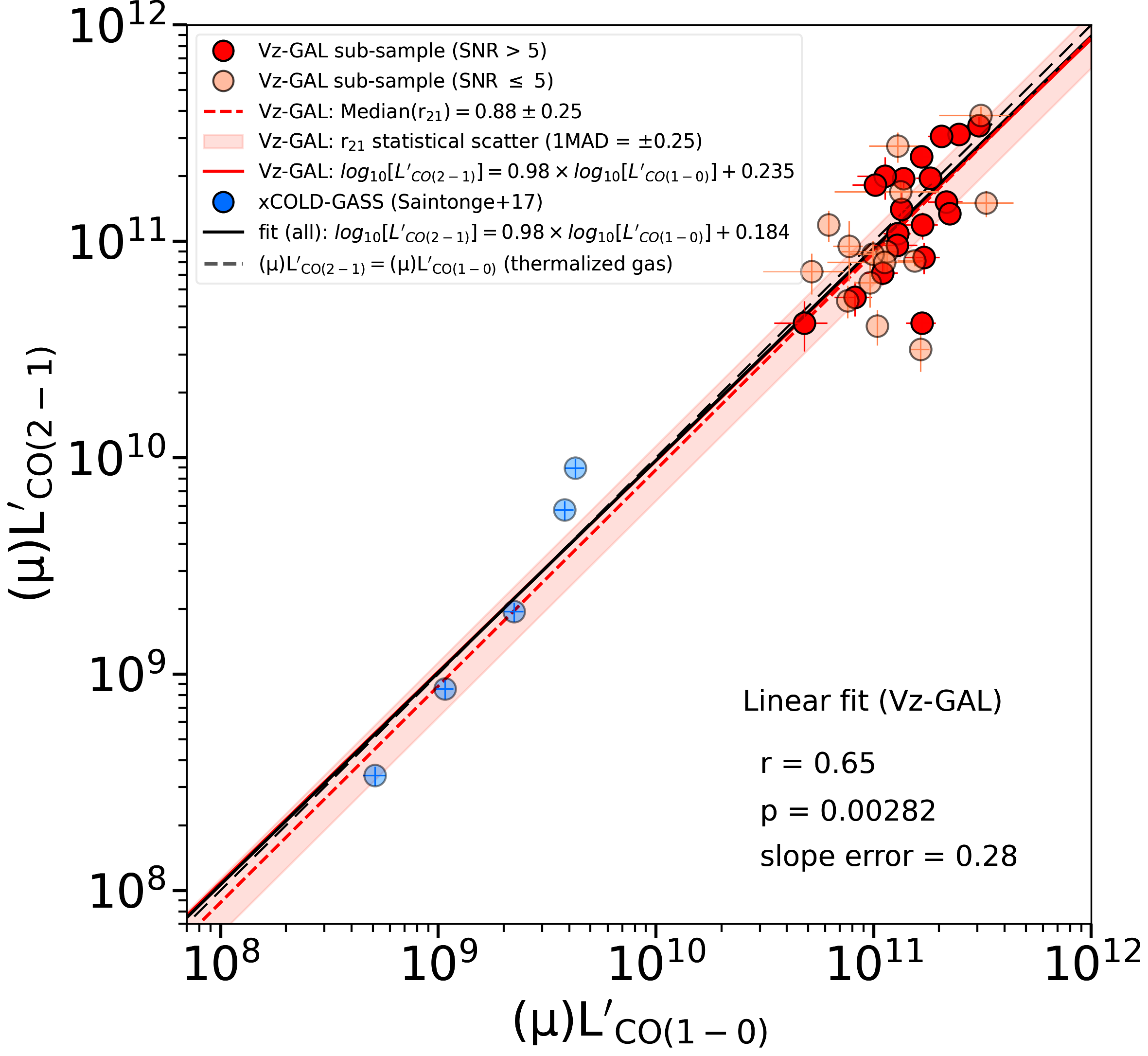}{0.50\textwidth}{}
    \fig{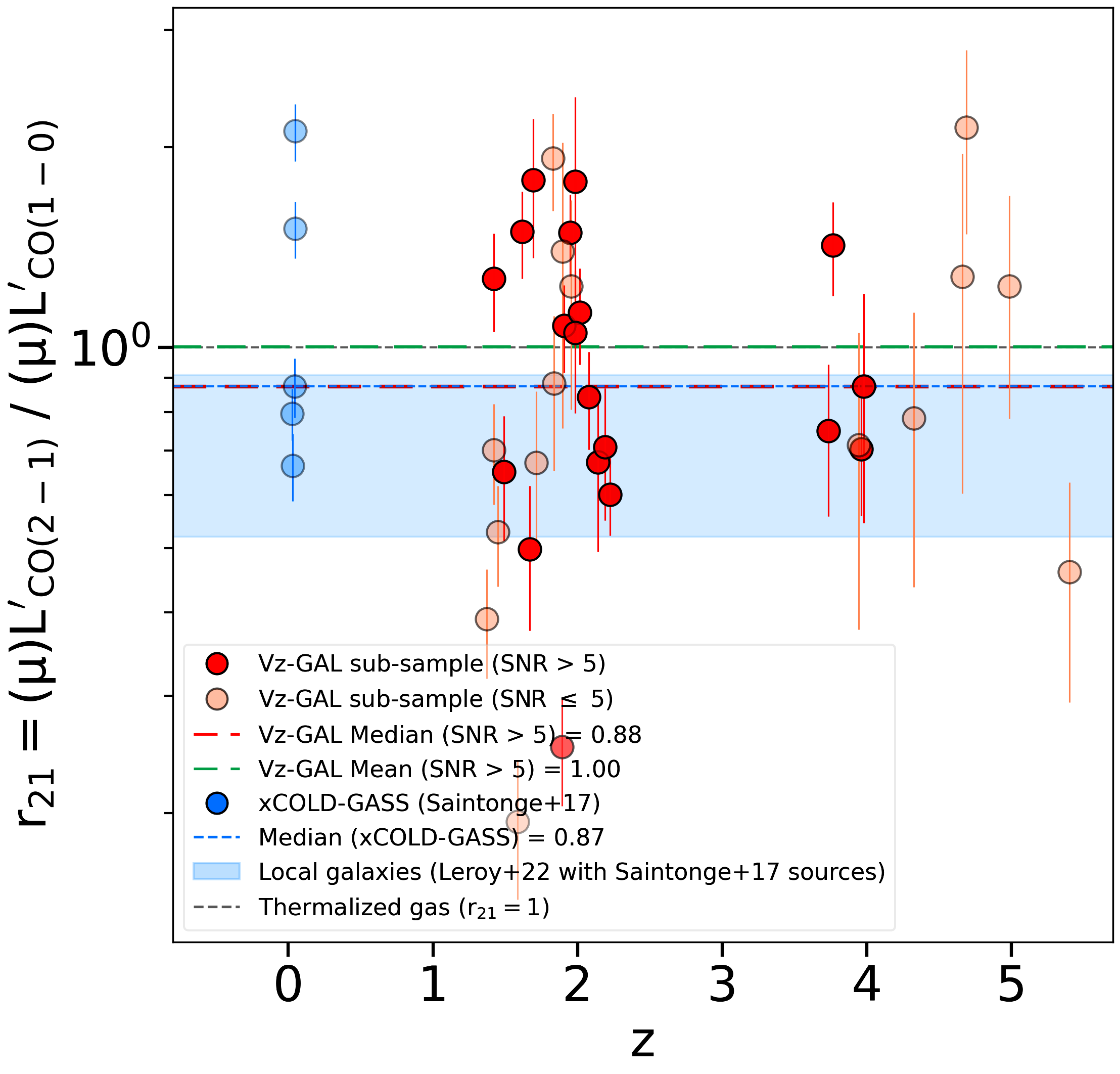}{0.485\textwidth}{}
  }
    \gridline{
    \fig{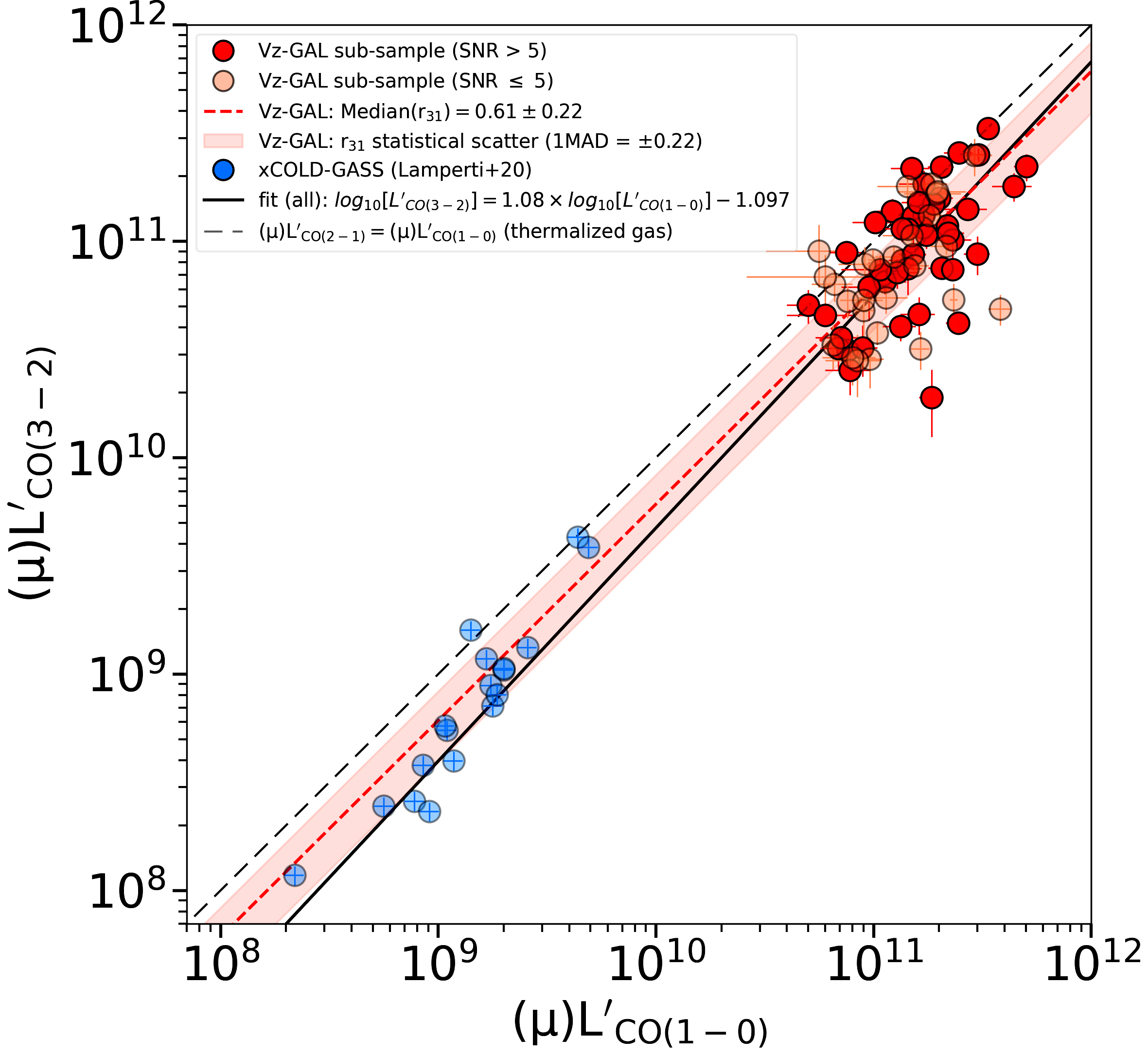}{0.50\textwidth}{}
    \fig{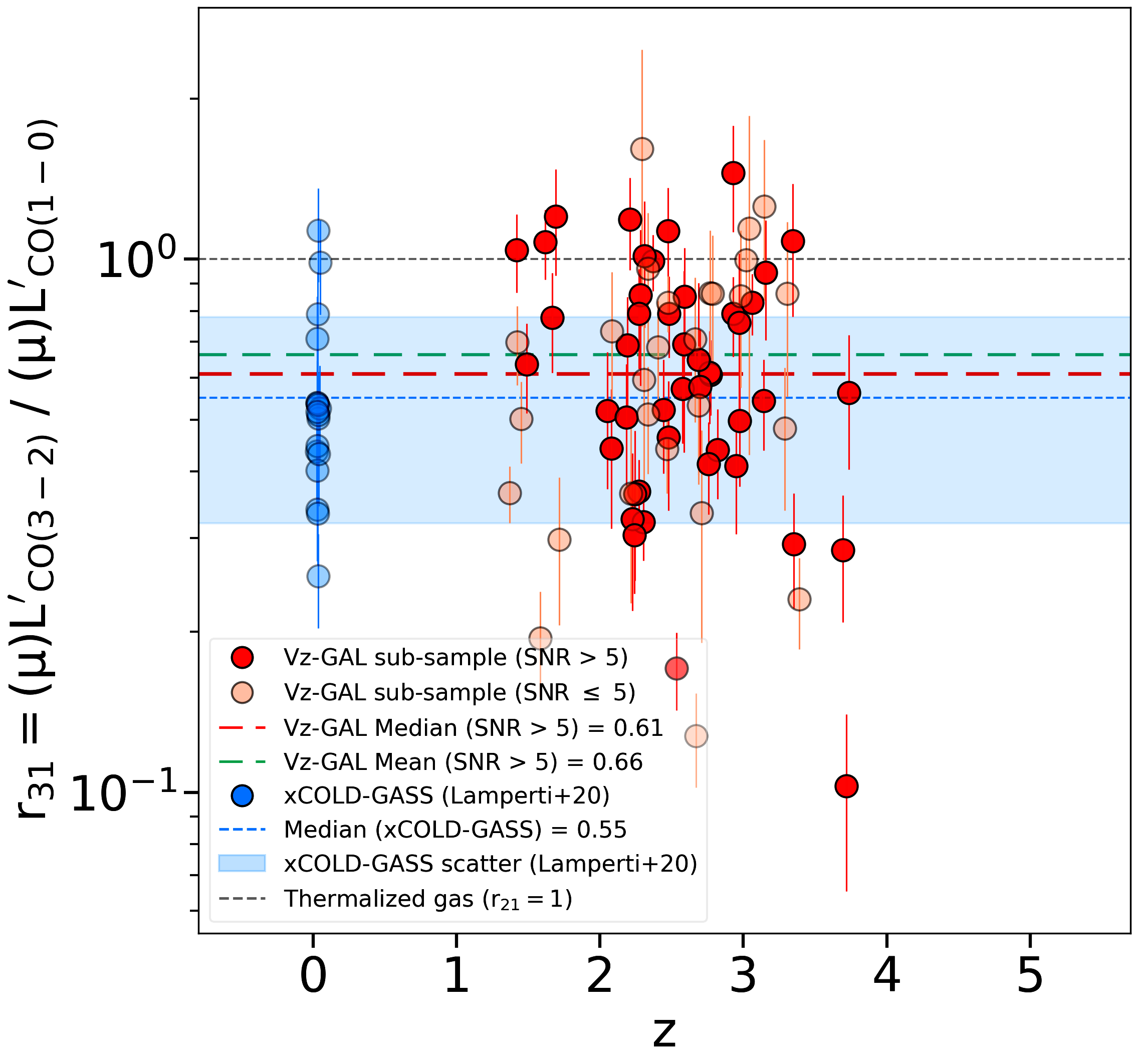}{0.485\textwidth}{}
  }
  \caption{\textit{({Upper panel})} Comparison of the \coonezero and \cotwoone line fluxes for the \vzgal DSFGs where both lines have been measured --- with \cotwoone from this work for the $z>3$ sources and the rest from \citet{cox+2023}. The data points with SNR $> 5$ are highlighted in dark red color, for which the linear fit and median/mean values are presented. The results for the high-$z$ DSFGs are compared to massive ($M_{\ast}>{10}^{10} \mathrm{M_{\odot}}$) local MS galaxies selected from the xCOLD-GASS survey \citep[][see text for details]{Saintonge+2017}. \textit{({Lower} panel)} Same as the {upper} panel, but with a comparison of CO(3$-$2) with CO(1$-$0). Here CO(3$-$2) data of the xCOLD-GASS sample is from \citet{lamperti2020}. In both upper and lower panels, the \textit{left panel} shows an empirical correlation between the higher-\textit{J} CO line and \coonezero luminosities. The slope of the best linear fits agrees with the sample median within the error-bars. Local galaxies are overplotted as blue data points. The \textit{right panel} shows the redshift distribution of the higher-\textit{J} CO line to \coonezero ratio (i.e., $r_{21}$). Median and mean values are shown with red and green dashed lines, respectively. xCOLD-GASS median is shown using a blue dashed line, in agreement with the \vzgal median within error-bars. The shaded region in blue represents the spread in $r_{21}$ or $r_{31}$ for local galaxies \citep{Saintonge+2017,lamperti2020,leroy2022}. In all the plots, black dashed line shows the optically thick and thermalized gas condition ($r_{J1}=1$).}
  \label{fig:co21co10_co31co10}
\end{figure*}

As shown in Figure~\ref{fig:lineratio_stat}, the median ratio of \cotwoone to \coonezero line luminosity is observed to be close to unity ($r_{21}=0.88$$\pm$$0.25$). This indicates that the molecular gas masses, that are derived based on \cotwoone observations, will be closer to the ones derived using Equation~\ref{eqn:gasmass} and are less uncertain than those derived using higher-\textit{J} CO lines that are mostly observed to have subthermal line ratios. To investigate a plausible relation between the line intensities of these two low-\textit{J} CO levels, we plot ${L}^{\prime}_{\rm CO(2-1)}$ versus ${L}^{\prime}_{\rm CO(1-0)}$ for 34 \vzgal sources for which both \coonezero and \cotwoone observations are available {from this study (V\textit{z}--GAL) and the \zgal survey} (upper-left panel of Figure~\ref{fig:co21co10_co31co10}). We derive a strong linear correlation in log-space with a slope of 0.98$\pm$0.28 and a Y-intercept of $0.24$. The statistical median- or mean-$r_{21}$ of our DSFGs (within the error-bars) agrees with the criteria posed by optically thick and thermalized gas conditions, suggesting $r_{21}\sim1$. 

{Similar comparisons for large galaxy-samples with available observations of both \coonezero and \cotwoone exist for local (U)LIRGs \citep{montoya2023} and nearby star-forming galaxies \citep{denbrok2021,leroy2022,keenan2025}, but are scarce at higher redshifts. However, \cotwoone has been used to derive gas mass properties of high-\textit{z} galaxies in the absence of \coonezero data \citep[see e.g.,][and references therein]{genzel2010,bothwell2013,silverman2018,valentino2018,tacconi2020}. Our homogeneous target selection now confirms the usability of \cotwoone as an alternative calibrator for gas mass derivation (Equation~\ref{eqn:gasmass}) for the to-date largest DSFG sample.} 

In the upper-right panel of Figure~\ref{fig:co21co10_co31co10}, we explore the redshift evolution of $r_{21}$, which shows no trend.  Our derived median for the \vzgal DSFGs is consistent with the value found for the massive, local MS galaxies with $M_{\ast}>{10}^{10} \mathrm{M_{\odot}}$ \citep[xCOLD-GASS;][]{Saintonge+2017}. Similar $r_{21}$ values have also been observed in the local Universe, particularly around galactic nuclei \citep[i.e., inner kiloparsec region][]{braine1992,weiss2005m82}. \citet{leroy2022} compiled the low-\textit{J} CO observations of local galaxies with different stellar masses to find a median $r_{21}$ of 0.64 with the inclusion of the xCOLD-GASS sample from \citet{Saintonge+2017}. They derive the 16th and 84th percentile values of this $r_{21}$ to be 0.52 and 0.91, respectively. \citet{leroy2022} attribute these results to biases due to sample selection. In particular, they expect higher $r_{21}$ values in the high-SFR galaxies as could be the case for massive galaxies from \citet{Saintonge+2017}. {This has been recently confirmed by \citet{keenan2025} using detailed \cotwoone measurements of the xCOLD-GASS galaxies. It is also consistent with the results for local (U)LIRGs \citep[see e.g.,][]{montoya2023} and our intensely star-forming high-\textit{z} DSFGs, having $r_{21}$ close to unity.}

In Figure~\ref{fig:co21co10_co31co10} (lower panel), we present the ${L}^{\prime}_{\rm CO(3-2)}-{L}^{\prime}_{\rm CO(1-0)}$ correlation for 73 \vzgal DSFGs with the CO(3$-$2) luminosities from \zgal. This plot primarily shows subthermal $r_{31}$ values for our targets, which aligns with the trends observed in massive, {high-SFR}, local main-sequence galaxies from the xCOLD-GASS sample \citep{lamperti2020}. {Our results for $r_{31}$ are also consistent with those explored for a sample of about 40 ULIRGs in the local Universe \citep{montoya2023}.} Further, we find no apparent redshift evolution here.

\subsection{Gas-to-dust Mass Ratios}
\label{subsec:gasdustratio}

\begin{figure}[h]
\centering
\includegraphics[width=0.45\textwidth]{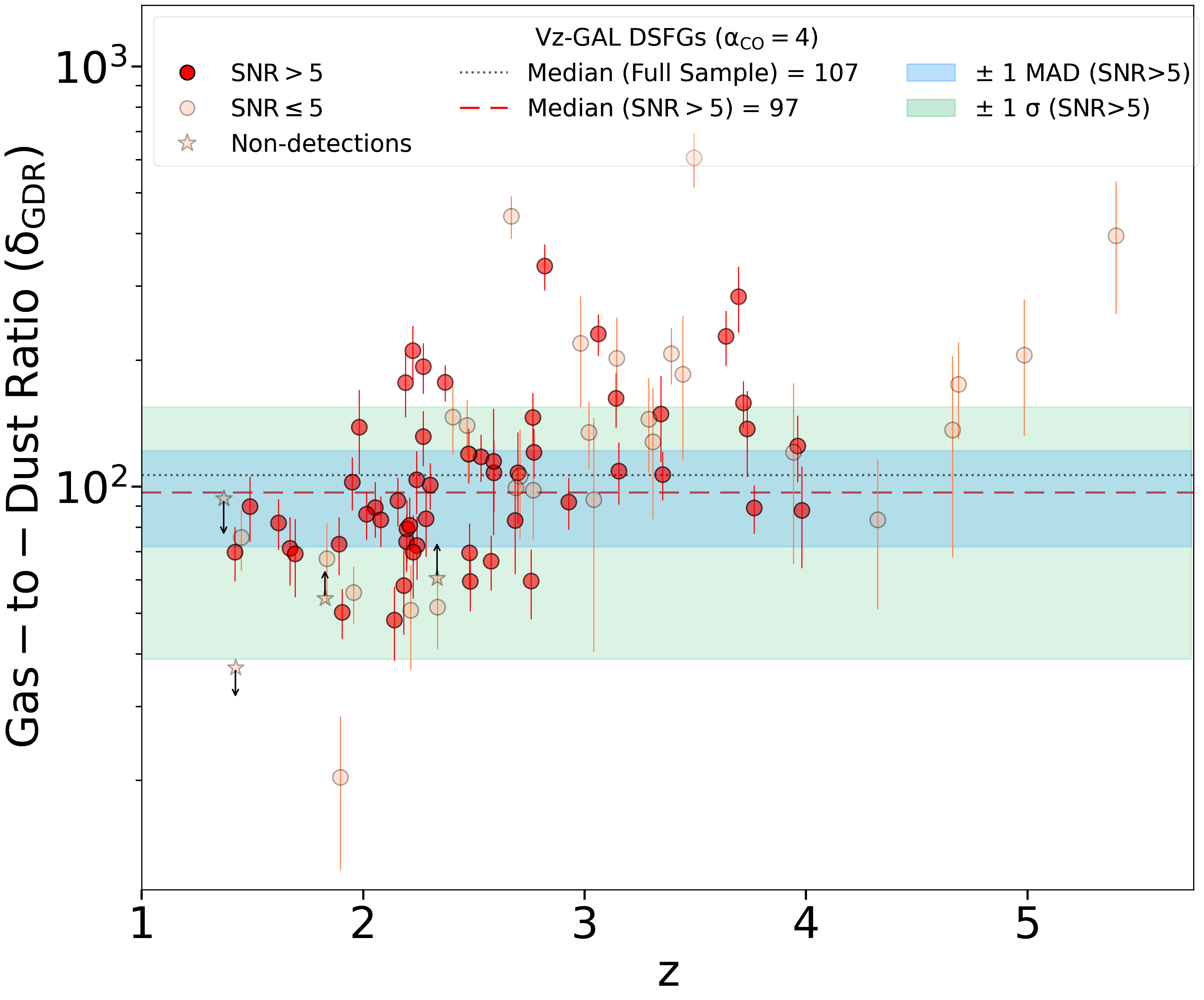}
\caption{Gas-to-Dust mass ratio (${\delta}_{\rm GDR}$) of the \vzgal sample using ${\alpha}_{\rm CO}=4$ $\mathrm{{M}_{\odot}~{\left(K~km~{s}^{-1}~{pc}^{2} \right)}^{-1}}$ and Milky Way-like $\kappa_{\nu}$. The median ${\delta}_{\rm GDR}$ of the full sample is shown with a dotted black line and the median for the targets with a signal-to-noise ratios (SNR) $> 5$ with a dashed red line. The shaded green and blue regions highlight the $1\sigma$ and (unscaled) 1 MAD (median absolute deviation) ranges for the sources with $\rm SNR>5$, respectively.}
\label{fig:gdr}
\end{figure}

Gas-to-dust mass ratios (${\delta}_{\rm GDR}={M}_{\rm H_2}/{M}_{\rm dust}$) provide crucial insights on the evolutionary state and star formation activity of galaxies. Figure~\ref{fig:gdr} presents the ${\delta}_{\rm GDR}$ of our high-\textit{z} dusty galaxies. Dust masses (${M}_{\rm dust}$) are from \citet{ismail+2023}, which were calculated using \citet{draine2014} templates with Milky Way-like absorption cross section per unit dust mass ($\kappa_{\nu}$). Assuming solar metallicity environments in our high-\textit{z} dusty galaxies, we adopt these dust masses, and also use Milky Way-like CO-$\mathrm{H_2}$ conversion factor to derive gas masses ($M_{\rm H_2}$) and subsequently ${\delta}_{\rm GDR}$. For the \vzgal DSFGs detected above $5\sigma$ significance, we find a median ${\delta}_{\rm GDR} \sim 100$ using ${\alpha}_{\rm CO}$=4 and Milky Way-like $\kappa_{\nu}$. Within the $\pm$1~MAD around the median ${\delta}_{\rm GDR}$, our ratios agree with the values explored in the literature for local/high-\textit{z} dusty starbursts \citep[e.g.,][]{herrero2019,harrington19,popping2023} and the values derived in the \zgal study \citep{berta+2023}, after adapting to their choice of ${\alpha}_{\rm CO}$. 

More importantly, our median gas-to-dust mass ratio of about 100 {for the chosen values of} $\alpha_{\rm CO}$=4 and $\kappa_{850}$=$0.047~\mathrm{m^2/kg}$ \citep{draine2014,ismail+2023}, leads to {a ratio} $\kappa_{\rm H}$=${\delta}_{\rm GDR}/\kappa_{850}$ of 2130~$\mathrm{kg/m^2}$, which is consistent with the mid-range of extragalactic determinations, $\kappa_{\rm H}$=1500$-$2200~$\mathrm{kg/m^2}$, used by \citet{dunne+2022} for metal-rich, dusty galaxies across redshifts. This is, in turn, also in agreement with measurements of the diffuse ISM in the Milky Way.

A broad distribution (${\delta}_{\rm GDR} \sim 50-300$) within this sample highlights inherent differences in the metallicities and star-forming environment of individual DSFGs as also observed in their gas depletion times (see Figure~\ref{fig:z_evol_sfe_tdepl}). Local LIRGs and non-LIRG spiral galaxies also show a similarly broad distribution in gas-to-dust mass ratios in the literature \citep[e.g.,][]{leroy2011,herrero2019}.

While we see a range in ${\delta}_{\rm GDR}$ \citep[hence, also in metallicities but not exclusively; see ][]{popping2023} for the \vzgal DSFGs, many of them are expected to have super-solar metallicity given their dusty nature \citep{dunne+2022,berta+2023}. We note that the ratios for the \vzgal sample are mostly consistent with the Milky Way and other local disks with similar metallicity (and ${\alpha}_{\rm CO}\sim 4$), though they tend to lie at the lower end of the typical ${\delta}_{\rm GDR}$ range known for these MS galaxies \citep[e.g.,][]{andreani1995,zubko2004,remyruyer2014}. In other words, about 50\% of our sample (with SNR$>$5) showing ${\delta}_{\rm GDR}<100$ (Figure~\ref{fig:gdr}) indicates a possibility of them having super-solar metallicity plausibly due to larger dust masses. However, there are caveats in derivations of dust masses that need to be addressed.

Although \coonezero observations help reduce the uncertainty related to the excitation ladder of the CO molecule by providing direct measurements of ${L}^{\prime}_{\rm CO(1-0)}$, interpreting the gas-to-dust mass ratio remains challenging due to two more sources of uncertainty: (1) the gas mass dependence on the choice of ${\alpha}_{\rm CO}$ and (2) the dust mass estimation from the SED fitting that depends on the dust temperature ($T_{\rm d}$) and an assumption on $\kappa_{\nu}$. For our 500~$\mu$m-selected \vzgal sample, \citet{ismail+2023} derived the dust masses using the optically thin approximation of the modified blackbody (MBB) to fit the cold dust component. This may lead to lower dust temperatures and higher ${M}_{\rm dust}$ as both are degenerate. As explained by \citet{ismail+2023}, this could be improved with spatially resolved observations from which one can estimate the sizes of the DSFGs, which will be useful in combination with more precise GMBB (general MBB, optically thick) models. In general, we expect the GMBB models to provide higher dust temperatures and lower $M_{\rm dust}$. A reasonable ${\delta}_{\rm GDR}$ range could still be obtained with our chosen $\alpha_{\rm CO}$ and $\kappa_{\nu}$, if the dust masses were lower. However, as such, we do not regard this finding as a strong evidence for a Milky Way-like ${\alpha}_{\rm CO}$.

\subsection{Comparison to [CI](1$-$0)} \label{subsec:CIcomparison}

\begin{figure}
\centering
\includegraphics[width=0.45\textwidth]{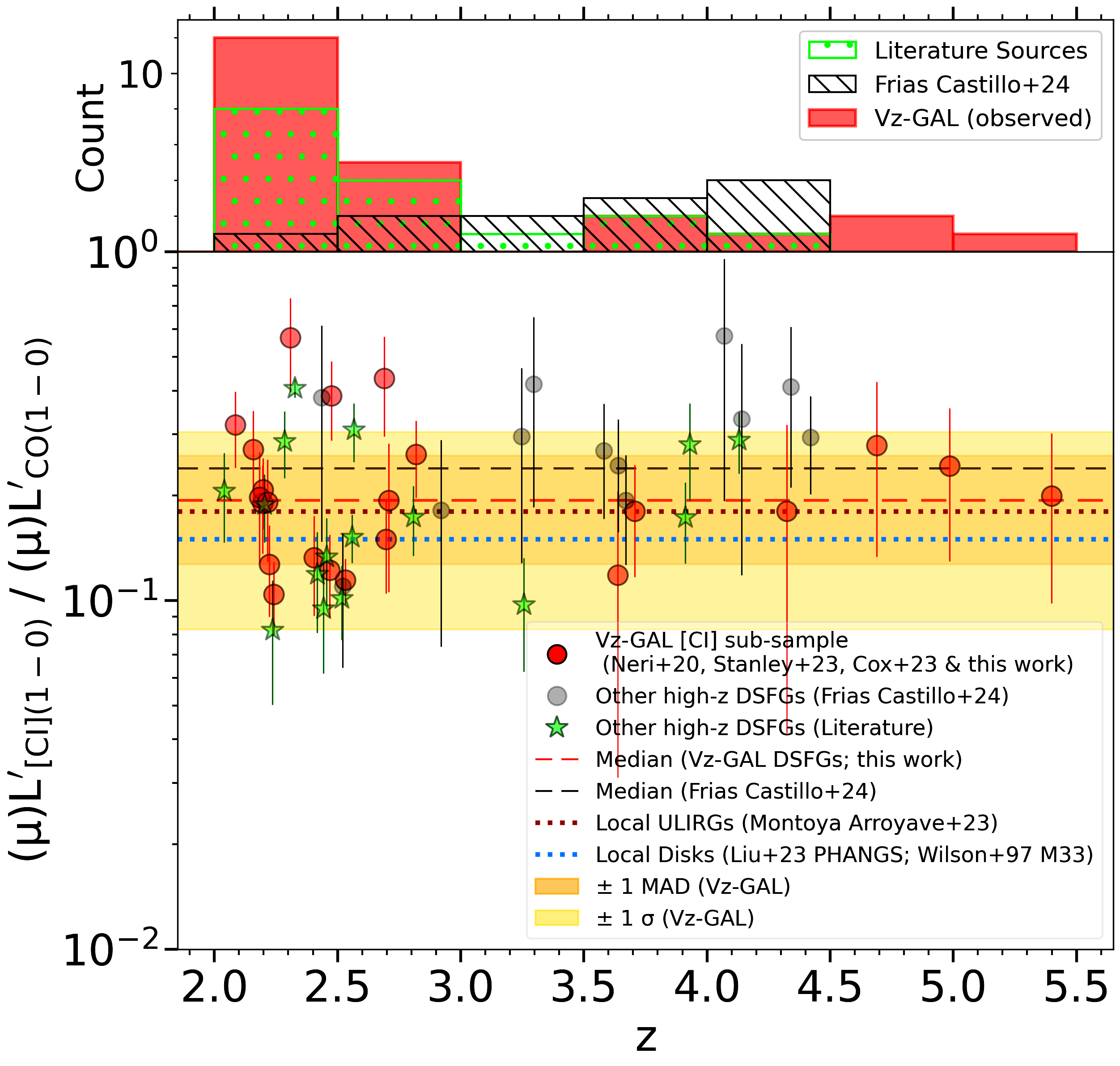}
\caption{Comparison of the \coonezero line luminosity with the luminosity of an alternative cold gas tracer, [CI](1$-$0), for 23 \vzgal targets. Other 12 high-$z$ DSFGs from \citet{friascastillo2024} are shown as gray points with detections of both the lines. More heterogeneous literature points for 16 dusty galaxies at $z>1$ are shown with green asterisk markers that have both \cionezero \citep{alaghbandzadeh2013,walter2011,dannerhr19,nesvedba2019,dunne+2022} and \coonezero \citep{riechers06,riechers09,lestrade11,riechers11,thomson12,alaghbandzadeh2013,dannerhr19,harrington19,dunne+2022} observations. The red and black dashed lines show the median ratios for the \vzgal and \citet{friascastillo2024} data points that are compared to the values derived for local ULIRGs \citep{montoya2023} and star-forming MS galaxies \citep{wilson1997,liu2023} in brown and blue dotted lines, respectively. The histogram above the main plot summarizes the redshift distribution and statistical significance of our subsample compared to the literature.}
\label{fig:ci10co10}
\end{figure}

The atomic carbon fine-structure emission line, [CI](1$-$0) ($\rm {\nu}_{rest} = 492.16~GHz$), 
has been explored as a promising tracer of extended cold gas in distant DSFGs due to its observed linear correlation with \coonezero \citep[e.g.,][]{israel1995,papadopoulos2004,weiss2005,dunne21,dunne+2022,Gururanjan2023,friascastillo2024}. As a three-level system, [CI] allows to determine $T_{\rm ex}$ of the gas by observing the [CI](1$-$0) and [CI](2$-$1) ($\rm {\nu}_{rest} = 809.34~GHz$) transitions. Importantly, the availability of both [CI] and \coonezero observations helps in narrowing down the range in the possible values of the physical properties (density and temperature) related to the gas excitation in DSFGs with well-sampled CO SLEDs.  

To date, most of the high-\textit{z} DSFGs detected in [CI] lines from the literature do not have \coonezero observations, leading to an uncertainty in their [CI]/CO ratios due to adopted CO line ratios ($r_{J,J-1}$; $J\geq2$) (see Section~\ref{subsec:colineratio}). We conduct a [CI](1$-$0)/\coonezero ratio comparison in a robust and statistically meaningful sample of 23 \vzgal DSFGs with [CI](1$-$0) from \cite{neri+2020} and \cite{cox+2023}. As depicted in Figure~\ref{fig:ci10co10}, we find a median ratio (with unscaled MAD error-bars) of log(${L}^{\prime}_{\rm [CI](1-0)}/{L}^{\prime}_{\rm CO(1-0)})=-0.71\pm0.12$, which is consistent with the values of $-0.62\pm0.14$ (other high-\textit{z} DSFGs) and $-0.74\pm0.12$ (local ULIRGs) found by \citet{friascastillo2024} and \citet{montoya2023}, respectively. The values presented by \citet{friascastillo2024} are based on twelve 850$\mu$m-selected unlensed DSFGs at $2<z<5$. In addition, our ratios also agree with local MS galaxies that show a global log(${L}^{\prime}_{\rm [CI](1-0)}/{L}^{\prime}_{\rm CO(1-0)})\sim -0.85$ \citep[e.g.,][]{wilson1997,liu2023} and other heterogeneous samples of 16 high-\textit{z} galaxies (see Figure~\ref{fig:ci10co10} and its caption), confirming the linear relation between [CI] and CO ground states across redshifts for various galaxy populations, irrespective of the expected differences in their excitation conditions.

 Furthermore, we find no redshift evolution of [CI]/CO ratio based on the available statistics (Figure~\ref{fig:ci10co10}). Similar ratios across populations at a wide range of redshifts therefore point to a common baseline state of the ISM, upon which different star formation modes, efficiencies, and scales are built --- highlighting that the [CI](1$-$0) is also a usable mass estimator.\footnote{Apparently, abundance variations of [CI] are expected to be minor across different source selections. It, therefore, still works in favor of \cionezero in cases where [CI] can be optically thin --- for which a line intensity is expected to vary linearly with abundance.} The similarity in the [CI](1$-$0)/CO(1$-$0) line {luminosity} ratio between \vzgal and local MS galaxies disfavors the usage of two different CO-$\mathrm{H_2}$ conversion factors (${\alpha}_{\rm CO}$) for these two galaxy populations (and in turn also for local ULIRGs); i.e., a bimodal behavior. This motivated our choice of ${\alpha}_{\rm CO}=4$ throughout the statistical analyses of our \vzgal large sample (see Section~\ref{subsec:lum} and Appendix~\ref{app:a}).

 The lack of redshift evolution is also interesting given that \coonezero (${T}_{\rm ex} \sim 5 \, \rm K$) has lower excitation temperature than \cionezero (${T}_{\rm ex} \sim 24 \, \rm K$), making it plausibly more susceptible to excitation by the increasingly warm CMB toward high-\textit{z} \citep[e.g.,][]{dacunha2013}. Furthermore, we observe only a marginal increase in the [CI]/CO ratios of ULIRGs and high-\textit{z} DSFGs compared to local star-forming MS galaxies (Figure~\ref{fig:ci10co10}), suggesting broadly similar ISM conditions, despite the fact that these systems exhibit significantly higher \coonezero luminosities (Figure~\ref{fig:KSrelation}), elevated star formation efficiencies, and shorter depletion timescales (Figure~\ref{fig:z_evol_sfe_tdepl}). Detailed investigations of these aspects needs the support from semi-empirical radiative transfer models to understand the star formation and gas excitation conditions.

\subsection{CO(1$-$0): the best gas mass tracer?}
\label{subsec:co10caveats}
 
 {The  molecular gas masses we derived in this work are best calibrated by the canonical tracer, CO(1--0), since $\alpha_{\rm CO}$ is calibrated on \coonezero line luminosity. However, in many local and high-\textit{z}, actively star-forming galaxies, lower-\textit{J} transitions like \coonezero and \cotwoone are found to be thermalized and saturated; i.e., their brightness temperature ratio is $r_{21}\sim1$ \citep[e.g.,][and this work]{Saintonge+2017,leroy2022,harrington2021}, owing to their high optical depth.} In such cases, the brightness temperature ($T_{\rm b}$) of these transitions closely approaches the kinetic temperature ($T_{\rm kin}$) of the gas, modulo corrections due to the CMB. Consequently, these low-\textit{J} CO lines become insensitive to the molecular gas columns for such extreme environments. In other words, although gas masses are robustly calibrated using \coonezero line luminosity as per Equation~\ref{eqn:gasmass}, this ground-state CO transition may not be the best column density tracer. 
 
 The linear dependence between the \cionezero and \coonezero line luminosities (Section~\ref{subsec:CIcomparison}) suggests that perhaps \cionezero is also optically thick in such extreme environments and is expected to have similar limitation as CO(1--0). However, the luminosity of \cionezero can be explored to complement CO observations in statistically large samples of actively star-forming galaxies. For each high-\textit{z} DSFG, the ``best calibration" of column density and total gas masses therefore will be obtained by modeling multiple lines (and continuum) observations together \citep[e.g.,][]{harrington2021}.

\section{Conclusions} 
\label{sec:conclusion}

We presented the first results from the \vzgal VLA Large Program. Out of the targeted 92 sub-millimeter/infrared bright (${L}_\mathrm{8-1000 \mu m}>$ 10$^{12}$$-$10$^{13}$ $\mathrm{L}_\mathrm{\odot}$), dusty star-forming galaxies (DSFGs), 90 are detected above a significance level above 2$\sigma$, including 57 with a signal-to-noise ratio greater than 5. Only two DSFGs remain undetected with signal levels below 2$\sigma$. When combined with the 14 DSFGs from the pilot survey \citep{stanley+2023}, our full sample comprises 106 galaxies, making \vzgal the largest \coonezero survey of DSFGs in the early Universe to date. 

We derive molecular gas masses based on CO(1$-$0), along with gas scaling relations, gas-to-dust mass ratios, and CO line ratios for this large, homogeneous sample --- quintupling the existing \coonezero statistics of \textit{Herschel}-selected galaxies in the early Universe. Effectively, our sample size doubles all the available high-redshift \coonezero measurements combined from the literature, which have rather heterogeneous selection methods. Additionally, we report \cotwoone line luminosities for 10 \vzgal DSFGs, and investigate [CI]/CO line ratios for 23 galaxies using \cionezero measurements from the NOEMA \zgal survey \citep{neri+2020,cox+2023}. We summarize the key findings as follows:
\vspace{2mm}

\begin{enumerate}[nosep]

    \item We obtain CO(1$-$0)-derived gas masses (using a Milky Way-like  CO-$\mathrm{H_2}$ conversion factor $\alpha_{\rm CO}$=4) for the \vzgal galaxies {of $M_{\rm H_2} = (2-20) \times {10}^{11}~\mathrm{M_{\odot}}$}, which are not corrected for gravitational lensing.

    \item {We find no evidence supporting a bimodal conversion factor ($\alpha_{\rm CO}$). Further, the data do not exclude the possibility that a single $\alpha_{\rm CO}$ applies to all galaxy populations.} 

    \item Both local ULIRGs and high-\textit{z} DSFGs are confirmed to exhibit more efficient star formation and/or have warmer dust temperatures than local MS galaxies. These two highly star-forming populations also show an internally self-regulated range in SFRs per unit gas mass.

    \item Gas depletion timescales derived using $\alpha_{\rm CO}$=4 for our sample are in the range $50-600$ Myr, confirming that such high-\textit{z} DSFGs are potentially a mixed population of starbursts and MS galaxies at those redshifts. Its confirmation however needs stellar masses to characterize the MS.

    \item {Robustly derived CO line ratios (Table~\ref{tab:lineratios}) 
    for the \vzgal sample confirm, with better statistics, the values obtained by \citet{carilliwalter2013,sharon+2016,yang2017} and \citet{harrington2021}.}

    \item Although the median CO line ladder of \vzgal galaxies peaks around $J$=5, a wide variety of gas excitation conditions are found across these high-\textit{z} DSFGs. 
    
    \item The median gas-to-dust mass ratio ($\delta_{\rm GDR}$=$100$) derived using Milky Way-like $\alpha_{\rm CO}$ and $\kappa_{\nu}$ is consistent with other extragalactic values explored for metal-rich, dusty galaxies \citep[e.g.,][]{dunne+2022}.

    \item The consistency in the [CI](1$-$0)/\coonezero luminosity ratio values across redshifts in distant DSFGs, local ULIRGs and star-forming MS galaxies suggests broadly similar conditions in their interstellar medium, despite the differences in \coonezero line luminosities and gas depletion times.

    \item CO(1$-$0) --- the most robustly calibrated gas mass tracer --- may not be the best indicator of the column density of gas. Therefore, the ``best calibration" will require excitation modeling of multiple CO lines, with a potential inclusion of \cionezero as a promising alternative tracer.  

\end{enumerate}

\vspace{2mm}

    To summarize, \vzgal is a unique large sample of comprehensive \coonezero measurements of homogeneously selected high-\textit{z} dusty galaxies. With available higher-\textit{J} CO lines along with continuum measurements from both \textit{Herschel} and NOEMA (\zgal survey), the \vzgal forms a basis for future studies involving the understanding of multi-phase interstellar medium of these galaxies in the early Universe. For example, more precise gas mass, temperature, density, opacity, and the $\alpha_{\rm CO}$ conversion factor can be obtained from a combination of CO, [CI], and continuum observations via semi-empirical, turbulent gas modeling \citep[e.g.,][]{harrington2021}, where \coonezero observations are essential for anchoring the CO SLEDs. Our forthcoming modeling efforts --- along with follow-up [CI] observations --- will not only facilitate a full-fledge characterization of these high-\textit{z} DSFGs but also allow the exploration of redshift evolution with respect to the varying temperature of the cosmic microwave background \citep[e.g.,][]{dacunha2013}. 
        
    The \vzgal \coonezero statistics will remain unrivaled in the northern sky until the advent of the next generation VLA (ngVLA\footnote{\href{https://ngvla.nrao.edu/}{https://ngvla.nrao.edu/}}), which will be capable in incorporating fainter and/or more distant DSFGs \citep{murphy2018}. Recent {and future} ALMA upgrades \citep{Carpenter2023}, however, have a potential to expand such \coonezero and \cotwoone studies to the southern fields. In particular, Band 1 of ALMA ($35-50~{\rm GHz}$) shares a common frequency range mainly with the VLA \texttt{Q} band, allowing to cover \coonezero at $1.3 < z < 2.3$ and \cotwoone at $3.6 < z < 5.6$ with better sensitivity.

\facilities{VLA/NRAO \citep{2009IEEEP..97.1448P}} 
\software{CASA \citep{McMullin+2007casa}, CARTA \citep{cartaref2024}, {Python v3.13 packages\footnote{\href{https://docs.python.org/3.13/reference/index.html}{https://docs.python.org/3.13/reference/index.html}}: Astropy \citep{astropy2022}, Matplotlib \citep{matplotlib2007}, Numpy \citep{numpy2020}, Pandas \citep{reback2020pandas}, Scipy \citep{scipy2020}}}

\section*{Acknowledgments}
{We thank the anonymous referee for their helpful suggestions.} We are grateful to the staff at National Radio Astronomy Observatory (NRAO) for making the observations of this \vzgal Large Program possible. The Karl G. Jansky Very Large Array (VLA) of NRAO is supported by the National Science Foundation (NSF). We thank Viral Parekh (Observatory Scientist, NRAO) for helping with a flux scaling process for the flaring calibrator 3C138. \vzgal project benefited from the \zgal (ANR-AAPG2019) of the French National Research Agency (ANR). PP is a member of the International Max Planck Research School (IMPRS) in Astronomy and Astrophysics. PP, DR, and AW acknowledge the Collaborative Research Center 1601 (SFB 1601 sub-project C2) funded by the Deutsche Forschungsgemeinschaft (DFG, German Research Foundation) -- 500700252. PP thanks N. Sulzenauer and D. Colombo, SFB 1601 collaborators from Bonn, for valuable discussions. LM acknowledges financial support from the South African Department of Science and Innovation’s National Research Foundation under the ISARP RADIOMAP Joint Research Scheme (DSI-NRF Grant Number 150551) and the CPRR Projects (DSI-NRF Grant Number SRUG2204254729).

\section*{Data Availability}

The full data release from the \vzgal VLA Large Program (LP), including calibrated data cubes, continuum-subtracted spectra, moment maps, and \coonezero line measurements for 106 DSFGs, is publicly available at the \vzgal website\footnote{\href{https://vzgal.uni-koeln.de/}{https://vzgal.uni-koeln.de/}}. The data are provided in standard FITS format with accompanying CSV tables for line fluxes and source properties. A README file with data structure and usage guidelines is included in the repository. Raw visibilities from the VLA observations are accessible via the NRAO Science Data Archive\footnote{\href{https://data.nrao.edu/}{https://data.nrao.edu/}} under project codes \texttt{VLA/20A-083} (pilot sample), \texttt{VLA/23B-169} (LP), and \texttt{VLA/25A-099} (LP). 

{In addition, all the figures depicted in this paper are available at \url{https://doi.org/10.5281/zenodo.17664164}. This also includes original files of integrated intensity maps and spectra of all the \vzgal targets from this work that are summarized in Figures~\ref{fig:image-grid_helms_v1}, \ref{fig:image-grid_helms_v2_hers}, and \ref{fig:image-grid_herbs} (Appendix~\ref{app:b}).}

\section*{Data Usage and Citation}

Researchers using the \vzgal data products in published work are kindly asked to cite this paper and acknowledge the \vzgal team. Suggested acknowledgment:

\textit{``This publication makes use of products from the Vz-GAL Large Program (VLA/20A-083, 23B-169, and 25A-099), which is based on data taken with the National Science Foundation (NSF)'s Karl G. Jansky Very Large Array (VLA/NRAO), and partially funded through the German Science Foundation (DFG) Collaborative Research Center SFB 1601. The National Radio Astronomy Observatory (NRAO) is a facility of the NSF operated under cooperative agreement by Associated Universities, Inc."}

\bibliography{vzgal_I}{}
\bibliographystyle{aasjournal}

\appendix 
\section{Appendix A} \label{app:a}

\subsection{Choice of CO-\texorpdfstring{${H}_{2}$}{H2} Conversion Factor} \label{subsec:massalphaco}

Although \coonezero observations enable us to derive robustly calibrated gas masses by directly providing ${{L}^{\prime}}_{\rm CO(1-0)}$ without a need of knowing the excitation of the CO line ladder, the derivation still depends on the choice of ${\alpha}_{\rm CO}$ (Equation~\ref{eqn:gasmass}). This conversion factor is speculated to be a function of the physical properties of galaxies (e.g., gas density, kinetic temperature, velocity dispersion, metallicity) in theoretical and semi-empirical models \citep[see][]{narayanan2011,harrington2021}. However, the accepted Galactic value of this conversion factor is ${\alpha}_{\rm CO}$ $\mathrm{\sim 4.3 ~{M}_{\odot}~{\left(K~km~{s}^{-1}~{pc}^{2} \right)}^{-1}}$ (hereafter without units attached for easy readability), which is also considered to be applicable for other main-sequence galaxies \citep[e.g.,][]{tacconi+2008,carilliwalter2013,sandstorm2014,hughes2017,Riechers2020,wang2022}. In contrast, assuming ${\alpha}_{\rm CO}\sim 0.8$ for local ULIRGs and high-\textit{z} DSFGs has been a common practice \citep[e.g.,][]{tacconi+2008,riechers+2011sled2,carilliwalter2013,riechers+2013,casey+2014,sharon+2016}. This is commonly motivated by the study \citet{downes_solomon_1998}, who derived this average conversion factor for a sample of 10 local ULIRGs based on a comparison of their enclosed dynamical masses to the results from radiative transfer models. Based on these results, \citet{genzel2010} and \citet{daddi2010} adopted a bimodal behavior of ${\alpha}_{\rm CO}$. In other words, the bimodality refers to two distinct values of ${\alpha}_{\rm CO}$ for normal star-forming disks on the SFR$-M_{\ast}$ main-sequence (MS) and for starbursts lying above the MS due to their typically shorter gas depletion timescales (and higher star formation efficiency).

The bimodality of ${\alpha}_{\rm CO}$ has been debated for the past decade and has also been countered by theoretical models predicting a rather smooth distribution of ${\alpha}_{\rm CO}$ based on physical conditions in galaxies \citep[see][]{narayanan2011,narayanan2012b,narayanan2012}. In addition, correlation analyses of $^{12}$\coonezero and $^{13}$\coonezero observations of the central molecular zone in the Milky Way shows lower conversion factors than those seen in the disk \citep{fixsen1999,kohno2024}, supporting ${\alpha}_{\rm CO}$ dependence on the physical conditions. However, its conclusive observational evidence with large statistics across a wide redshift range has been unavailable. \citet{dunne+2022} also discard the bimodality of ${\alpha}_{CO}$ due to its inferred requirement for a dust conversion factor bimodality, which has never been observed.

Overall, there is no clear consensus in the literature about what generic value of the ${\alpha}_{\rm CO}$ conversion factor fits all the observations, especially at earlier cosmic epochs in the relative lack of resolved observations and knowledge about the physical properties of the environment within DSFGs. Although these dusty systems are expected to have Milky Way like metallicities, a number of factors may lower the ${\alpha}_{\rm CO}$ values, such as: (1) high gas temperature and velocity dispersions leading to enhanced CO emission, (2) diffuse or overlapping gas leading to more luminous CO per unit $\mathrm{H_2}$ mass, and (3) starburst modes like local ULIRGs. Studying the ${\alpha}_{\rm CO}$ conversion factor that is determined by a balance between all these different physical conditions within high-\textit{z} galaxies requires semi-empirical models \citep[e.g.,][]{harrington2021,weiss2005} that also show different ${\alpha}_{\rm CO}$ values for cold and warm gas within a galaxy. 

However, we caution the reader that there could be a plausible discrepancy with the ${\alpha}_{\rm CO} \sim 4$ in regard to dynamical masses ($M_{\rm dyn}=M_{\rm gas}+M_{\ast}$) of the local ULIRGs and high-\textit{z} systems. A few studies \citep[e.g.,][]{downes_solomon_1998,bolatto2013} have already pointed out that the Milky Way like conversion factor often leads to an inconsistency with observed dynamical masses. However, assumptions in terms of the ``virial coefficient" in dynamical mass calculations also remain uncertain depending on the galaxy structure/geometry and kinematics \citep{natascha2009,angello2014}. Detailed investigation of galaxies with known stellar masses and resolved gas dynamics is a way forward to further characterize the rotation curves and dynamical masses of such systems in the early Universe. This also helps further constraining the derived gas masses (hence, the CO-$\mathrm{H_2}$ conversion factors), but is beyond the scope of this work. 

Due to a lack of conclusive evidence against the simplest assumption of a Milky Way like conversion factor, we use ${\alpha}_{\rm CO}=4$ (including the contribution from Helium; see Table~\ref{tab:co10}) for the purpose of statistical analyses of our large sample presented in this paper. In addition to the \vzgal DSFGs, we also apply the same conversion factor to all different populations; e.g., other high-\textit{z} DSFGs, local MS galaxies and ULIRGs, that are used for a comparison in Section~\ref{sec:results}. This assumption, while averaging over the variation in physical conditions across these galaxies, helps to highlight any additional trends within these populations. This ${\alpha}_{\rm CO}$ gives gas-to-dust mass ratios similar to those seen in local sources with solar and super-solar metallicity (discussed in Section~\ref{subsec:gasdustratio}) that are consistent given that high-\textit{z} DSFGs are dusty and already metal-rich.

\subsection{\vzgal Continuum Measurements} \label{subsec:cont}

 Although our observations were not designed to detect continuum emission, $\sim$40\% of the targets show radio continuum emission above 2${\sigma}_{c}$. In Appendix~\ref{app:b}, we show continuum contours (in blue) overlaid on the CO moment-0 maps. Nine sources (HeLMS-17 W, HeLMS-19, HeLMS-24, HeLMS-27, HeLMS-41, HerBS-72, HerBS-108, HerBS-183, HerBS-188) are detected above 5${\sigma}_{c}$ significance. Five of them (HeLMS-19, HeLMS-41, HerBS-183, HerBS-185, and HerBS-188) show confirmed foreground radio sources along the line of sight in VLASS\footnote{\href{https://science.nrao.edu/vlass}{https://science.nrao.edu/vlass}}. Several other DSFGs appear to have their continuum potentially contaminated by an overlapping radio source on the plane of the sky. 
  In the case of HeLMS-51, the continuum emission is significantly offset by $4''$ to the west from the \coonezero emitting DSFGs, in line with a similar component seen in the \zgal data \citep{cox+2023}, possibly revealing an unrelated source in the observed field. In addition, we detect multiple other serendipitous sources in various target fields (see the table below and observed radio maps in Appendix~\ref{app:b}). Here we have colored the continuum contours in magenta for all the cases of unrelated radio sources within the field.

\renewcommand{\arraystretch}{1.2}

\begin{longdeluxetable}{cccccc}
\tabletypesize{\scriptsize} 
\tablecaption{Serendipitous Continuum Detections of sources that are not related to the targeted \vzgal DSFGs. 
\label{tab:extra_continuum}}
\tablehead{
    \colhead{Target} & Radio &
     \colhead{Position} &
    \colhead{${\nu}_\mathrm{cont}$} &
      \colhead{${S}_\mathrm{cont}$} &
      \colhead{Signal-to-noise}\\
    Field & Source & [RA,DEC]  & (GHz)  &  (mJy) & Ratio (SNR) \\
}
\startdata
HeLMS-49  & RC & [23:37:22.09, -06:47:51.76] &35   &  0.10±0.02  & 5.8   \\
HeLMS-51  & RC  & [23:26:17.26, -2:53:17.87] &28   &  0.11±0.03  & 4.1   \\
HeLMS-57  & RC1  &  [00:35:19.53, 07:28:14.08]  & 30   &  0.09±0.02 & 3.4 \\
   &  RC2 &   [00:35:19.81, 07:27:55.17]   & 30   &  0.14±0.02  & 2.9 \\
    &  &   & 38   &  0.19±0.05   & 2.1 \\
HerS-20  &  RC & [01:02:45.52, 01:05:54.65]  & 29    & 0.09±0.02   & 5.7  \\
HerBS-191  &  RC  &  [12:47:53.32, 32:25:00.90] &  25   &  0.09±0.02   & 2.5 \\
HerBS-204  & RC  &  [13:29:09.81, 30:10:16.49] &  25   &  0.11±0.02  & 4.4 \\
\enddata
\end{longdeluxetable}

\subsection{\vzgal Spectral Line Measurements} \label{subsec:colines}

\setlength{\tabcolsep}{2pt}

\renewcommand{\arraystretch}{1.15}

\begin{longdeluxetable}{ccccccccccccc}
\tabletypesize{\scriptsize}
\tablecaption{\vzgal (this work): \coonezero Line and Underlying Continuum Properties, Molecular Gas Masses, and Gas Depletion Times. 
\label{tab:co10}}
\tablehead{
    \colhead{Source} & 
    \colhead{\rule{0pt}{2.05em}\zgal} & 
    \colhead{$\rm {S}_{peak}$} & 
    \colhead{$\Delta v_{FWHM}$}& 
    \colhead{$\rm {I}_{CO(1-0)}$} & 
    \colhead{$\rm \mu {L}^{\prime}_{CO(1-0)}$$^{a}$} & 
    \colhead{$\rm \left[\frac{4.0}{\alpha_{CO}}\right] \mu {M}_{\rm H_2}$} & 
    \colhead{$\rm {\tau}_{dep}$} & 
    \multicolumn{2}{c}{${\nu}_\mathrm{cont}$ (GHz)$^{\dag}$} & 
    \multicolumn{2}{c}{${S}_\mathrm{cont}$ (mJy)$^{\ddag}$} & \colhead{Line} \\
    &$\rm {\textit{z}}_{spec}$ &(mJy)&($\rm km~{s}^{-1}$)&($\rm Jy~km~{s}^{-1}$)&(${10}^{11}$ $L_l$)&($\rm {10}^{11}~{M}_{\odot}$)&(Myr)&
    \colhead{IF1} & 
    \colhead{IF2} & 
    \colhead{IF1} & 
    \colhead{IF2} & \colhead{Significance$^{b}$} \\
}
\startdata
HeLMS-1 &   1.9047  &   2.42±0.21 &   372±38    &   0.96±0.13 &   1.82±0.25$^{l}$ &   7.27±1.00$^{l}$ &   142.70±21.70  & 30.7  & 39.5  & 0.03±0.01   & $<$0.13 & 6$\sigma$ \\   
HeLMS-3 &   1.4199  &   3.63±0.39 &   576±58    &   2.22±0.33 &   2.47±0.37$^{l}$ &   9.87±1.46$^{l}$ &   233.73±41.50  &  47.1  &   & $<$0.33  &  & 5$\sigma$ \\   
HeLMS-11 &  2.4834  &   0.84±0.10 &   745±74    &   0.67±0.10 &   2.01±0.30$^{l}$ &   8.03±1.21$^{l}$ &   231.14±36.33  &  32.5  &   & $<$0.12  &  & 7$\sigma$ \\   
HeLMS-12 &  2.3699  &   2.05±0.13 &   554±41    &   1.21±0.12 &   3.36±0.33$^{l}$ &   13.4±1.34$^{l}$ &  365.08±41.20  &  27.1  & 34.1  & $<$0.10  &  $<$0.27 & 11$\sigma$ \\   
HeLMS-14 &  1.6168  &   2.65±0.25 &   516±52    &   1.45±0.20 &   2.05±0.28$^{l}$ &   8.21±1.13$^{l}$ &   384.49±59.68   &  43.5  &   & $<0.39$  &  & 5$\sigma$ \\   
HeLMS-16 &  2.8187  &   1.12±0.08 &   1136±114    &   1.35±0.17 &   5.05±0.64 &   20.2±2.54 &   436.02±58.28  &  30.0  &  38.5 & 0.08±0.02  & $<$0.21 & 11$\sigma$ \\   
HeLMS-17 WE & -- &   -- &   --   &  0.88±0.10 & 2.32±0.38 & 9.28±0.92 & 259.59±38.27  &27.1 & 34.9 & 0.16±0.03  & $<$0.11 & -- \\ 
W & 2.2972 &   0.56±0.07 &   816±117   &   0.49±0.09 &   1.29±0.24$^{l}$ &   5.16±0.96$^{l}$ &   --  &  --  & --  &  -- &  -- & 5$\sigma$ \\   
E & 2.2983 &   0.28±0.05 &   1312±273  &   0.39±0.11 &   1.03±0.29 &   4.12±1.16 &  -- & -- & -- & --  & -- & 3$\sigma$ \\
HeLMS-19   & 4.6885 &   0.43±0.07 &   337±64    &   0.15±0.04 &   1.29±0.34$^{l}$ &   5.18±1.34$^{l}$ &   57.18±15.18  &  20.7  &   & 1.18±0.06  &  & 3$\sigma$ \\    
HeLMS-20   & {2.1947} &   0.56±0.07 &   990±100  &   0.59±0.09 &   1.44±0.22$^{l}$ &   5.75±0.90$^{l}$ &   197.14±31.98  &  27.9  & 35.9  & $<$0.11  & $<$0.18 & 6$\sigma$  \\   
HeLMS-21   & 2.7710 & 1.11±0.10  &   315±35    &   0.37±0.05 &   1.35±0.18$^{l}$ &   5.40±0.73$^{l}$ &   99.59±14.35  &  30.4  & 38.5  & 0.06±0.02  &  $<$0.08 & 6$\sigma$ \\   
HeLMS-23   & 1.4888 & 2.18±0.26   &   391±54    &   0.91±0.16 &   1.10±0.19$^{l}$ &   4.39±0.78$^{l}$ &   143.92±31.12  &  45.6  &   & $<$0.13  &  &  6$\sigma$ \\   
HeLMS-24 &   4.9841    &   0.53±0.12 &   596±153   &   0.34±0.12 &   3.10±1.10$^{l}$ &   12.4±4.42$^{l}$ &  147.12±53.35   &  19.7  &   & 0.08±0.02  & &  4$\sigma$\\   
HeLMS-25   & 2.1408 &   0.30±0.05 &   1095±110   &   0.35±0.07 &   0.82±0.16 &   3.26±0.65 &   112.08±23.51   &  28.4  & 36.5  & 0.04±0.01  &  $<$0.07 & 5$\sigma$  \\    
HeLMS-26 EW & -- &  -- &  -- &    0.26±0.03  &    0.90±0.10  &   3.59±0.36  &  123.91±31.19 & 31.1  & 38.5 & $<$0.05  & $<$0.10 & -- \\
E & 2.6899 &  0.65±0.12 &  216±46 & 0.15±0.02  &   0.52±0.05  &    2.07±0.21  &   --   & --  & -- & --& -- & 4$\sigma$  \\
W & 2.6875 &  0.24±0.07 &  442±156 & 0.11±0.01  &   0.38±0.04  &    1.52±0.15  &   --  &  -- &  -- & --  & -- & 2$\sigma$  \\
HeLMS-27   & 3.7652 &   0.39±0.03 &   554±48    &   0.23±0.03 &   1.37±0.18$^{l}$ &   5.49±0.72$^{l}$ &   83.79±11.60  &  21.5$^{\ast}$  &   &  0.05±0.01 & & 8$\sigma$ \\   
HeLMS-28   & 2.5327 &   1.06±0.08 &   697±70    &   0.79±0.10 &   2.45±0.31$^{l}$ &   9.80±1.25$^{l}$ &  328.89±44.78  &  32.1  &   & $<$0.08  & & 13$\sigma$ \\ 
{HeLMS-31}  &   1.9495  &     1.48±0.15   &   532±53   &   0.84±0.08 &   1.66±0.17$^{l}$ &   6.64±0.66$^{l}$ &  284.56±45.81  &  30.2  & 38.9  & $<$0.12  & $<$0.26 & 6$\sigma$ \\
HeLMS-32 CN1N2  &   --  &   --&  -- &   0.60±0.10   & 0.96±0.15 &   3.84±0.35 &  25.72±3.00  &  41.9  &   & $<$0.06  & & -- \\
C  &   1.7153  &   0.47±0.10 &   493±49 &   0.25±0.07  &   0.40±0.11  &   1.60±0.44 &  --  &  --  &  -- & --  &  -- & 3$\sigma$  \\ 
{N1}    &   1.7153 &   0.58±0.10 &   346±71  &   0.21±0.06 &   0.34±0.09 &   1.36±0.36 &   --    &    -- & --  & --  &  -- & 3$\sigma$  \\
{N2}  &   1.7153 &   0.89±0.10 &   148±19    &   0.14±0.04 &   0.22±0.06 &   0.88±0.24 &  --    &   -- & --  & --  &  -- &  2$\sigma$ \\
HeLMS-34 & 2.2715 & 1.82±0.16 &   411±41    &   0.80±0.11 &   2.06±0.28$^{l}$ &   8.24±1.14$^{l}$ &   183.11±28.18  &  27.3  & 35.0  & 0.09 ± 0.03  & $<$0.22 & 10$\sigma$  \\   
HeLMS-35 & 1.6684 & 1.71±0.21 &   627±90    &   1.14±0.21 &   1.70±0.31$^{l}$  &   6.82±1.25$^{l}$  &  605.63±117.32  &  42.6  &   & $<$0.18  & &  5$\sigma$ \\   
HeLMS-36 & 3.9802 & 0.21±0.04 &   330±70    &   0.07±0.02 &   0.48±0.13 &   1.92±0.52 &   33.07±9.10  & 18.2   & 23.0  &  $<$0.05 & $<$0.04 & 5$\sigma$   \\   
HeLMS-37 & 2.7576 & 0.48±0.06 &   416±60    &   0.21±0.03 &   0.76±0.08 &   3.04±0.30 &  89.19±9.00  &  30.5  & 38.5  & $<$0.05  & $<$0.07 & 6$\sigma$  \\   
HeLMS-38 & 2.1898 & 1.27±0.19 &   513±51    &   0.69±0.12 &   1.68±0.29$^{l}$ &   6.73±1.17$^{l}$ &  216.66±39.89  &   28.0 & 36.0  & 0.06±0.02  &  $<$0.09 & 6$\sigma$   \\   
HeLMS-39 & 2.7658 & 0.79±0.08 &   575±58    &   0.48±0.07 &   1.75±0.25 &   6.99±1.01 &   256.93±39.50  &   30.4 & 38.5  & $<$0.07  & $<$0.08 &  6$\sigma$ \\   
HeLMS-40 W1W2 & -- &  -- &  --   &   0.53±0.05 &   2.18±0.22 &   8.72±0.88 & 19.71±2.00 &  27.6  & 32.5  & 0.05±0.01  & $<$0.10 & --   \\ 
 W1 & 3.1445 &  0.32±0.05 &   786±79   &   0.27±0.05 &   1.01±0.22 &   4.04±0.88 &   --  &  -- & --  & --  & -- & 4$\sigma$   \\
 W2 & 3.1395 &  0.39±0.05 &   629±63   &   0.26±0.05 &   1.17±0.22$^{l}$ &   4.68±0.88$^{l}$ &--  & --   & --  &  -- & -- &  5$\sigma$   \\   
HeLMS-41 & 2.3353 & 0.42±0.06 &   547±86    &  0.24±0.05  &   0.66±0.14$^{l}$ &   2.65±0.54$^{l}$ &   99.96±21.54   & 30.7$^{\ast}$   &   & 0.20±0.02  & & 4$\sigma$ \\  
HeLMS-42 & 1.9558 &  1.46±0.15 &  249±25 & 0.39±0.06  &   0.77±0.12$^{l}$  &    3.08±0.48$^{l}$  &   149.63±25.15   &  30.2 & 38.8  & $<$0.06  & $<$0.12 &  2$\sigma$  \\
HeLMS-43 & 2.2912 &  0.44±0.12 &  460±146 & 0.22±0.09  &   0.56±0.24  &    2.26±0.94  &  78.66±33.29   & 27.1 & 34.8 & $<$0.07  & $<$0.07 & 4$\sigma$   \\
HeLMS-44 & 1.3700 & $<$2.3 & 385±16$^{i}$ & $<$0.89 & $<$1.04$^{l}$ & $<$4.16$^{l}$ & $<$341.40   &  48.0  &   & $<$0.19  & &  $<$2$\sigma$  \\   
HeLMS-45 & 5.3998& 0.35±0.08 & 847±227 & 0.32±0.11  & 3.28±1.10$^{l}$  & 13.1±4.57$^{l}$  & 143.78±50.81  &  15.9  &  & 0.06 ± 0.01  & & 3$\sigma$   \\
HeLMS-46 & 2.5772 & 0.66±0.07 &   673±67    &   0.47±0.07 &   1.52±0.22$^{l}$ &   6.07±0.90$^{l}$ &   250.42±39.51  & 32.5   &   & $<$0.07  & &  7$\sigma$   \\   
HeLMS-47 NS & -- &  -- &  -- & 0.90±0.07  &   2.24±0.16  &    8.97±0.66  &   276.59±44.29   & 27.7  & 35.6  & $<$0.07  & $<$0.10 & --  \\
N & 2.2232 &  0.64±0.10 &  834±135 & 0.57±0.06  &   1.42±0.14  &    5.68±0.57  &   --   & --  & --  & --  & -- &  5$\sigma$  \\
S & 2.2232 &  0.71±0.14 &  439±100 & 0.33±0.03  &   0.82±0.08  &    3.29±0.33  &  --   & --  & --  & --  & -- &  4$\sigma$  \\
HeLMS-48 & 3.3514 & 0.88±0.08 &   647±64    &   0.61±0.08 &   3.00±0.40$^{l}$ &   12.0±1.59$^{l}$ &   329.52±46.56  &  23.2$^{\ast}$  &   & 0.05±0.01  & &  8$\sigma$  \\ 
HeLMS-49 CS & -- &  -- &  -- & 0.54±0.03  &   1.34±0.09$^{l}$  &    5.35±0.38$^{l}$  &   176.55±26.14   & 35.3 &   & $<$0.05  & &  --  \\
C & 2.2154 &  0.66±0.1 &  426±74 & 0.30±0.03  &   0.74±0.07  &    2.97±0.30  &  --   & -- &   & --  & & 5$\sigma$ \\
S & 2.2154 &  0.42±0.1 &  548±144 & 0.24±0.03  &   0.59±0.06  &    2.38±0.24  &  --  & -- &   & --  & & 3$\sigma$   \\
HeLMS-50 & {2.0532} & 0.45±0.05 &   1387±139  &   0.66±0.10 &   1.44±0.22 &   5.76±0.87 &  331.50±53.86  &  29.2  & 37.6  &  $<$0.04 & $<$0.08 & 5$\sigma$   \\   
HeLMS-51 & 2.1559 & 0.64±0.06 &   888±89    &   0.60±0.08 &   1.43±0.19$^{l}$ &   5.72±0.76$^{l}$ &   361.01±51.70  &  28.3  & 36.3  & $<$0.03   & $<$0.07 &  8$\sigma$  \\  
HeLMS-52 & 2.2092 &  1.22±0.14 &  235±30 & 0.31±0.05  &   0.75±0.12$^{l}$   &    3.01±0.49$^{l}$   &   99.59±17.72   &  27.8 & 35.7  & 0.09±0.02  & $<$0.06 & 7$\sigma$   \\ 
HeLMS-54 & 2.7070 & 0.40±0.08 &   564±130   &   0.24±0.07  &   0.84±0.24 &   3.35±0.98 &   154.66±46.98  &  30.9  & 38.5  & $<$0.05  & $<$0.08 & 4$\sigma$  \\   
HeLMS-55 & 2.2834 & 1.28±0.21 &   428±43    &   0.58±0.11 &   1.52±0.29$^{l}$ &   6.09±1.15$^{l}$ &  304.62±59.80  & 27.2   & 34.9  & 0.05±0.02  & $<$0.13 & 7$\sigma$  \\    
HeLMS-56 CNW & -- &  -- &  -- & 0.46±0.05  &   2.33±0.10  &    9.32±0.50  &   231.38±37.20   &  25.7 &   & 0.02±0.01  & & --  \\
C & 3.3909 &  0.26±0.03 &  726±75 & 0.20±0.02  &   1.01±0.10$^{l}$   &    4.04±0.40$^{l}$   &   --   & -- &   & -- & & 2$\sigma$  \\
N & 3.3909 &  0.36±0.13 &  353±152 & 0.14±0.02  &   0.71±0.07  &    2.83±0.28  &   --   &  -- &   & --  & & 2$\sigma$  \\
W & 3.3909 &  0.52±0.16 &  212±77 & 0.12±0.01  &   0.61±0.06  &    2.43±0.24  &   --   &  -- &   & -- & &  2$\sigma$ \\
HeLMS-57 & 1.9817 &  0.68±0.10 &  915±154 & 0.66±0.08  &   1.34±0.15  &   5.38±0.55  &   302.70±32.22   & 29.9  &  38.5 & 0.16±0.03  & $<$0.10 & 5$\sigma$  \\
HerS-2 & 2.0149 &   1.99±0.17 &   684±68    &   1.45±0.19 &   3.04±0.40$^{l}$ &   12.2±1.59$^{l}$ &   341.05±48.49  &  29.6  &  38.0 & $<$0.16  & $<$0.35 & 9$\sigma$   \\   
HerS-3 & 3.0608 &   2.02±0.14 &   330±27    &   0.71±0.08 &   3.03±0.34$^{l}$ &   12.1±1.37$^{l}$  &  149.77±41.86 &  28.2  & 32.5  & 0.10±0.03  &  $<$0.17 &  7$\sigma$  \\   
HerS-5 & 1.4491 &   1.74±0.23 &   721±72    &   1.33±0.22 &   1.54±0.25$^{l}$ &   6.15±1.01$^{l}$  &  346.64±66.31  &  46.4  &   & $<$0.37  & & 4$\sigma$ \\   
HerS-7 & 1.9838 &   0.82±0.14 &   636±124   &   0.55±0.14 &   1.13±0.29 &   4.53±1.14 &  193.19±50.56  &  29.9  & 38.4  & $<$0.05  & $<$0.10 & 5$\sigma$   \\   
HerS-8 & 2.2431 &   0.48±0.05 &   687±89    &   0.35±0.06 &   0.89±0.15 &   3.55±0.61 &  133.52±24.19  &  27.5  & 35.4  &  $<$0.04  & $<$0.06 &  6$\sigma$ \\   
HerS-10 & 2.4690 & 1.43±0.15  &   475±48 &  0.72±0.11 &   2.16±0.33$^{l}$  &   8.63±1.31$^{l}$ &   332.94±53.67   & 32.5   &   & $<$0.13  & & 4$\sigma$   \\   
HerS-11 & 4.6618 & 0.22±0.07    & 681±269 &   0.16±0.08 &   1.33±0.67 &   5.31±2.66 &   73.57±37.17  &  19.8  &   & $<$0.09  & & 3$\sigma$  \\   
HerS-12 & 2.2707 & 0.58±0.06  &   1187±119    &   0.73±0.11 &   1.89±0.28$^{l}$  &   7.57±1.14$^{l}$  &  275.14±45.15  & 27.3   & 35.1  & $<$0.06 &  $<$0.08 &  6$\sigma$  \\   
HerS-13 & 2.4759 & 0.87±0.08  &   439±50    &   0.41±0.06 &   1.22±0.18$^{l}$ &   4.88±0.72$^{l}$  &  130.58±21.04  &  32.5  &   & 0.09±0.02  & & 7$\sigma$ \\   
HerS-14 & 3.3441 & 0.35±0.05  &   926±174   &   0.34±0.08 &   1.70±0.40$^{l}$ &   6.81±1.58$^{l}$ &   126.56±30.02  &  27.0  & 32.5  & 0.04±0.01  & $<$0.04 &  5$\sigma$   \\   
HerS-15 & 2.3019 & 1.10±0.09  &   743±74    &   0.87±0.11 &   2.30±0.29$^{l}$ &   9.21±1.16$^{l}$ &   433.83±59.07  &  27.0  & 34.7  & $<$0.12  & $<$0.20 &  7$\sigma$  \\   
{HerS-16} & 2.1971  &  1.08±0.14 & 724±72    &   0.83±0.14 &   2.04±0.34$^{l}$ &   8.16±1.36$^{l}$ &   386.29±69.04  &  27.9  & 35.9  & 0.06±0.02  & $<$0.23 & 9$\sigma$  \\ 
HerS-17 & 3.0186 &  0.61±0.09 &   682±68    &   0.44±0.08  &   1.85±0.33$^{l}$ &   7.40±1.34$^{l}$ &   184.76±34.73   &  28.5  & 32.5  & $<$0.08  & $<$0.08 &  4$\sigma$  \\    
HerS-18 E & 1.6926  &  1.12±0.16 &   558±92    &   0.66±0.14 &   1.02±0.22$^{l}$  &   4.08±0.86$^{l}$  & --   &  42.1  &   &  $<$0.12 & & 5$\sigma$ \\   
HerS-20 & 2.0792 &  1.61±0.14 & 339±34  &   0.58±0.08 &   1.29±0.18$^{l}$  &   5.14±0.71$^{l}$  &   300.15±45.65 &  29.0  & 37.2  & 0.06±0.01  & $<$0.16 & 10$\sigma$   \\   
HerBS-46 & 1.8349 &  0.56±0.11 &  938±94 & 0.56±0.12  &   0.99±0.21  &    3.97±0.85  &   276.96±61.78   &  41.2 &   & $<$0.10  & & 4$\sigma$ \\
HerBS-48 & 3.1438 & 0.42±0.07 &   713±144   &   0.32±0.08 &   1.43±0.39 &   5.70±1.43 &   120.95±30.84  &  27.6  & 32.5  & $<$0.08  &  $<$0.07 &  4$\sigma$  \\   
{HerBS-50} & 2.9280 & 0.54±0.05 & 621±62 & 0.36±0.05  &   1.44±0.20$^{l}$ &   5.76±0.80$^{l}$ &   401.50±20.85  & 29.2   & 32.5  &  $<$0.07 &  $<$0.06 & 5$\sigma$   \\ 
HerBS-51 & 2.1827 &  0.62±0.10 &  448±87 & 0.30±0.07  &   0.71±0.17  &    2.85±0.68  &   116.70±28.42   &  36.6 &   & $<$0.05  & & 5$\sigma$ \\
HerBS-53 EW & 1.4229 &  $<$1.2  &  543±87$^{i}$ &  $<$0.65  &  $<$0.76$^{l}$  &  $<$3.03$^{l}$ &   $<$342.65   &  47.1 &   & $<$0.18  & & $<$2$\sigma$   \\
HerBS-65 & 2.6858 &  0.49±0.08 &  529±103 & 0.28±0.07  &   0.95±0.24  &    3.80±0.96  &   134.36±35.20   &  34.7$^{\ast}$ &   & $<$0.06  & & 5$\sigma$   \\
HerBS-72 & 3.6380 &  0.57±0.06 &  550±56 & 0.33±0.05  &   1.89±0.28$^{l}$  &    7.55±1.13$^{l}$  &  173.91±33.26   &  24.3 &   & 0.04±0.01  & & 8$\sigma$  \\
HerBS-78 & 3.7344 &  0.40±0.06 &  508±88 & 0.22±0.05  &   1.28±0.30  &    5.10±1.18  &  107.36±26.91   &  23.8 &   & $<$0.03  & & 5$\sigma$   \\
HerBS-91 CE &  2.4047 & 0.72±0.08    &   567±78    &   0.43±0.08 &   1.24±0.23$^{l}$ &   4.96±0.91$^{l}$ &  172.40±33.57   &  33.3  &   & $<$0.12  & & 4$\sigma$  \\   
{HerBS-105} & 2.6695 & 2.35±0.14 & 448±45 & 0.18±0.04   &   0.62±0.14$^{l}$ &   2.48±0.56$^{l}$ &   704.17±93.80  &   32.1 &   & $<$0.06  & & 2$\sigma$ \\ 
HerBS-108 & 3.7168 &    0.53±0.05 &   561±56    &   0.32±0.04 &   1.85±0.23$^{l}$ &   7.41±0.94$^{l}$ &   152.26±20.74  &  21.6$^{\ast}$  &   & 0.07±0.01  & & 8$\sigma$  \\   
HerBS-109 SNW & -- & --&--& --  &   --  &    --  &  --  &44.0 &  & $<$0.21  & &  --  \\
S & 1.5843 & 1.52±0.16&750±74& 1.21±0.13  &   1.64±0.16  &    6.56±0.66  &  --  &-- &  & -- & &  2$\sigma$ \\
NW & 1.5850 &$<$0.81 &662±84$^{i}$&$<$0.57  &   $<$0.78  &    $<$3.10  &   --  & -- &  & -- & & $<$2$\sigma$  \\
HerBS-116 & 3.1547 &  0.42±0.06 &  799±80 & 0.36±0.06  &   1.61±0.27  &    6.42±1.08  &   203.76±35.35   & 30.2$^{\ast}$  &   & $<$0.04  & & 7$\sigma$   \\
HerBS-126 & 2.5875 &  0.33±0.07 &  941±238 & 0.33±0.11  &   1.07±0.36  &    4.27±1.42  &   196.91±66.57   &  32.6 &   & $<$0.05  & & 5$\sigma$  \\
HerBS-129 & 3.3077 &  0.53±0.12 & 720±186 & 0.41±0.14  &   1.97±0.68$^{l}$  &    7.88±2.72$^{l}$  &   245.27±85.47   & 26.8 & 34.8 & $<$0.09  & $<$0.08 & 4$\sigma$   \\
HerBS-136 & 3.2884 &    0.38±0.06 &   589±116   &   0.24±0.06 &   1.14±0.29 &   4.58±1.15 &   137.99±35.68  &  27.1  & 32.5  &  $<$0.06 & $<$0.05 & 4$\sigma$  \\   
HerBS-137 & 3.0408 &  0.31±0.11 &  429±180 & 0.14±0.08  &   0.60±0.34$^{l}$  &    2.40±1.35$^{l}$  &   76.25±43.31   & 28.5 & 34.5 & $<$0.08  & $<$0.08 & 2$\sigma$  \\ 
HerBS-143 & 2.2406 &  0.46±0.05 &  1077±148 & 0.53±0.09  &   1.33±0.23  &    5.33±0.91  &   340.73±61.08   &  36.0 &   & $<$0.04  & & 7$\sigma$   \\
HerBS-157 & 1.8971 & {{0.33±0.13}}    &   {{787±79}}   &   {{0.28±0.11}}   &   {{0.52±0.21}}   &   {{2.08±0.83}}   &  {{189.09±76.21}}  &  30.8  & 39.5  & $<$0.07  & $<$0.15 & 2$\sigma$  \\  
 HerBS-162 SWNE & -- &  -- &  -- & 0.33±0.02  &   0.99±0.05  &    3.96±0.30  &   --   &  33.6 &   & $<$0.04  & & --   \\
SW & 2.4739 &  0.47±0.12 &  305±94 & 0.15±0.02  &   0.45±0.05  &    1.80±0.20  &  --   &  -- &   & --  & & 3$\sigma$   \\
NE & 2.4742 &  0.42±0.10 &  408±116 & 0.18±0.02  &   0.54±0.05  &    2.16±0.22  &   --  &  -- &   & --  & &   3$\sigma$  \\
HerBS-165 & 2.2251 & 0.57±0.08    &   516±83 &  0.31±0.07 &   0.78±0.18 &   3.12±0.70 &   228.60±53.40   &  27.7  & 35.7  & $<$0.05  & $<$0.11 & 5$\sigma$   \\   
HerBS-167 & 2.2144 & 0.32±0.06    &  952±196 &  0.32±0.09 &   0.80±0.22 &   3.21±0.89 &   273.64±77.46  &  27.8  & 35.7  &  $<$0.05 & $<$0.17 & 3$\sigma$  \\   
HerBS-169 & 2.6977 & 0.57±0.09    &   535±99 &  0.32±0.08  &   1.13±0.28 &   4.50±1.11 &   171.30±43.65  &  32.2  &   & $<$0.04  & &  5$\sigma$ \\   
HerBS-171 & 2.4793 & 0.48±0.05    &   451±59 &  0.23±0.04 &   0.69±0.12 &   2.77±0.48 &   151.11±27.69  &  32.6  &   & $<$0.07  & & 5$\sigma$   \\   
HerBS-176 & 2.9805 &  0.50±0.10 &  894±197 & 0.48±0.14  &   1.95±0.57$^{l}$  &    7.79±2.29$^{l}$  &   288.20±86.14   &  30.9$^{\ast}$ &   & $<$0.06  & & 4$\sigma$   \\
HerBS-177 & 3.9625 &  0.55±0.06 &  571±76 & 0.33±0.06  &   2.16±0.39$^{l}$  &    8.66±1.55$^{l}$  &   176.11±32.96   & 22.7  &   & 0.02±0.01 & & 9$\sigma$  \\
HerBS-179 & 3.9423 &  0.53±0.16 &  310±109 & 0.17±0.08  &   1.12±0.51  &    4.49±2.06  &   106.95±49.23   & 22.8 &   & 0.05±0.01  & &  2$\sigma$ \\
HerBS-183 & 1.8919 & 1.34±0.16    &   624±62 &  0.89±0.14 &   1.67±0.26$^{l}$ &   6.67±1.05$^{l}$ &   590.04±100.73  &  35.2$^{\ast}$  &   & 0.39±0.02  & &  7$\sigma$ \\   
HerBS-185 & 4.3238 & 0.23±0.06    &  632±178 &  0.15±0.06 &   1.15±0.44$^{l}$ &   4.59±1.78$^{l}$ &  93.92±36.96   &  21.1  &   & 0.02±0.01  & & 4$\sigma$   \\   
HerBS-187 EW & -- & --  & -- & -- &   -- &  -- &   $>$210.24  & 41.2  &   & $<$0.13  & &  -- \\
E & 1.8285    & $<$0.25  &591±76$^{i}$&$<$0.16&$<$0.28 &$<$1.13 & --  &    --  &  &  -- & &  $<$2$\sigma$  \\  
W & 1.8274 & 0.64±0.11  &  517±106 &  0.35±0.09 &   0.62±0.16 &   2.48±0.64 &   --  & --  &   &--  & & 3$\sigma$  \\
HerBS-188 & 2.7675 &    0.24±0.04 &  987±192 &  0.25±0.06 &   0.91±0.22 &   3.65±0.87 &   141.75±34.77  &  34.5$^{\ast}$   &   &  0.31±0.03  & &  4$\sigma$ \\   
HerBS-190 & 2.5890 &    0.52±0.06 &  755±107 &  0.42±0.08 &   1.35±0.26 &   5.41±1.04 &   249.12±49.96  &  32.7  &   &  $<$0.09 & & 5$\sigma$  \\   
HerBS-191 & 3.4428 &  0.20±0.05 &  1751±490 & 0.37±0.14  &   1.93±0.72  &    7.72±2.90  &   246.26±93.48   & 25.4  &   & 0.04±0.01  & & 3$\sigma$  \\
HerBS-193 & 3.6951 &    0.31±0.03 &  849±110 &  0.28±0.05 &   1.62±0.29 &   6.49±1.16  &   144.84±26.94   &  24.1  &   & 0.02±0.01  & &6$\sigma$  \\   
HerBS-194 NS & -- & --&  -- &  -- &  -- &   -- &   $>$135.24   & 27.0  &  34.4 & $<$0.05  & $<$0.07   & -- \\
N & 2.3335 &  0.26±0.05 &  862±194 &  0.24±0.07 &   0.65±0.19 &   2.60±0.76 &   --   & --  &  -- & --  & -- & 3$\sigma$  \\
S & 2.3316  &  $<$0.20   &538±140$^{i}$ &$<$0.12&$<$0.32 &$<$1.30 &  -- &  --  &  -- & --  & -- &  $<$2$\sigma$ \\  
HerBS-204 EW & -- & -- & -- & 0.40±0.08  &  2.12±0.23  &    8.48±0.90  &   190.78±20.52   & 25.1   &   & $<$0.04  & & --  \\
E & 3.4937 &  0.31±0.08 &  581±163 & 0.19±0.03  &   1.01±0.11  &    4.04±0.41  &   --   & --  &   & -- & &  3$\sigma$  \\
W & 3.4933 &  0.37±0.08 &  536±131 & 0.21±0.07  &   1.11±0.13  &    4.44±0.45  &   --   & --  &   & -- & & 3$\sigma$ \\
\enddata
\end{longdeluxetable}

\renewcommand{\arraystretch}{1.15}

\begin{longdeluxetable}{ccccccccccc}
\tabletypesize{\scriptsize} 
\tablecaption{\vzgal (this work): \cotwoone Line and Underlying Continuum Properties
\label{tab:co21}}
\tablehead{
    \colhead{Source} & 
    \colhead{\rule{0pt}{2.05em}$\rm {\textit{z}}_{spec}$} & 
    \colhead{$\rm {S}_{peak}$} & 
    \colhead{$\Delta v_{FWHM}$}& 
    \colhead{$\rm {I}_{CO(2-1)}$} & 
    \colhead{$\rm \mu {L}^{\prime}_{CO(2-1)}$$^{a}$} & 
    \multicolumn{2}{c}{${\nu}_\mathrm{cont}$ (GHz)$^{\dag}$} & 
    \multicolumn{2}{c}{${S}_\mathrm{cont}$ (mJy)$^{\ddag}$} & \colhead{Line}  \\
    &&(mJy)&($\rm km~{s}^{-1}$)&($\rm Jy~km~{s}^{-1}$)&(${10}^{11}$ $L_l$)&
    \colhead{IF1} & 
    \colhead{IF2} & 
    \colhead{IF1} & 
    \colhead{IF2} & \colhead{Significance$^{b}$} \\
}
\startdata
HeLMS-19    & 4.6885    &   4.29±0.45 &   288±35    &   1.31±0.21 &   2.76±0.44$^{l}$ &  41.1  &   & 0.87±0.10  & & 5$\sigma$ \\
HeLMS-24    & 4.9841    &   2.26±0.14 &   691±50    &   1.66±0.16 &   3.83±0.37 &  34.4$^{\ast}$  &   & 0.09±0.02  & & 11$\sigma$  \\
HeLMS-27 & 3.7652 &   2.07±0.56 &   599±189    &   1.32±0.12 &   1.95±0.18$^{l}$ & 43.0  & 48.2  &  $<$0.14 & $<$0.35 & 2$\sigma$ \\
HeLMS-36 & 3.9802   &   0.78±0.15 &   309±67    &   0.26±0.07 &   0.42±0.11 & 45.6   &   & $<$0.15  & & 3$\sigma$ \\
HeLMS-45 & 5.3998   &   1.06±0.10 &   515±56    &   0.58±0.08 &   1.51±0.21 & 28.2   & 36.1  & $<$0.09  & $<$0.18 & 8$\sigma$ \\
HerS-11  & 4.6618   &   1.66±0.17 &   462±55    &   0.82±0.13 &   1.70±0.27 & 41.2  &   & $<$0.19 & & 7$\sigma$ \\
HerBS-78 & 3.7344  & 0.95±0.29 &    639±224 & 0.65±0.06  &  0.96±0.10  &  48.1  &   & $<$0.18  & & 3$\sigma$  \\ 
HerBS-177 & 3.9625  & 1.49±0.23 &  593±104   & 0.94±0.10  &  1.52±0.15$^{l}$   &  45.9  &   & $<$0.13  & & 5$\sigma$ \\ 
 HerBS-179 & 3.9423  & 1.23±0.38 &   385±137  & 0.50±0.05  &  0.80±0.10  &  46.1  &   & $<$0.19 &   & 2$\sigma$  \\ 
HerBS-185 & 4.3238  &   0.97±0.15 &   469±83    &   0.48±0.11 &   0.90±0.20$^{l}$ &  42.7  &   & $<$0.17  & & 4$\sigma$  \\  
\enddata
\tablecomments{Tables~\ref{tab:co10} and \ref{tab:co21}\vspace{-7mm}
\tablenotetext{a}{Given in units of ${L}_{l}$=$\rm K~km~{s}^{-1}~{pc}^{2}$.}
\vspace{-1.00mm}
\tablenotetext{\dag}{For the targets with the values given only for IF1, it represents the central frequency of a continuous 2~GHz wide band, where the IF1 and IF2 are adjacent to each other. {Please refer to Section~\ref{sec:vlaobs} for more details.}}
\vspace{-1.00mm}
\tablenotetext{\ddag}{{For continuum non-detections (SNR $<2$), we provide the 3${\sigma}_{c}$ upper-limits on the continuum flux densities (assuming an unresolved emission within a beam), where ${\sigma}_{c}$ is the rms noise of a given continuum map.}}
\vspace{-1.00mm}
\tablenotetext{{b}}{{\coonezero or \cotwoone line detection level in terms of $\sigma$, the rms noise of the integrated intensity (moment-0) map.}}
\vspace{-1.00mm}
\tablenotetext{*}{\texttt{CASA tclean} \texttt{mfs}-mode-based net central frequency for a combined continuum of individual IF1 and IF2 detections at two different central frequencies.}
\vspace{-1.00mm}
\tablenotetext{i}{{For non-detections (SNR $<2$) in CO(1$-$0), FWHM values are from \textit{z}$-$GAL mid-\textit{J} CO-line results. The upper-limit on the \coonezero line flux of HeLMS-44 has been derived using the linewidth from the \zgal \cotwoone spectrum.}}
\vspace{-1.00mm}
\tablenotetext{l}{Line luminosities and gas masses of lensed DSFG candidates. Many of these targets have known lensing factor $\mu>1$ from \citet{Borsato2024}.}
}
\end{longdeluxetable}

\renewcommand{\arraystretch}{1.15}

\begin{longdeluxetable}{ccccccccccccc}
\tabletypesize{\scriptsize} 
\tablecaption{\vzgal Pilot Program DSFGs from \citet{stanley+2023}: \coonezero Line and Underlying Continuum Properties, Molecular Gas Masses, and Gas Depletion Times adapted to ${\alpha}_{CO}=4$ $\mathrm{{M}_{\odot}~{\left(K~km~{s}^{-1}~{pc}^{2} \right)}^{-1}}$. 
\label{tab:co10pilot}}
\tablehead{
    \colhead{Source} & 
    \colhead{\rule{0pt}{2.05em}$\rm {\textit{z}}_{spec}$} & 
    \colhead{$\rm {S}_{peak}$} & 
    \colhead{$\Delta v_{FWHM}$}& 
    \colhead{$\rm {I}_{CO(1-0)}$} & 
    \colhead{$\rm \mu {L}^{\prime}_{CO(1-0)}$$^{a}$} & 
    \colhead{$\rm \left[\frac{4.0}{\alpha_{CO}}\right] \mu {M}_{\rm H_2}$} & 
    \colhead{$\rm {\tau}_{dep}$} & 
    \multicolumn{2}{c}{${\nu}_\mathrm{cont}$ (GHz)} & 
    \multicolumn{2}{c}{${S}_\mathrm{cont}$ (mJy){$^{\ddag}$}} & \colhead{Line}  \\
    &&(mJy)&($\rm km~{s}^{-1}$)&($\rm Jy~km~{s}^{-1}$)&(${10}^{11}$ $L_l$)&($\rm {10}^{11}~{M}_{\odot}$)&(Myr)&
    \colhead{IF1} & 
    \colhead{IF2} & 
    \colhead{IF1} & 
    \colhead{IF2} & \colhead{Significance$^{b}$} \\
}
\startdata
HerBS-34 &  2.633 &   0.7±0.12 &   593±112    &   0.44±0.11 &   1.5±0.4 &   6.0±1.6 &   150±40  & 32.0 &  & $<$0.03   & & 4$\sigma$ \\   
HerBS-43a & 3.212  &   0.3±0.06 &   1166±249    &   0.37±0.10 &   1.7±0.5 &   6.8±2.0 &   200±70  & 22.2 &  & 0.024±0.007   & & 5$\sigma$  \\  
HerBS-43b &  4.054 &   0.08±0.03 &   744±290    &   0.064±0.033 &   0.4±0.2 &   1.6±0.8 &   110±50  & 22.2 &  & 0.019±0.003   & & 5$\sigma$ \\   
HerBS-44 &  2.927 &   0.95±0.14 &   377±63    &   0.38±0.08 &   1.5±0.3 &   6.0±1.2 &   50±10  &  29.2 & 38.5  & $<$0.05   & $<$0.08 & 6$\sigma$ \\   
HerBS-54 & 2.442  &   0.8±0.11 &   1087±176    &   0.92±0.19 &   2.7±0.6 &   10.8±2.4 &   450±110  & 33.1 & & 0.072±0.13   & & 5$\sigma$ \\   
HerBS-58 &  2.084 &   2.12±0.32 &   363±64    &   0.82±0.19 &   1.8±0.4 &   7.2±1.6 &   410±90  &  28.8 & 37.3 & $<$0.04  & $<$0.06 & 4$\sigma$ \\   
HerBS-70E & 2.308  &   0.49±0.09 &   622±130    &   0.32±0.09 &   0.9±0.2 &   3.6±0.8 &   90±30  & 27.1 & 34.8 & 0.60±0.02   & 0.47±0.02 & 4$\sigma$ \\   
HerBS-70W &  2.311 &   0.83±0.15 &   197±39    &   0.17±0.04 &   0.5±0.1 &   2.0±0.4 &   230±120  &  27.1 & 34.8  & $<$0.03  & $<$0.03 & 5$\sigma$  \\   
HerBS-79 & 2.078  &   1.24±0.17 &   787±125    &   1.04±0.22 &   2.3±0.5$^{l}$ &   9.2±2.0$^{l}$ &   510±120  & 28.7 & 37.5 &  $<$0.03   & $<$0.04 & 5$\sigma$  \\   
HerBS-89a & 2.949  &   0.64±0.09 &   1586±247    &   1.08±0.22 &   4.4±0.9$^{l}$ &   17.6±3.6$^{l}$ &   610±130  & 29.2 & 38.5 & $<$0.06   & $<$0.15 & 5$\sigma$  \\   
HerBS-95E &  2.972 &   0.2±0.04 &   658±137    &   0.14±0.04 &   0.6±0.2 &   2.4±0.8 &   210±80  & 29.0 & 38.5 &  $<$0.02  & $<$0.06 & 5$\sigma$ \\   
HerBS-95W & 2.973  &   0.95±0.12 &   522±76    &   0.52±0.10 &   2.2±0.4 &   8.8±1.6 &   550±120  & 29.0 & 38.5 &  $<$0.02   & $<$0.06 & 7$\sigma$  \\   
HerBS-113 & 2.787  &   1.46±0.21 &   497±81    &   0.77±0.16 &   2.9±0.6$^{l}$ &   11.6±2.4$^{l}$ &   390±90  & 30.4 & 38.5 & $<$0.05   & $<$0.07 & 4$\sigma$ \\   
HerBS-154 &  3.707 &   0.95±0.11 &   384±54    &   0.39±0.07 &   2.3±0.4$^{l}$ &   9.2±1.6$^{l}$ &   120±30  & 19.2 & 24.4 &  $<$0.07  & $<$0.03 & 5$\sigma$ \\     
\enddata
\tablecomments{
\tablenotetext{a}{Given in units of ${L}_{l}$=$\rm K~km~{s}^{-1}~{pc}^{2}$.}
\vspace{-1.00mm}
\tablenotetext{{\ddag}}{{{For continuum non-detections (SNR $<2$), we provide the 3${\sigma}_{c}$ upper-limits on the continuum flux densities (assuming an unresolved emission within a beam), where ${\sigma}_{c}$ is the rms noise of a given continuum map.}}}
\vspace{-1.00mm}
\tablenotetext{{b}}{{\coonezero line detection level in terms of $\sigma$, the rms noise of the integrated intensity (moment-0) map.}}
\vspace{-1.00mm}
\tablenotetext{l}{Line luminosities and gas masses of lensed DSFG candidates. The signatures of their lensed nature can be seen in \citet{berta2021, stanley+2023}.}
}
\end{longdeluxetable}

\section{Appendix B} \label{app:b}

\subsection{Comparison of the $L_{\rm IR}$ (hence, dust temperature) and the SFE proxy of galaxy populations at cosmic noon} \label{subsec:cosmic_noon_SFEs}

\begin{figure}[h]
\centering
\includegraphics[width=0.65\textwidth]{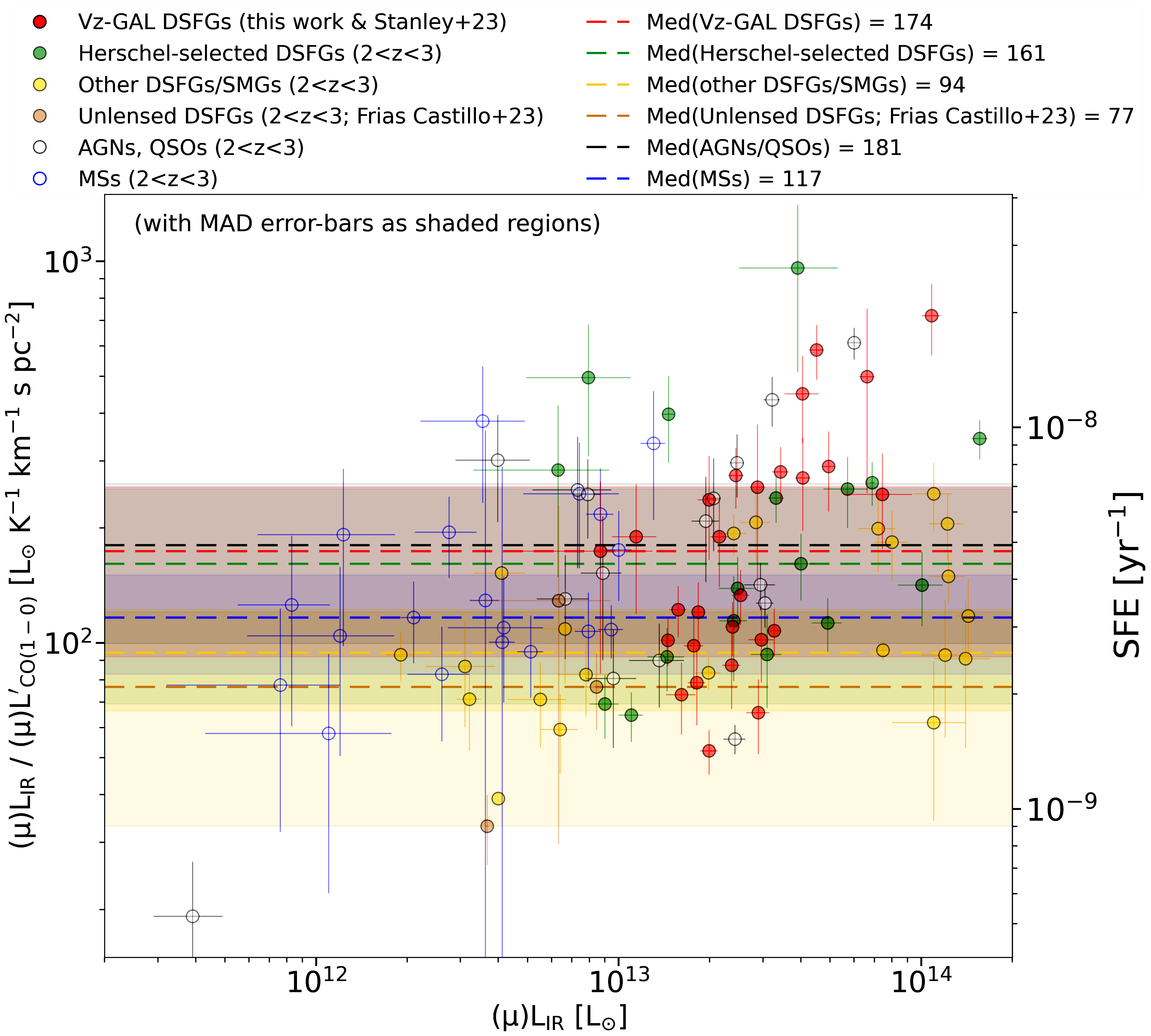}
\caption{A comparison of $L_{\rm IR}/{{L}^{\prime}}_{\rm CO(1-0)}$ ratio, a proxy for SFE, with respect to the $8-1000 \mu m$ luminosity ($L_{\rm IR}$) for various galaxy populations around cosmic noon ($z=2-3$). The median ratios for these galaxy populations are comparable within the (unscaled) median absolute deviation (MAD) error-bars, modulo a few outliers from each populations.}
\label{fig:lprimecolir_Lir_z23}
\end{figure}

\newpage
\subsection{\vzgal sources with extreme CO line ratios} \label{subsec:line_ratios_extreme}

\renewcommand{\arraystretch}{1.15}

\begin{longdeluxetable}{ccccccccccccc}
\tabletypesize{\scriptsize} 
\tablecaption{\vzgal DSFGs with Extreme Line Ratios. Most of the superthermal ratios are consistent with being $r_{J1} \leq 1$ (subthermal) within the $1\sigma$ error-bars and given that the total uncertainty due to the flux calibration is $10-15\%$ (see Section~\ref{subsec:datareduction}).
\label{tab:extreme_lineratios}}
\tablehead{
\rule{0pt}{2.05em}&\multicolumn{6}{c}{SNR $>$ 5}&\multicolumn{6}{c}{SNR $\leq$ 5}\\
&\multicolumn{3}{c}{Superthermal}&\multicolumn{3}{c}{Subthermal}& \multicolumn{3}{c}{Superthermal}&\multicolumn{3}{c}{Subthermal}\\
&\multicolumn{3}{c}{(with ${{r}}_{{J1}}>1$)}&\multicolumn{3}{c}{(with ${{r}}_{{J1}}<0.2$)}& \multicolumn{3}{c}{(with ${{r}}_{{J1}}>1$)}&\multicolumn{3}{c}{(with ${{r}}_{{J1}}<0.2$)}\\
    \colhead{Ratio} & 
    \colhead{DSFG} &
     \colhead{Value} &
       \colhead{Detection} &
      \colhead{DSFG} &
      \colhead{Value} &
           \colhead{Detection} &
    \colhead{DSFG} &
     \colhead{Value} &
       \colhead{Detection} &
      \colhead{DSFG} &
      \colhead{Value} &
           \colhead{Detection} \\
}
\startdata
${{r}}_{{21}}$  & HeLMS-1  &  1.08±0.16 &  $ 6 \sigma$ &  &   &   & HeLMS-19 &   2.14±0.66 &  $ 3 \sigma$   & HerBS-109 S &    0.19±0.10 &  $ 2 \sigma$  \\
& HeLMS-3  &  1.27±0.21 &  $ 5 \sigma$ &  &  &  & HeLMS-24 &   1.24±0.45 &  $ 4 \sigma$  &  &   &  \\
& HeLMS-14  &  1.49±0.22 &  $ 5 \sigma$ &  &  &  & HeLMS-42 &   1.24±0.43 &  $ 2 \sigma$  &  &  &  \\
& HeLMS-27  &  1.42±0.23 &  $ 8 \sigma$ &  &  & & HerS-11 &   1.28±0.68 &  $3 \sigma$  &  &   &  \\
& HeLMS-31  &  1.49±0.21 &  $ 6 \sigma$ &  &  & & HerBS-157 &   1.39±0.64 &  $ 2 \sigma$  &  &   & \\
& HeLMS-57  &  1.05±0.25 &  $ 5 \sigma$ &  &   &  &  &   &  &  &  &  \\
& HerS-2  &  1.13±0.19 &  $ 9 \sigma$ &  &   & &  &   &  &  &  & \\
& HerS-7  &  1.78±0.60 &  $ 5 \sigma$ &  &   &  &  &   & &  &   &  \\
& HerS-18E  &  1.79±0.42 &  $ 5 \sigma$ &  &  & &  &  &  &  &   & \\ 
\midrule
${{r}}_{{31}}$& HeLMS-3  &  1.04±0.17 &  $ 5 \sigma$ & HeLMS-28 &   0.17±0.03 &  $ 13 \sigma$  & HeLMS-43 &   1.61±0.86 &  $ 4 \sigma$  & HerBS-105 &   0.13±0.08 &  $ 2 \sigma$  \\
& HeLMS-14  &  1.08±0.16 &  $ 5 \sigma$ & HerBS-108 &   0.10±0.04 &  $ 8 \sigma$  & HerBS-48 &   1.26±0.42 &  $ 4 \sigma$  & HerBS-109 S &   0.19±0.10 &  $ 2 \sigma$  \\
& HeLMS-52  &  1.19±0.23 &  $ 7 \sigma$ &  &  &  & HerBS-137 &   1.14±0.71 &  $ 2 \sigma$  &  &    &   \\
& HerS-13  &  1.13±0.23 &  $ 7 \sigma$ &  &   &   &  &  & &  &    &   \\
& HerS-14  &  1.08±0.16 &  $ 5 \sigma$ &  &   &  &  &   & &  &    &    \\
& HerS-18E  &  1.20±0.27 &  $ 5 \sigma$ &  &  &  &  &   & &  &   &   \\
\hline
${{r}}_{{41}}$& HerS-14  &  1.13±0.30 &  $ 5 \sigma$ & HerBS-126 &   0.15±0.06 &  $ 5 \sigma$  & HeLMS-19 &   1.33±0.42 &  $ 3 \sigma$  & HerBS-167 &   0.19±0.09 &  $ 3 \sigma$  \\
& HerS-20  &  1.27±0.27 &  $ 10 \sigma$ &  &   &   & HeLMS-24 &   1.14±0.42 &  $ 4 \sigma$  &  &    &    \\
\midrule
${{r}}_{{51}}$&   &  &   & HeLMS-48 &   0.20±0.04 &  $ 8 \sigma$  &  &    &   & HerBS-105 &   0.06±0.04 &  $ 2 \sigma$  \\
\midrule
${{r}}_{{61}}$&   &  &  & HerBS-72 &   0.14±0.03 &  $ 8 \sigma$  &  &    &   &  &   &  \\
&   &   &   & HerBS-193 &   0.15±0.03 &  $ 6 \sigma$  &  &   &   &  &    &   \\
\enddata
\end{longdeluxetable}

\subsection{Comments on individual DSFGs with novel spectral and spatial features} \label{subsec:interesting_targets}

In this section, we comment on individual sources of the \vzgal survey, highlighting new aspects revealed by the VLA \coonezero or the continuum observations, and commenting on any characteristics found in the fields where they are located. The sources here discussed are listed in their order of appearance in the tables and figures.

\begin{itemize}[leftmargin=*, label={}, itemsep=3pt]

\item {\bf HeLMS-1} - This AGN-starburst galaxy is amplified by a foreground galaxy group at $z \sim 0.5$ \citep[seen in Hubble Space Telescope (HST) imaging;][]{Borsato2024}, into a bright northeast arc and a weaker image $\sim 8''$ to the southwest \citep{cox+2023,Borsato2024}. Only the northern image is clearly detected in the \coonezero emission line, and has $3{\sigma}_{c}$ detection at 30~GHz in the continuum emission  ($\rm 0.03\pm0.01 \, mJy$). The southern image has only a $3{\sigma}$ upper limit in \coonezero of $\rm 0.43 \, Jy \, km \, s^{-1}$, that remains compatible with the relatively weak CO(4$-$3) emission line $\rm (5.10\pm0.45 \, Jy \, km \, s^{-1})$.

\item {\bf HeLMS-3} - This source has the strongest CO(3$-$2) and \cotwoone emission lines in the \zgal survey ($\rm 11.24\pm0.86$ and $\rm 20.72\pm1.55 \, Jy \, km \, s^{-1}$, respectively) and also the strongest \coonezero emission line of the entire \vzgal sample with $\rm 2.22\pm0.33 \, Jy \, km \, s^{-1}$. HeLMS-3 is gravitationally amplified by a foreground galaxy group seen in HST F110W imaging \citep{Borsato2024}. The north-south extension, that was apparent in the NOEMA data, is clearly seen in the \vzgal \coonezero moment-0 map, which shows a strong image to the south and a weaker one to the north.

\item {\bf HeLMS-14} - The \coonezero emission of HeLMS-14 is resolved into two arc-like structures located east and west with a separation of $\sim 4''$. This source is gravitationally lensed by a foreground galaxy at $z=0.26$ seen in the HST F110W image that also reveals an almost complete ring of the background source with a radius of $r \sim 4''$ with clear sub-structures \citep{Borsato2024}. The two arcs seen in the \coonezero emission follow the extent of the rest-frame optical ring structure.

\item {\bf HeLMS-17} - The field of HeLMS-17 contains a single source to the East with a spectroscopic redshift of $z=2.2983$ and a double source to the West, with two close-by ($\sim 4''$) components (W1 and W2) that have the same redshift $z=2.2972$ \citep{cox+2023}. The Western source is aligned with a galaxy seen in the HST F110W image and is therefore probably lensed. Both the Eastern and Western sources are detected in the \coonezero emission line with broad profiles that are comparable to the higher-$J$ CO lines ($\rm \Delta v=1312\pm273$ and $\rm 816\pm117 \, km \, s^{-1}$ for the E and W source, respectively). The Western source shows a distinct, although weak, \coonezero extension to the East that corresponds to the position of the second component (W2). The continuum emission is detected at 27~GHz in the Western source with a flux density of $\rm 0.16\pm0.03 \, mJy$, but not at 35~GHz with an upper limit of $\sigma_{\rm c}=0.11 \, \rm mJy$.

\item {\bf HeLMS-19} - This source is lensed by a foreground $z$=0.14 galaxy into a large Einstein ring with two main arcs, separated by $\sim 5''$, that are partially traced in {\it HST} F110W imaging \citep{Borsato2024} . The two arcs were detected in multiple emission lines (CO(4$-$3), (5$-$4) and (7$-$6) and in the two [CI] transitions) and in the dust continuum with NOEMA \cite[][]{cox+2023}. Both arcs are also seen in \coonezero and CO(2$-$1). The strong continuum emission, that is detected at 41.1~GHz ($\rm 0.87\pm0.10 \, mJy$) and 20.7~GHz ($\rm 1.18\pm0.06 \, mJy$), peaks at the center of the Einstein ring and most likely traces the radio continuum of the foreground galaxy.

\item {\bf HeLMS-20} - The \coonezero emission line of this source shows a well-defined, broad ($\rm 990\pm100 \, km \, s^{-1}$) double Gaussian profile. Similar profiles are found in the CO(4$-$3) and [CI](1$-$0) emission lines from the \zgal survey, although the weaker red-shifted component was not identified in \cite{cox+2023}. As a result, we revise the spectroscopic redshift of this source ($z=2.1947$) reported by \citet{cox+2023} to $z=2.1975$.

\item {\bf HeLMS-26} - The two sources that were detected in the higher-$J$ CO lines to the East and the West around the \textit{Herschel} position, separated by $\sim 12''$ (corresponding to a linear distance of $\rm \sim 100 \, kpc$), are clearly detected in \coonezero displaying the similar broad and narrow line widths as in the high-$J$ CO lines for the West and East sources, respectively.

\item {\bf HeLMS-32} - The weak central source, that was detected in \cotwoone and CO(3$-$2), is also seen in the \coonezero emission line. At the same redshift, to the north, two other sources, N1 and N2, are tentatively detected in \coonezero emission. None of these sources were detected in the \zgal survey \citep{cox+2023}.

\item {\bf HeLMS-40} - Two sources were detected in the NOEMA field, W1 \& W2, separated by $\sim 4''$, at slightly different redshifts of $z=3.1445$ and $z=3.1395$ \citep{cox+2023}. We detect both in \coonezero emission with similar profiles as in the higher-$J$ CO transitions. The source W2 is aligned with a visible galaxy ($z=0.821$) seen in the HST F110W imaging \citep{nayyeri+2016lensed,Borsato2024}. Both sources have also been been observed in the $\rm 873 \, \mu m$ continuum with ALMA \citep{Amvrosiadis2018} and both W1 and W2 have recently each been resolved into two components (Bakx et al. in prep.).

\item {\bf HeLMS-47} - This source is resolved in the \coonezero emission line showing two strong peaks to the north-east and the south-west separated by $\sim 4''$, in line with the extension that was seen in the NOEMA data \citep[][]{cox+2023}.

\item {\bf HeLMS-49} - The source is resolved in \coonezero, revealing an emission peak at the center and a slightly weaker peak to the southwest, separated by $\sim 5''$. Aligned with a galaxy seen in the {\it HST} F110W image \citep[][]{Borsato2024}, HeLMS-49 is likely gravitationally amplified. The spectroscopic redshift of HeLMS-49 was measured based on the detection of water and atomic carbon emission lines \citep[][]{cox+2023}, preventing to date an analysis of CO for this galaxy.

\item {\bf HeLMS-50} - The \coonezero emission line profile of this source shows a well-defined, broad ($\rm 1837\pm139 \, km \, s^{-1}$) double Gaussian. A close comparison with the profiles of the CO(3$-$2) and (4$-$3) emission lines from the \zgal survey shows that the profiles of all three lines are comparable, although the weaker blue-shifted component was not identified by \cite{cox+2023}. As a result, the spectroscopic redshift of this source is revised from $z=2.0532$ to $z=2.0515$.

\item {\bf HeLMS-51} - This source is lensed as it is aligned with a galaxy seen in {\it HST} F110W imaging \citep[][]{Borsato2024}. To the west of HeLMS-51, a source is detected in the 28~GHz continuum with a flux density of $\rm 0.11\pm0.03 \, mJy$. Aligned with another galaxy seen in the {\it HST} F110W image, this source is unrelated to HeLMS-51.

\item {\bf HeLMS-56} - The main central source at $z=3.391$, that was detected in \zgal, is clearly detected in \coonezero emission, displaying a similar double Gaussian line profile as the high-$J$ CO transitions. This central source is aligned with a galaxy seen in {\it HST} F110W imaging \citep[][]{Borsato2024} and is therefore likely lensed. Two additional sources in the field of HeLMS-56, that are at the same redshift as the central source, are detected in \coonezero -- one $12''$ to the north and the other $13''$ to the west, corresponding to linear distances of $\rm \sim 100 \, kpc$. Both sources have comparable line fluxes and display single Gaussian profiles, contrary to the central source.

\item {\bf HerS-3} - This source is an Einstein cross with a fifth central image that is described in detail by \citet{cox2025hers3}. The \coonezero emission line clearly reveals the strong northern and southern images, while the three images along the east-west axis are barely resolved apart. The continuum is also detected, mostly along the north-south elongation, at 28~GHz but remain undetected at 32.5~GHz.

\item {\bf HerS-7} - The \coonezero emission of this source is dominated by a strong central component, which was also detected in the \cotwoone and CO(4$-$3) emission lines \citep{cox+2023}. The second weaker component, $\rm \sim 3.5''$ to the East, is also detected in \coonezero with a slightly red-shifted velocity of $\rm \sim 250 \, km \, s^{-1}$. The two other sources in the field that were only detected in the millimeter continuum emission, one  $\rm \sim 11''$ to the North and a weaker source  $\rm \sim 12''$ to the West, remain undetected in the current \vzgal data.

\item {\bf HerS-10} - This source is aligned with a bright galaxy seen in {\it HST} F110W imaging \citep{Borsato2024} and is gravitationally amplified. The \coonezero emission displays the same double-peaked profile as CO(3$-$2) and shows an extension in the east-west direction, that could be compatible with an arc-like structure or two sides of an incomplete Einstein ring, as is seen in ALMA 350~GHz imaging.

\item {\bf HerS-13} - This source is spatially unresolved in CO(1$-$0). Its \coonezero line shape is interesting, given that it shows signatures of a potential differential magnification across the blue and red sides of the spectrum. \textit{HST} F110W imaging \citep{Borsato2024} of the field shows a foreground galaxy that is lensing HerS-13, which is also consistent with an incomplete Einstein ring seen in ALMA 350~GHz imaging.

\item {\bf HerS-18} - This source also has a spatial component in the west \citep[see][]{cox+2023}, which is at $z=0.56$; hence, outside our observational range for the \coonezero transition.

\item {\bf HerBS-91} - The \coonezero moment-0 map reveals two east and central sources that were also detected in the higher-$J$ CO lines at the same redshift \citep[$z=2.4047$][]{cox+2023}. To enhance the S/N, the spectra of both components have been combined. 

\item {\bf HerBS-109} - Three distinct components were detected for this source in the $z$-GAL survey: the north-west and south components, separated by $12''$ and having similar redshifts of $z=1.5850$ and 1.5843, respectively; and a component to the north-east at $z=2.8385$ that is unrelated to the binary system \citep{cox+2023}. The \coonezero emission line is only detected in the southern component and remained undetected in the north-west component. Further, the receiver band used for these observations did not include the frequency of the red-shifted \coonezero emission line of the north-east component. 

\item {\bf HerBS-116} - This source was resolved into two components in the 2mm continuum emission, located in the east and west, and separated by $\sim 4''$, that are both at the same redshift $z=3.1573$ \citep{cox+2023}. The \coonezero  emission peaks at the stronger east component and shows an extension towards the west that coincides with the second weaker component of this system. Since no galaxy is visible in \textit{HST} FW110 imaging or in the SDSS \citep{Borsato2024}, this system is unlikely to be lensed.

\item {\bf HerBS-162} - The \coonezero emission traces two components to the southwest and northeast, separated by $\sim 7''$, which were also detected in CO(3$-$2) and (4$-$3). They have slightly different redshifts of $z=2.4739$ and 2.4742, respectively \citep{cox+2023}.

\item {\bf HerBS-183} - This source is located at the western edge of a bright galaxy visible in {\it HST} F110W imaging \citep{Borsato2024}. The \coonezero emission line has a broad double-Gaussian profile that is similar to those of \cotwoone and CO(4$-$3). The strong radio continuum at 35~GHz, with a flux density of $\rm 0.39\pm0.02 \, mJy$, is shifted to the east and aligned with the central region of the foreground galaxy, indicating that the continuum emission is most likely related to the lensing galaxy.

\item {\bf HerBS-187} - Two sources separated by $\sim~4''$, eastern and western, were detected in this field \citep{cox+2023} in the \cotwoone and (4$-$3) emission lines. They also detect an additional weaker southern source in the millimeter continuum. Of these, we only detect the western source in \coonezero.

\item {\bf HerBS-188} - This source displays a broad \coonezero emission line with a shape comparable to the CO(3$-$2) and CO(5$-$4) profiles \citep{cox+2023}. Interestingly, the strong radio continuum, with a flux density of $\rm 0.33\pm0.03 \, mJy$, is offset from the \coonezero peak by $\sim 3''$. Although there is no visible galaxy seen in {\it HST} F110W imaging, the VLASS image shows a 3~GHz radio source in the foreground that is aligned with this line of sight.

\item {\bf HerBS-194} - Of the two sources that were detected in CO(3$-$2) and CO(4$-$3), only the northern source with stronger CO emission lines, is detected in CO(1--0). The southern source has an upper $3\sigma$ limit that is compatible with the finding that the CO(3$-$2) line flux is twice weaker than the northern source \citep{cox+2023}.

\item {\bf HerBS-204} - This system is the only case in the \zgal sample that shows two galaxies that are separated by $\sim 6''$ and are linked by a bridge of matter, with all components detected in the dust continuum and the emission lines CO(4$-$3) and (5$-$4) \citep{neri+2020,cox+2023}. The total extent of this system is $\rm \sim 45 \, kpc$. The \coonezero emission line is clearly detected in both the eastern and western galaxies, with profiles that are comparable to the higher-$J$ CO transitions. However, no \coonezero emission could be traced in the region in between the two galaxies. 

\end{itemize}

\newpage

\subsection{Integrated intensity (moment-0) maps and spectra of \coonezero and CO(2$-$1) of the \vzgal DSFGs} \label{subsec:mom0_spec_vzgal}

{We here provide an overview of all \coonezero and \cotwoone integrated intensity maps and spectra for the newly detected \vzgal DSFGs from this work. Sources in the HeLMS and HerS fields are shown in Figures~\ref{fig:image-grid_helms_v1}~and~\ref{fig:image-grid_helms_v2_hers}, and those in the H-ATLAS fields are shown with their HerBS IDs in Figure~\ref{fig:image-grid_herbs}. An overview of the pilot sources is shown in Figure~1 of \citet{stanley+2023}.}

\begin{figure*}[!htbp]
\centering

\begin{minipage}{0.235\textwidth}
    \centering
    \includegraphics[width=\textwidth]{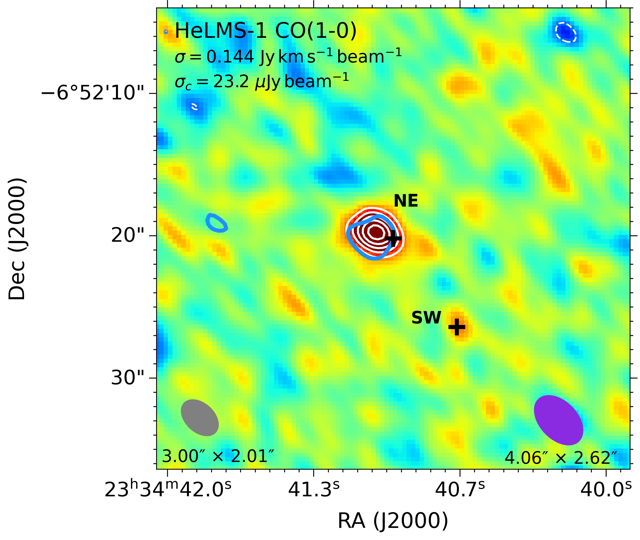}
\end{minipage}
\hfill
\begin{minipage}{0.235\textwidth}
    \centering
    \includegraphics[width=\textwidth]{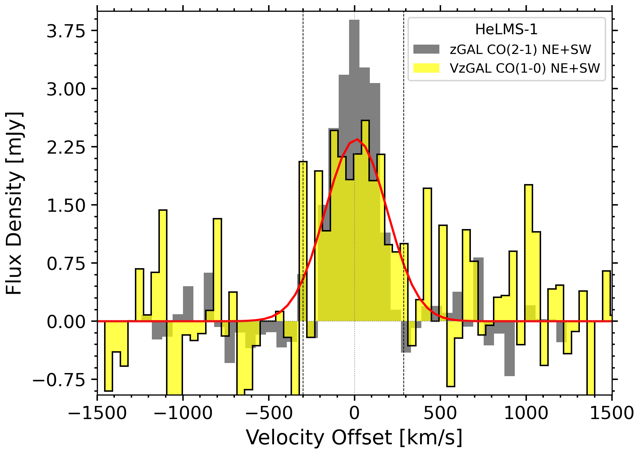}
\end{minipage}
\hfill
\begin{minipage}{0.235\textwidth}
    \centering
    \includegraphics[width=\textwidth]{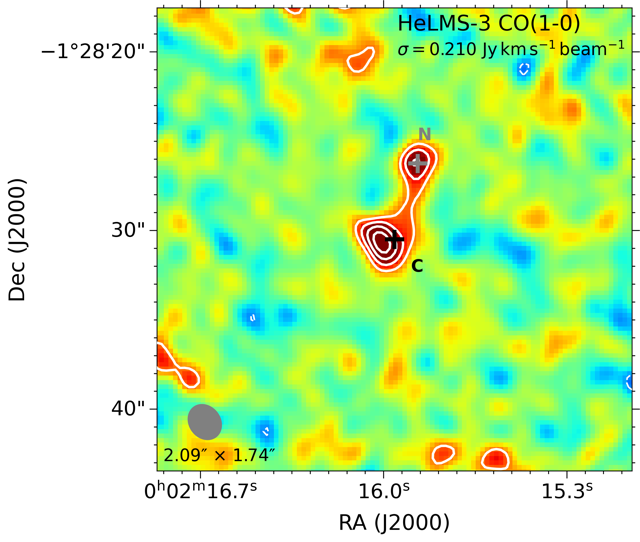}
\end{minipage}
\hfill
\begin{minipage}{0.235\textwidth}
    \centering
    \includegraphics[width=\textwidth]{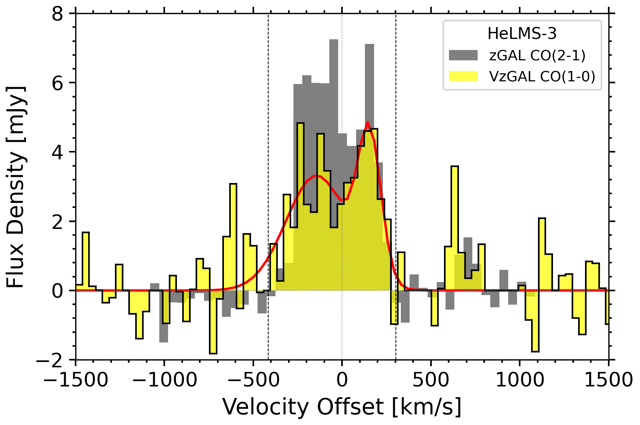}
\end{minipage}

                                        \vspace{1em}

\begin{minipage}{0.235\textwidth}
    \centering
    \includegraphics[width=\textwidth]{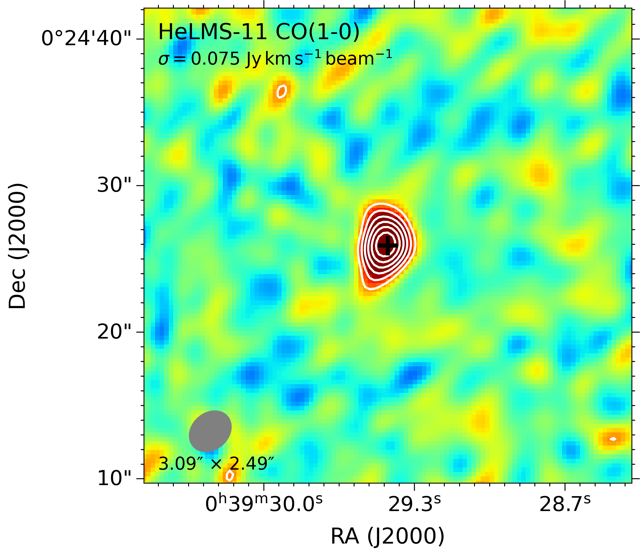}
\end{minipage}
\hfill
\begin{minipage}{0.235\textwidth}
    \centering
    \includegraphics[width=\textwidth]{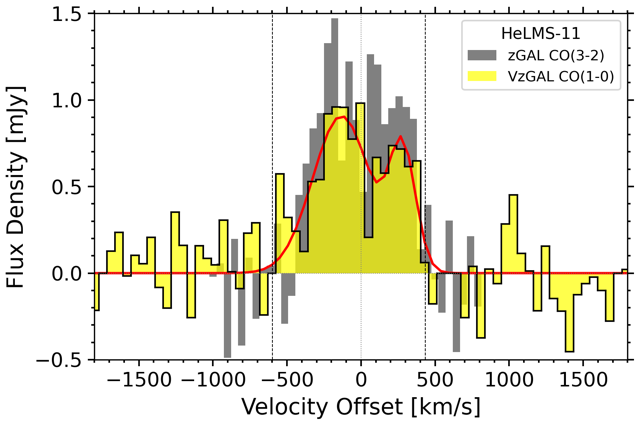}
\end{minipage}
\hfill
\begin{minipage}{0.235\textwidth}
    \centering
    \includegraphics[width=\textwidth]{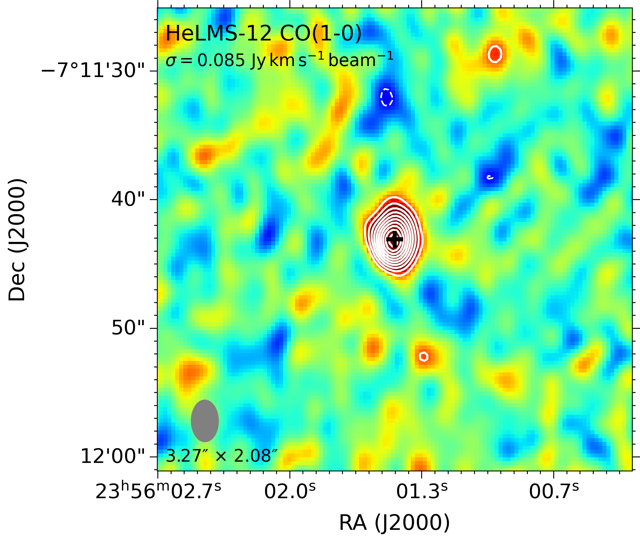}
\end{minipage}
\hfill
\begin{minipage}{0.235\textwidth}
    \centering
    \includegraphics[width=\textwidth]{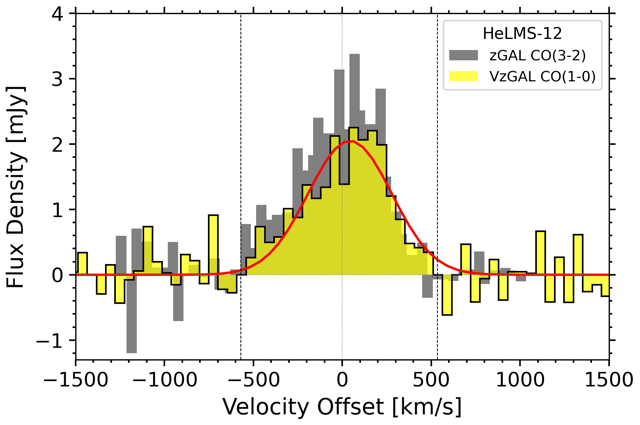}
\end{minipage}

                                        \vspace{1em}

\begin{minipage}{0.235\textwidth}
    \centering
    \includegraphics[width=\textwidth]{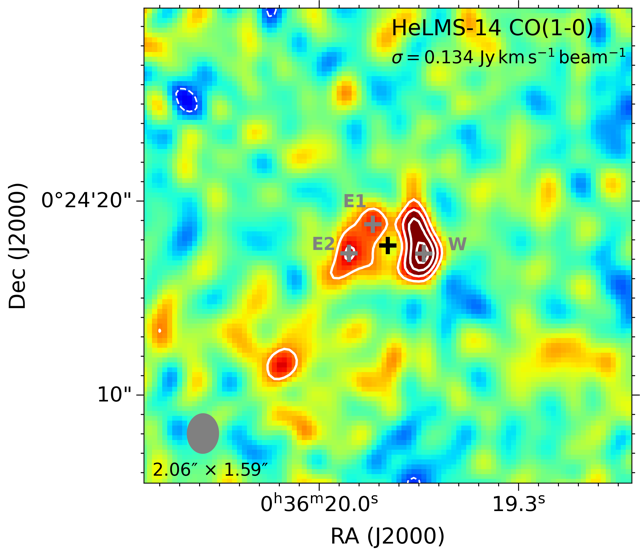}
\end{minipage}
\hfill
\begin{minipage}{0.235\textwidth}
    \centering
    \includegraphics[width=\textwidth]{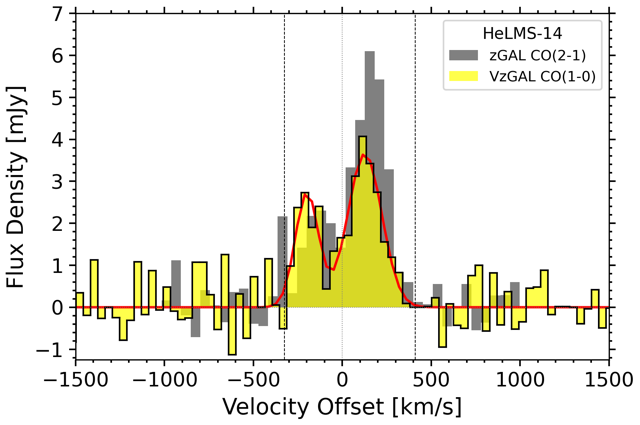}
\end{minipage}
\hfill
\begin{minipage}{0.235\textwidth}
    \centering
    \includegraphics[width=\textwidth]{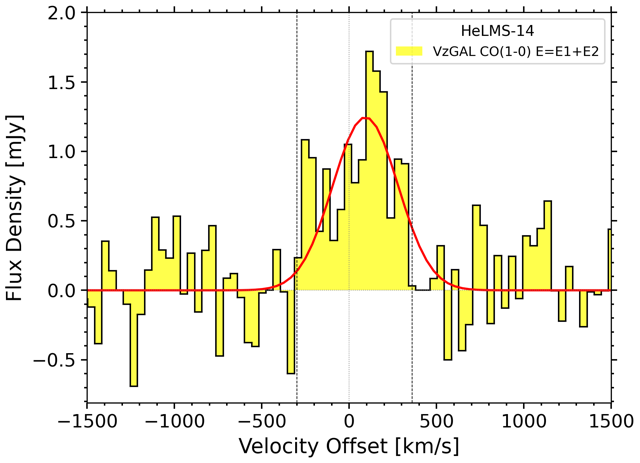}
\end{minipage}
\hfill
\begin{minipage}{0.235\textwidth}
    \centering
    \includegraphics[width=\textwidth]{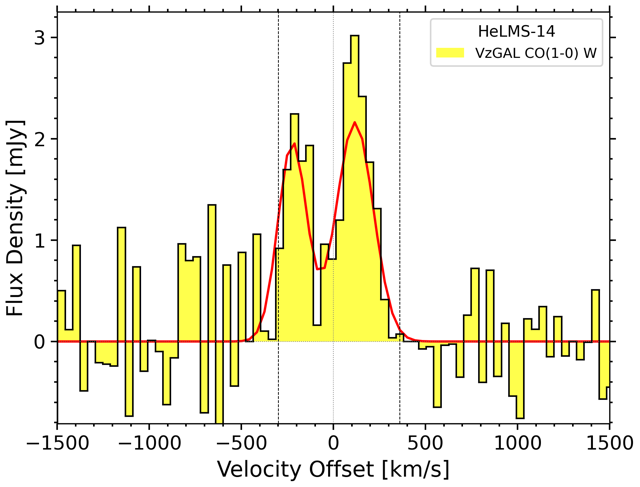}
\end{minipage}

    \addtocounter{figure}{-1}

\caption{Moment-0 maps and spectra of low-\textit{J} CO emission, namely \coonezero and CO(2--1), of the \vzgal galaxies (HeLMS-1 to HeLMS-48 but HeLMS-47 here). We provide the rms noise ($\sigma$) of each moment-0 map along with the synthesized beam size (gray ellipses) in the image itself. For targets with lensed images or multiplicity, we distinctly add cross-markers identifying different spatial components on these moment-0 maps. Otherwise the cross markers represent the central position of the DSFG as found in the \zgal survey \citep{cox+2023}. The CO emission is shown using solid/dashed white (positive/negative) contours, starting at $\pm 2\sigma$. These contours vary in the steps of $1 \sigma$. For objects with detected continuum emission above $2{\sigma}_{c}$, we show continuum flux density levels as blue contours. In the cases where we detect a serendipitous radio source along the line of sight or in the field, we show its continuum with magenta contours instead. Please refer to Appendix~\ref{app:a} (Section~\ref{subsec:cont}) for more details. The synthesized beam (purple ellipses) and noise level ($\sigma_{c}$) of continuum maps are depicted in the images. The extracted spectra of our \coonezero observations (see Section~\ref{subsec:spec}), depicted in yellow, are compared to higher-\textit{J} CO lines shown in gray \citep{cox+2023}. For ten \vzgal DSFGs with additional \cotwoone detection, we also present the \cotwoone spectra alongside the \coonezero and higher-\textit{J} CO lines. Each line spectra from \vzgal detections is fitted with Gaussian profile(s) as red curves, with their fitting parameters shown in Appendix~\ref{app:a} (Tables~\ref{tab:co10} and \ref{tab:co21}). The spectra in gray are scaled to $1.5\times$ the peak of the yellow spectrum. Black vertical dashed lines over the extracted spectra indicate the velocity range over which the moment-0 map of the CO line was integrated. Spectra of individual spatial components of lensed/multiple DSFGs are also depicted, wherever applicable.} 
\label{fig:image-grid_helms_v1}
\end{figure*}

\clearpage

\begin{figure*}[!htbp]
\centering

\begin{minipage}{0.235\textwidth}
    \centering
    \includegraphics[width=\textwidth]{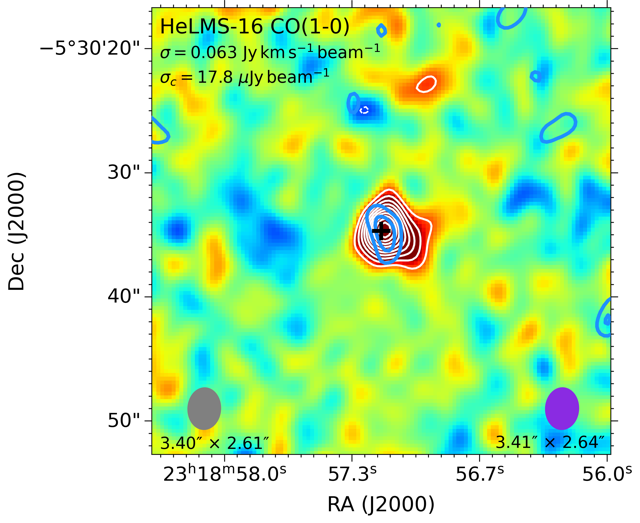}
\end{minipage}
\hfill
\begin{minipage}{0.235\textwidth}
    \centering
    \includegraphics[width=\textwidth]{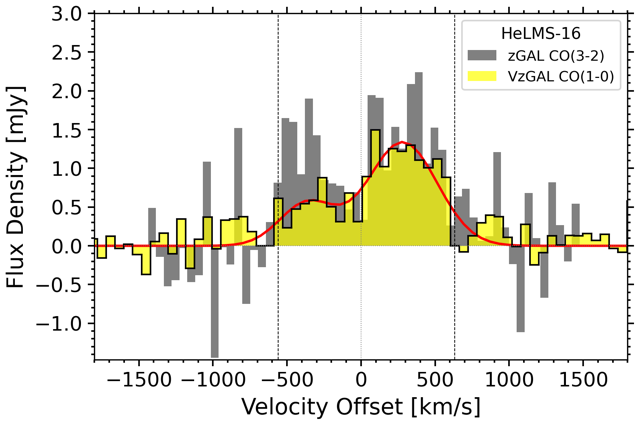}
\end{minipage}
\hfill
\begin{minipage}{0.235\textwidth}
    \centering
    \includegraphics[width=\textwidth]{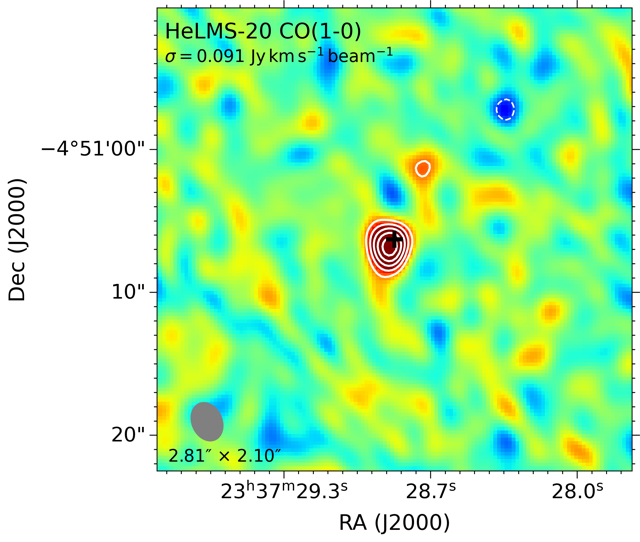}
\end{minipage}
\hfill
\begin{minipage}{0.235\textwidth}
    \centering
    \includegraphics[width=\textwidth]{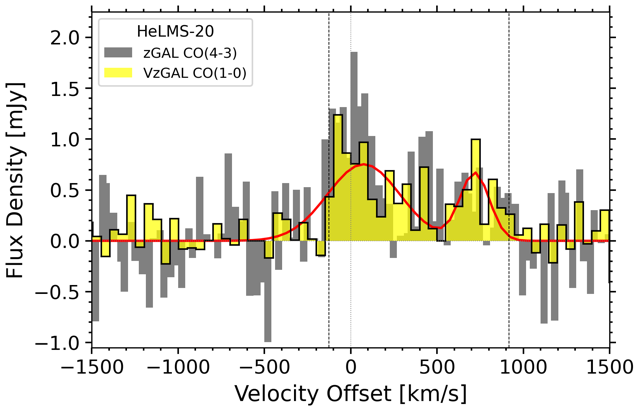}
\end{minipage}

                    \vspace{1em}

\begin{minipage}{0.235\textwidth}
    \centering
    \includegraphics[width=\textwidth]{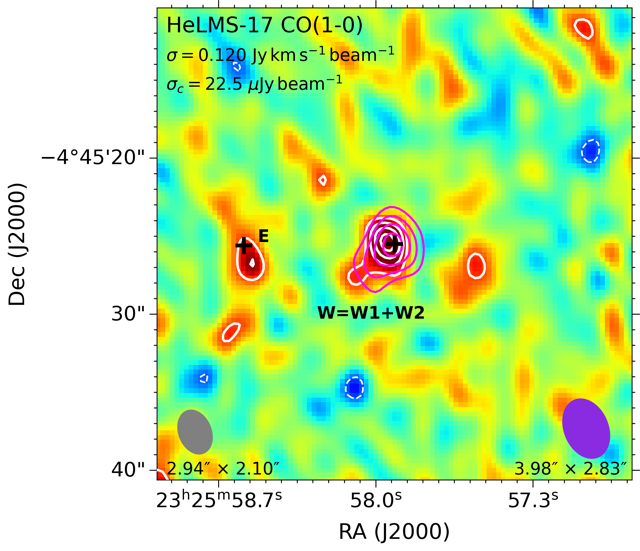}
\end{minipage}
\hfill
\begin{minipage}{0.235\textwidth}
    \centering
    \includegraphics[width=\textwidth]{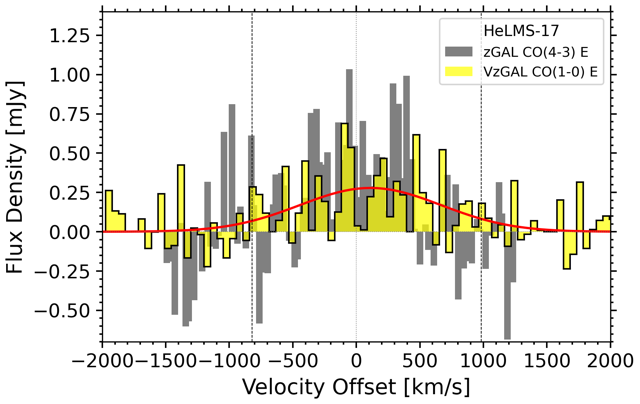}
\end{minipage}
\hfill
\begin{minipage}{0.235\textwidth}
    \centering
    \includegraphics[width=\textwidth]{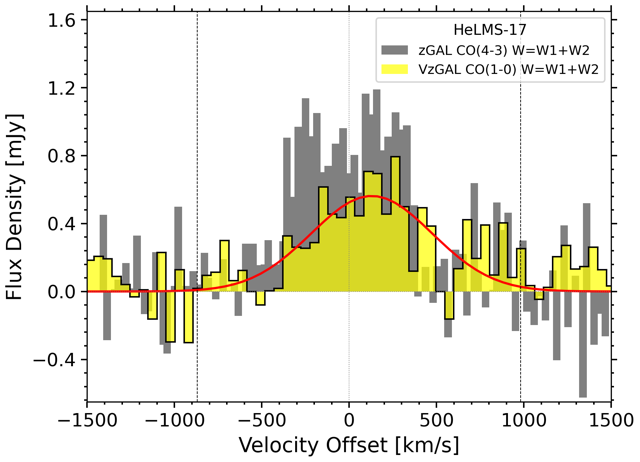}
\end{minipage}

                                        \vspace{1em}

\begin{minipage}{0.235\textwidth}
    \centering
    \includegraphics[width=\textwidth]{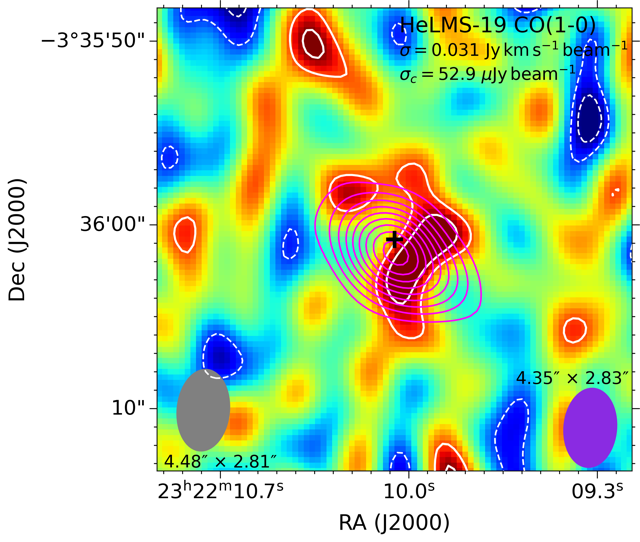}
\end{minipage}
\hfill
\begin{minipage}{0.235\textwidth}
    \centering
    \includegraphics[width=\textwidth]{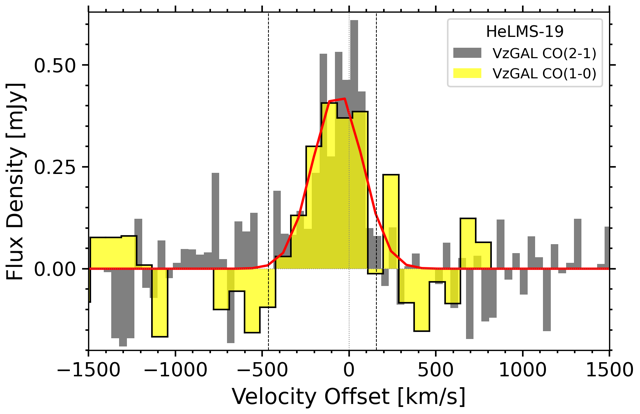}
\end{minipage}
\hfill
\begin{minipage}{0.235\textwidth}
    \centering
    \includegraphics[width=\textwidth]{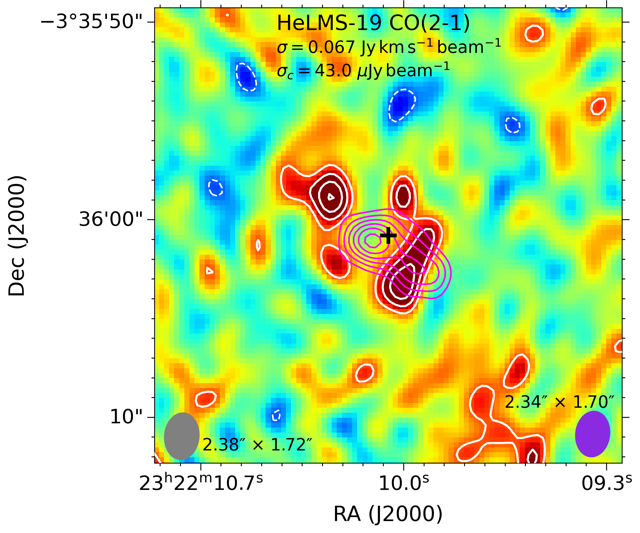}
\end{minipage}
\hfill
\begin{minipage}{0.235\textwidth}
    \centering
    \includegraphics[width=\textwidth]{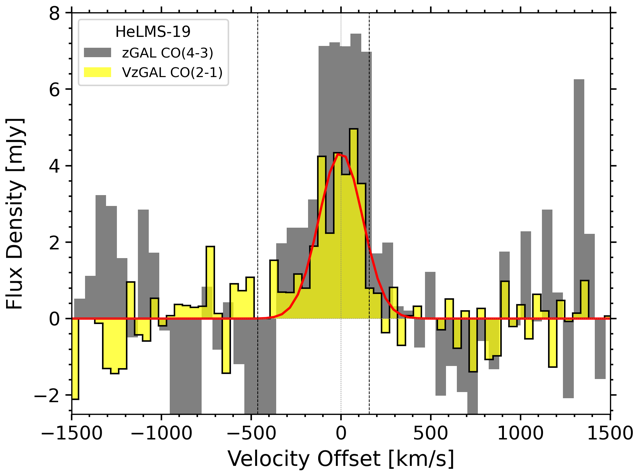}
\end{minipage}

                                        \vspace{1em}

\begin{minipage}{0.235\textwidth}
    \centering
    \includegraphics[width=\textwidth]{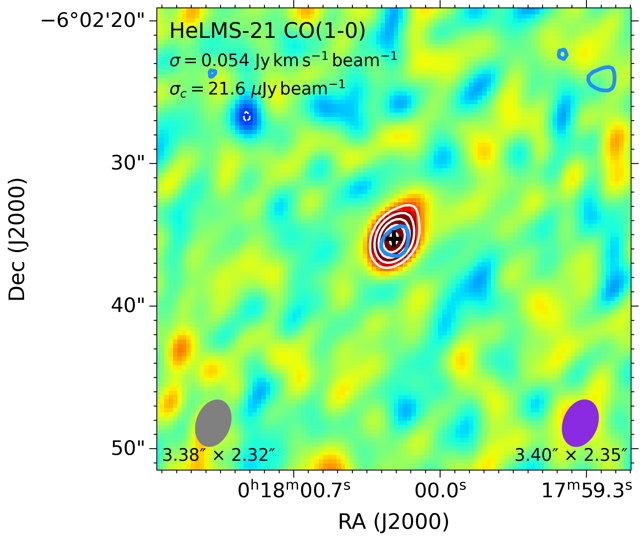}
\end{minipage}
\hfill
\begin{minipage}{0.235\textwidth}
    \centering
    \includegraphics[width=\textwidth]{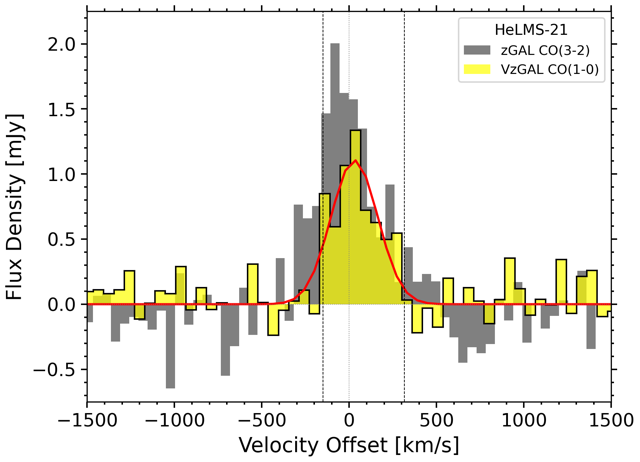}
\end{minipage}
\hfill
\begin{minipage}{0.235\textwidth}
    \centering
    \includegraphics[width=\textwidth]{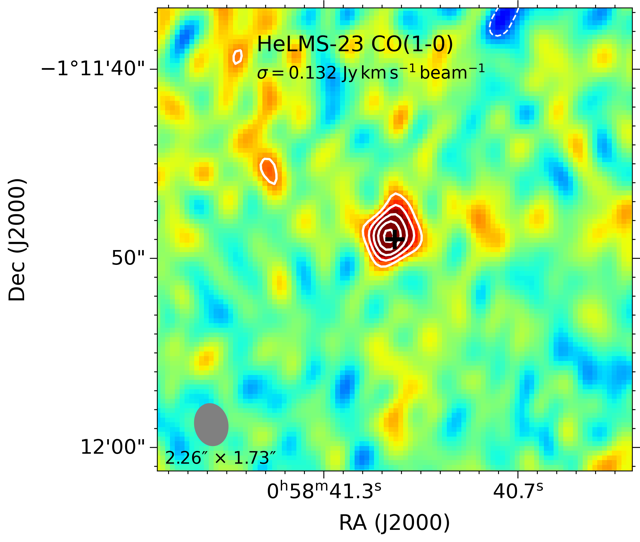}
\end{minipage}
\hfill
\begin{minipage}{0.235\textwidth}
    \centering
    \includegraphics[width=\textwidth]{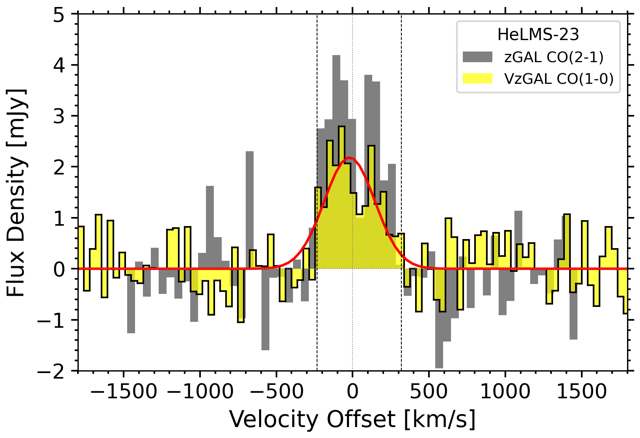}
\end{minipage} 

                                        \vspace{1em}

\begin{minipage}{0.235\textwidth}
    \centering
    \includegraphics[width=\textwidth]{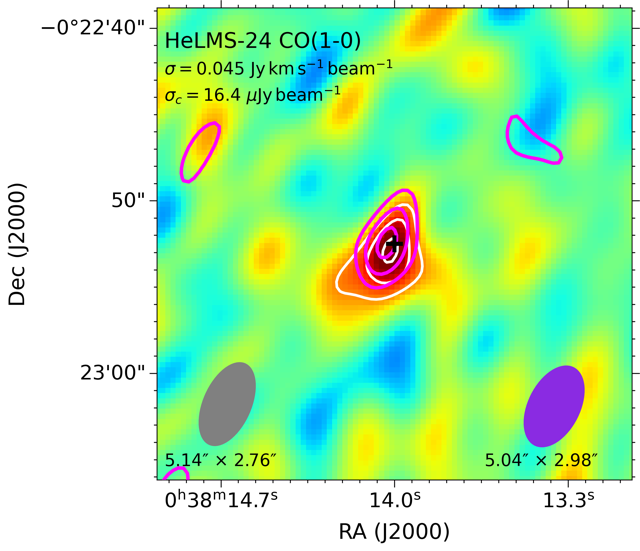}
\end{minipage}
\hfill
\begin{minipage}{0.235\textwidth}
    \centering
    \includegraphics[width=\textwidth]{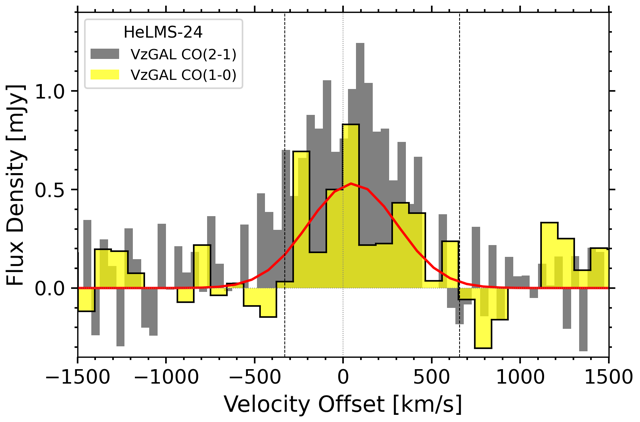}
\end{minipage}
\hfill
\begin{minipage}{0.235\textwidth}
    \centering
    \includegraphics[width=\textwidth]{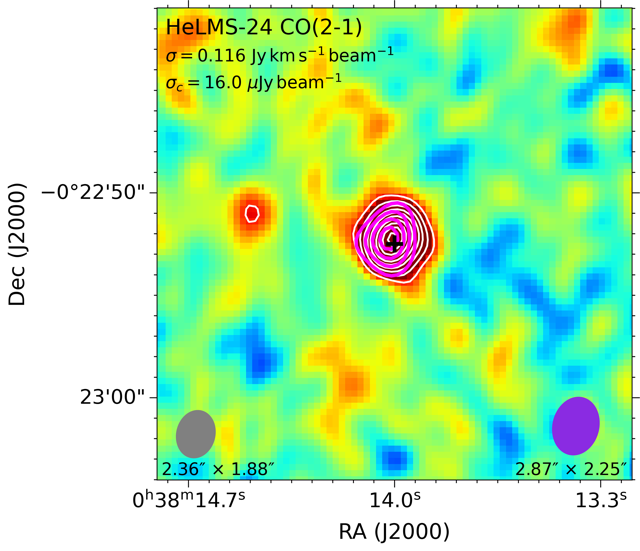}
\end{minipage}
\hfill
\begin{minipage}{0.235\textwidth}
    \centering
    \includegraphics[width=\textwidth]{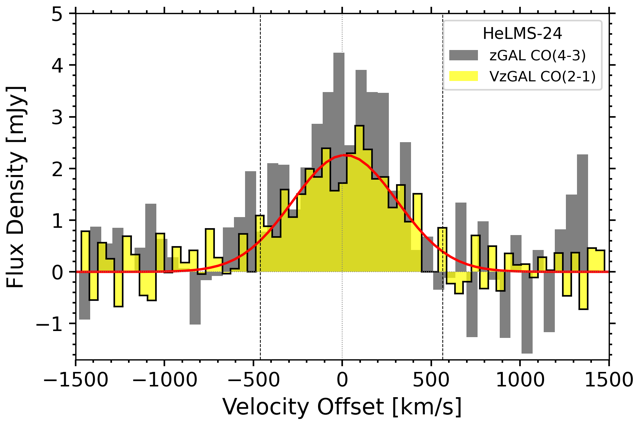}
\end{minipage}

                                        \vspace{1em}

\begin{minipage}{0.235\textwidth}
    \centering
    \includegraphics[width=\textwidth]{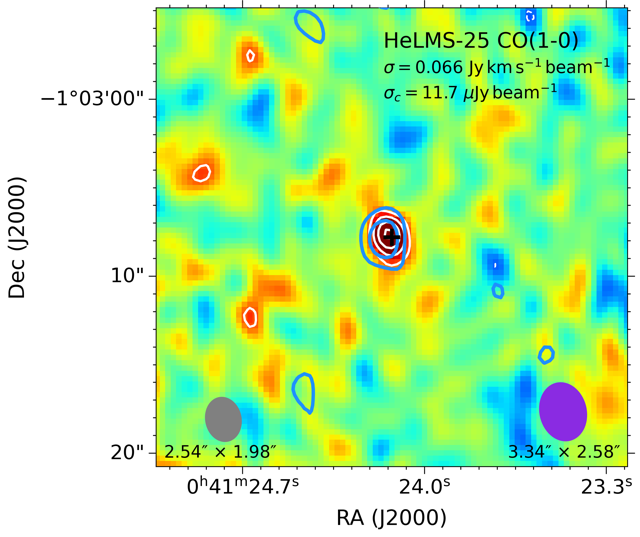}
\end{minipage}
\hfill
\begin{minipage}{0.235\textwidth}
    \centering
    \includegraphics[width=\textwidth]{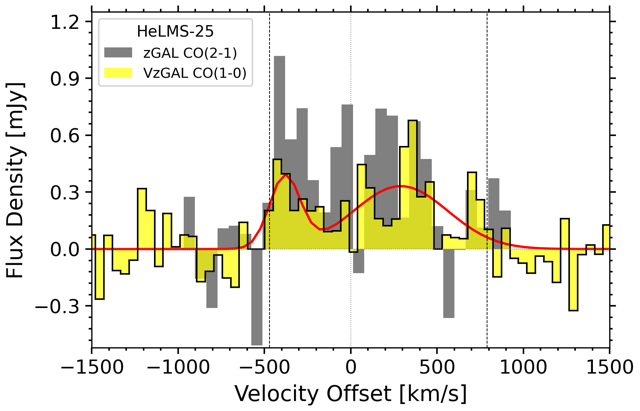}
\end{minipage}
\hfill
\begin{minipage}{0.235\textwidth}
    \centering
    \includegraphics[width=\textwidth]{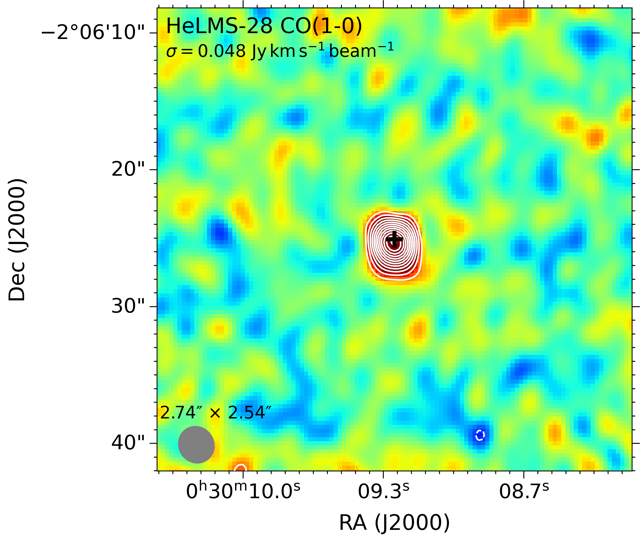}
\end{minipage}
\hfill
\begin{minipage}{0.235\textwidth}
    \centering
    \includegraphics[width=\textwidth]{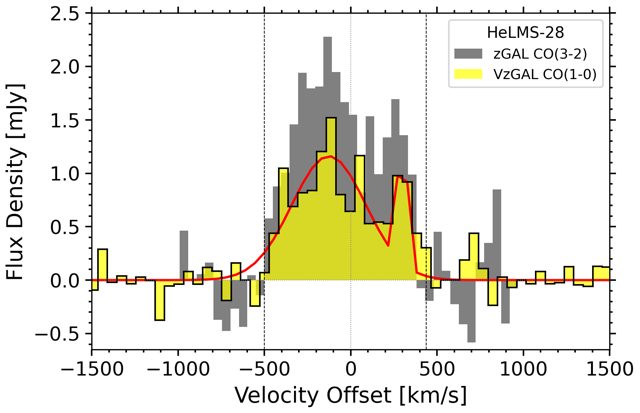}
\end{minipage}

    \addtocounter{figure}{-1}
\caption{(continued)}
\end{figure*}

\clearpage

\begin{figure*}[!htbp]
\centering

\begin{minipage}{0.235\textwidth}
    \centering
    \includegraphics[width=\textwidth]{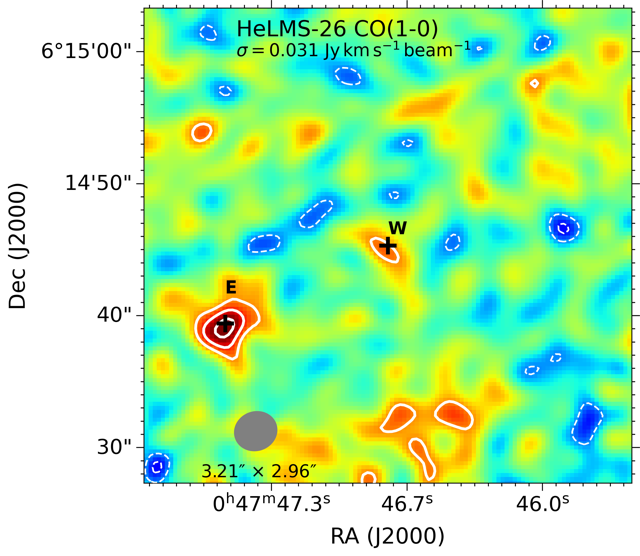}
\end{minipage}
\hfill
\begin{minipage}{0.235\textwidth}
    \centering
    \includegraphics[width=\textwidth]{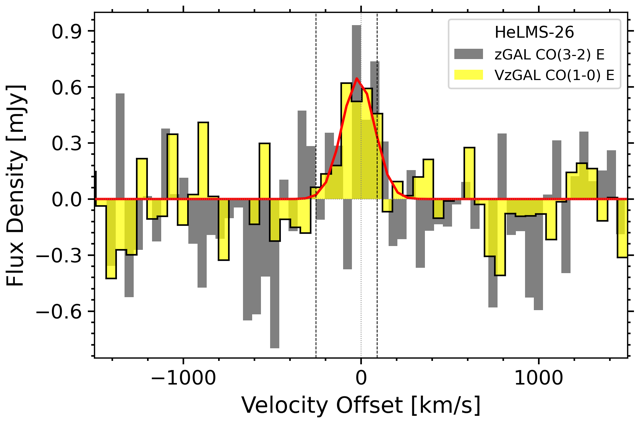}
\end{minipage}
\hfill
\begin{minipage}{0.235\textwidth}
    \centering
    \includegraphics[width=\textwidth]{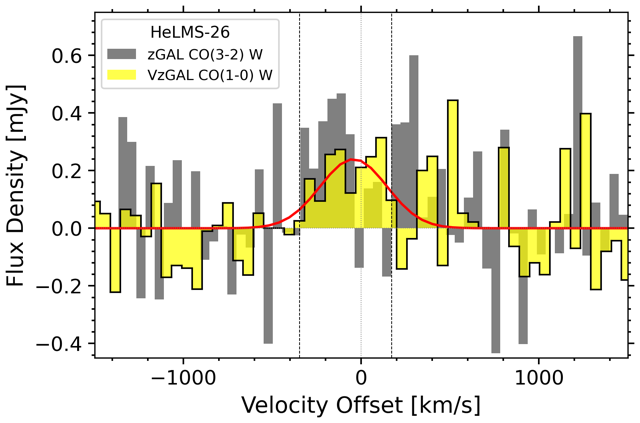}
\end{minipage}

                \vspace{1em}

\begin{minipage}{0.235\textwidth}
    \centering
    \includegraphics[width=\textwidth]{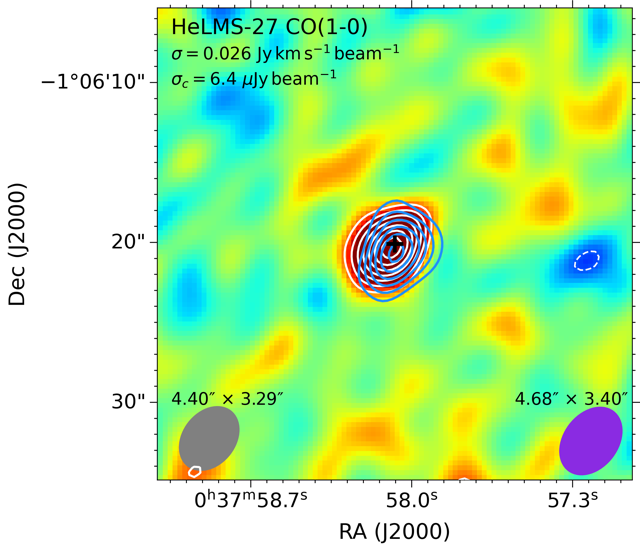}
\end{minipage}
\hfill
\begin{minipage}{0.235\textwidth}
    \centering
    \includegraphics[width=\textwidth]{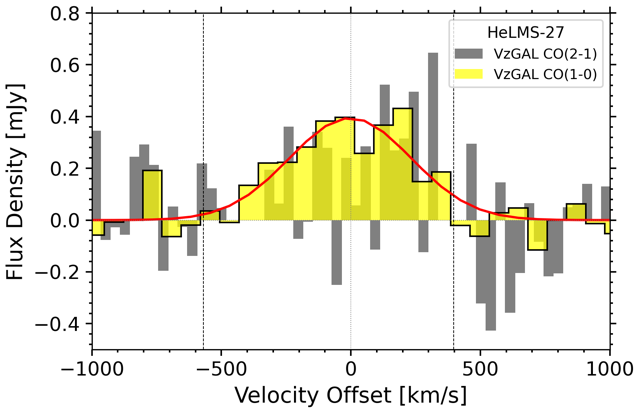}
\end{minipage}
\hfill
\begin{minipage}{0.235\textwidth}
    \centering
    \includegraphics[width=\textwidth]{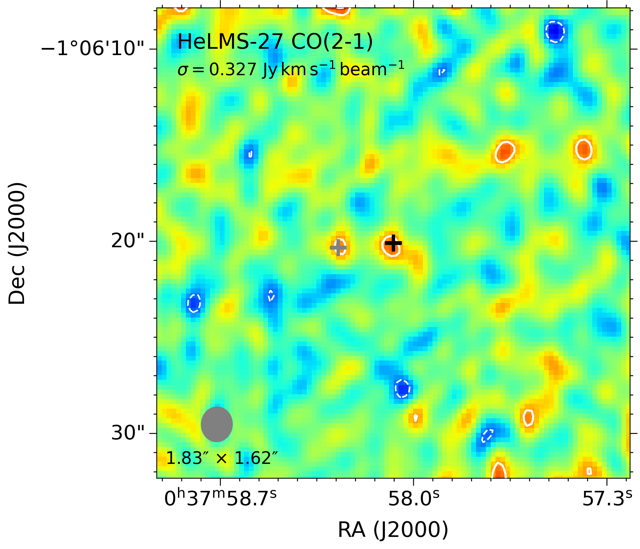}
\end{minipage}
\hfill
\begin{minipage}{0.235\textwidth}
    \centering
    \includegraphics[width=\textwidth]{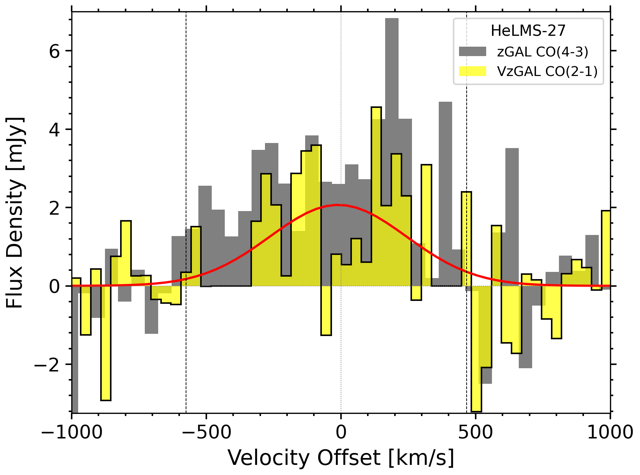}
\end{minipage}

                                        \vspace{1em}

\begin{minipage}{0.235\textwidth}
    \centering
    \includegraphics[width=\textwidth]{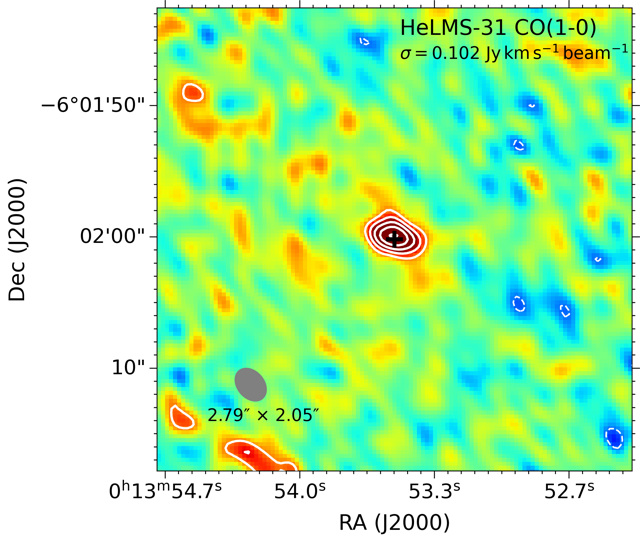}
\end{minipage}
\hfill
\begin{minipage}{0.235\textwidth}
    \centering
    \includegraphics[width=\textwidth]{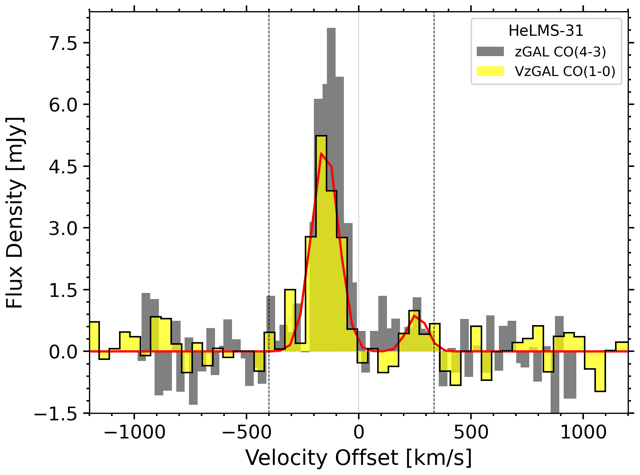}
\end{minipage}
\hfill
\begin{minipage}{0.235\textwidth}
    \centering
    \includegraphics[width=\textwidth]{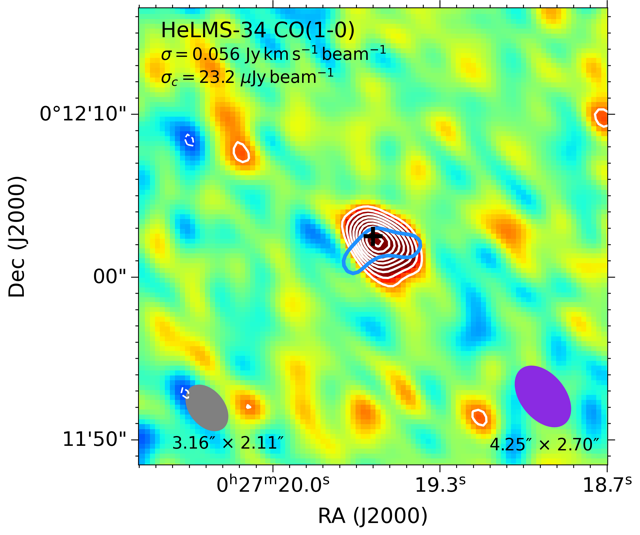}
\end{minipage}
\hfill
\begin{minipage}{0.235\textwidth}
    \centering
    \includegraphics[width=\textwidth]{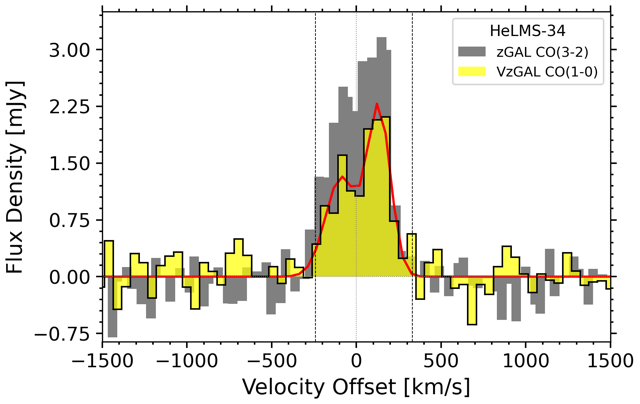}
\end{minipage}

                                        \vspace{1em}

\begin{minipage}{0.235\textwidth}
    \centering
    \includegraphics[width=\textwidth]{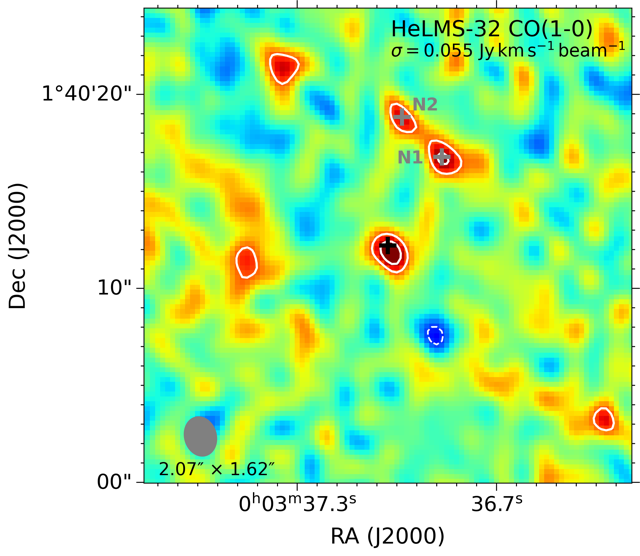}
\end{minipage}
\hfill
\begin{minipage}{0.235\textwidth}
    \centering
    \includegraphics[width=\textwidth]{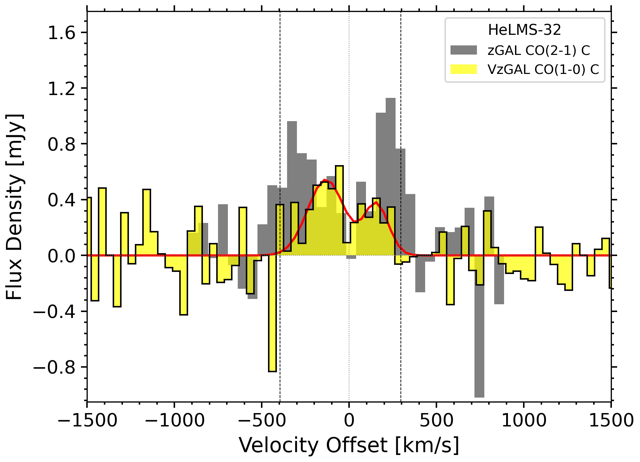}
\end{minipage}
\hfill
\begin{minipage}{0.235\textwidth}
    \centering
    \includegraphics[width=\textwidth]{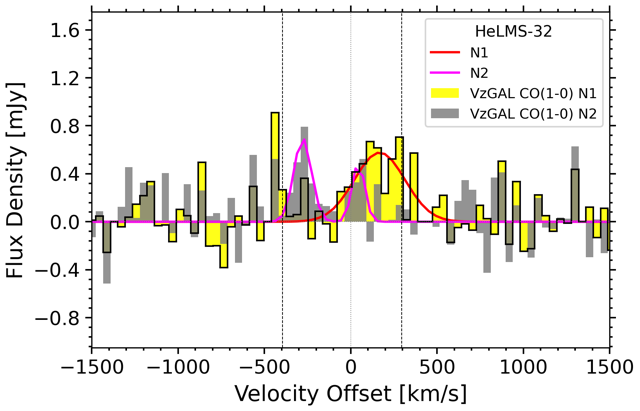}
\end{minipage}

                                        \vspace{1em}
                                        
\begin{minipage}{0.235\textwidth}
    \centering
    \includegraphics[width=\textwidth]{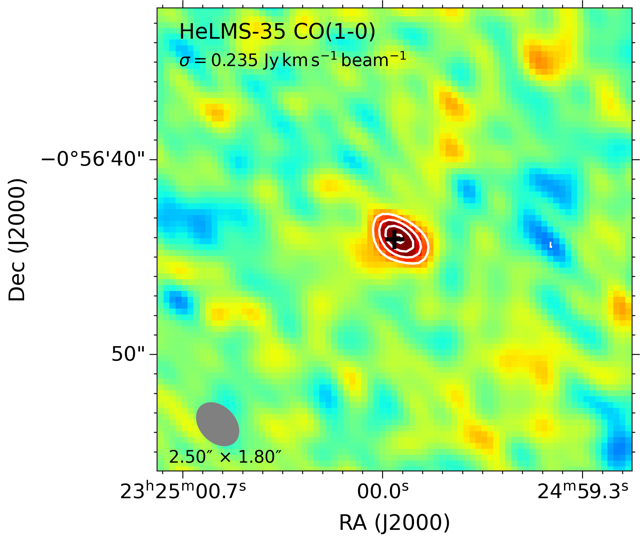}
\end{minipage}
\hfill
\begin{minipage}{0.235\textwidth}
    \centering
    \includegraphics[width=\textwidth]{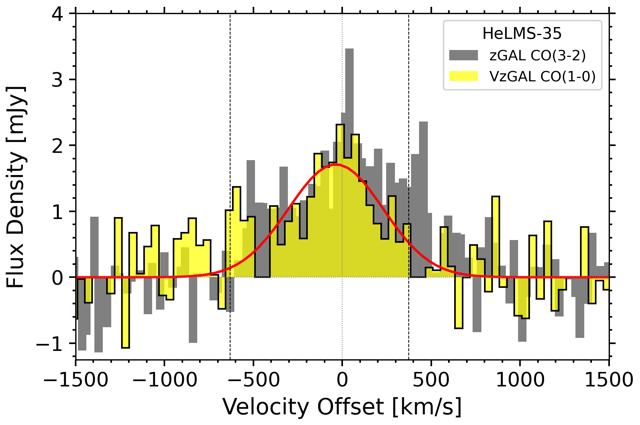}
\end{minipage}
\hfill
\begin{minipage}{0.235\textwidth}
    \centering
    \includegraphics[width=\textwidth]{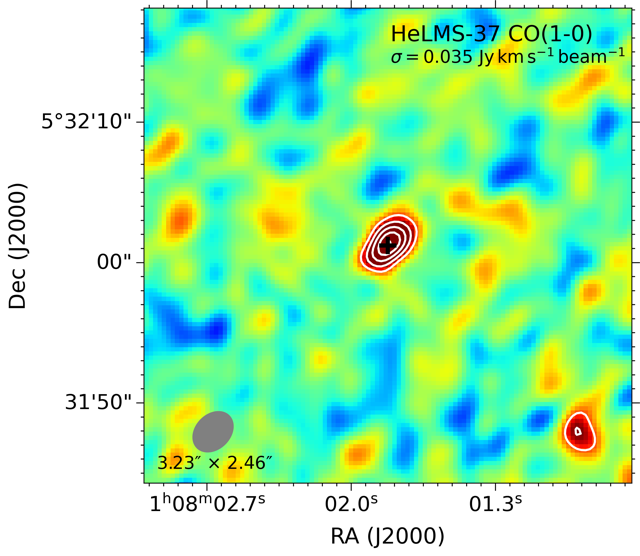}
\end{minipage}
\hfill
\begin{minipage}{0.235\textwidth}
    \centering
    \includegraphics[width=\textwidth]{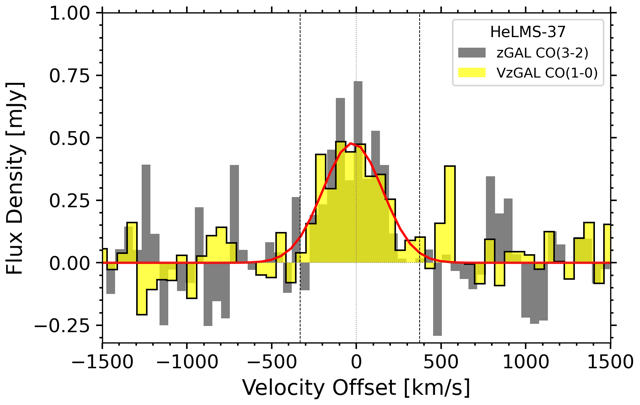}
\end{minipage}

                                        \vspace{1em}
                                        
\begin{minipage}{0.235\textwidth}
    \centering
    \includegraphics[width=\textwidth]{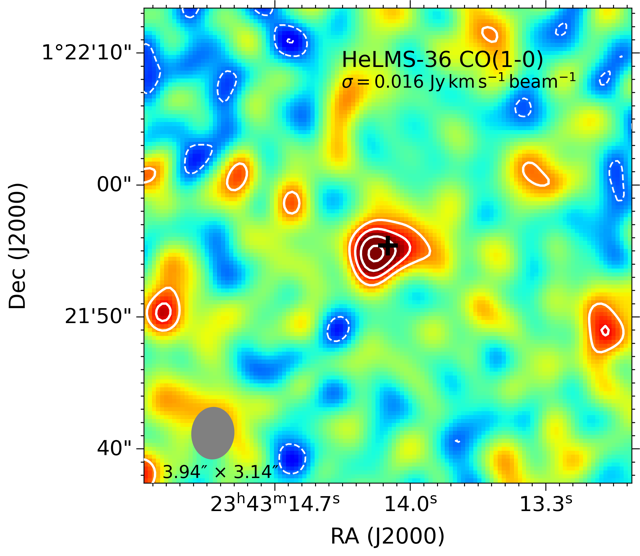}
\end{minipage}
\hfill
\begin{minipage}{0.235\textwidth}
    \centering
    \includegraphics[width=\textwidth]{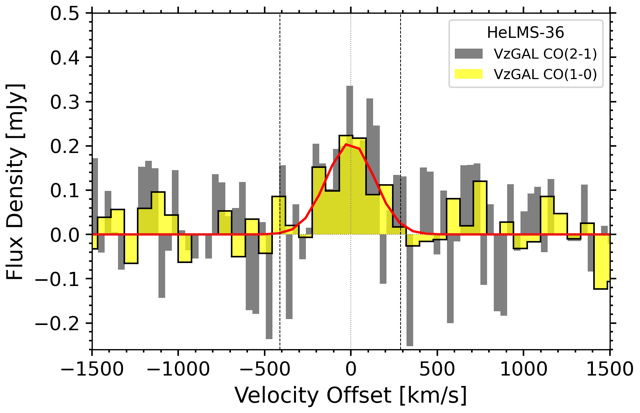}
\end{minipage}
\hfill
\begin{minipage}{0.235\textwidth}
    \centering
    \includegraphics[width=\textwidth]{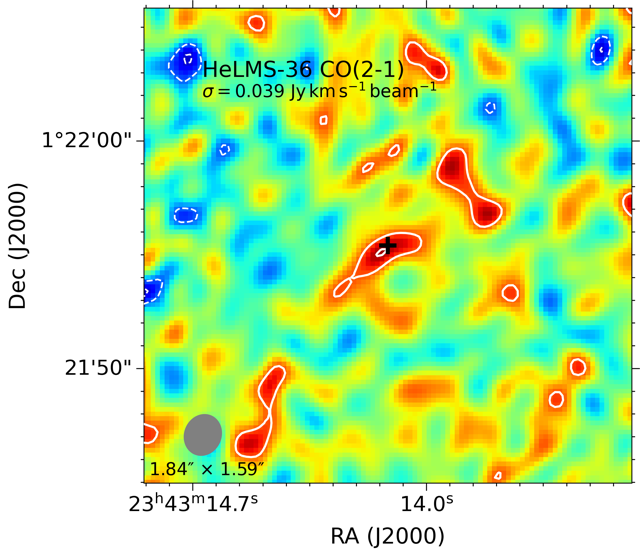}
\end{minipage}
\hfill
\begin{minipage}{0.235\textwidth}
    \centering
    \includegraphics[width=\textwidth]{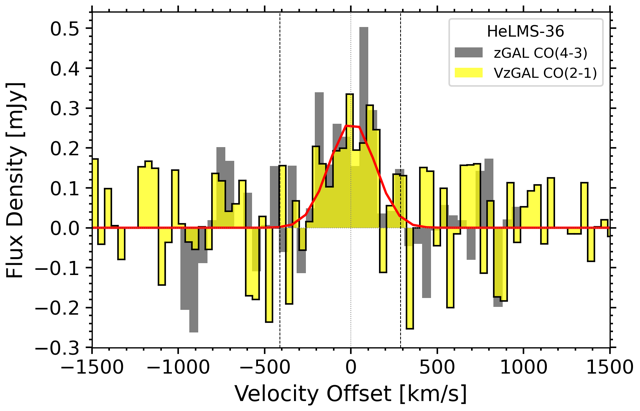}
\end{minipage}

    \addtocounter{figure}{-1}
\caption{(continued)}
\end{figure*}

\clearpage

\begin{figure*}[!htbp]
\centering

\begin{minipage}{0.235\textwidth}
    \centering
    \includegraphics[width=\textwidth]{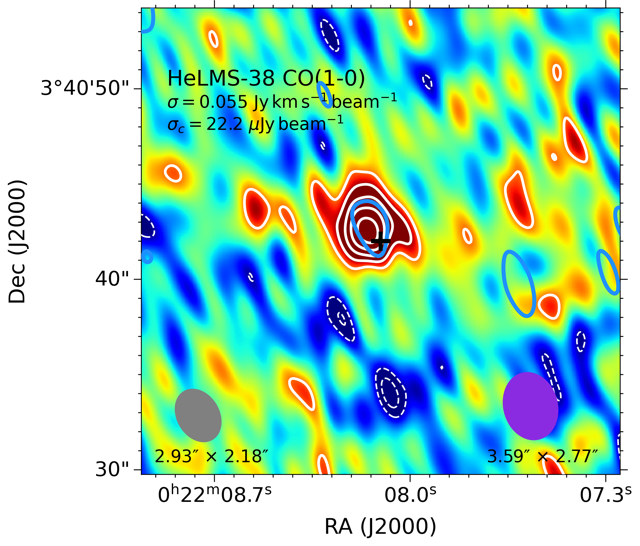}
\end{minipage}
\hfill
\begin{minipage}{0.235\textwidth}
    \centering
    \includegraphics[width=\textwidth]{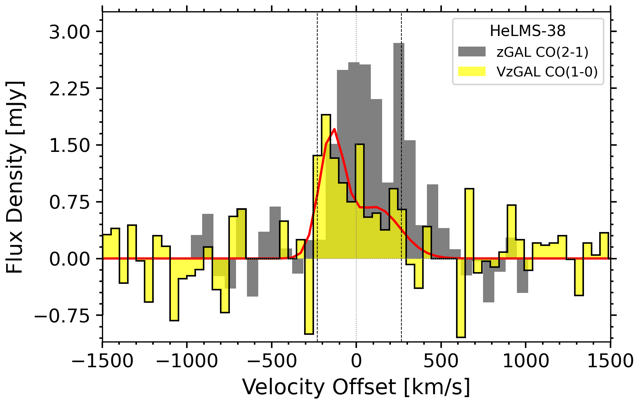}
\end{minipage}
\hfill
\begin{minipage}{0.235\textwidth}
    \centering
    \includegraphics[width=\textwidth]{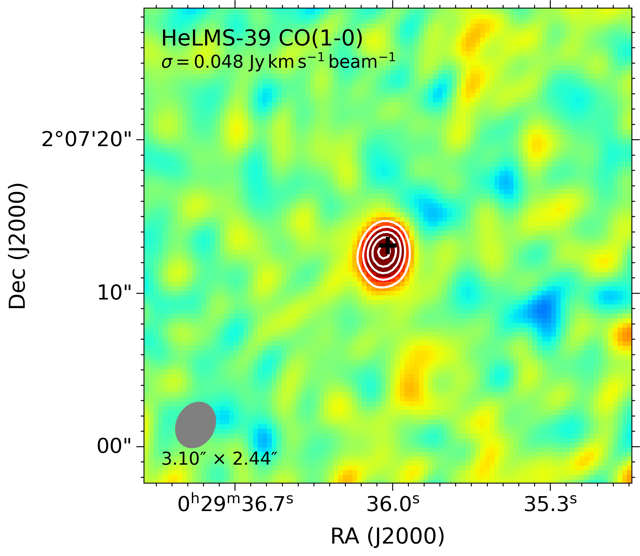}
\end{minipage}
\hfill
\begin{minipage}{0.235\textwidth}
    \centering
    \includegraphics[width=\textwidth]{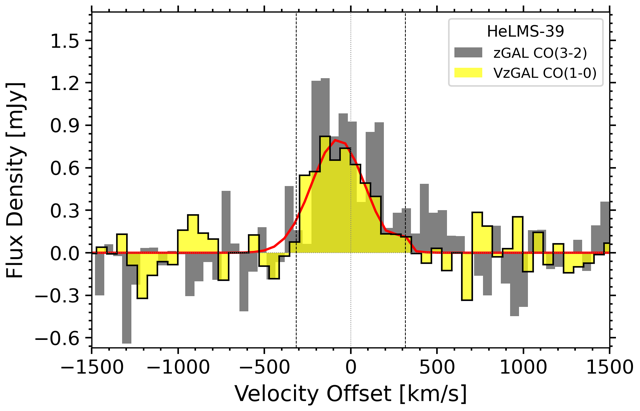}
\end{minipage}

                \vspace{1em}

\begin{minipage}{0.235\textwidth}
    \centering
    \includegraphics[width=\textwidth]{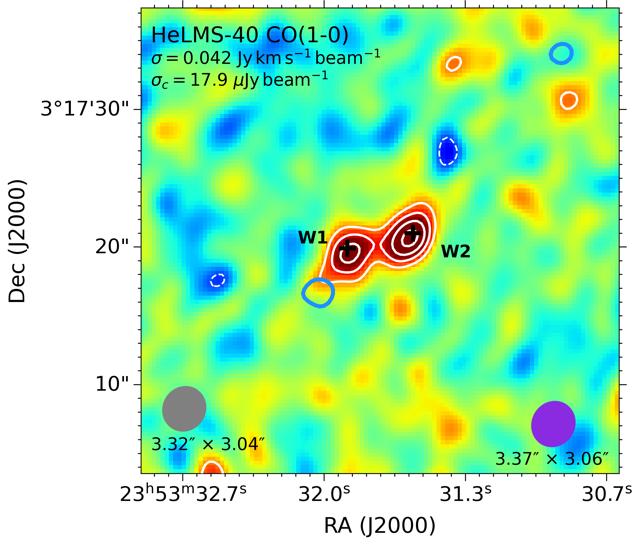}
\end{minipage}
\hfill
\begin{minipage}{0.235\textwidth}
    \centering
    \includegraphics[width=\textwidth]{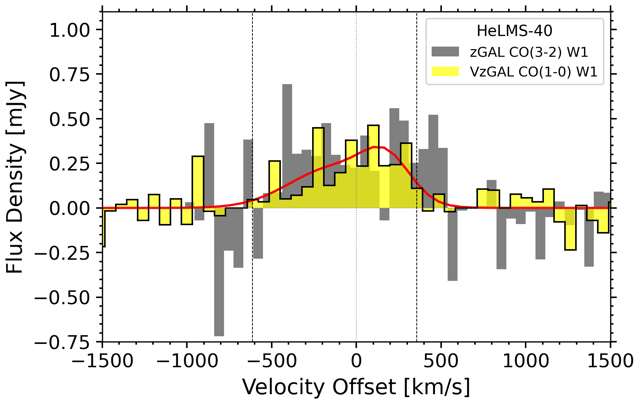}
\end{minipage}
\hfill
\begin{minipage}{0.235\textwidth}
    \centering
    \includegraphics[width=\textwidth]{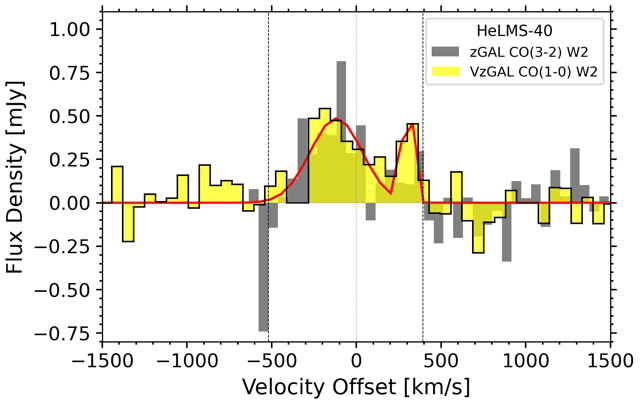}
\end{minipage}

                                        \vspace{1em}                  
                                        
\begin{minipage}{0.235\textwidth}
    \centering
    \includegraphics[width=\textwidth]{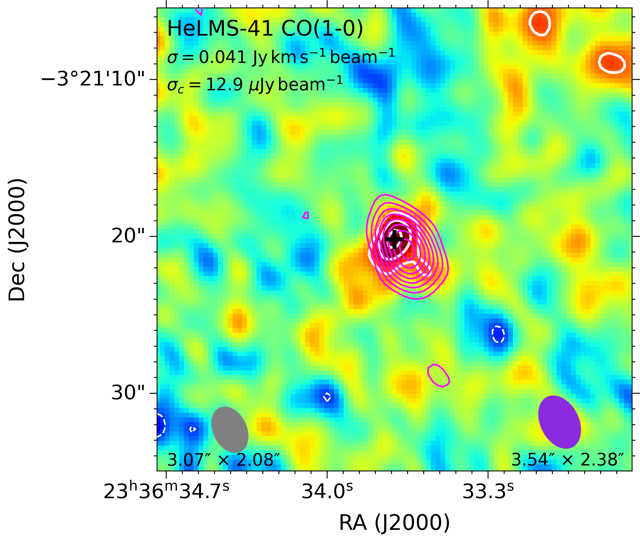}
\end{minipage}
\hfill
\begin{minipage}{0.235\textwidth}
    \centering
    \includegraphics[width=\textwidth]{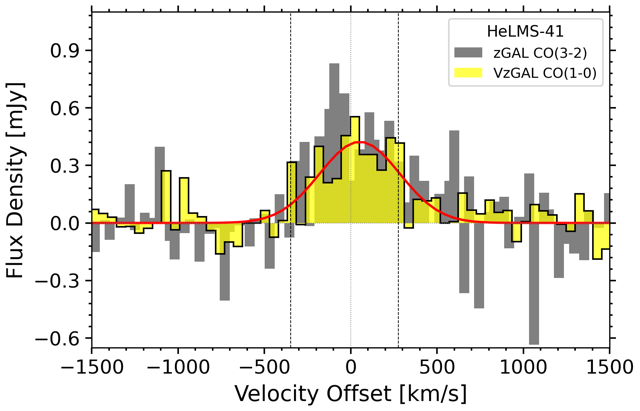}
\end{minipage}
\hfill
\begin{minipage}{0.235\textwidth}
    \centering
    \includegraphics[width=\textwidth]{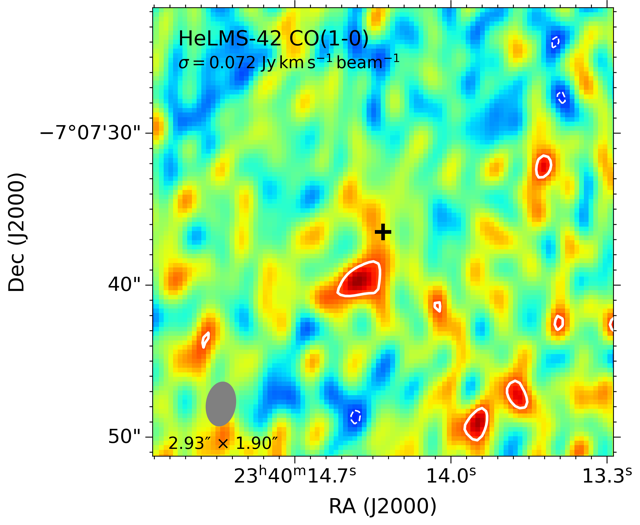}
\end{minipage}
\hfill
\begin{minipage}{0.235\textwidth}
    \centering
    \includegraphics[width=\textwidth]{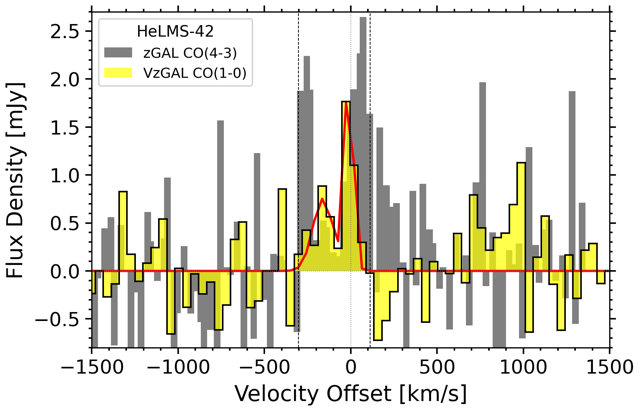}
\end{minipage}

                                        \vspace{1em}

\begin{minipage}{0.235\textwidth}
    \centering
    \includegraphics[width=\textwidth]{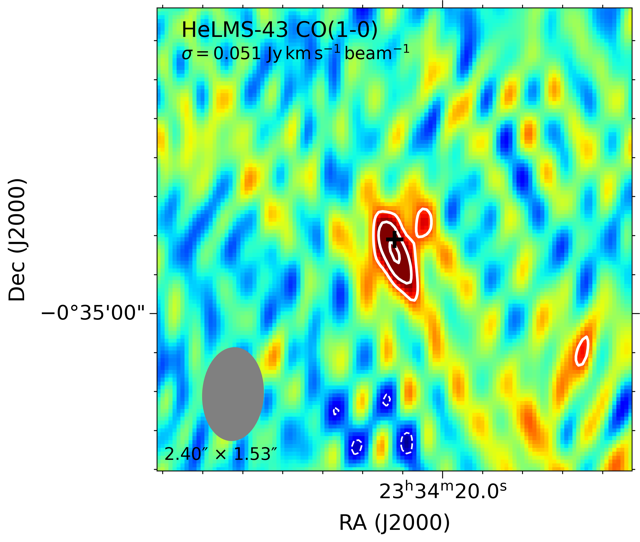}
\end{minipage}
\hfill
\begin{minipage}{0.235\textwidth}
    \centering
    \includegraphics[width=\textwidth]{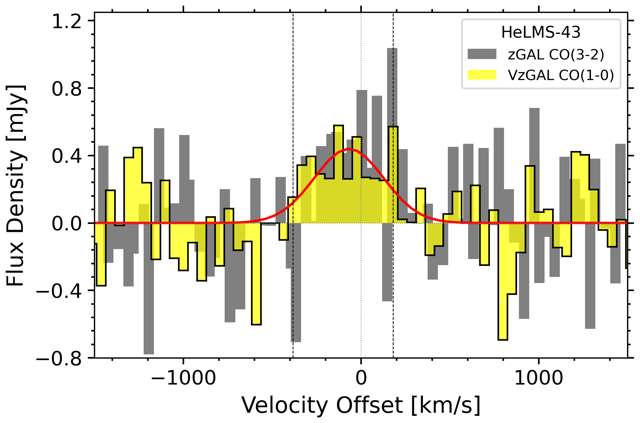}
\end{minipage}
\hfill
\begin{minipage}{0.235\textwidth}
    \centering
    \includegraphics[width=\textwidth]{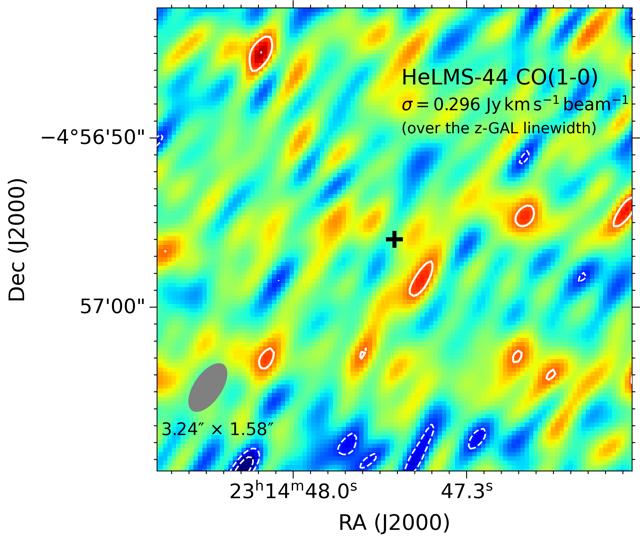}
\end{minipage}
\hfill
\begin{minipage}{0.235\textwidth}
    \centering
    \includegraphics[width=\textwidth]{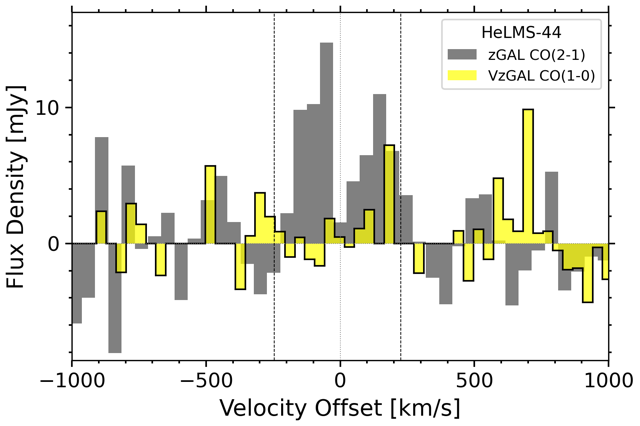}
\end{minipage}

                                        \vspace{1em}
                                                                 
\begin{minipage}{0.235\textwidth}
    \centering
    \includegraphics[width=\textwidth]{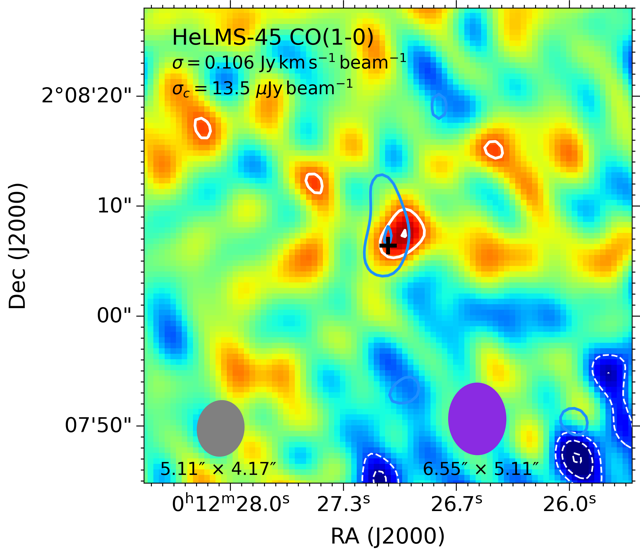}
\end{minipage}
\hfill
\begin{minipage}{0.235\textwidth}
    \centering
    \includegraphics[width=\textwidth]{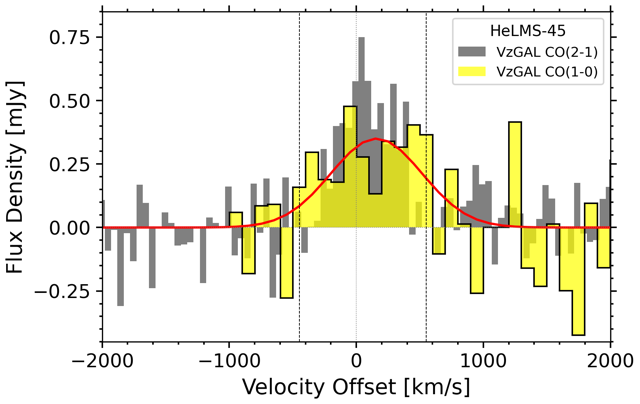}
\end{minipage}
\hfill
\begin{minipage}{0.235\textwidth}
    \centering
    \includegraphics[width=\textwidth]{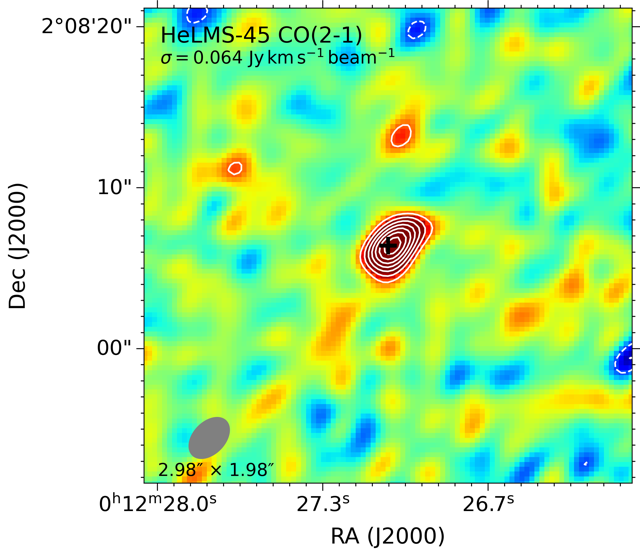}
\end{minipage}
\hfill
\begin{minipage}{0.235\textwidth}
    \centering
    \includegraphics[width=\textwidth]{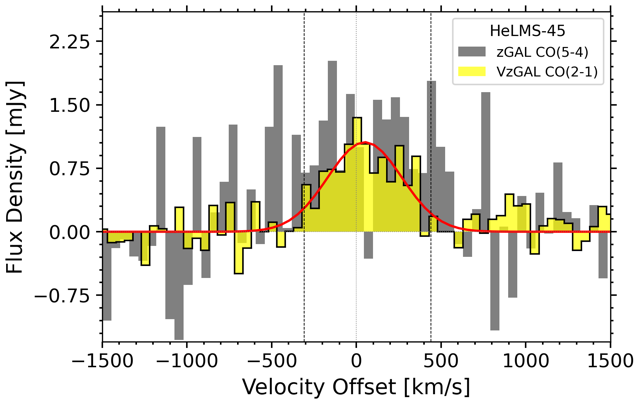}
\end{minipage}

                                        \vspace{1em}
                                        
\begin{minipage}{0.235\textwidth}
    \centering
    \includegraphics[width=\textwidth]{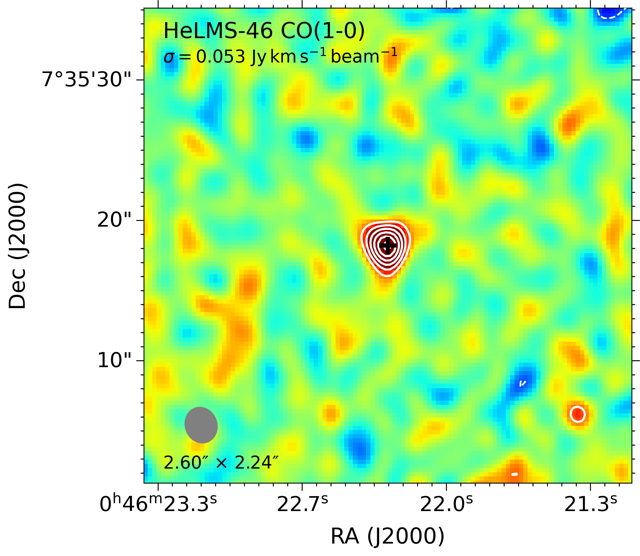}
\end{minipage}
\hfill
\begin{minipage}{0.235\textwidth}
    \centering
    \includegraphics[width=\textwidth]{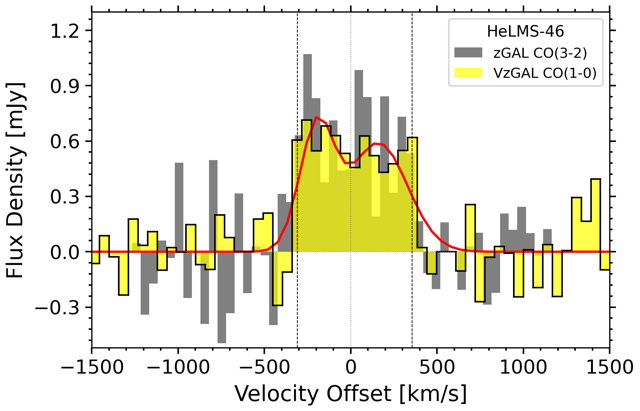}
\end{minipage}
\hfill
\begin{minipage}{0.235\textwidth}
    \centering
    \includegraphics[width=\textwidth]{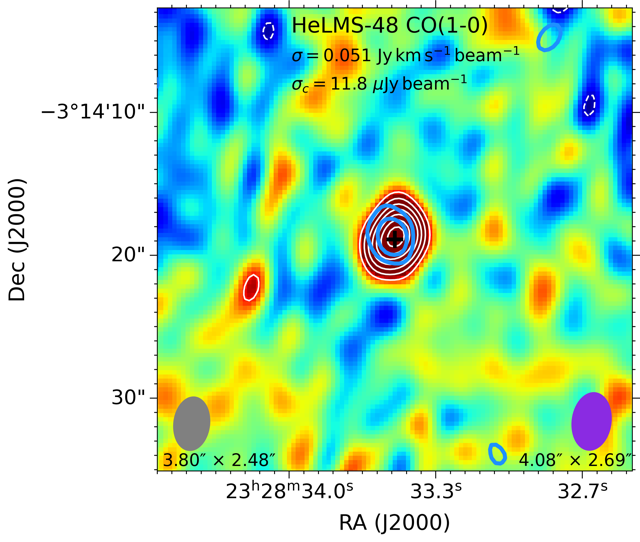}
\end{minipage}
\hfill
\begin{minipage}{0.235\textwidth}
    \centering
    \includegraphics[width=\textwidth]{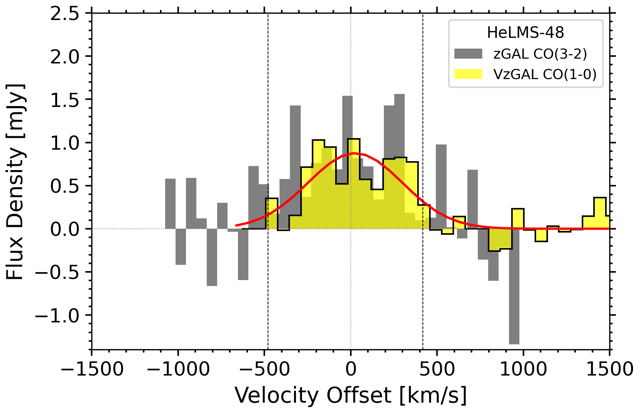}
\end{minipage}

    \addtocounter{figure}{-1}
\caption{(continued)}
\end{figure*}

\clearpage

\begin{figure*}[!htbp]
\centering

\begin{minipage}{0.235\textwidth}
    \centering
    \includegraphics[width=\textwidth]{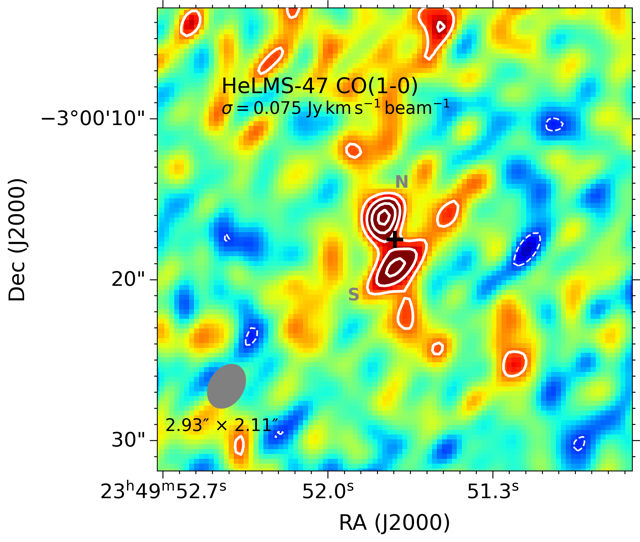}
\end{minipage}
\hfill
\begin{minipage}{0.235\textwidth}
    \centering
    \includegraphics[width=\textwidth]{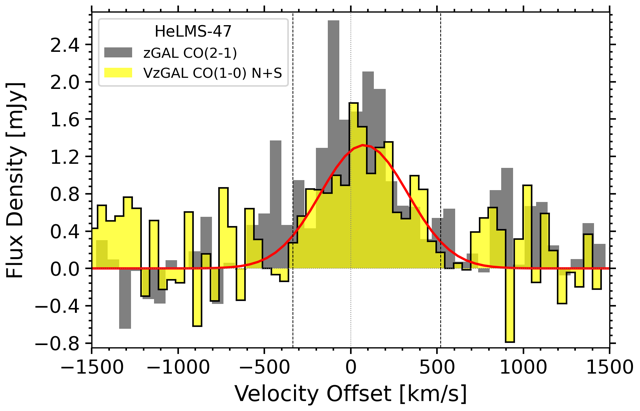}
\end{minipage}
\hfill
\begin{minipage}{0.235\textwidth}
    \centering
    \includegraphics[width=\textwidth]{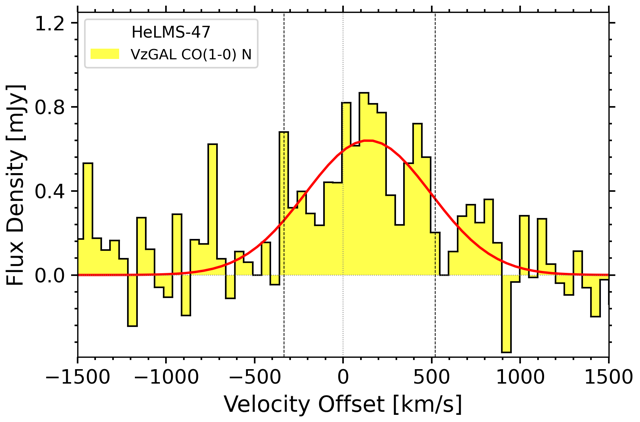}
\end{minipage}
\hfill
\begin{minipage}{0.235\textwidth}
    \centering
    \includegraphics[width=\textwidth]{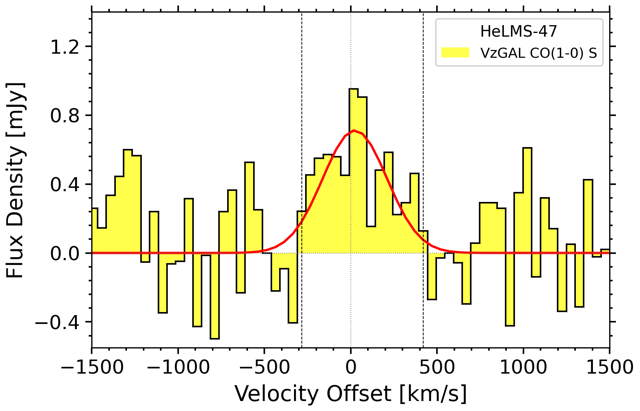}
\end{minipage}

            \vspace{1em}

\begin{minipage}{0.235\textwidth}
    \centering
    \includegraphics[width=\textwidth]{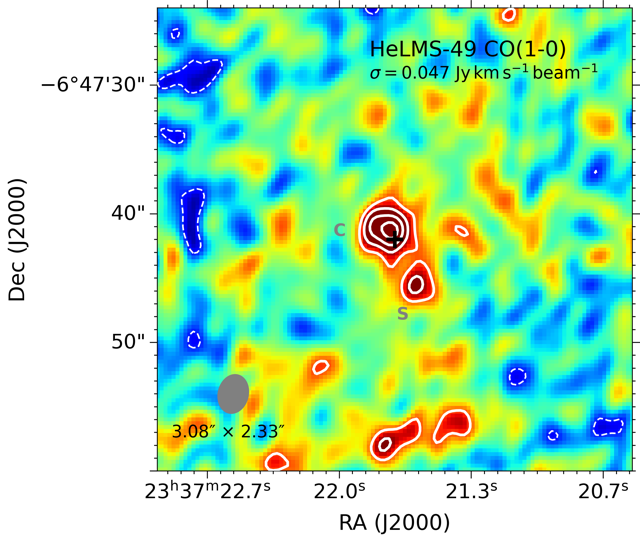}
\end{minipage}
\hfill
\begin{minipage}{0.235\textwidth}
    \centering
    \includegraphics[width=\textwidth]{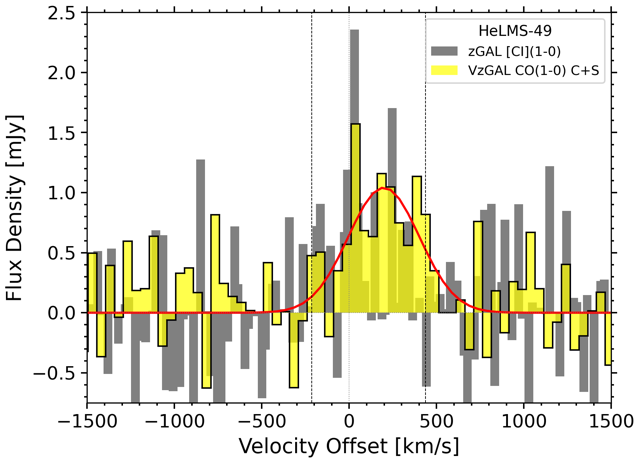}
\end{minipage}
\hfill
\begin{minipage}{0.235\textwidth}
    \centering
    \includegraphics[width=\textwidth]{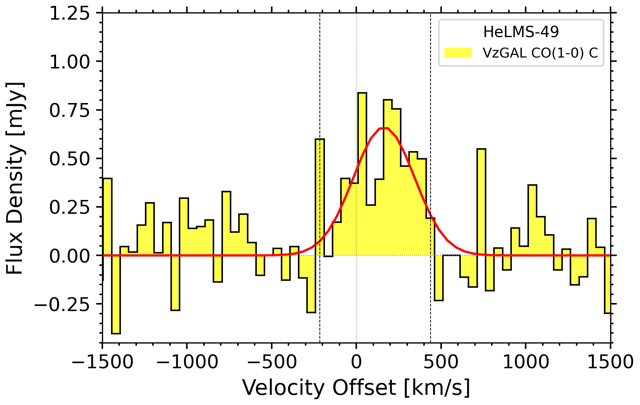}
\end{minipage}
\hfill
\begin{minipage}{0.235\textwidth}
    \centering
    \includegraphics[width=\textwidth]{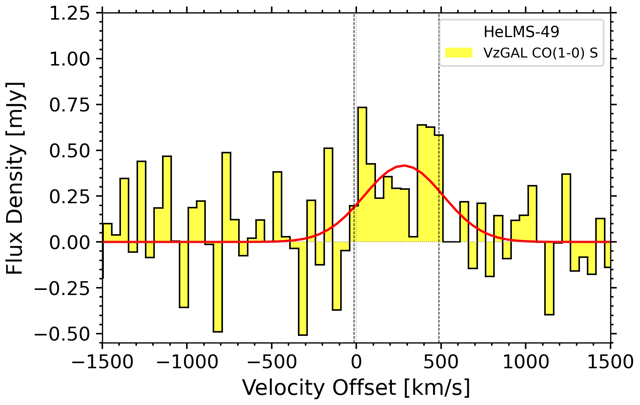}
\end{minipage}

                                        \vspace{1em}
                                        
\begin{minipage}{0.235\textwidth}
    \centering
    \includegraphics[width=\textwidth]{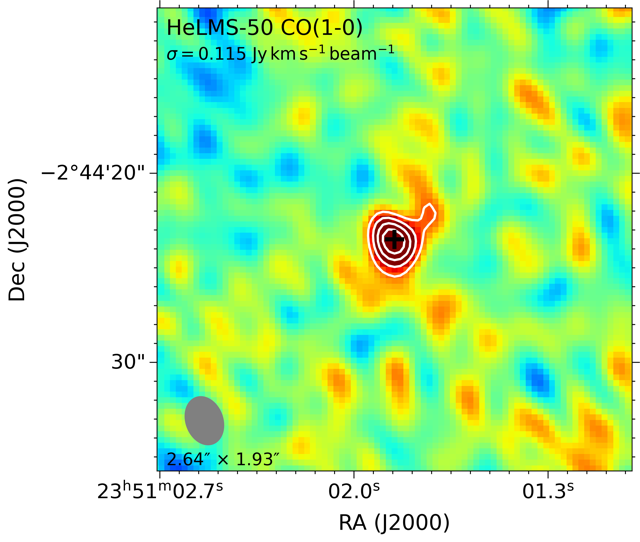}
\end{minipage}
\hfill
\begin{minipage}{0.235\textwidth}
    \centering
    \includegraphics[width=\textwidth]{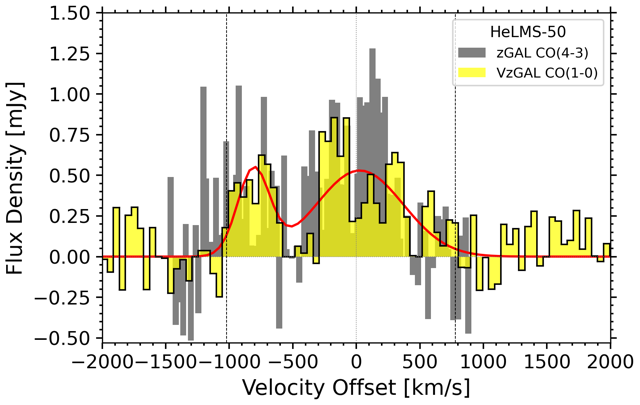}
\end{minipage}
\hfill
\begin{minipage}{0.235\textwidth}
    \centering
    \includegraphics[width=\textwidth]{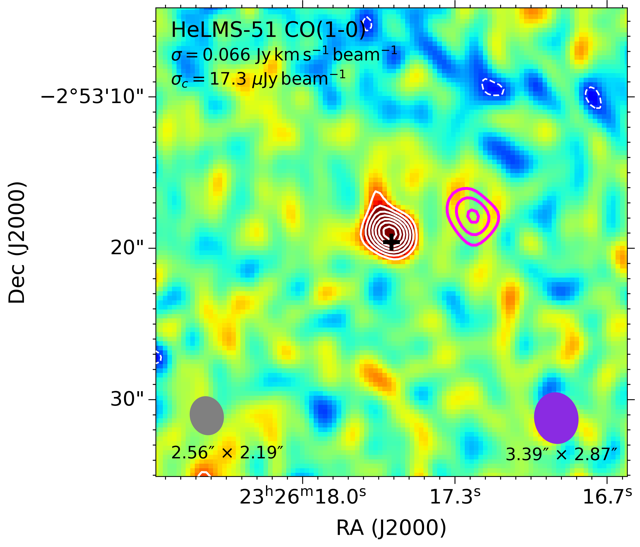}
\end{minipage}
\hfill
\begin{minipage}{0.235\textwidth}
    \centering
    \includegraphics[width=\textwidth]{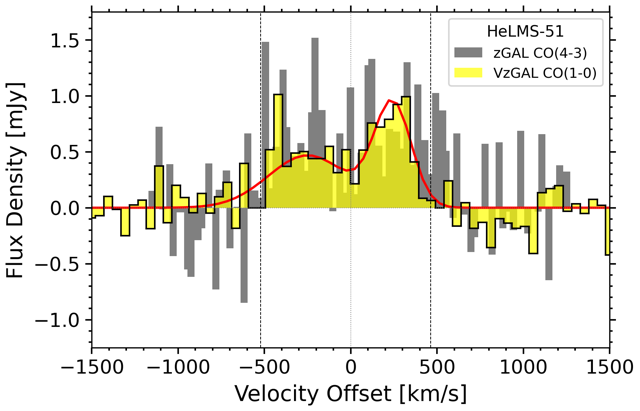}
\end{minipage}

                                        \vspace{1em}
                                        
\begin{minipage}{0.235\textwidth}
    \centering
    \includegraphics[width=\textwidth]{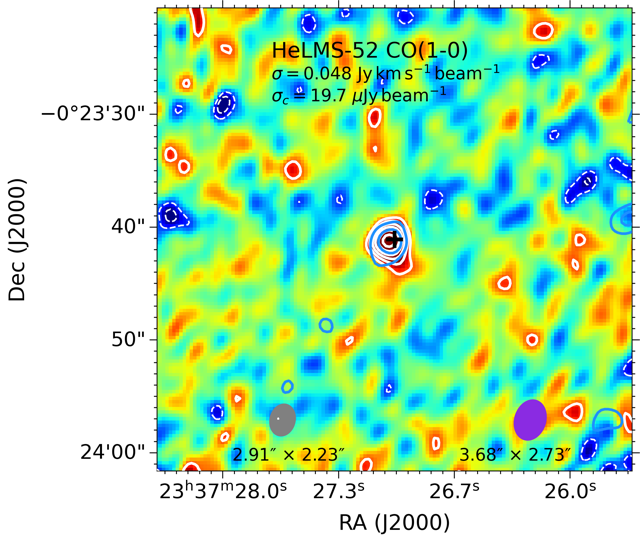}
\end{minipage}
\hfill
\begin{minipage}{0.235\textwidth}
    \centering
    \includegraphics[width=\textwidth]{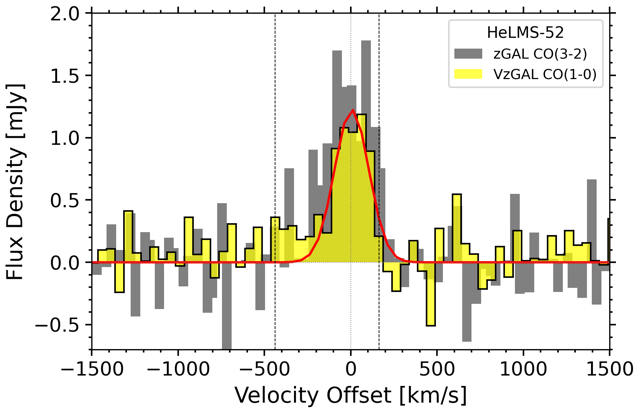}
\end{minipage}
\hfill
\begin{minipage}{0.235\textwidth}
    \centering
    \includegraphics[width=\textwidth]{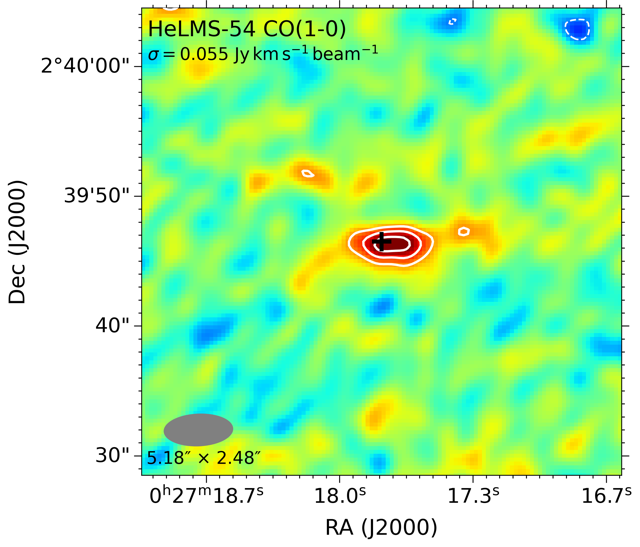}
\end{minipage}
\hfill
\begin{minipage}{0.235\textwidth}
    \centering
    \includegraphics[width=\textwidth]{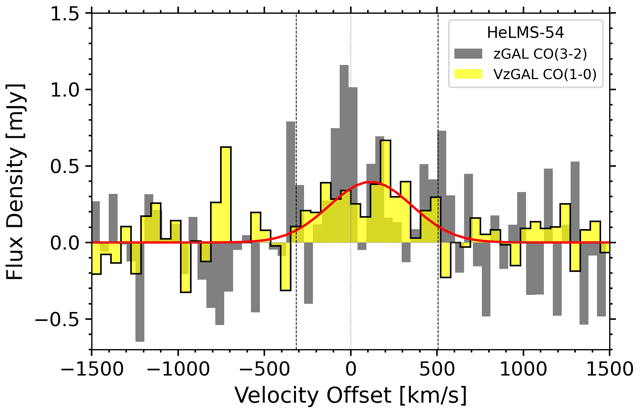}
\end{minipage}

                                \vspace{1em}
     
\begin{minipage}{0.235\textwidth}
    \centering
    \includegraphics[width=\textwidth]{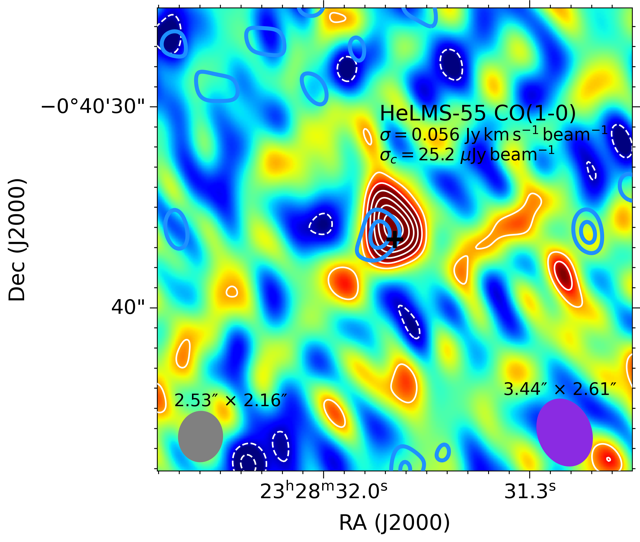}
\end{minipage}
\hfill
\begin{minipage}{0.235\textwidth}
    \centering
    \includegraphics[width=\textwidth]{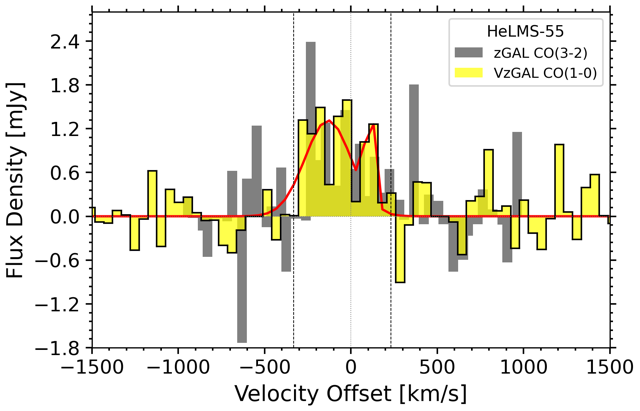}
\end{minipage}
\hfill
\begin{minipage}{0.235\textwidth}
    \centering
    \includegraphics[width=\textwidth]{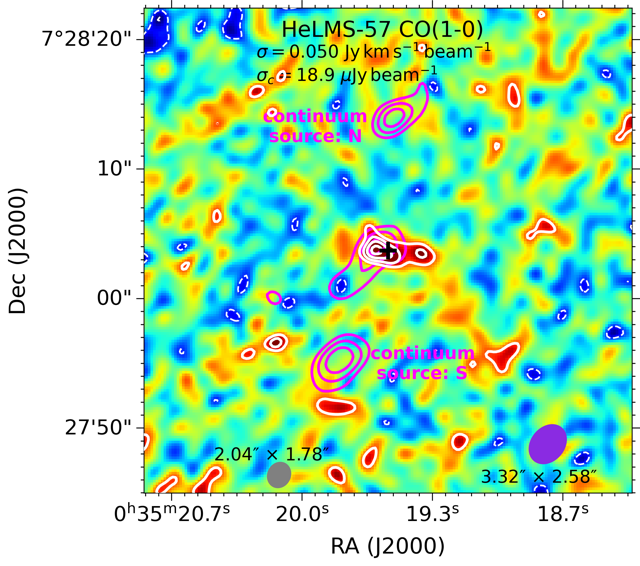}
\end{minipage}
\hfill
\begin{minipage}{0.235\textwidth}
    \centering
    \includegraphics[width=\textwidth]{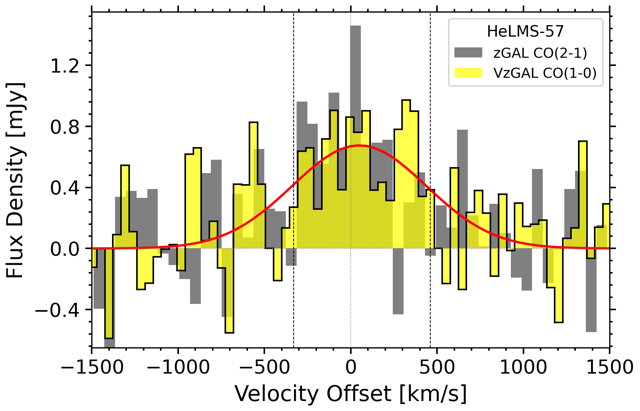}
\end{minipage}

                                        \vspace{1em}
                                        
\begin{minipage}{0.235\textwidth}
    \centering
    \includegraphics[width=\textwidth]{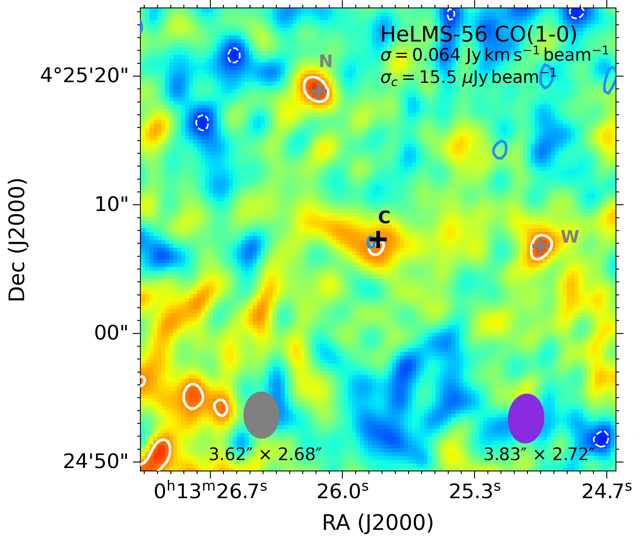}
\end{minipage}
\hfill
\begin{minipage}{0.235\textwidth}
    \centering
    \includegraphics[width=\textwidth]{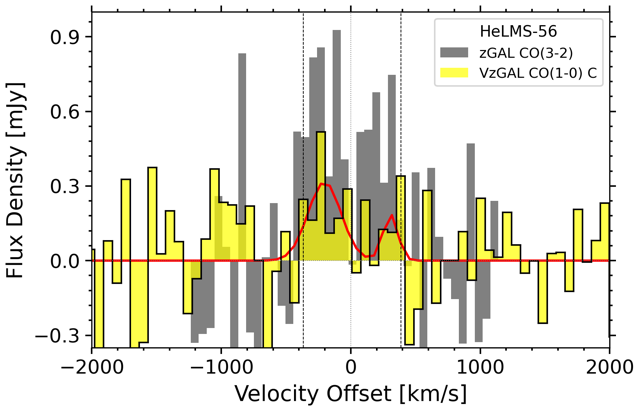}
\end{minipage}
\hfill
\begin{minipage}{0.235\textwidth}
    \centering
    \includegraphics[width=\textwidth]{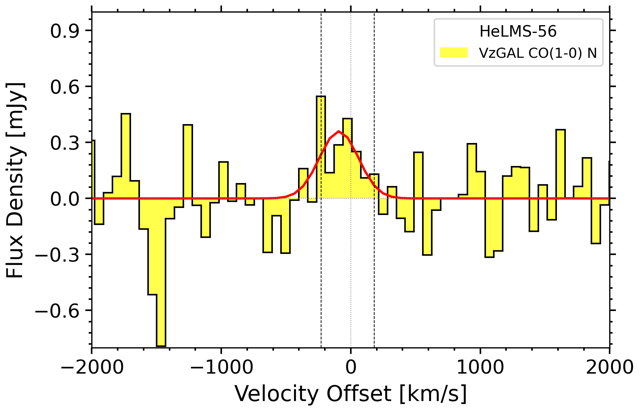}
\end{minipage}
\hfill
\begin{minipage}{0.235\textwidth}
    \centering
    \includegraphics[width=\textwidth]{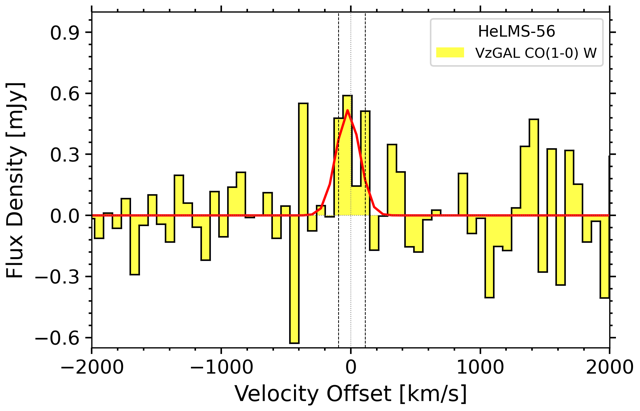}
\end{minipage}


\caption{Same as Figure~\ref{fig:image-grid_helms_v1} with the remaining HeLMS sources (HeLMS-47 and HeLMS-49 to HeLMS-56) and all the HerS galaxies.} 
\label{fig:image-grid_helms_v2_hers}
\end{figure*}

\clearpage

\begin{figure*}[!htbp]
\centering

\begin{minipage}{0.235\textwidth}
    \centering
    \includegraphics[width=\textwidth]{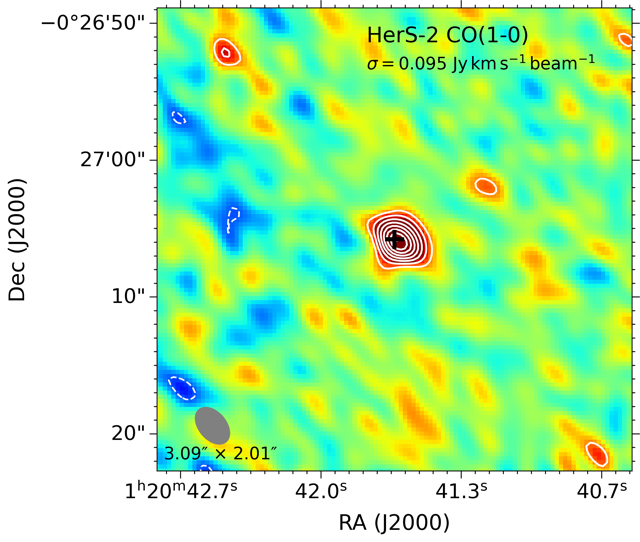}
\end{minipage}
\hfill
\begin{minipage}{0.235\textwidth}
    \centering
    \includegraphics[width=\textwidth]{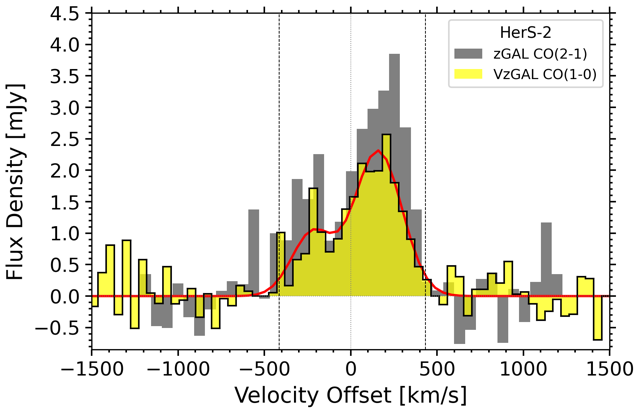}
\end{minipage}
\hfill                         
\begin{minipage}{0.235\textwidth}
    \centering
    \includegraphics[width=\textwidth]{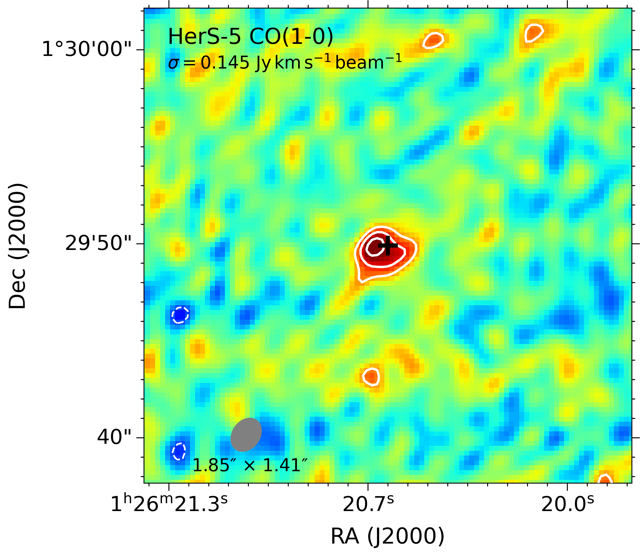}
\end{minipage}
\hfill
\begin{minipage}{0.235\textwidth}
    \centering
    \includegraphics[width=\textwidth]{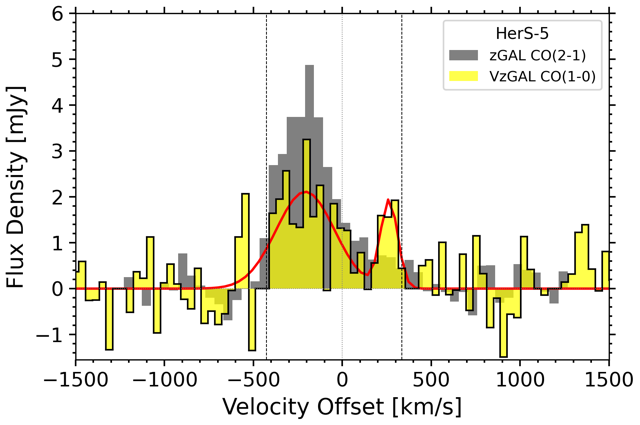}
\end{minipage}

                          \vspace{1em}

\begin{minipage}{0.235\textwidth}
    \centering
    \includegraphics[width=\textwidth]{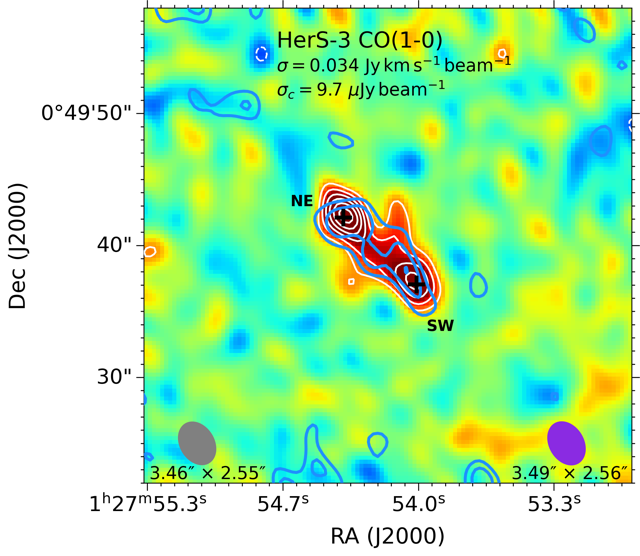}
\end{minipage}
\hfill
\begin{minipage}{0.235\textwidth}
    \centering
    \includegraphics[width=\textwidth]{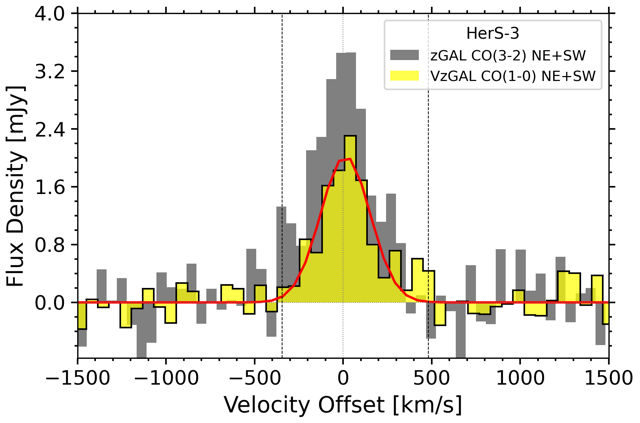}
\end{minipage}
\hfill
\begin{minipage}{0.235\textwidth}
    \centering
    \includegraphics[width=\textwidth]{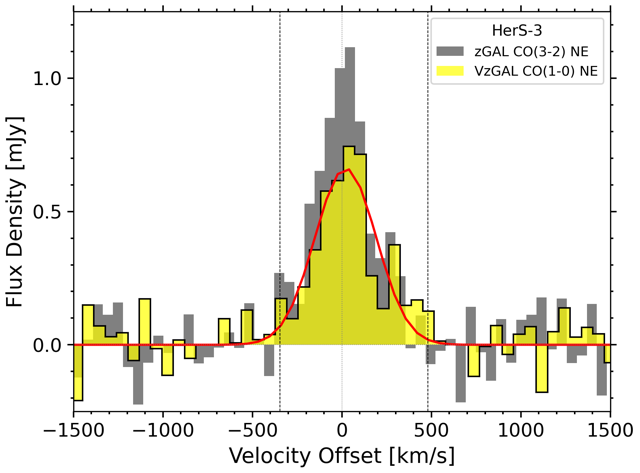}
\end{minipage}
\hfill
\begin{minipage}{0.235\textwidth}
    \centering
    \includegraphics[width=\textwidth]{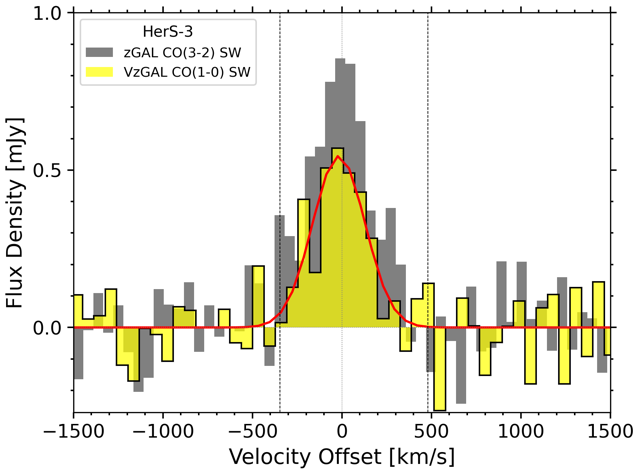}
\end{minipage}

                                        \vspace{1em}

\hfill
\begin{minipage}{0.235\textwidth}
    \centering
    \includegraphics[width=\textwidth]{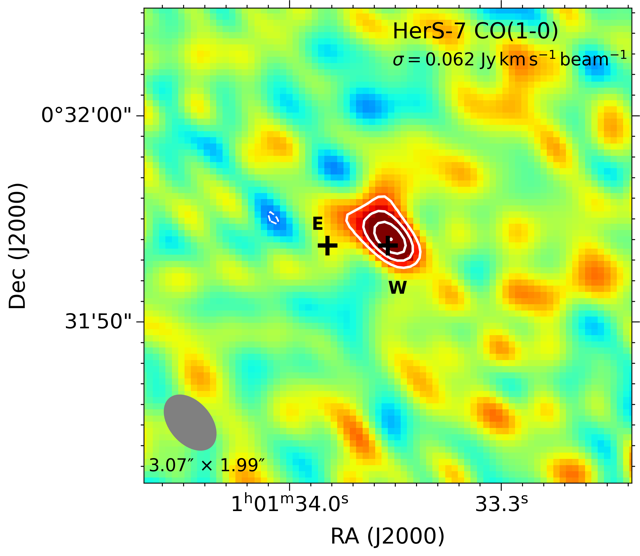}
\end{minipage}
\hfill
\begin{minipage}{0.235\textwidth}
    \centering
    \includegraphics[width=\textwidth]{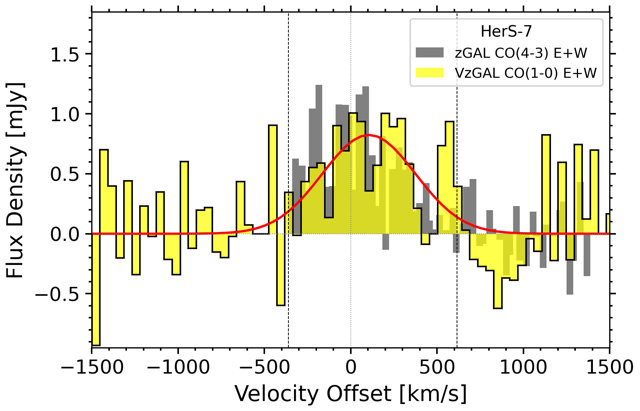}
\end{minipage}
\hfill
\begin{minipage}{0.235\textwidth}
    \centering
    \includegraphics[width=\textwidth]{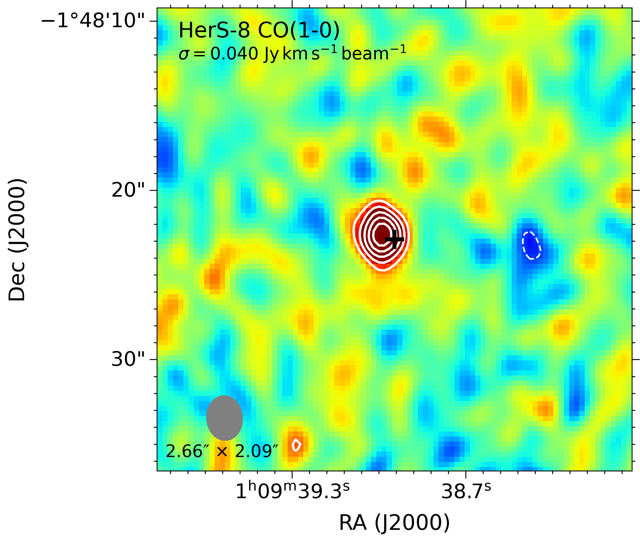}
\end{minipage}
\hfill
\begin{minipage}{0.235\textwidth}
    \centering
    \includegraphics[width=\textwidth]{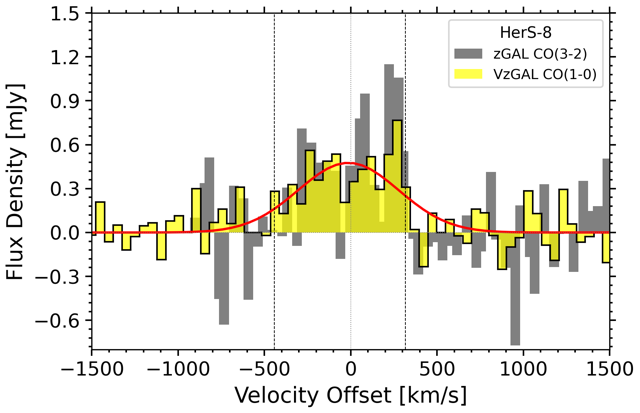}
\end{minipage}

                                    \vspace{1em}

\begin{minipage}{0.235\textwidth}
    \centering
    \includegraphics[width=\textwidth]{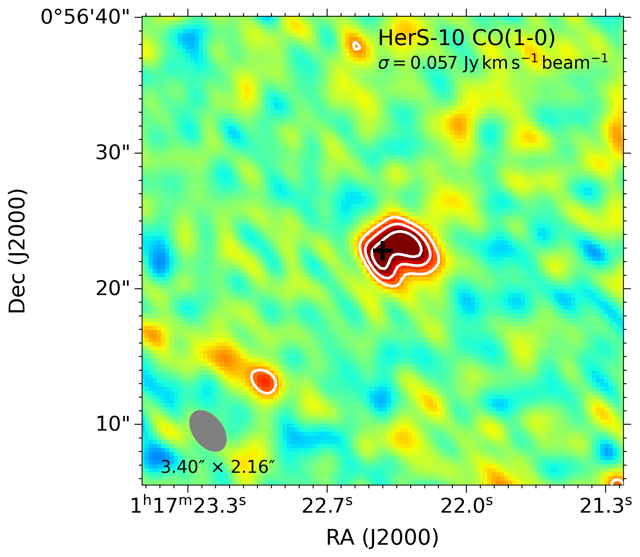}
\end{minipage}
\hfill
\begin{minipage}{0.235\textwidth}
    \centering
    \includegraphics[width=\textwidth]{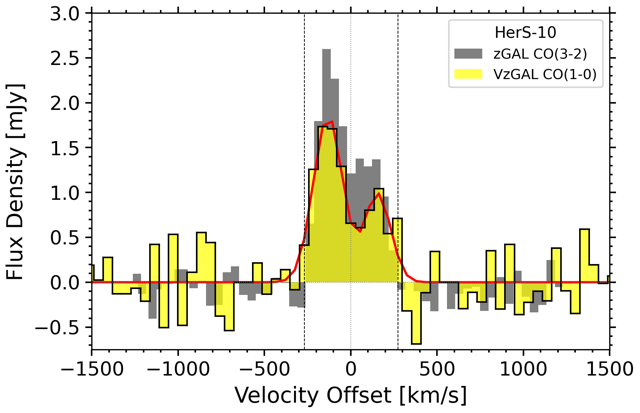}
\end{minipage}
\hfil
\begin{minipage}{0.235\textwidth}
    \centering
    \includegraphics[width=\textwidth]{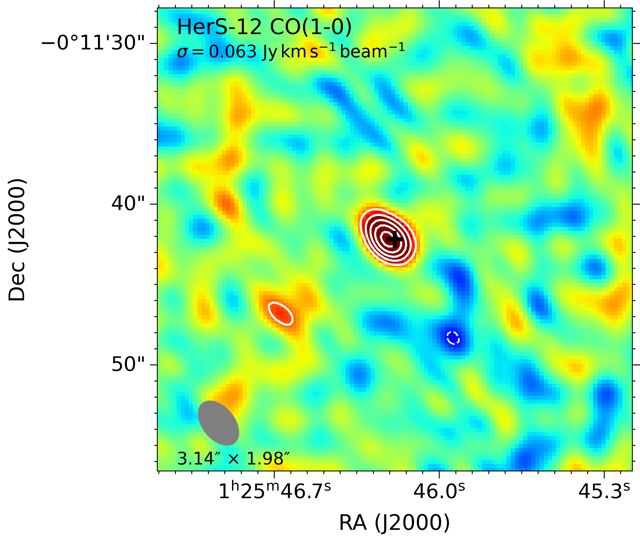}
\end{minipage}
\hfill
\begin{minipage}{0.235\textwidth}
    \centering
    \includegraphics[width=\textwidth]{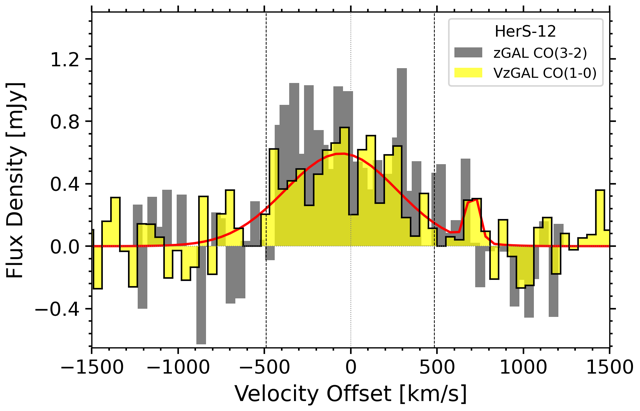}
\end{minipage}

                                        \vspace{1em}
                                        
\begin{minipage}{0.235\textwidth}
    \centering
    \includegraphics[width=\textwidth]{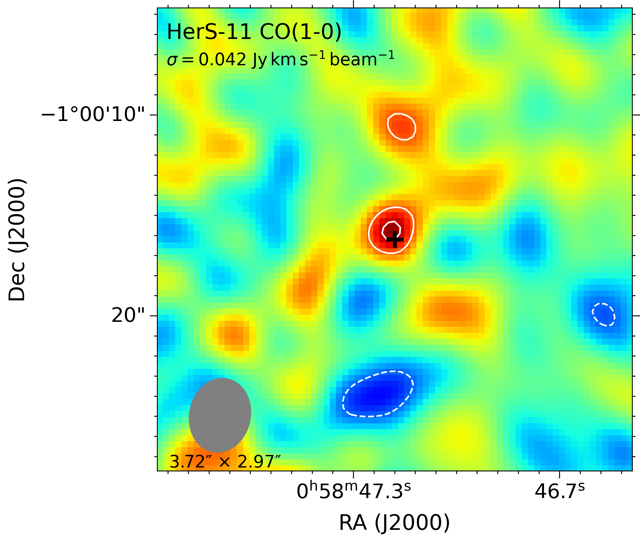}
\end{minipage}
\hfill
\begin{minipage}{0.235\textwidth}
    \centering
    \includegraphics[width=\textwidth]{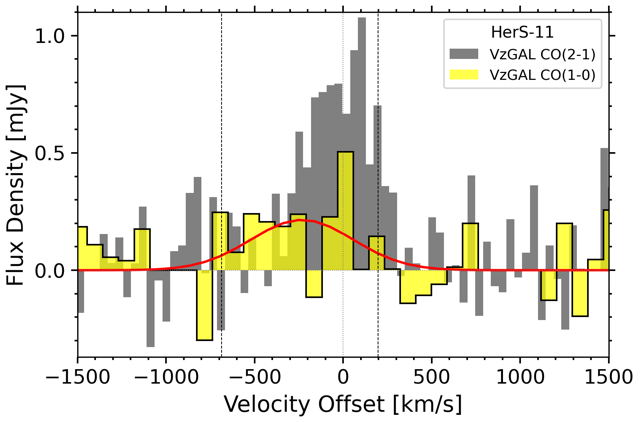}
\end{minipage} 
\hfill
\begin{minipage}{0.235\textwidth}
    \centering
    \includegraphics[width=\textwidth]{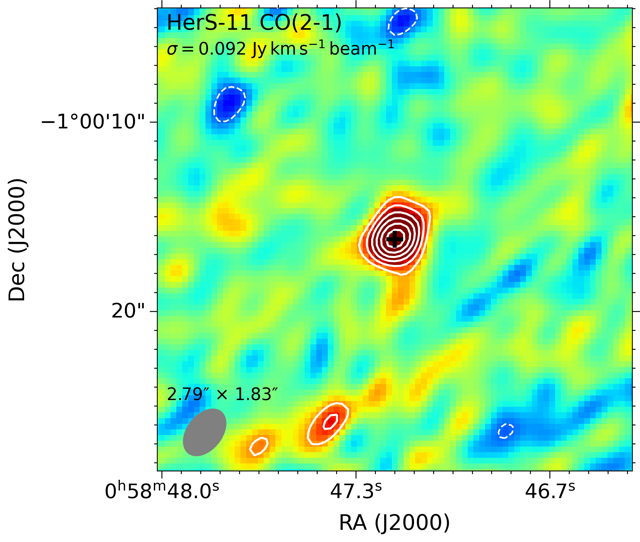}
\end{minipage}
\hfill
\begin{minipage}{0.235\textwidth}
    \centering
    \includegraphics[width=\textwidth]{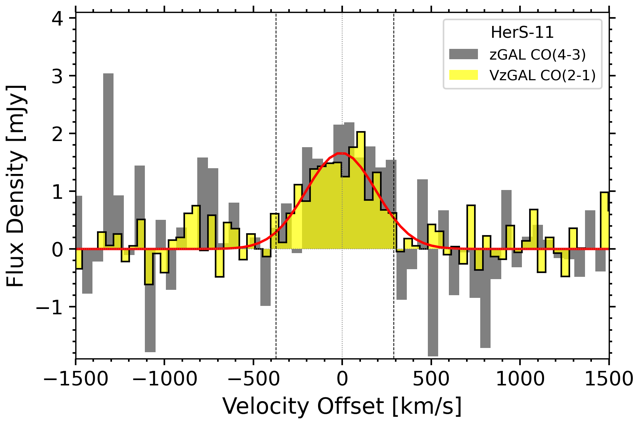}
\end{minipage}    

                                        \vspace{1em}

    \begin{minipage}{0.235\textwidth}
    \centering
    \includegraphics[width=\textwidth]{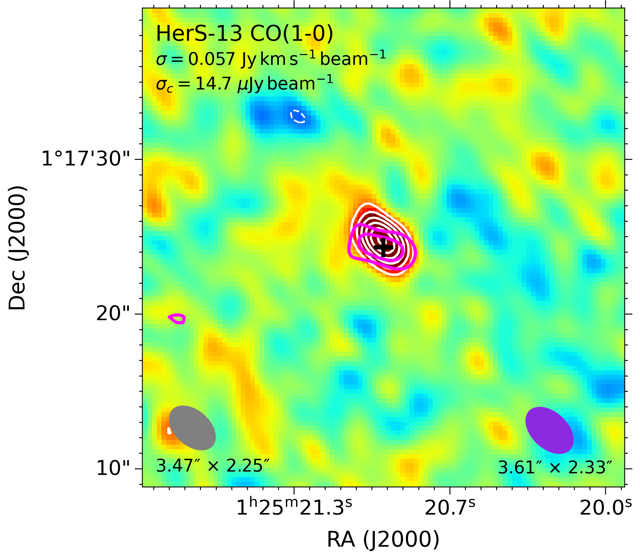}
\end{minipage}
\hfill
\begin{minipage}{0.235\textwidth}
    \centering
    \includegraphics[width=\textwidth]{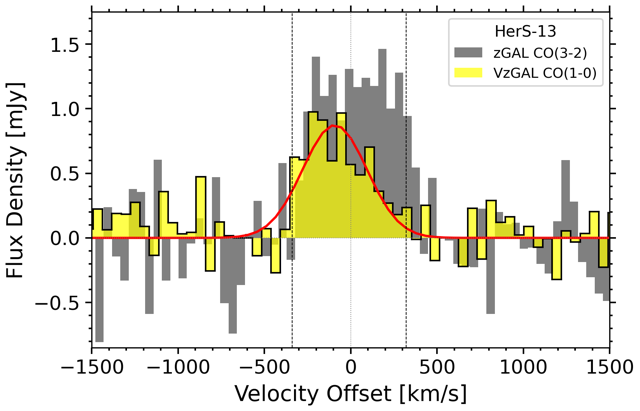}
\end{minipage}
\hfill                                        
\begin{minipage}{0.235\textwidth}
    \centering
    \includegraphics[width=\textwidth]{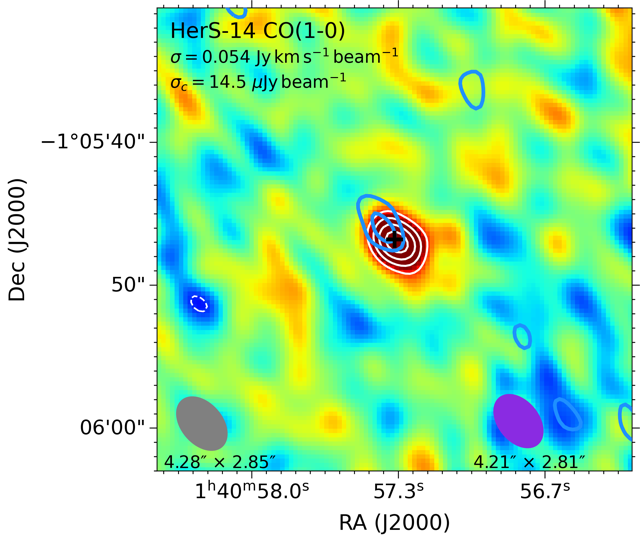}
\end{minipage}
\hfill
\begin{minipage}{0.235\textwidth}
    \centering
    \includegraphics[width=\textwidth]{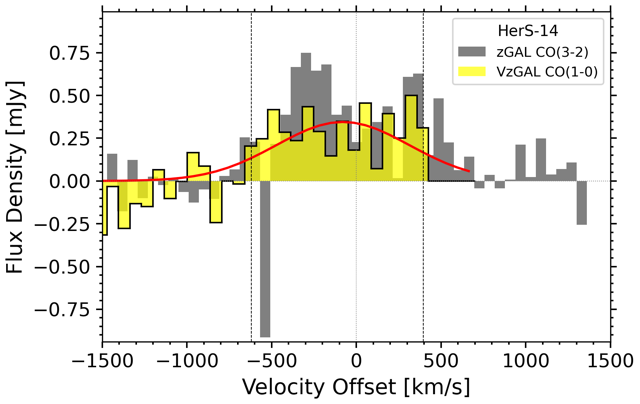}
\end{minipage}

    \addtocounter{figure}{-1}
\caption{(continued)}
\end{figure*}

\clearpage

\begin{figure*}[!htbp]
\centering

\begin{minipage}{0.235\textwidth}
    \centering
    \includegraphics[width=\textwidth]{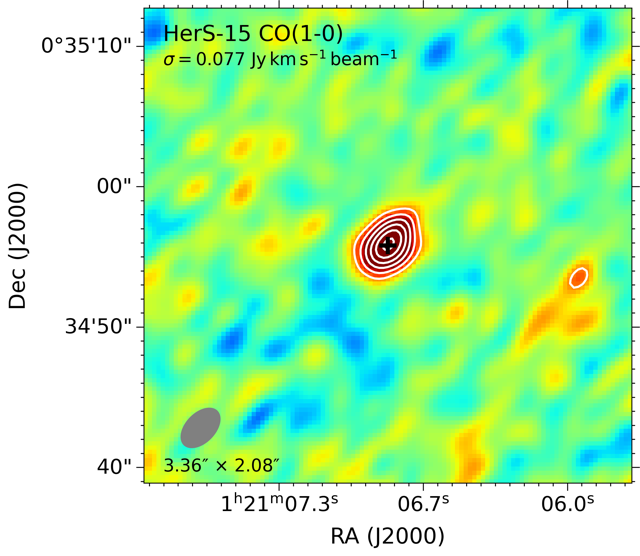}
\end{minipage}
\hfill
\begin{minipage}{0.235\textwidth}
    \centering
    \includegraphics[width=\textwidth]{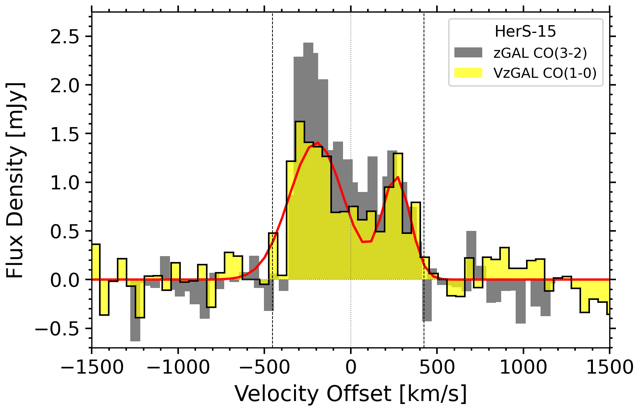}
\end{minipage}
\hfill
\begin{minipage}{0.235\textwidth}
    \centering
    \includegraphics[width=\textwidth]{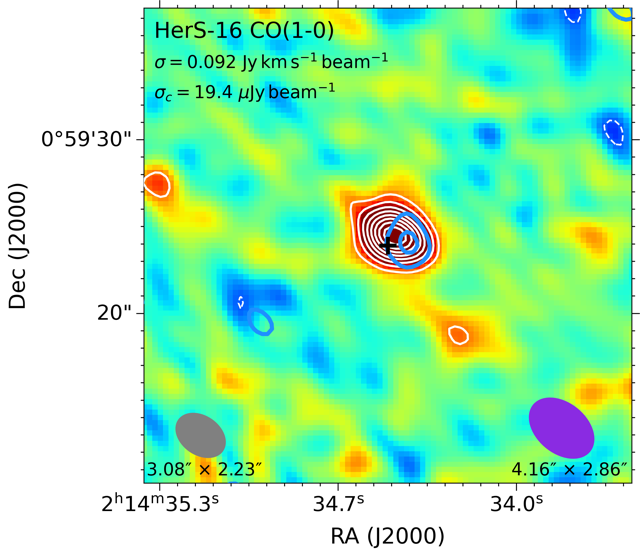}
\end{minipage}
\hfill
\begin{minipage}{0.235\textwidth}
    \centering
    \includegraphics[width=\textwidth]{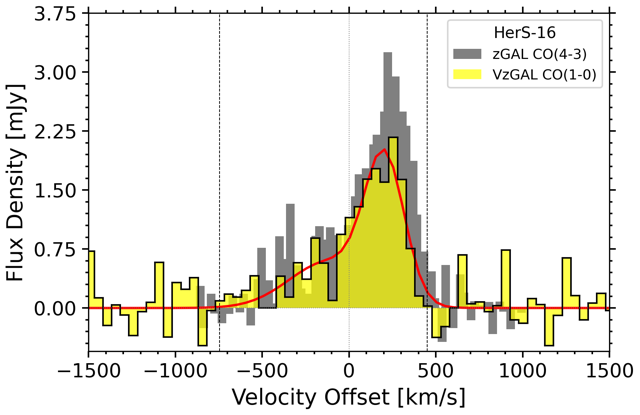}
\end{minipage}

                                \vspace{1em}

\begin{minipage}{0.235\textwidth}
    \centering
    \includegraphics[width=\textwidth]{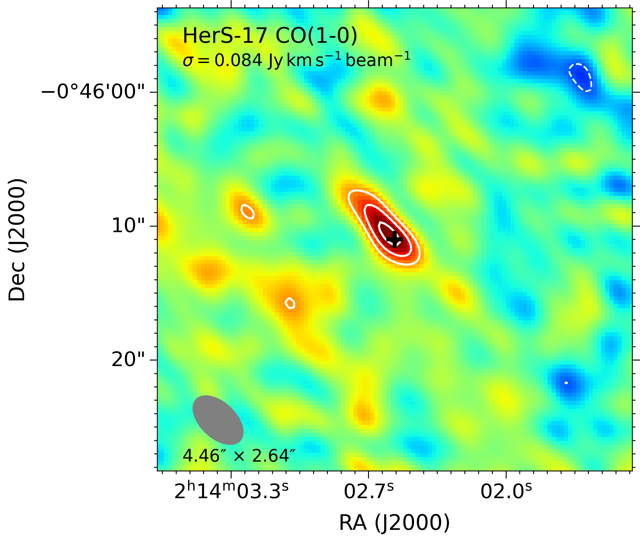}
\end{minipage}
\hfill
\begin{minipage}{0.235\textwidth}
    \centering
    \includegraphics[width=\textwidth]{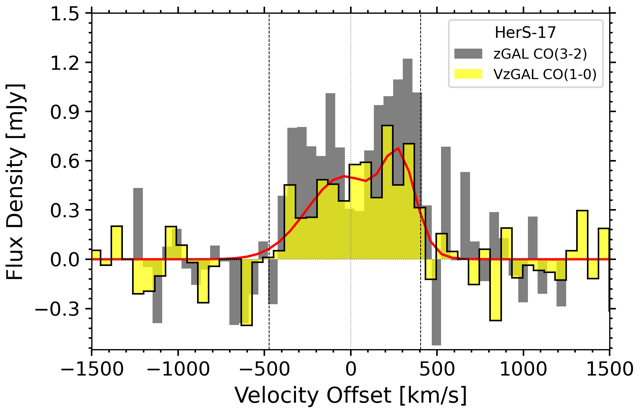}
\end{minipage}
\hfill
\begin{minipage}{0.235\textwidth}
    \centering
    \includegraphics[width=\textwidth]{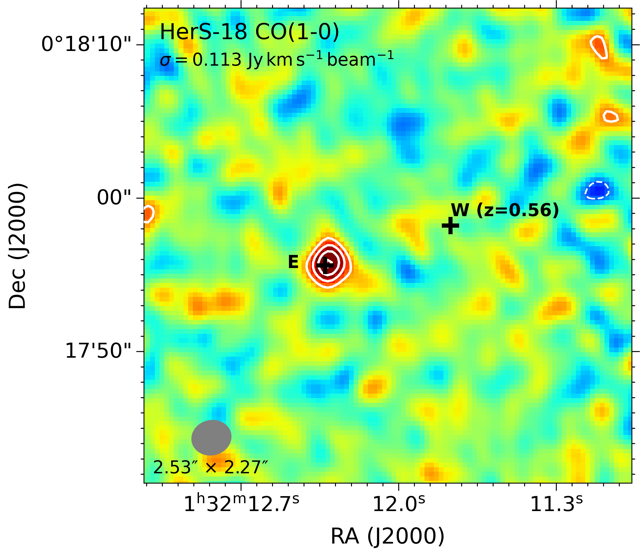}
\end{minipage}
\hfill
\begin{minipage}{0.235\textwidth}
    \centering
    \includegraphics[width=\textwidth]{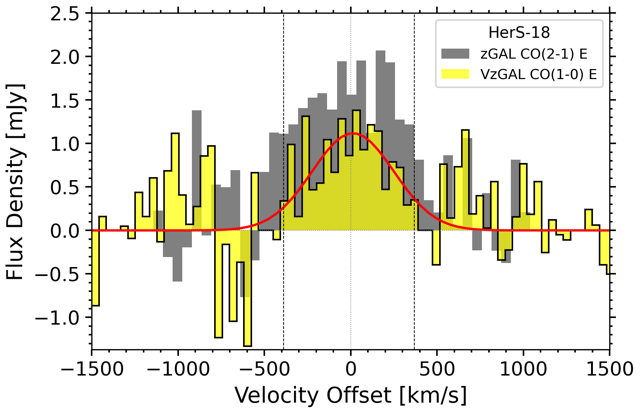}
\end{minipage}

                                        \vspace{1em}

\begin{minipage}{0.235\textwidth}
    \centering
    \includegraphics[width=\textwidth]{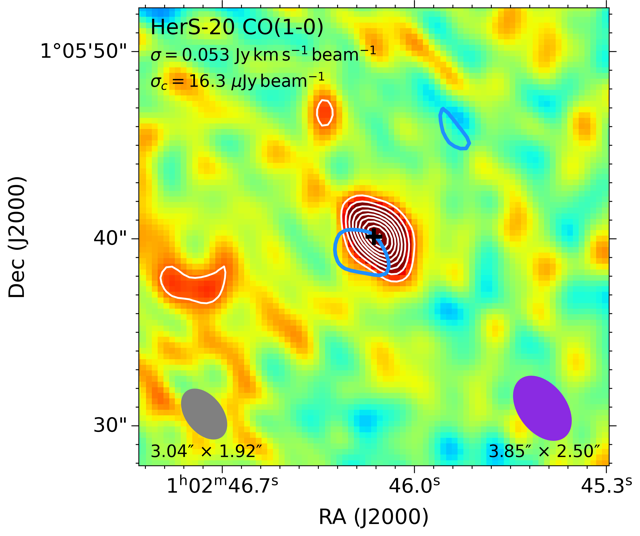}
\end{minipage}
\hfill
\begin{minipage}{0.235\textwidth}
    \centering
    \includegraphics[width=\textwidth]{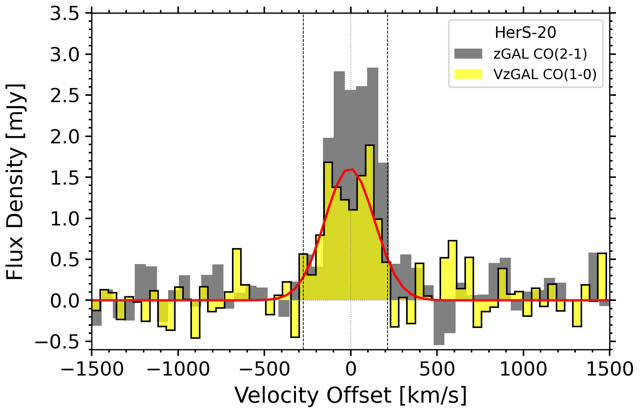}
\end{minipage}

    \addtocounter{figure}{-1}
\caption{(continued)}
\end{figure*}

\clearpage

\begin{figure*}[!htbp]
\centering

\begin{minipage}{0.235\textwidth}
    \centering
    \includegraphics[width=\textwidth]{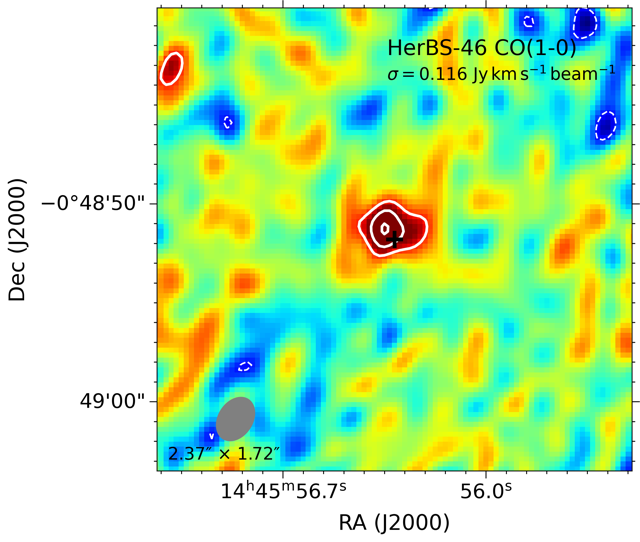}
\end{minipage}
\hfill
\begin{minipage}{0.235\textwidth}
    \centering
    \includegraphics[width=\textwidth]{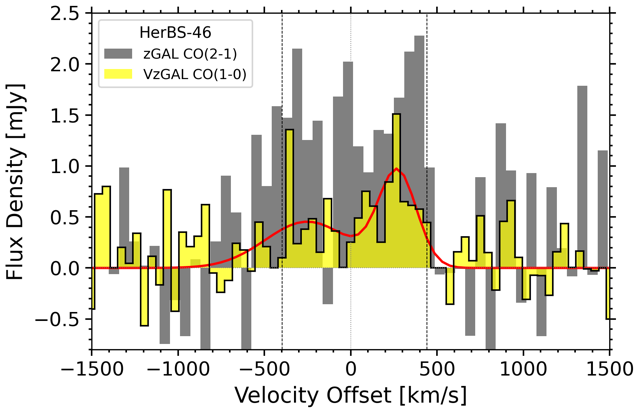}
\end{minipage}
\hfill
\begin{minipage}{0.235\textwidth}
    \centering
    \includegraphics[width=\textwidth]{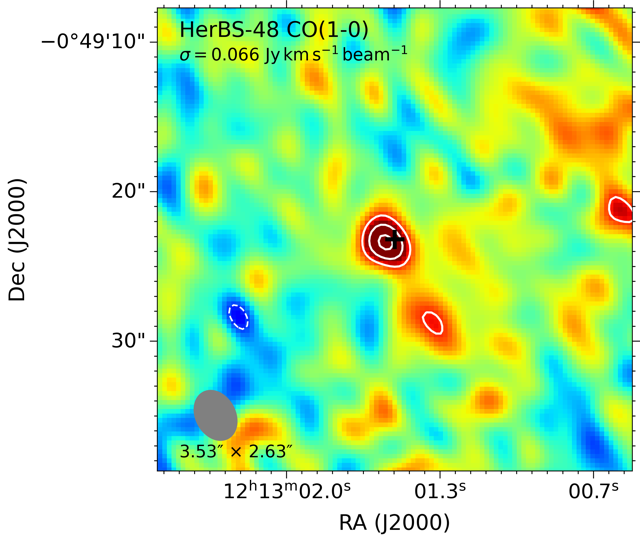}
\end{minipage}
\hfill
\begin{minipage}{0.235\textwidth}
    \centering
    \includegraphics[width=\textwidth]{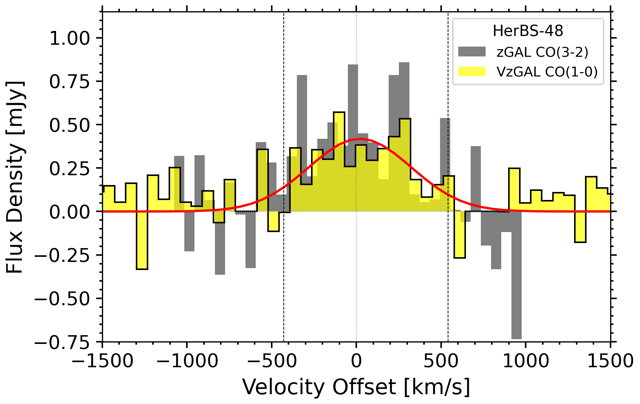}
\end{minipage}

                                            \vspace{1em}

\begin{minipage}{0.235\textwidth}
    \centering
    \includegraphics[width=\textwidth]{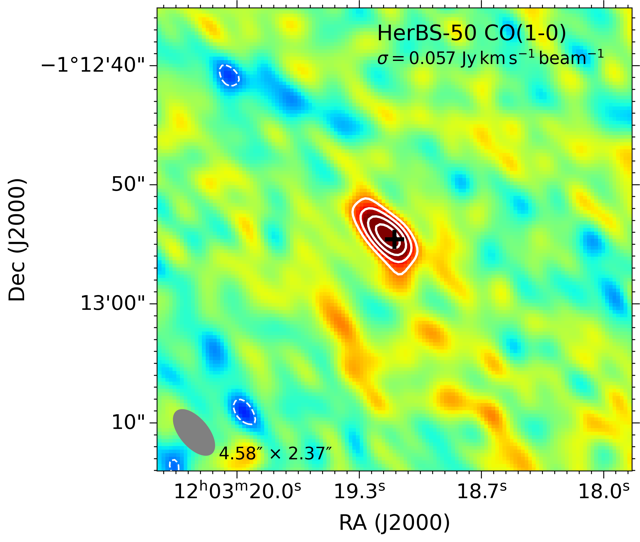}
\end{minipage}
\hfill
\begin{minipage}{0.235\textwidth}
    \centering
    \includegraphics[width=\textwidth]{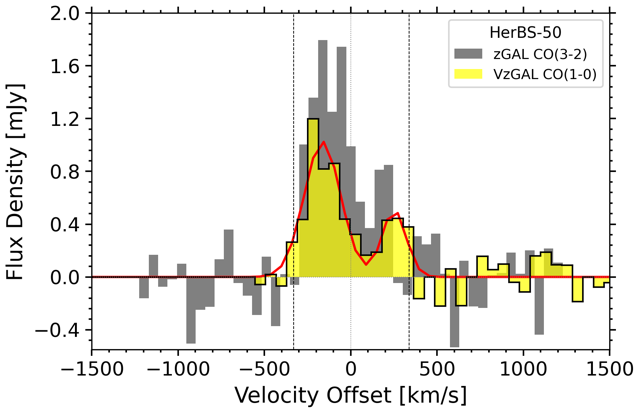}
\end{minipage}
\hfill
\begin{minipage}{0.235\textwidth}
    \centering
    \includegraphics[width=\textwidth]{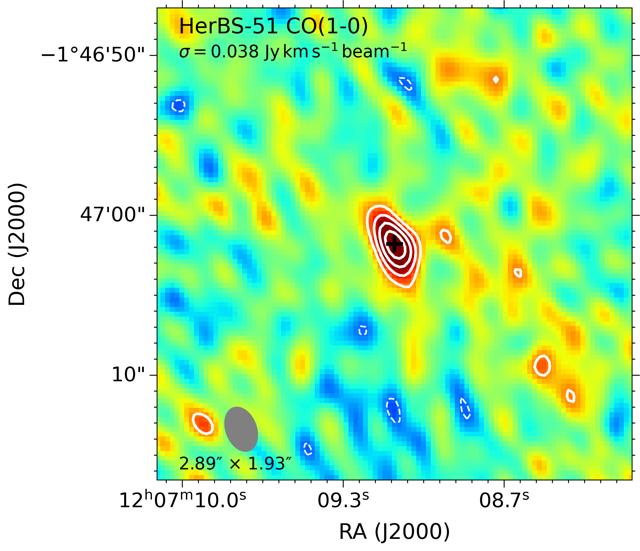}
\end{minipage}
\hfill
\begin{minipage}{0.235\textwidth}
    \centering
    \includegraphics[width=\textwidth]{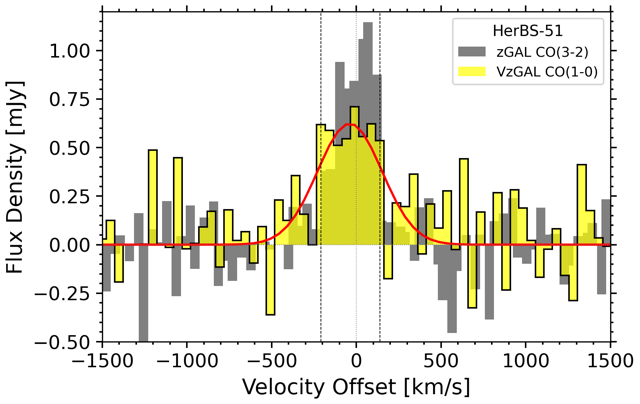}
\end{minipage}

                                            \vspace{1em}

\begin{minipage}{0.235\textwidth}
    \centering
    \includegraphics[width=\textwidth]{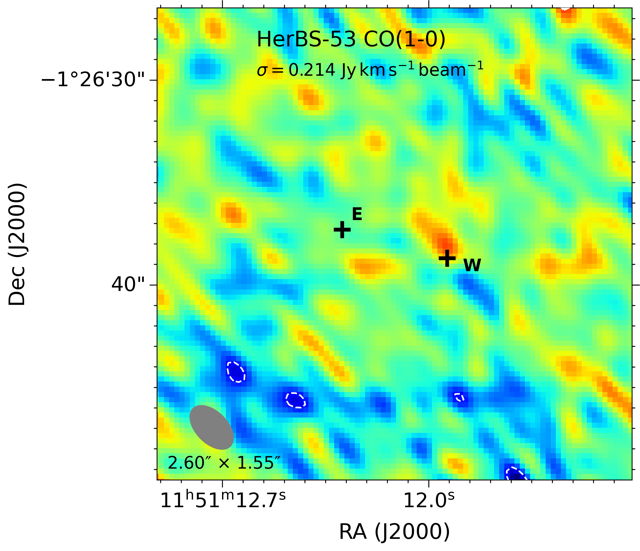}
\end{minipage}
\hfill
\begin{minipage}{0.235\textwidth}
    \centering
    \includegraphics[width=\textwidth]{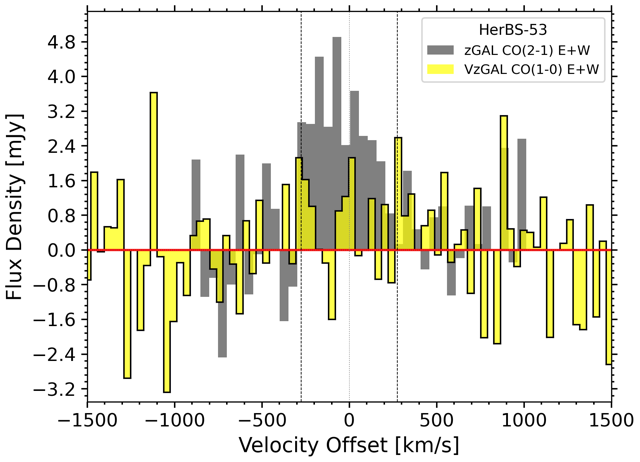}
\end{minipage}
\hfill
\begin{minipage}{0.235\textwidth}
    \centering
    \includegraphics[width=\textwidth]{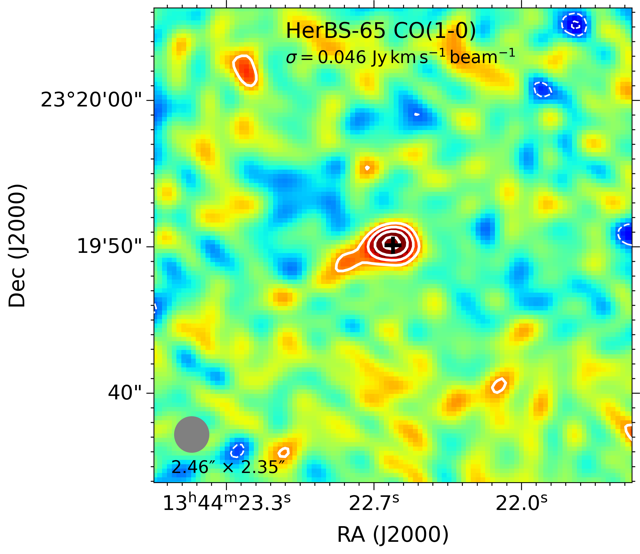}
\end{minipage}
\hfill
\begin{minipage}{0.235\textwidth}
    \centering
    \includegraphics[width=\textwidth]{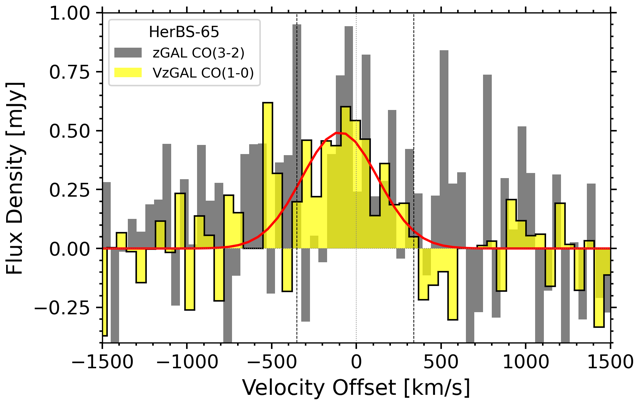}
\end{minipage}

                                            \vspace{1em}

\begin{minipage}{0.235\textwidth}
    \centering
    \includegraphics[width=\textwidth]{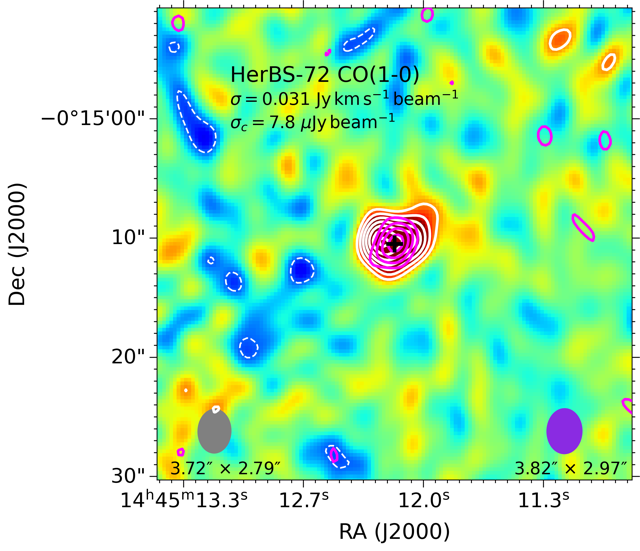}
\end{minipage}
\hfill
\begin{minipage}{0.235\textwidth}
    \centering
    \includegraphics[width=\textwidth]{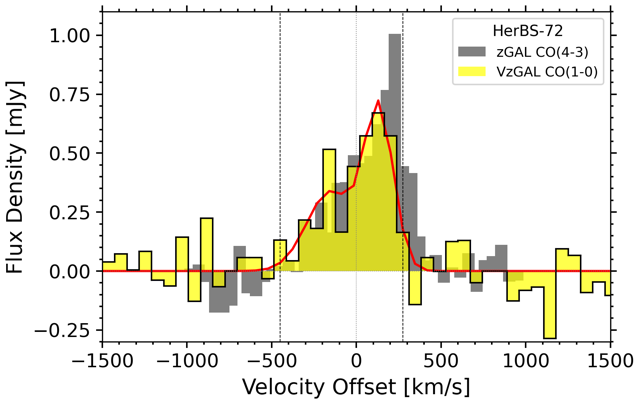}
\end{minipage}

                                            \vspace{1em}

\begin{minipage}{0.235\textwidth}
    \centering
    \includegraphics[width=\textwidth]{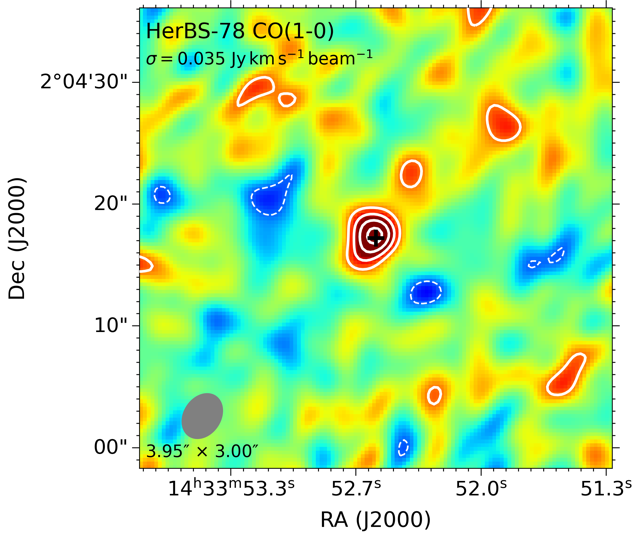}
\end{minipage}
\hfill
\begin{minipage}{0.235\textwidth}
    \centering
    \includegraphics[width=\textwidth]{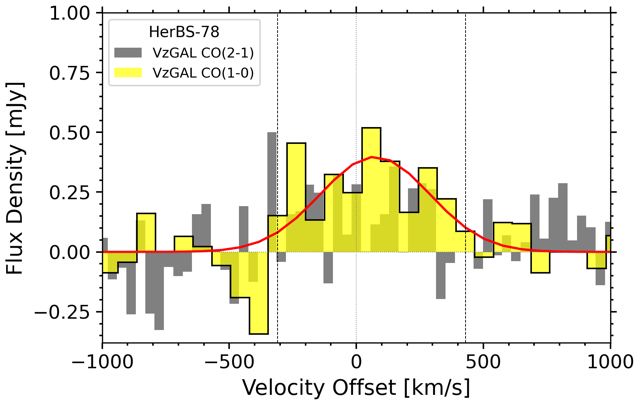}
\end{minipage}
\hfill
\begin{minipage}{0.235\textwidth}
    \centering
    \includegraphics[width=\textwidth]{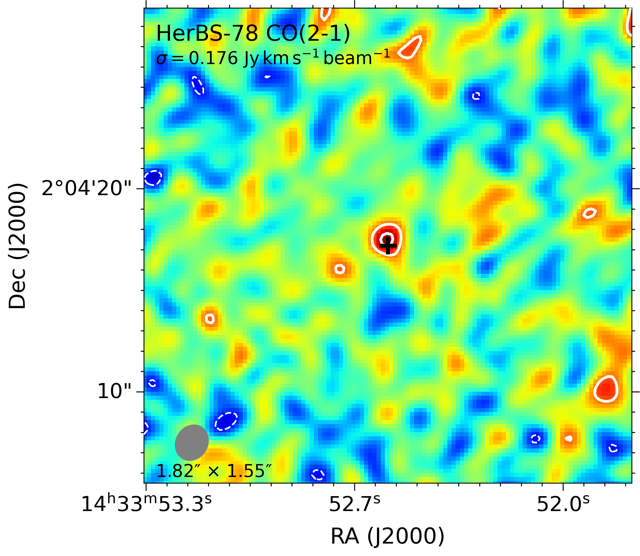}
\end{minipage}
\hfill
\begin{minipage}{0.235\textwidth}
    \centering
    \includegraphics[width=\textwidth]{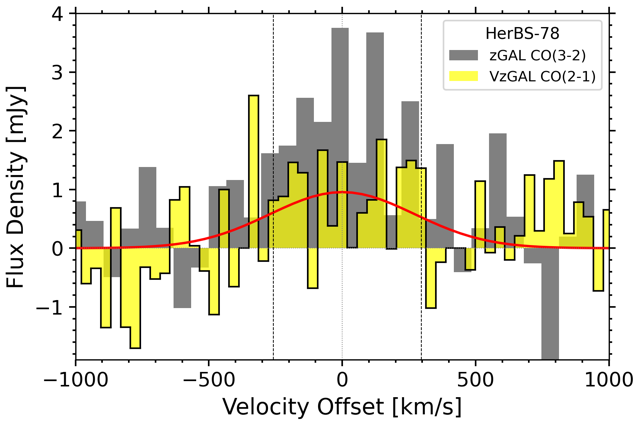}
\end{minipage}


\caption{Same as Figure~\ref{fig:image-grid_helms_v1} with DSFGs from the HerBS fields.} 
\label{fig:image-grid_herbs}
\end{figure*}

\clearpage

\begin{figure*}[!htbp]
\centering

\begin{minipage}{0.235\textwidth}
    \centering
    \includegraphics[width=\textwidth]{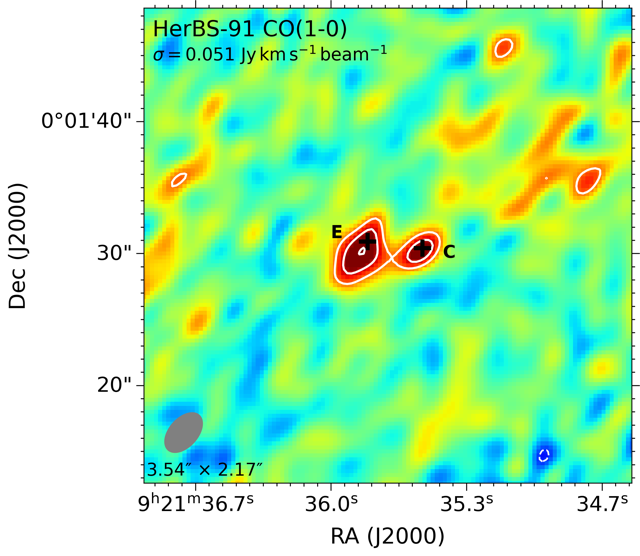}
\end{minipage}
\hfill
\begin{minipage}{0.235\textwidth}
    \centering
    \includegraphics[width=\textwidth]{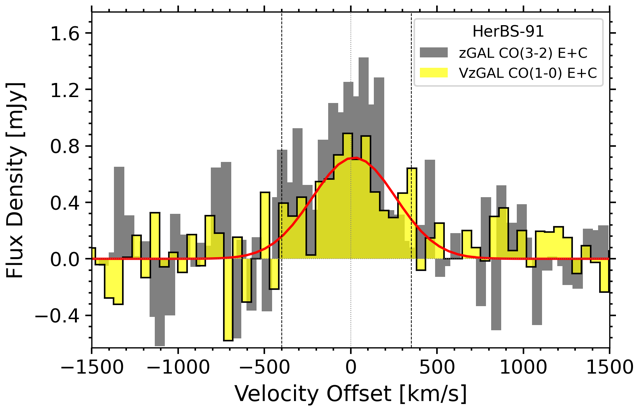}
\end{minipage}
\hfill
\begin{minipage}{0.235\textwidth}
    \centering
    \includegraphics[width=\textwidth]{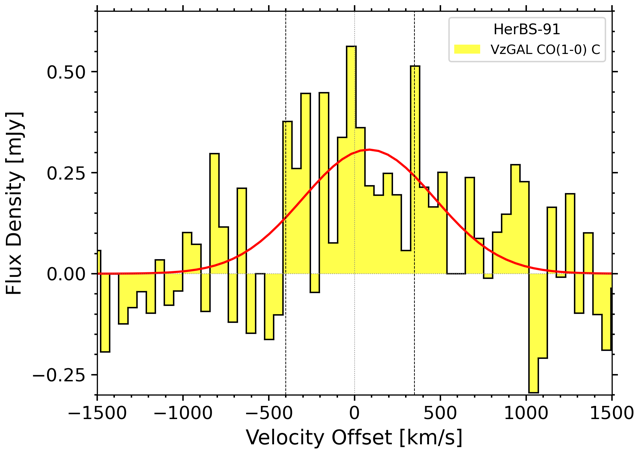}
\end{minipage}
\hfill
\begin{minipage}{0.235\textwidth}
    \centering
    \includegraphics[width=\textwidth]{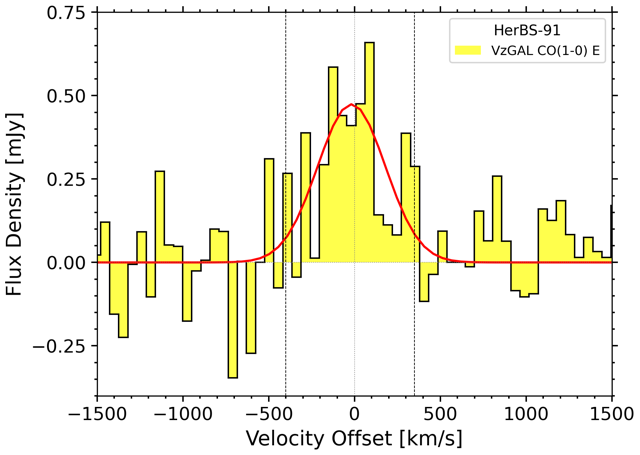}
\end{minipage}

                                        \vspace{1em}
    
\begin{minipage}{0.235\textwidth}
    \centering
    \includegraphics[width=\textwidth]{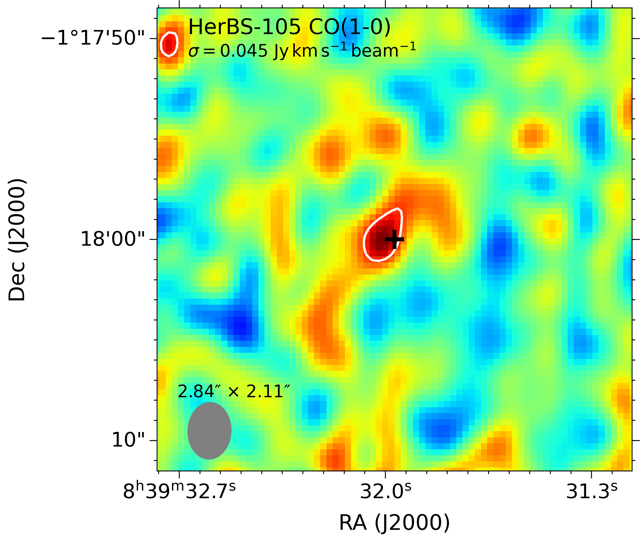}
\end{minipage}
\hfill
\begin{minipage}{0.235\textwidth}
    \centering
    \includegraphics[width=\textwidth]{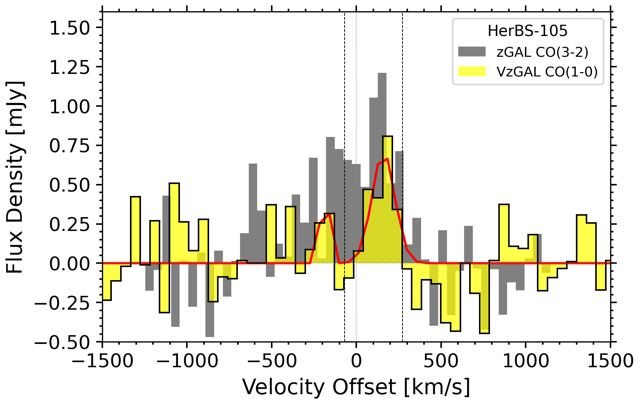}
\end{minipage}
\hfill
\begin{minipage}{0.235\textwidth}
    \centering
    \includegraphics[width=\textwidth]{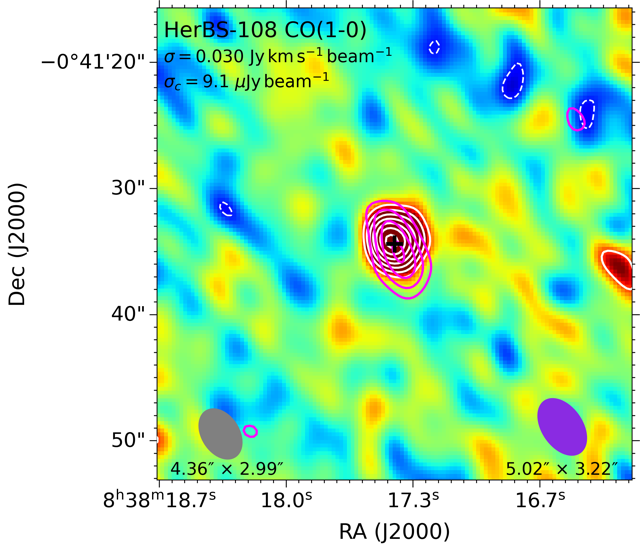}
\end{minipage}
\hfill
\begin{minipage}{0.235\textwidth}
    \centering
    \includegraphics[width=\textwidth]{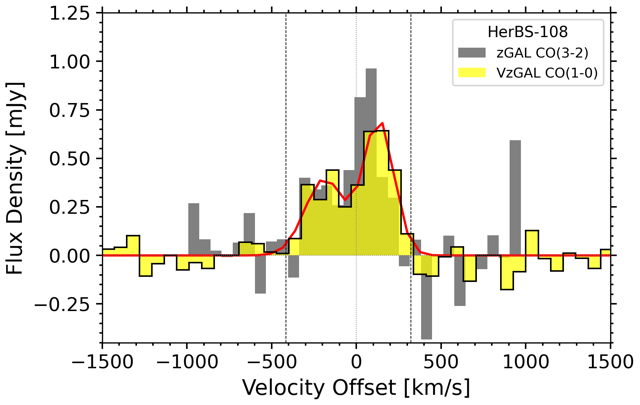}
\end{minipage}

                                            \vspace{1em}

\begin{minipage}{0.235\textwidth}
    \centering
    \includegraphics[width=\textwidth]{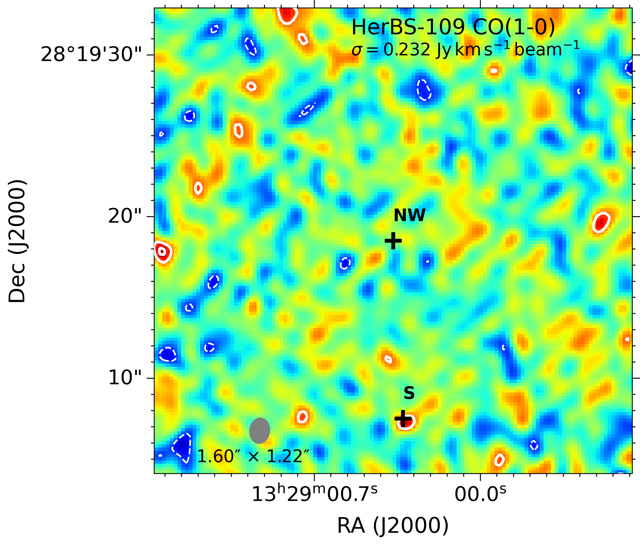}
\end{minipage}
\hfill
\begin{minipage}{0.235\textwidth}
    \centering
    \includegraphics[width=\textwidth]{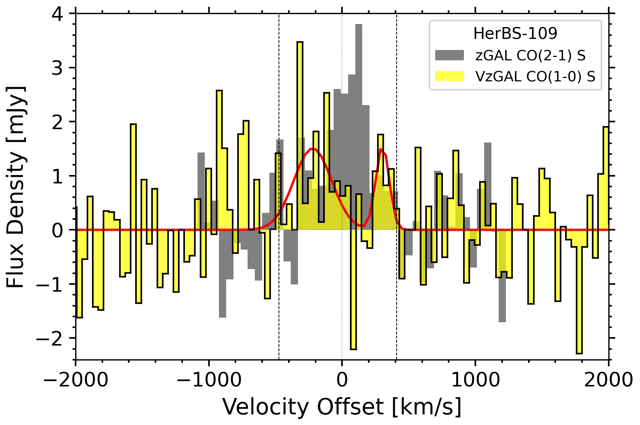}
\end{minipage}
\hfill
\begin{minipage}{0.235\textwidth}
    \centering
    \includegraphics[width=\textwidth]{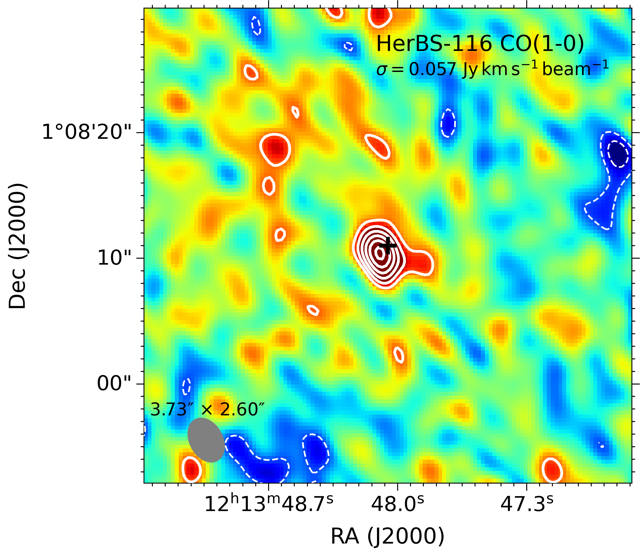}
\end{minipage}
\hfill
\begin{minipage}{0.235\textwidth}
    \centering
    \includegraphics[width=\textwidth]{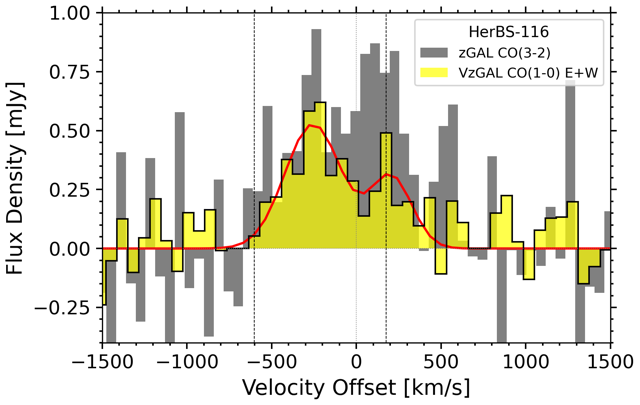}
\end{minipage}

                                            \vspace{1em}

\begin{minipage}{0.235\textwidth}
    \centering
    \includegraphics[width=\textwidth]{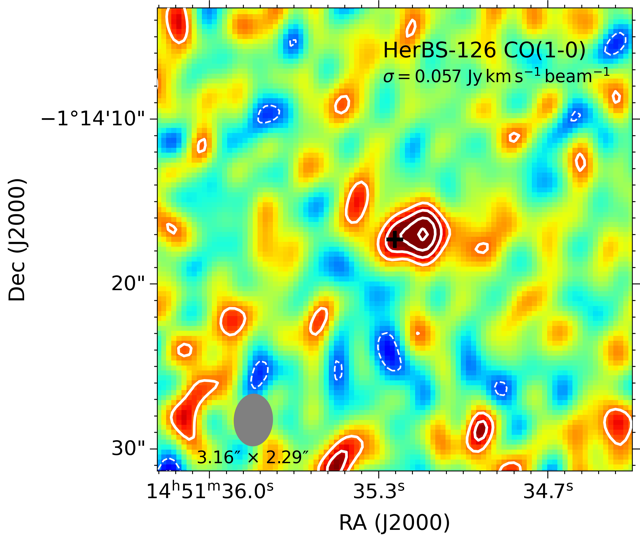}
\end{minipage}
\hfill
\begin{minipage}{0.235\textwidth}
    \centering
    \includegraphics[width=\textwidth]{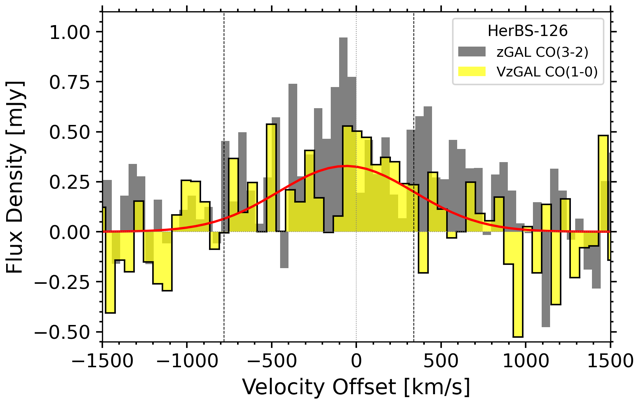}
\end{minipage}
\hfill
\begin{minipage}{0.235\textwidth}
    \centering
    \includegraphics[width=\textwidth]{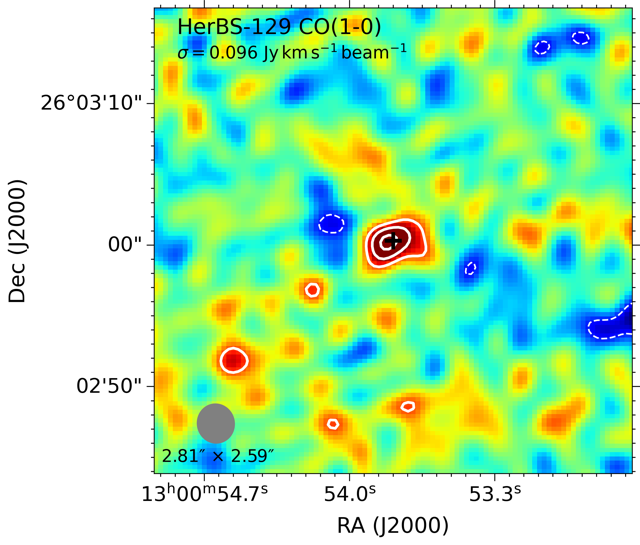}
\end{minipage}
\hfill
\begin{minipage}{0.235\textwidth}
    \centering
    \includegraphics[width=\textwidth]{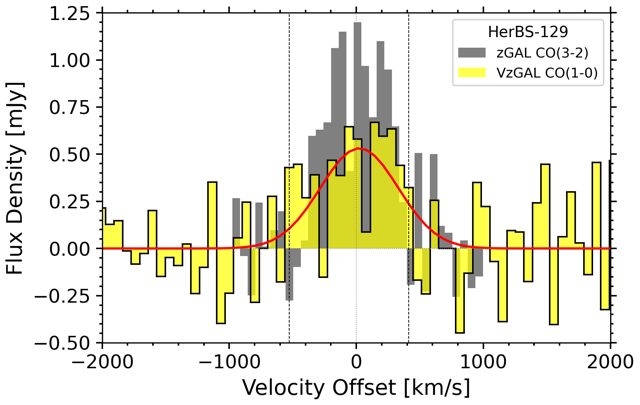}
\end{minipage}
 
                            \vspace{1em}
                                      
\begin{minipage}{0.235\textwidth}
    \centering
    \includegraphics[width=\textwidth]{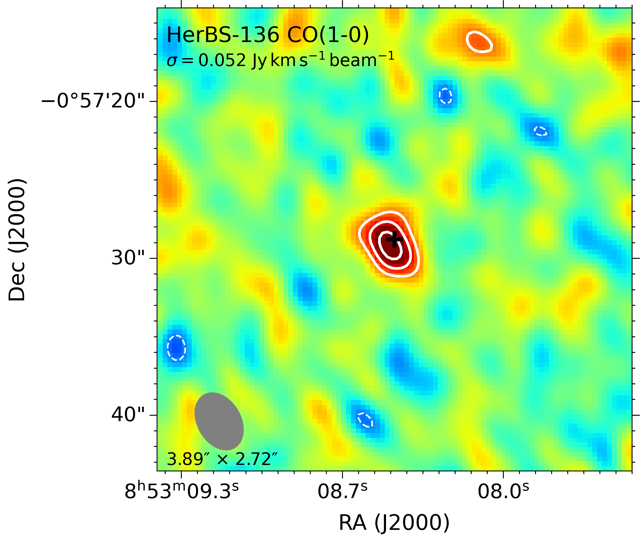}
\end{minipage}
\hfill
\begin{minipage}{0.235\textwidth}
    \centering
    \includegraphics[width=\textwidth]{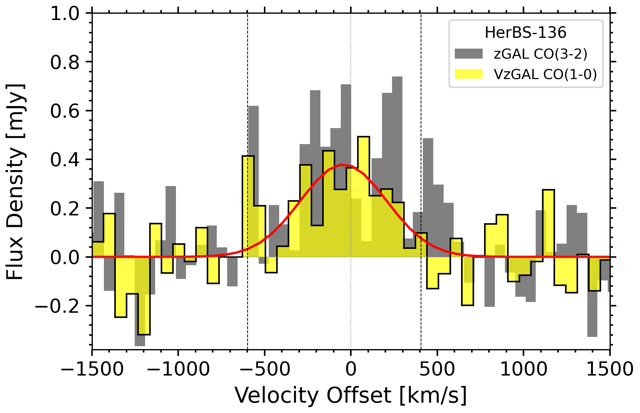}
\end{minipage}
\hfill
\begin{minipage}{0.235\textwidth}
    \centering
    \includegraphics[width=\textwidth]{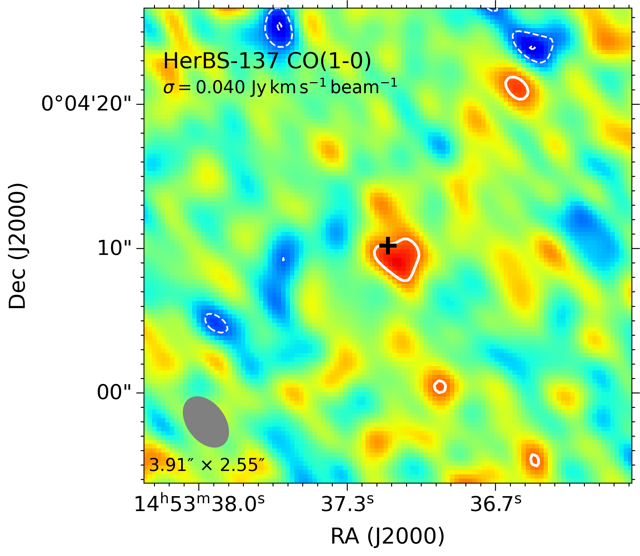}
\end{minipage}
\hfill
\begin{minipage}{0.235\textwidth}
    \centering
    \includegraphics[width=\textwidth]{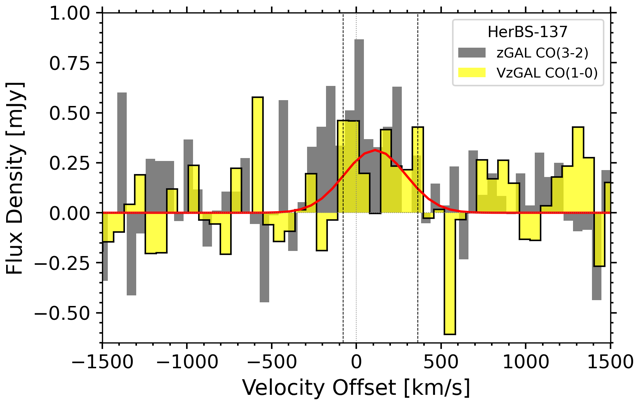}
\end{minipage}
 
                                            \vspace{1em}
                                      
\begin{minipage}{0.235\textwidth}
    \centering
    \includegraphics[width=\textwidth]{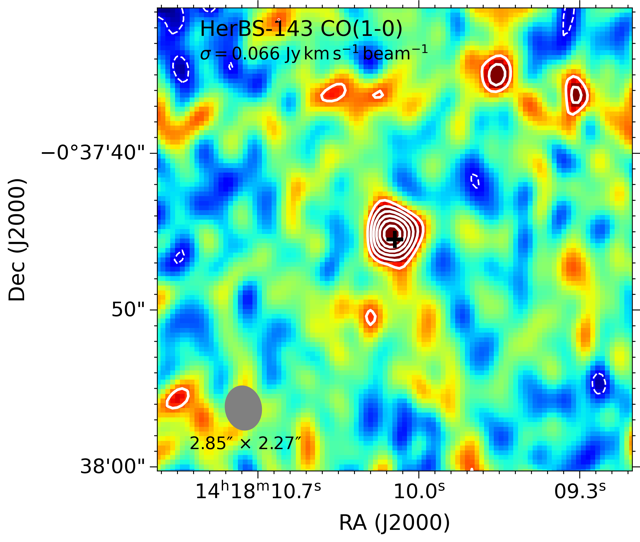}
\end{minipage}
\hfill
\begin{minipage}{0.235\textwidth}
    \centering
    \includegraphics[width=\textwidth]{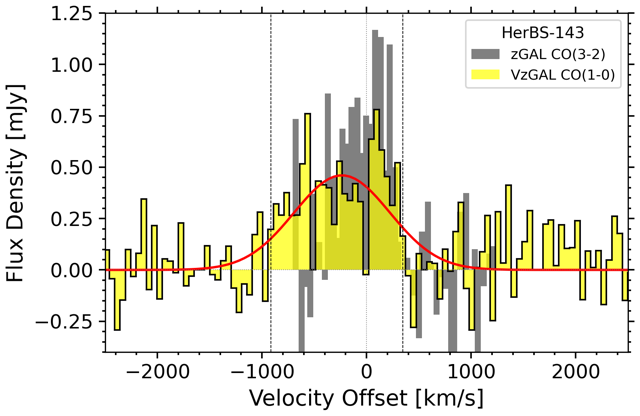}
\end{minipage}
\hfill
\begin{minipage}{0.235\textwidth}
    \centering
    \includegraphics[width=\textwidth]{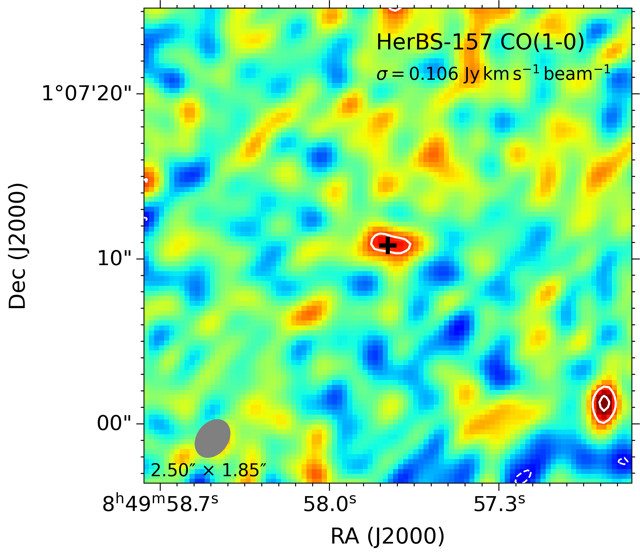}
\end{minipage}
\hfill
\begin{minipage}{0.235\textwidth}
    \centering
    \includegraphics[width=\textwidth]{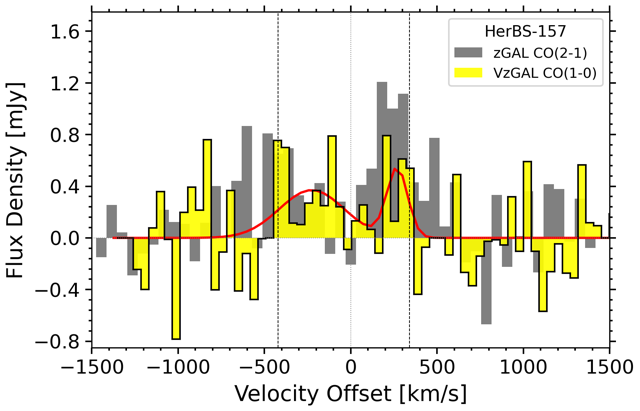}
\end{minipage}

    \addtocounter{figure}{-1}
\caption{(continued)}
\end{figure*}

\clearpage

\begin{figure*}[!htbp]
\centering

\begin{minipage}{0.235\textwidth}
    \centering
    \includegraphics[width=\textwidth]{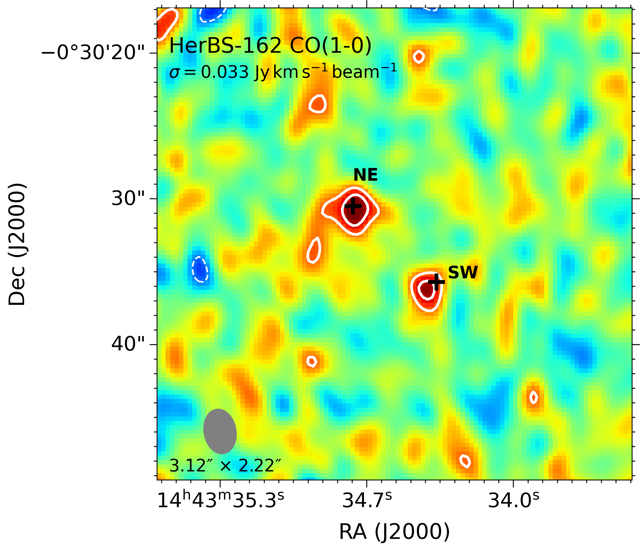}
\end{minipage}
\hfill
\begin{minipage}{0.235\textwidth}
    \centering
    \includegraphics[width=\textwidth]{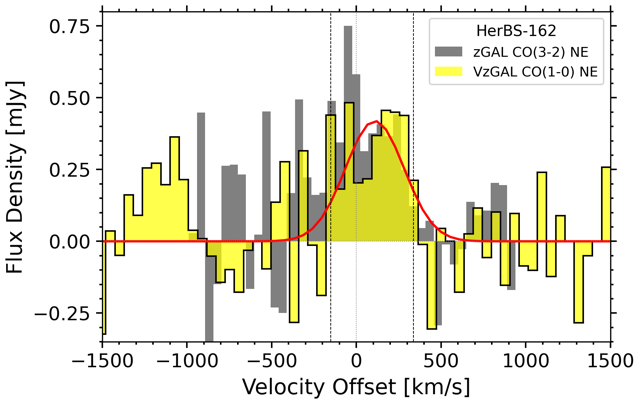}
\end{minipage}
\hfill
\begin{minipage}{0.235\textwidth}
    \centering
    \includegraphics[width=\textwidth]{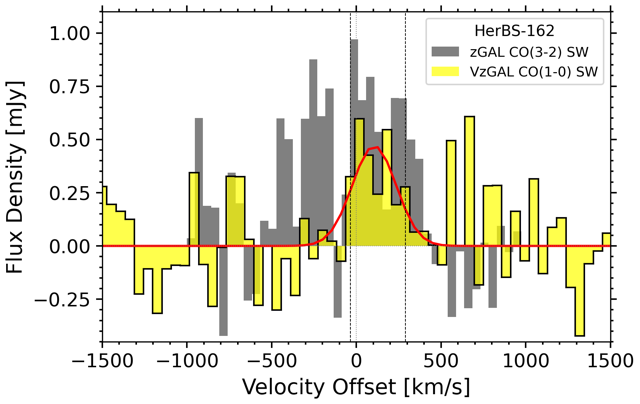}
\end{minipage}

                                \vspace{1em}
    
\begin{minipage}{0.235\textwidth}
    \centering
    \includegraphics[width=\textwidth]{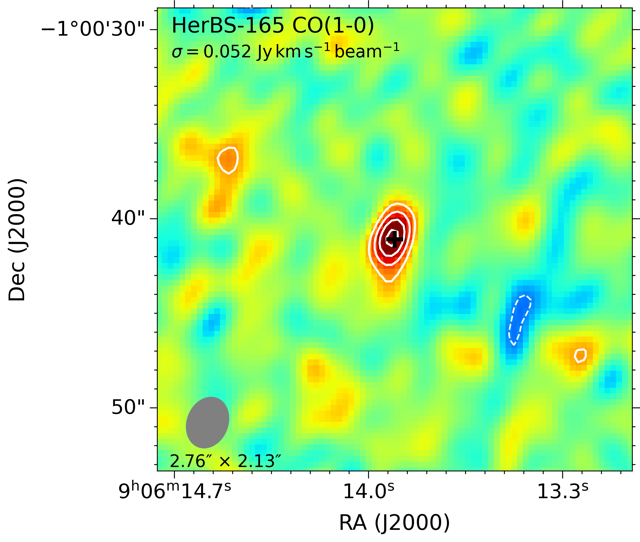}
\end{minipage}
\hfill
\begin{minipage}{0.235\textwidth}
    \centering
    \includegraphics[width=\textwidth]{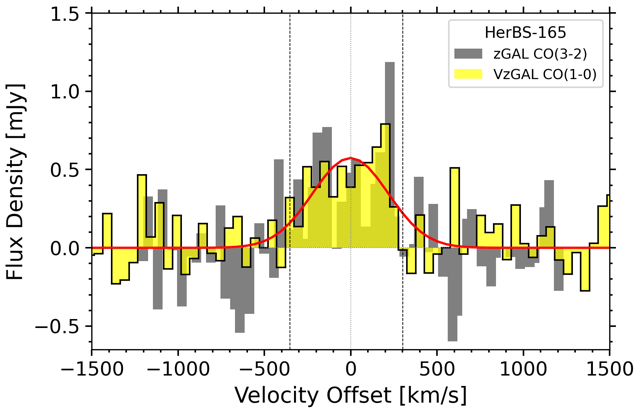}
\end{minipage}
\hfill
\begin{minipage}{0.235\textwidth}
    \centering
    \includegraphics[width=\textwidth]{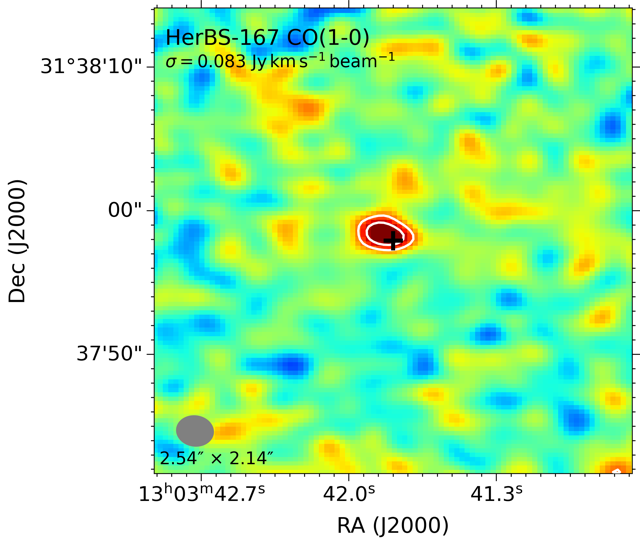}
\end{minipage}
\hfill
\begin{minipage}{0.235\textwidth}
    \centering
    \includegraphics[width=\textwidth]{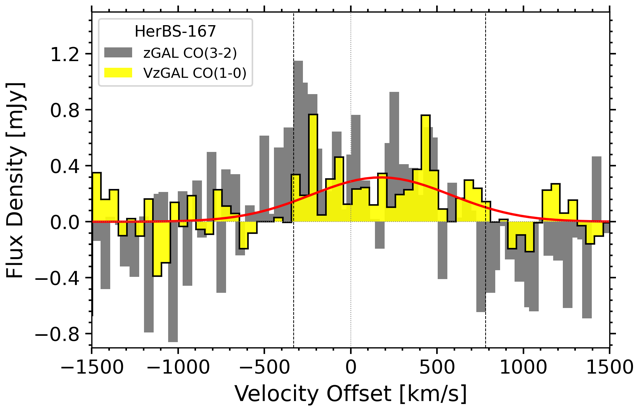}
\end{minipage}

                                           \vspace{1em}
                              
\begin{minipage}{0.235\textwidth}
    \centering
    \includegraphics[width=\textwidth]{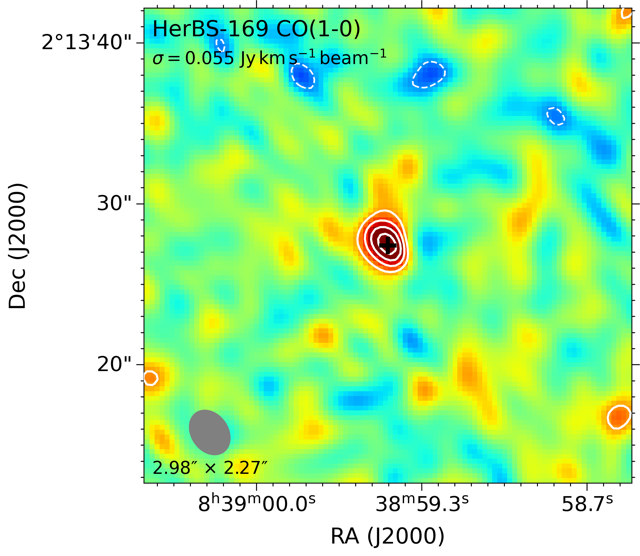}
\end{minipage}
\hfill
\begin{minipage}{0.235\textwidth}
    \centering
    \includegraphics[width=\textwidth]{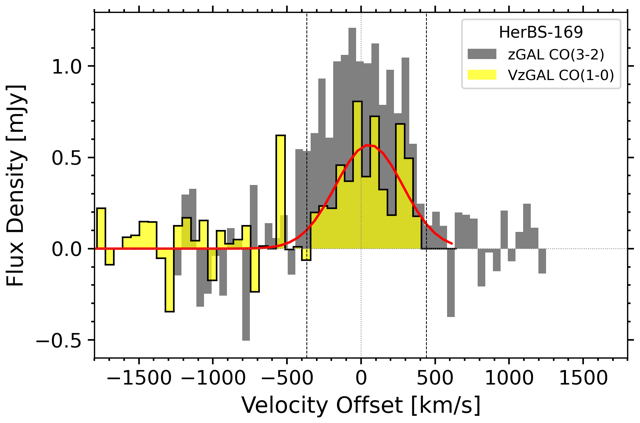}
\end{minipage}
\hfill
\begin{minipage}{0.235\textwidth}
    \centering
    \includegraphics[width=\textwidth]{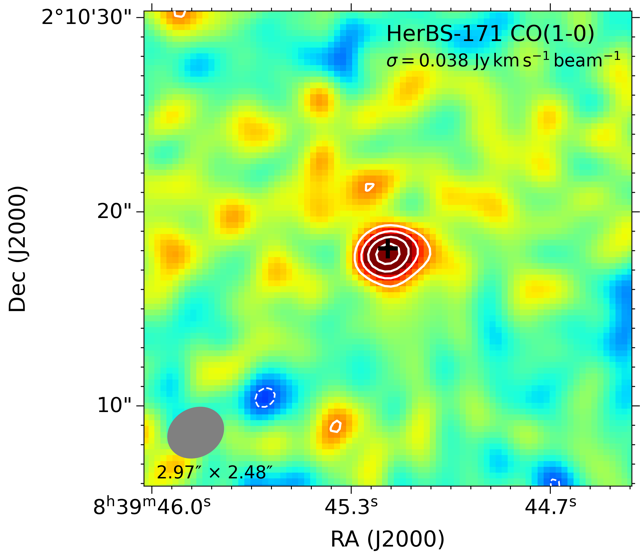}
\end{minipage}
\hfill
\begin{minipage}{0.235\textwidth}
    \centering
    \includegraphics[width=\textwidth]{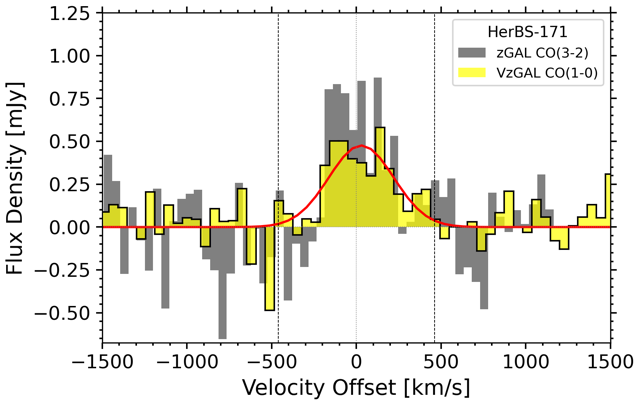}
\end{minipage}

                                           \vspace{1em}
                              
\begin{minipage}{0.235\textwidth}
    \centering
    \includegraphics[width=\textwidth]{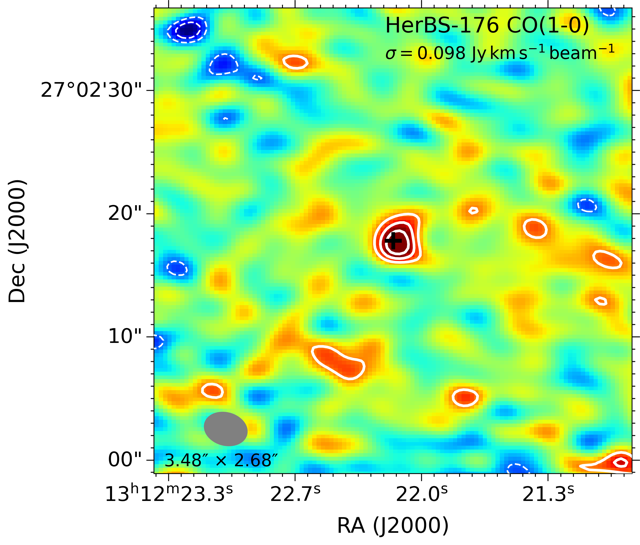}
\end{minipage}
\hfill
\begin{minipage}{0.235\textwidth}
    \centering
    \includegraphics[width=\textwidth]{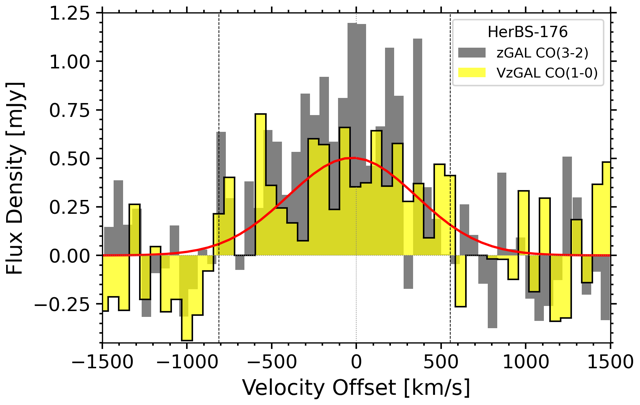}
\end{minipage}
\hfill
\begin{minipage}{0.235\textwidth}
    \centering
    \includegraphics[width=\textwidth]{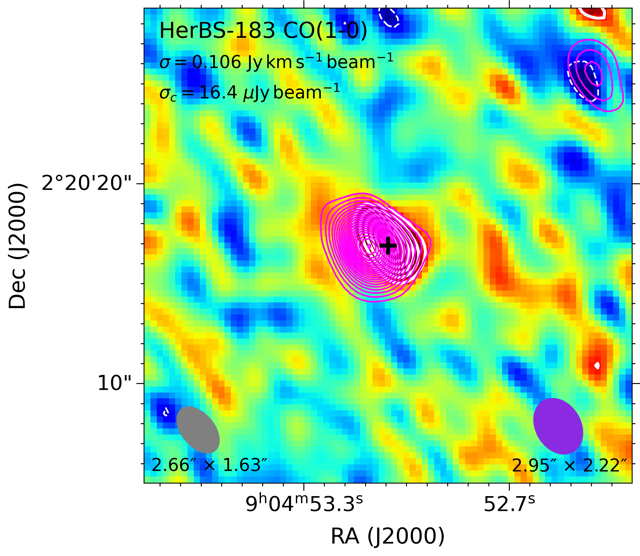}
\end{minipage}
\hfill
\begin{minipage}{0.235\textwidth}
    \centering
    \includegraphics[width=\textwidth]{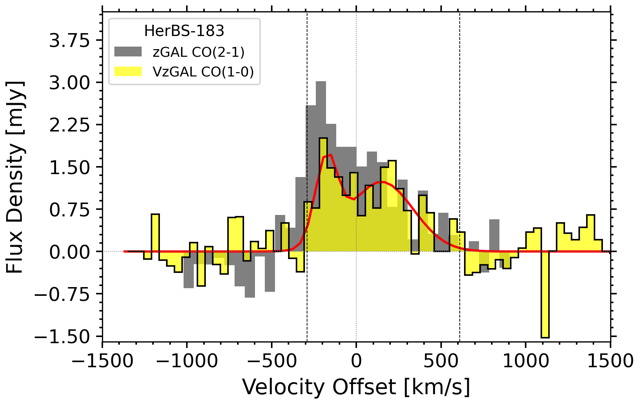}
\end{minipage}
                        \vspace{1em}
                                        
\begin{minipage}{0.235\textwidth}
    \centering
    \includegraphics[width=\textwidth]{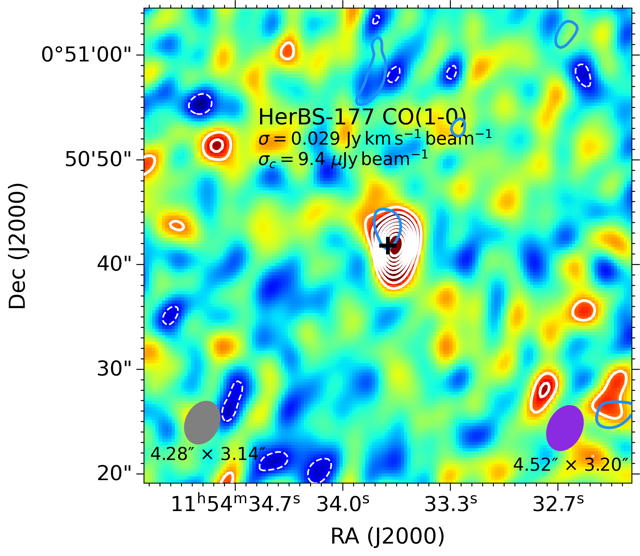}
\end{minipage}
\hfill
\begin{minipage}{0.235\textwidth}
    \centering
    \includegraphics[width=\textwidth]{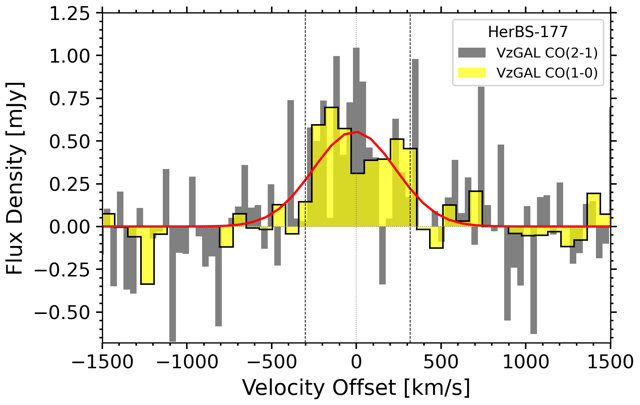}
\end{minipage}
\hfill
\begin{minipage}{0.235\textwidth}
    \centering
    \includegraphics[width=\textwidth]{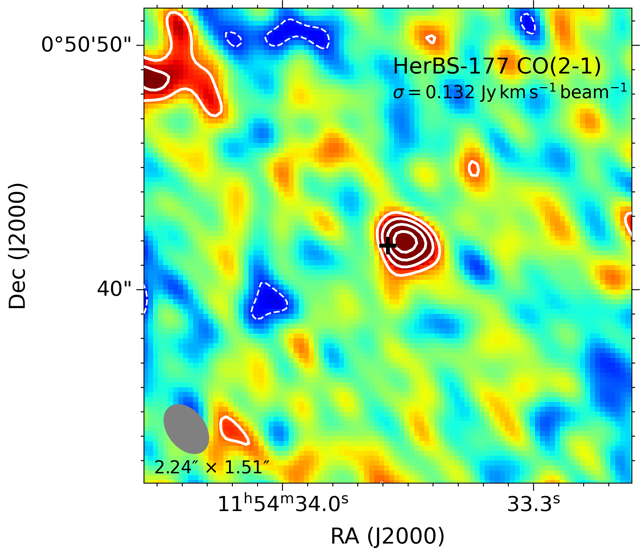}
\end{minipage}
\hfill
\begin{minipage}{0.235\textwidth}
    \centering
    \includegraphics[width=\textwidth]{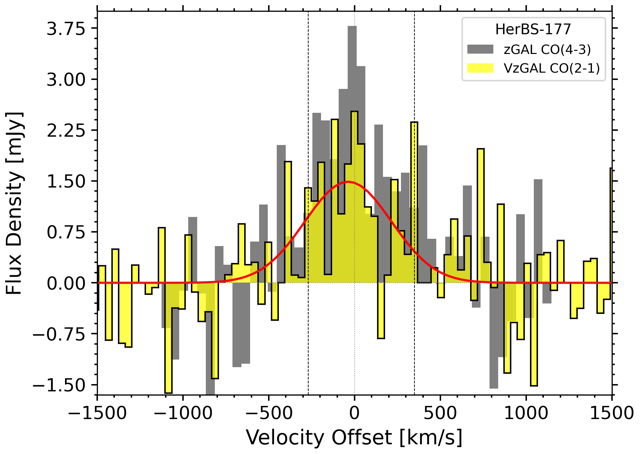}
\end{minipage}

                        \vspace{1em}
                                      
\begin{minipage}{0.235\textwidth}
    \centering
    \includegraphics[width=\textwidth]{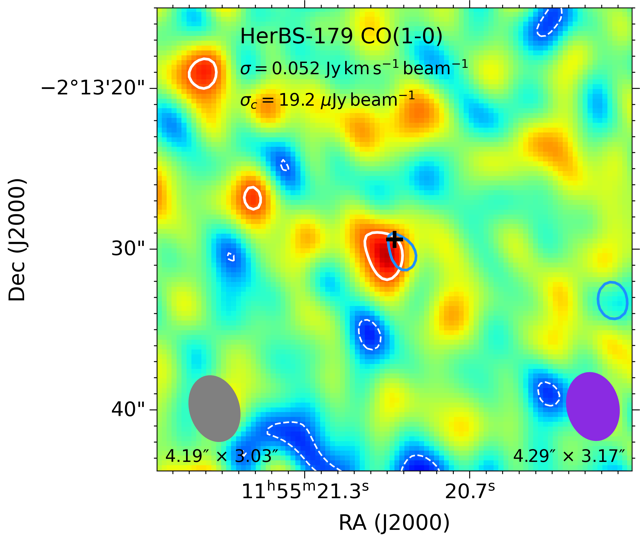}
\end{minipage}
\hfill
\begin{minipage}{0.235\textwidth}
    \centering
    \includegraphics[width=\textwidth]{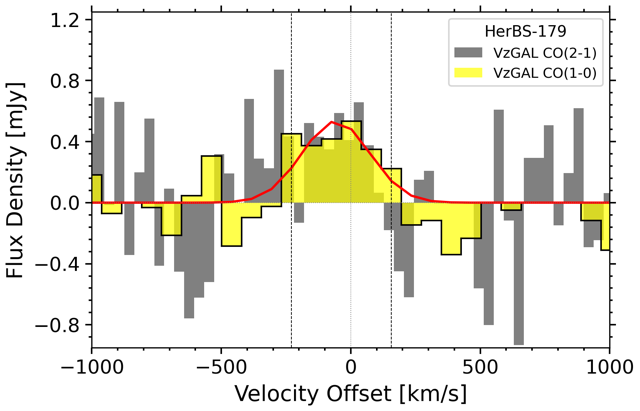}
\end{minipage}
\hfill
\begin{minipage}{0.235\textwidth}
    \centering
    \includegraphics[width=\textwidth]{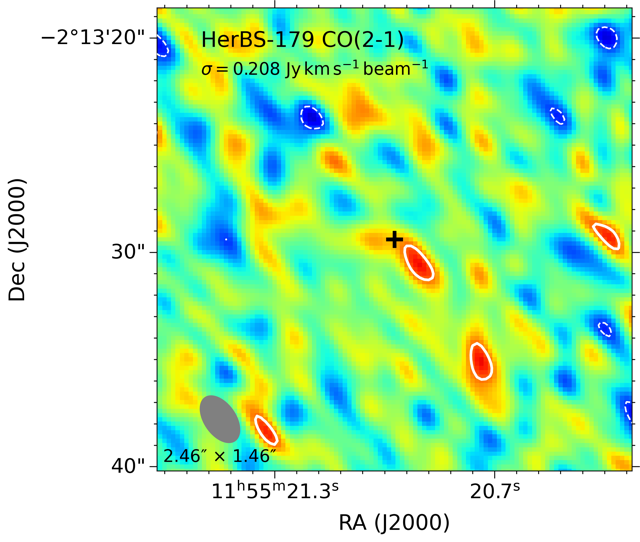}
\end{minipage}
\hfill
\begin{minipage}{0.235\textwidth}
    \centering
    \includegraphics[width=\textwidth]{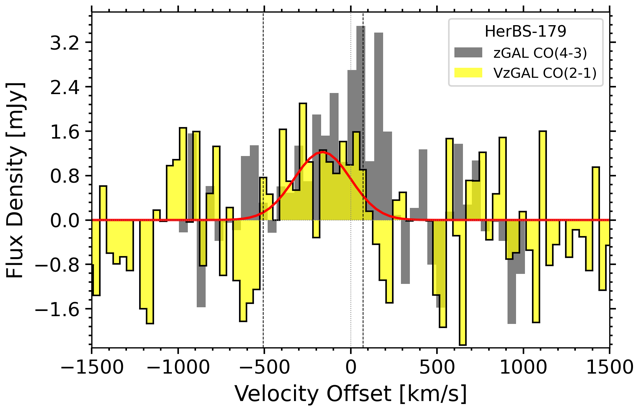}
\end{minipage}

    \addtocounter{figure}{-1}
\caption{(continued)}
\end{figure*}

\clearpage

\begin{figure*}[!htbp]
\centering

\begin{minipage}{0.235\textwidth}
    \centering
    \includegraphics[width=\textwidth]{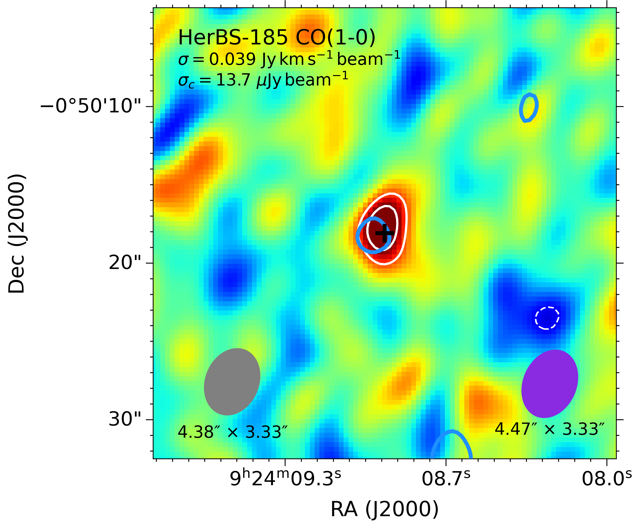}
\end{minipage}
\hfill
\begin{minipage}{0.235\textwidth}
    \centering
    \includegraphics[width=\textwidth]{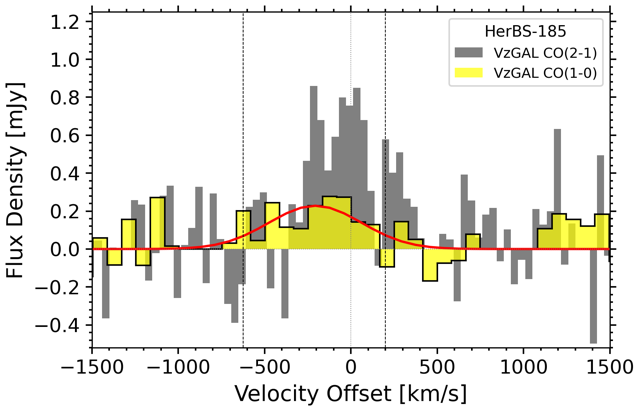}
\end{minipage}
\hfill
\begin{minipage}{0.235\textwidth}
    \centering
    \includegraphics[width=\textwidth]{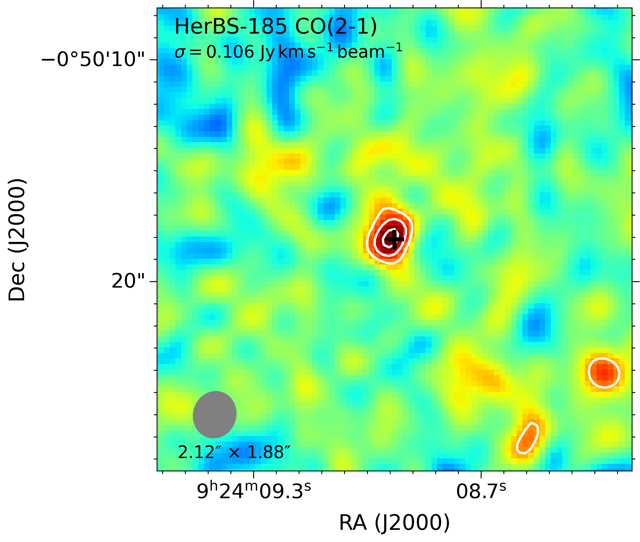}
\end{minipage}
\hfill
\begin{minipage}{0.235\textwidth}
    \centering
    \includegraphics[width=\textwidth]{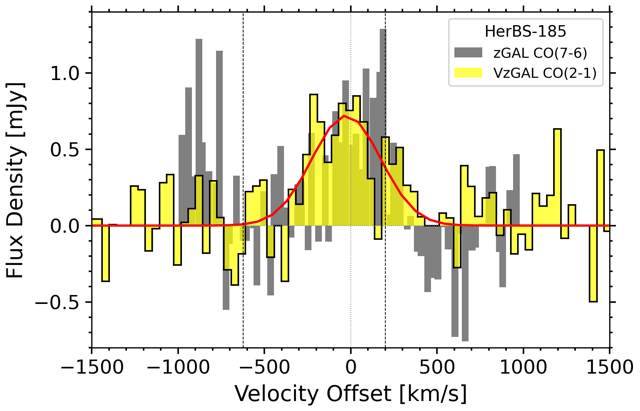}
\end{minipage} 

                                        \vspace{1em}                          

\begin{minipage}{0.235\textwidth}
    \centering
    \includegraphics[width=\textwidth]{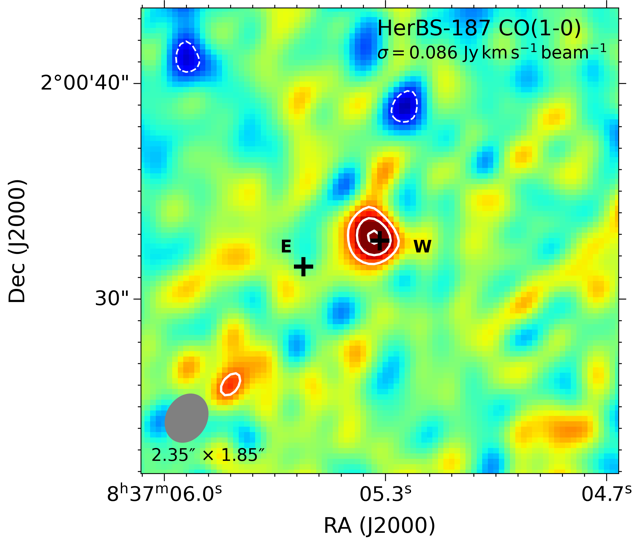}
\end{minipage}
\hfill
\begin{minipage}{0.235\textwidth}
    \centering
    \includegraphics[width=\textwidth]{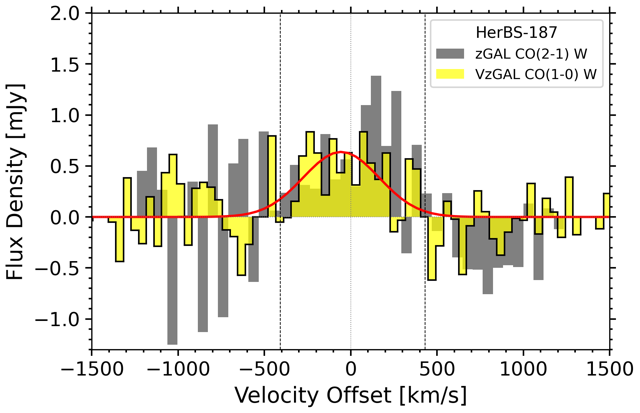}
\end{minipage}
\hfill
\begin{minipage}{0.235\textwidth}
    \centering
    \includegraphics[width=\textwidth]{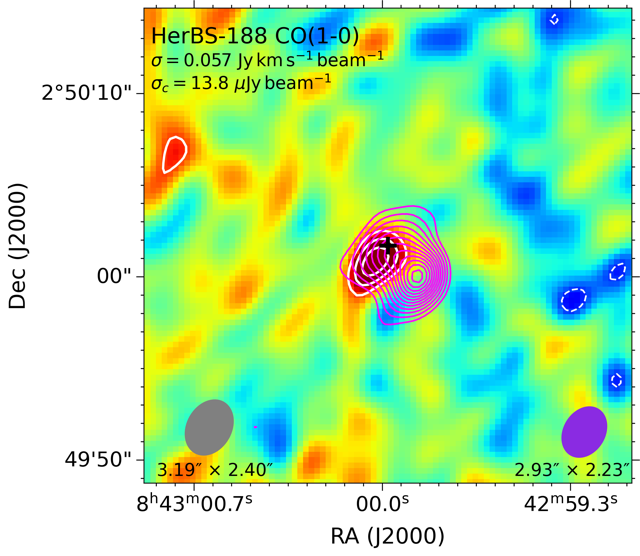}
\end{minipage}
\hfill
\begin{minipage}{0.235\textwidth}
    \centering
    \includegraphics[width=\textwidth]{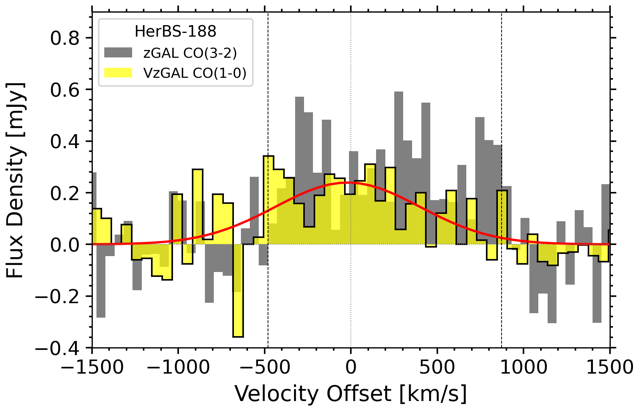}
\end{minipage}

                        \vspace{1em}
  
 \begin{minipage}{0.235\textwidth}
    \centering
    \includegraphics[width=\textwidth]{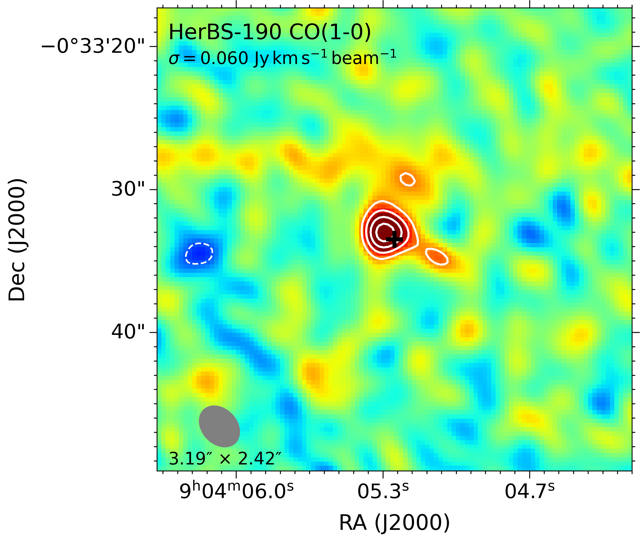}
\end{minipage}
\hfill
\begin{minipage}{0.235\textwidth}
    \centering
    \includegraphics[width=\textwidth]{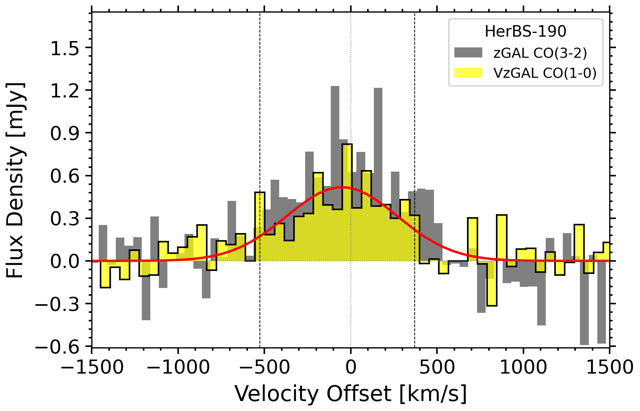}
\end{minipage}
\hfill
\begin{minipage}{0.235\textwidth}
    \centering
    \includegraphics[width=\textwidth]{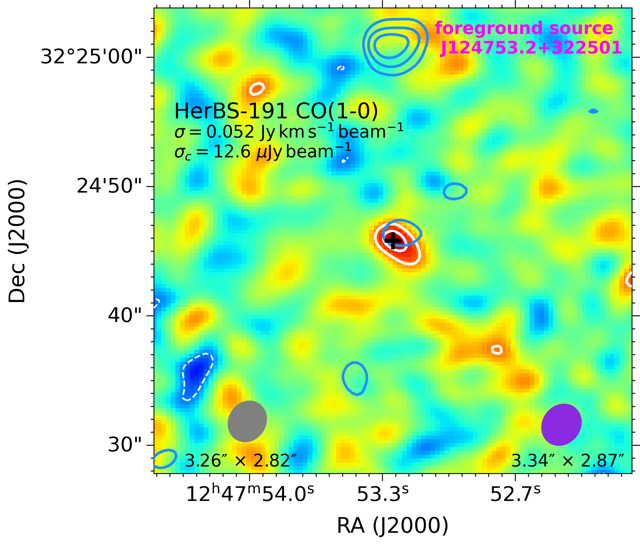}
\end{minipage}
\hfill
\begin{minipage}{0.235\textwidth}
    \centering
    \includegraphics[width=\textwidth]{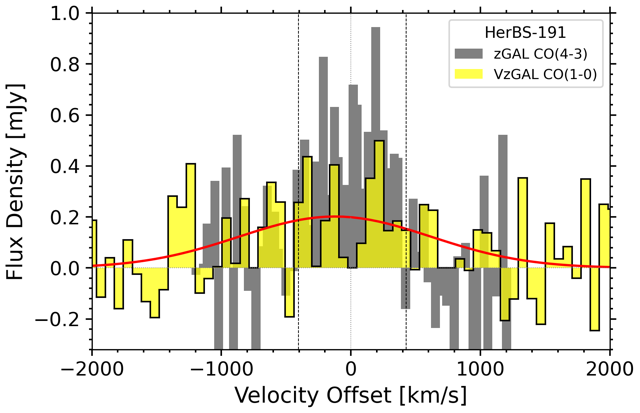}
\end{minipage}

                                            \vspace{1em}

\hfill
\begin{minipage}{0.235\textwidth}
    \centering
    \includegraphics[width=\textwidth]{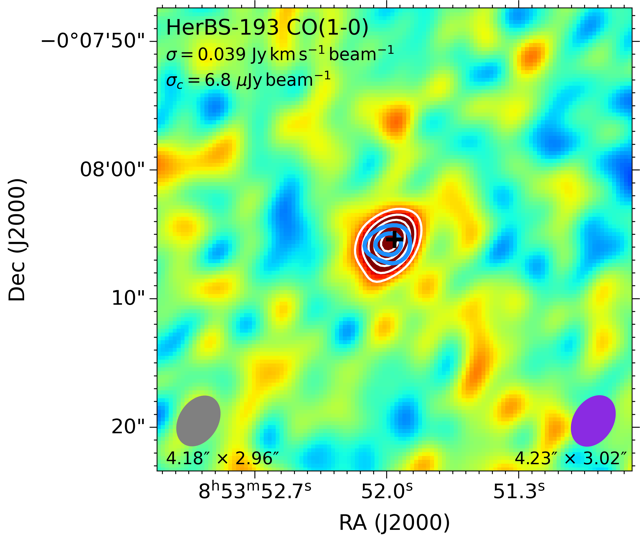}
\end{minipage}
\hfill
\begin{minipage}{0.235\textwidth}
    \centering
    \includegraphics[width=\textwidth]{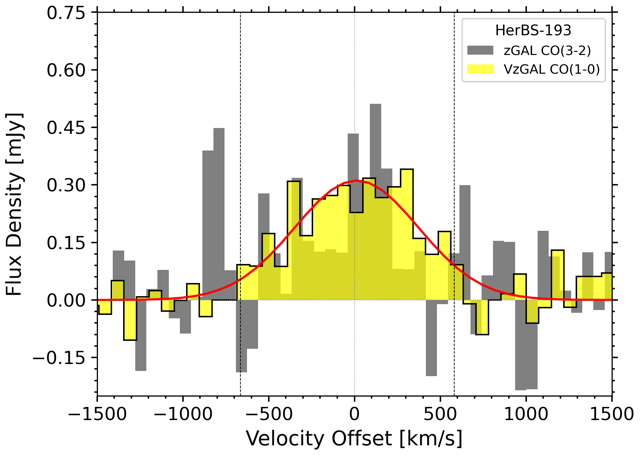}
\end{minipage}
\hfill
\begin{minipage}{0.235\textwidth}
    \centering
    \includegraphics[width=\textwidth]{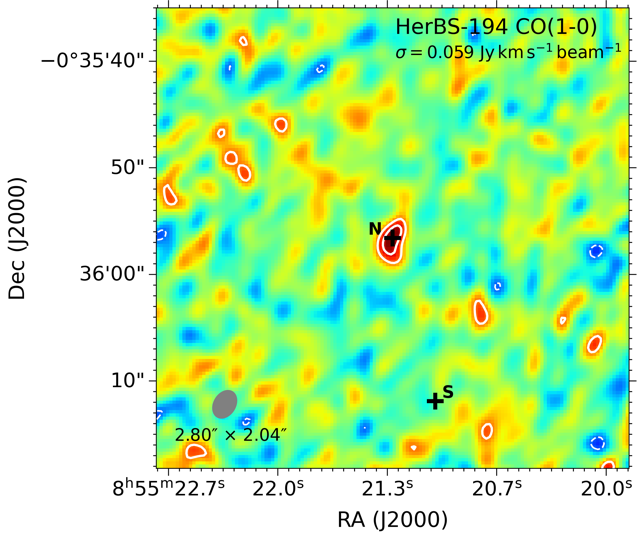}
\end{minipage}
\hfill
\begin{minipage}{0.235\textwidth}
    \centering
    \includegraphics[width=\textwidth]{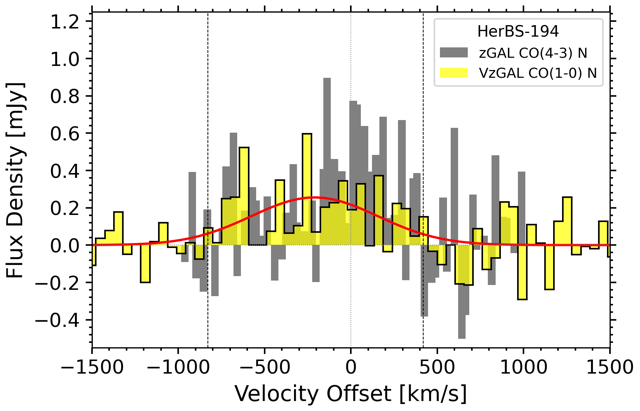}
\end{minipage}
  
                                            \vspace{1em}
                                      
\begin{minipage}{0.235\textwidth}
    \centering
    \includegraphics[width=\textwidth]{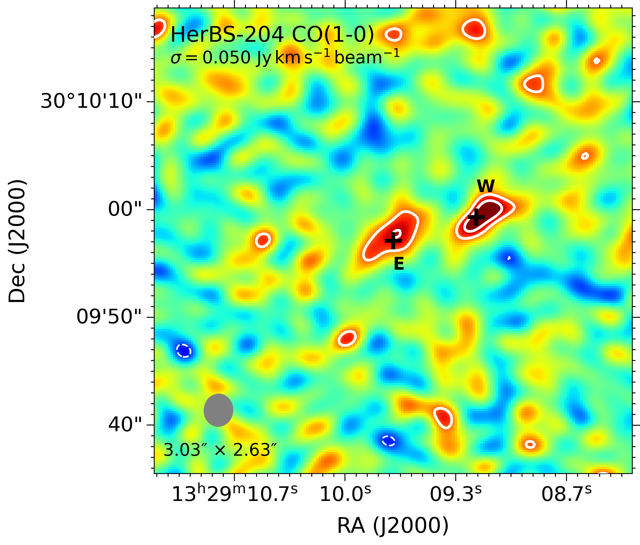}
\end{minipage}
\hfill
\begin{minipage}{0.235\textwidth}
    \centering
    \includegraphics[width=\textwidth]{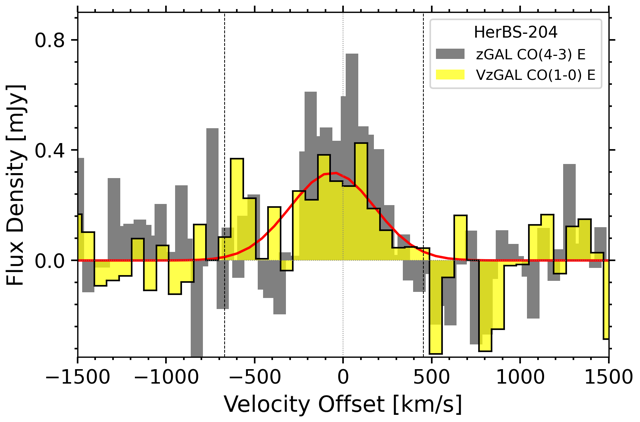}
\end{minipage}
\hfill
\begin{minipage}{0.235\textwidth}
    \centering
    \includegraphics[width=\textwidth]{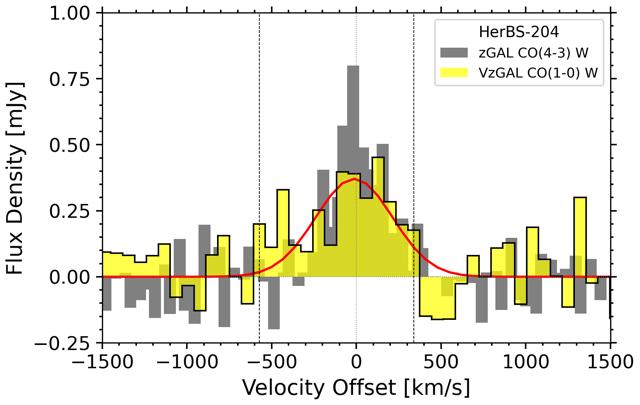}
\end{minipage}

    \addtocounter{figure}{-1}
\caption{(continued)}
\end{figure*}

\end{document}